
\let\em=\it

\font\sixrm=cmr6 \font\sixi=cmmi6 

\font\eightrm=cmr8  \let\smallrm=\eightrm
\font\eighti=cmmi8  \let\smalli=\eighti
\skewchar\eighti='177
\font\eightsy=cmsy8
\skewchar\eightsy='60
\font\eightit=cmti8
\font\eightsl=cmsl8
\font\eightbf=cmbx8
\font\eighttt=cmtt8

\font\ninerm=cmr9
\font\ninei=cmmi9
\skewchar\ninei='177
\font\ninesy=cmsy9
\skewchar\ninesy='60
\font\nineit=cmti9
\font\ninesl=cmsl9
\font\ninebf=cmbx9
\font\ninett=cmtt9

\def\eightpoint{\textfont0=\eightrm \scriptfont0=\fiverm 
                \def\rm{\fam0\eightrm}\relax
                \textfont1=\eighti \scriptfont1=\fivei 
                \def\mit{\fam1}\def\oldstyle{\fam1\eighti}\relax
                \textfont2=\eightsy \scriptfont2=\fivesy 
                \def\cal{\fam2}\relax
                \textfont3=\tenex \scriptfont3=\tenex 
                \def\it{\fam\itfam\eightit}\let\em=\it
                \textfont\itfam=\eightit
                \def\sl{\fam\slfam\eightsl}\relax
                \textfont\slfam=\eightsl
                \def\bf{\fam\bffam\eightbf}\relax
                \textfont\bffam=\eightbf \scriptfont\bffam=\fivebf
                \def\tt{\fam\ttfam\eighttt}\relax
                \textfont\ttfam=\eighttt
                \setbox\strutbox=\hbox{\vrule
                     height7pt depth2pt width0pt}\baselineskip=9.2pt
                \let\smallrm=\sixrm \let\smalli=\sixi
                \rm}
\def\ninepoint{\textfont0=\eightrm \scriptfont0=\fiverm
                \def\rm{\fam0\ninerm}\relax
                \textfont1=\ninei \scriptfont1=\fivei 
                \def\mit{\fam1}\def\oldstyle{\fam1\ninei}\relax
                \textfont2=\ninesy \scriptfont2=\fivesy 
                \def\cal{\fam2}\relax
                \textfont3=\tenex \scriptfont3=\tenex 
                \def\it{\fam\itfam\nineit}\let\em=\it
                \textfont\itfam=\nineit
                \def\sl{\fam\slfam\ninesl}\relax
                \textfont\slfam=\ninesl
                \def\bf{\fam\bffam\ninebf}\relax
                \textfont\bffam=\ninebf \scriptfont\bffam=\fivebf
                \def\tt{\fam\ttfam\ninett}\relax
                \textfont\ttfam=\ninett
                \setbox\strutbox=\hbox{\vrule
                     height7pt depth2pt width0pt}\baselineskip=10.4pt
                \let\smallrm=\sixrm \let\smalli=\sixi
                \rm}

\font\titlefont=cmss17 scaled\magstep1  
\font\secfont=cmss12 scaled\magstep1   

\font\fivebmi=cmmib6
\font\sixbmi=cmmib6     \skewchar\sixbmi='177
\font\ninebmi=cmmib10 at 9pt    \skewchar\ninebmi='177
\newfam\bmifam
\textfont\bmifam=\ninebmi
\scriptfont\bmifam=\sixbmi
\scriptscriptfont\bmifam=\fivebmi
\def\bmi{\fam\bmifam\ninebmi}
\def\b#1{{\bmi#1}}

\mathchardef\alpha="710B
\mathchardef\beta="710C
\mathchardef\gamma="710D
\mathchardef\delta="710E
\mathchardef\epsilon="710F
\mathchardef\zeta="7110
\mathchardef\eta="7111
\mathchardef\theta="7112
\mathchardef\iota="7113
\mathchardef\kappa="7114
\mathchardef\lambda="7115
\mathchardef\mu="7116
\mathchardef\nu="7117
\mathchardef\xi="7118
\mathchardef\pi="7119
\mathchardef\rho="711A
\mathchardef\sigma="711B
\mathchardef\tau="711C
\mathchardef\upsilon="711D
\mathchardef\phi="711E
\mathchardef\chi="711F
\mathchardef\psi="7120
\mathchardef\omega="7121
\mathchardef\varepsilon="7122
\mathchardef\vartheta="7123
\mathchardef\varpi="7124
\mathchardef\varrho="7125
\mathchardef\varsigma="7126
\mathchardef\varphi="7127


\newif\ifauxfile \auxfiletrue
\newcount\noundef \noundef=0

\newread\auxin
\newwrite\auxout
\ifauxfile
   \openin\auxin=\jobname.aux
   \ifeof\auxin\else\input\jobname.aux\fi
   \closein\auxin
   \immediate\openout\auxout=\jobname.aux
\fi

\def\ifundefined#1{%
   \expandafter\ifx\csname#1\endcsname\relax
}
\def\ref#1{\ifundefined{#1} $\bullet$#1$\bullet$%
  \immediate\write-1{Undefined reference #1}%
  \global\advance\noundef by1
\else\hbox{\csname#1\endcsname}\fi
}
\def\nameplace#1 {\ifauxfile\immediate\write\auxout%
{\def\string#1{\currentplace}}\fi%
\xdef#1{\currentplace}%
}

\def\bye{
   \ifnum\noundef>0 \immediate
     \write16{Undefined macro references: \number\noundef.}
   \fi
   \ifauxfile\closeout\auxout\fi
   \vfill\supereject\end
}

\newcount\fignum \fignum=0
\input psfig
\def\figure#1#2{%
  \topinsert#1
  \noindent\eightpoint#2
  \endinsert
}

\newcount\eqnum \eqnum=1
\everydisplay{\puteqnum}
\def\puteqnum#1$${#1\eqno(\the\eqnum)\global\advance\eqnum by 1$$}
\def\eqname#1{%
   \ifauxfile\immediate\write\auxout{\def\string#1{(\the\eqnum)}}\fi
   \xdef#1{(\the\eqnum)}
}
\def\refeq#1{%
 \advance\eqnum by -#1
 \hbox{(\the\eqnum)}%
 \advance\eqnum by #1
}

\newcount\secnum \newcount\ssecnum \secnum=0
\newif\ifappendix \appendixfalse

\def\tskip{\vskip12pt plus4pt minus4pt}
\def\secskip{\vskip24pt plus4pt minus4pt}
\def\ssecskip{\vskip18pt plus4pt minus4pt}

\def\maybebreak#1{\vskip0pt plus #1\hsize \penalty-500
                  \vskip0pt plus -#1\hsize}
\def\asecnum{\ifappendix
   {\ifcase\secnum $bullet$\or A\or B\or C\or D\or E\or F\or G\or H
      \else !!!\fi}\else\the\secnum\fi
}
\def\title#1\par{
   \secnum=0
   \xdef\currentplace{\asecnum}
   \par\vskip0pt plus.1\vsize\penalty-250
   \vskip0pt plus-.1\vsize\secskip\vskip\parskip
   {\noindent\titlefont\centerline{#1}}%
}
\def\titleline#1{\secnum=0\noindent{\titlefont\centerline{#1}}\tskip}
\def\abstract{\begingroup
\parskip=0pt\parindent=2em
\advance\leftskip by10em\noindent{\bf Abstract:}\smallskip\ninepoint\noindent}
\def\endabstract{\endgroup\secskip}

\def\section#1\par {
   \global\advance\secnum by 1
   \ssecnum=0
   \xdef\currentplace{\asecnum}
   \par\vskip0pt plus.2\vsize\penalty-250
   \vskip0pt plus-.2\vsize\secskip\vskip\parskip
   {\noindent\secfont\currentplace\ #1}\par\noindent}
\def\subsection#1\par{%
   \global\advance\ssecnum by 1
   \xdef\currentplace{\asecnum.\the\ssecnum}
   \par\vskip0pt plus.1\vsize\penalty-100
   \vskip0pt plus-.1\vsize\ssecskip\vskip\parskip
   {\noindent\secfont\currentplace\ #1}%
   \nobreak\par\noindent}
\def\appendices{
   \appendixtrue
   \secnum=0
   \ssecnum=0
   \vfill\eject
}

\def\biblio{\secskip\maybebreak{0.1}\noindent{\secfont References}\
\parskip=2pt  \begingroup\frenchspacing\ninepoint
\raggedright
}
\def\endbiblio{\endgroup}
\def\bibitem{
  \par\parindent=1.5em\hangindent=1.5em\hangafter=1
  \everypar={\hangindent=1.5em\hangafter=1\ignorespaces}
  \noindent\ignorespaces
}


\def\today{\ifcase\month\or
 January\or February\or March\or April\or May\or June\or
 July\or August\or September\or October\or November\or December\fi
 \space\number\day, \number\year}


\def\i{\relax\ifmmode{\rm i}\else\char16\fi}

\def\frac#1#2{{\textstyle{#1\over#2}}}

\def\d{{\rm d}}
\def\dddot#1{\ddot#1\kern-1.4pt\dot{\phantom{#1}}\kern-3pt}



\def\spose#1{\hbox to 0pt{#1\hss}}

\def\=#1{\overline{#1}}

\def\pr{\mathop{\smash{\rm p}\vphantom{\sin}}}

\def\lta{\mathrel{\spose{\lower 3pt\hbox{$\mathchar"218$}}
     \raise 2.0pt\hbox{$\mathchar"13C$}}}
\def\gta{\mathrel{\spose{\lower 3pt\hbox{$\mathchar"218$}}
     \raise 2.0pt\hbox{$\mathchar"13E$}}}

\def\kms{{\rm\,km\,s^{-1}}}

\def\Mpc{{\rm\,Mpc}}

\def\aa{A\&A}
\def\aj{AJ}
\def\apj{ApJ}
\def\apjs{ApJS}

\def\mn{MNRAS}

\def\newfigure#1{\global\advance\fignum by 1\ifauxfile%
\immediate\write\auxout{\def\string#1{\the\fignum}}%
\fi\xdef#1{\the\fignum}}

\def\newfig{\global\advance\figno by1}
\def\currentfigure{\the\figno}

\def\currenteq{(\the\eqnumber)}
\def\newe{\global\advance\eqnumber by 1
 \hbox{\currenteq}
}

\outer\long\def\crap#1{}

\parindent=0pt
\parskip=10pt

\def\title#1\par{
  \centerline{\Large#1}
  \vskip18pt
  \noindent
}

\newdimen\hhsize \hhsize=.5\hsize
\def\boxit#1{\vbox{\hrule\hbox{\vrule\kern3pt
        \vbox{\kern3pt#1\kern3pt}\kern3pt\vrule}\hrule}}

\titleline{The Demography of Massive Dark Objects}
\titleline{in Galaxy Centres}

\bigskip
\begingroup\ninepoint
\centerline{ John Magorrian}
\centerline{Canadian Institute for Theoretical Astrophysics, 
University of Toronto,}
\centerline{60 St. George St., Toronto M5S 3H8, Canada}
\centerline{Electronic mail: magorrian@cita.utoronto.ca}

\medskip
\centerline{ Scott Tremaine}
\centerline{CIAR Cosmology and Gravity Program, 
Canadian Institute for Theoretical Astrophysics,} 
\centerline{University of Toronto, 60 St. George St., Toronto M5S 3H8, Canada}
\centerline{Electronic mail: tremaine@cita.utoronto.ca}

\medskip
\centerline{ Douglas Richstone}
\centerline{Department of Astronomy, University of Michigan, 
Ann Arbor, MI 48109}
\centerline{Electronic mail: dor@astro.lsa.umich.edu}


\medskip
\centerline{ Ralf Bender}
\centerline{Universit\"ats-Sternwarte,}
\centerline{Scheinerstrasse 1,}
\centerline{M\"unchen 81679, Germany}
\centerline{Electronic mail: bender@usm.uni-muenchen.de}

\medskip
\centerline{ Gary Bower}
\centerline{Kitt Peak National Observatory, National Optical 
Astronomy Observatories\footnote{$^1$}{Operated by AURA under 
cooperative agreement with the U. S. National Science Foundation.},}
\centerline{P.O. Box 26732, Tucson, AZ 85726}
\centerline{Electronic mail: gbower@noao.edu}

\medskip
\centerline{ Alan Dressler}
\centerline{The Observatories of the Carnegie Institution, 
813 Santa Barbara St., Pasadena, CA 91101}
\centerline{Electronic mail: dressler@ociw.edu}

\medskip
\centerline{ S. M. Faber}
\centerline{UCO/Lick Observatory, Board of Studies in 
Astronomy and Astrophysics,}
\centerline{ University of California, Santa Cruz, CA 95064}
\centerline{Electronic mail: faber@ucolick.org}

\medskip
\centerline{ Karl Gebhardt}
\centerline{Department of Astronomy, University of Michigan, 
Ann Arbor, MI 48109}
\centerline{Electronic mail: gebhardt@astro.lsa.umich.edu}

\medskip
\centerline{ Richard Green}
\centerline{Kitt Peak National Observatory, National Optical 
Astronomy Observatories$^1$,}
\centerline{P.O. Box 26732, Tucson, AZ 85726}
\centerline{Electronic mail: green@noao.edu}

\medskip
\centerline{ Carl Grillmair}
\centerline{Jet Propulsion Laboratory,}
\centerline{Mail Stop 183-900, 4800 Oak Grove Drive,}
\centerline{Pasadena, CA 91109}
\centerline{Electronic mail: carl@grandpa.jpl.nasa.gov}

\medskip
\centerline{ John Kormendy}
\centerline{Institute for Astronomy, 
University of Hawaii, 2680 Woodlawn Dr., Honolulu, HI 96822}
\centerline{Electronic mail: kormendy@oort.ifa.hawaii.edu}

\medskip
\centerline{ Tod R. Lauer}
\centerline{Kitt Peak National Observatory, National Optical 
Astronomy Observatories$^1$,}
\centerline{P.O. Box 26732, Tucson, AZ 85726}
\centerline{Electronic mail: lauer@noao.edu}
\endgroup
\vfill\eject

\abstract
We construct dynamical models for a sample of 36 nearby galaxies with
Hubble Space Telescope photometry and ground-based kinematics.  The
models assume that each galaxy is axisymmetric, with a two-integral
distribution function, arbitrary inclination angle, a
position-independent stellar mass-to-light ratio~$\Upsilon$, and a
central massive dark object (MDO) of arbitrary mass~$M_\bullet$.  They
provide acceptable fits to 32 of the galaxies for some value of
$M_\bullet$ and $\Upsilon$; the four galaxies that cannot be fit have
kinematically decoupled cores.  The mass-to-light ratios inferred for
the 32 well-fit galaxies are consistent with the fundamental plane
correlation $\Upsilon\propto L^{0.2}$, where $L$ is galaxy luminosity.
In all but six galaxies the models require at the 95\% confidence
level an MDO of mass $M_\bullet\sim 0.006 M_{\rm
bulge}\equiv0.006\Upsilon L$.  Five of the six galaxies consistent
with $M_\bullet=0$ are also consistent with this correlation.  The
other (NGC~7332) has a much stronger upper limit on $M_\bullet$.  We
consider various parameterizations for the probability distribution
describing the correlation of the masses of these MDOs with other
galaxy properties.  One of the best models can be summarized thus: a
fraction $f\simeq0.97$ of galaxies have MDOs, whose masses are well
described by a Gaussian distribution in $\log (M_\bullet/M_{\rm
bulge})$ of mean $-2.27$ and width $\sim0.07$.

\endabstract

\section Introduction

The evidence that massive dark objects (MDOs) are present in the
centers of nearby galaxies is reviewed by Kormendy \& Richstone (1995;
hereafter KR95). Further evidence that post-dates this review is
described by Bender, Kormendy \& Dehnen (1997), van der Marel et
al. (1997) and Kormendy et al.\ (1997a).  The MDOs are probably black
holes, since star clusters of the required mass and size are difficult
to construct and maintain, and since black-hole quasar remnants are
expected to be common in galaxy centers; however, this identification
is not important for the purposes of this paper.  Following Kormendy
(1993a), KR95 suggest that at least 20\% of nearby hot galaxies
(ellipticals and spiral bulges) have MDOs and point out that the
observed MDO masses exhibit the correlation
$M_\bullet\simeq0.003M_{\rm bulge}$, where $M_{\rm bulge}$ is the mass
of the hot stellar component of the galaxy.  (Throughout this paper we
use the word ``bulge'' to refer to the hot stellar component of a
galaxy, whether elliptical or spiral.)  For a ``bulge'' with constant
mass-to-light ratio $\Upsilon$ and luminosity $L$, $M_{\rm
bulge}\equiv\Upsilon L$.

The machinery for modelling the kinematics of hot galaxies to
determine whether MDOs are present has increased steadily in
sophistication over the past two decades. The earliest models
(e.g. Young et al.~1978) fitted only the line-of-sight dispersion of
spherical galaxies and assumed that the stellar distribution function
was isotropic. Modern programs (e.g. Rix et al.~1997; Gebhardt et
al.~1997) fit the entire line-of-sight velocity distribution for
arbitrary axisymmetric galaxy models. While the most general and
accurate possible models, and the highest resolution spectroscopic
observations, were needed to establish the presence of the first few
MDOs, we have learned with experience that estimates of the MDO mass
based on cruder models and observations are usually fairly
accurate. An example is the MDO in M87: Young et al.~(1978) estimated
the mass to be $\sim 5\times10^9M_\odot$ from spherical, isotropic
models, very close to the $3\times10^9M_\odot$ determined by Harms et
al. (1994) from HST spectra of a ring of ionized gas at 20 pc from the
center.

This experience suggests that it is worthwhile to estimate MDO masses
using relatively simple models applied to a large sample of
galaxies. We cannot yet insist on HST spectroscopy for our sample,
since this is still available only for a few galaxies; on the other
hand HST photometry is available for over 60 hot galaxies.  In this
paper we examine a sample of~36 hot galaxies for which both HST
photometry and reasonable quality, ground-based, long-slit
spectroscopy are available. We look for evidence of MDOs among these
by fitting two-integral axisymmetric dynamical models to the data for
each galaxy.  These are not the most general types of models, but they
are quick to compute and will guide us towards galaxies to which we
should apply more precise (and expensive) observations and models. Our
results also provide a first look at the statistical distribution of
MDOs as a function of galaxy luminosity and other parameters. They do
{\it not} establish unambiguously that an MDO is present in any
individual galaxy.

The paper is organized as follows.  The next section gives a brief
outline of the data we use.  This is followed by a detailed
description of our modelling procedure and the assumptions that go in
to it.  Section~4 presents results for individual galaxies.  What
these tell us about the MDO mass distribution is tackled in Section~5.
Finally Section~6 sums up.

\section Data

Our sample consists of all reasonably dust-free hot galaxies with HST
photometry and ground-based velocity dispersion and rotation velocity
profiles.  The sample contains 36 galaxies, listed in Table~1 along
with details of the sources of the observations we use.  The Appendix
contains comments about some of the galaxies.  Most of the objects
were observed prior to the first HST servicing mission with the
Planetary Camera (0.043'' per pixel) through filter F555W (roughly
Johnson V). The sample includes galaxies observed in a number of HST
programs (Lauer et al.~1992a,b, Grillmair et al.~1994, Jaffe et
al.~1994, Forbes et al.~1995, Lauer et al.~1995); the reduction
procedures are described by Lauer et al.\ (1995), Byun et al.\ (1996),
and Faber et al.\ (1997).  Some of the galaxies show evidence for
nuclear activity.  For each of these the table lists a radius $R_{\rm
min}$, inside which non-stellar radiation probably makes a significant
contribution to the observed light.

The HST data extend only to about 10'' from the centre.  For most of
the galaxies we take published ground-based photometry and join it
smoothly to the HST data to obtain a global photometric profile.  For
the remaining galaxies, we assume that the outer parts are well
described by an $R^{1/4}$ profile with the same flattening as the
outermost HST isophote.  We take the effective radius of the $R^{1/4}$
profile from the literature, if available; otherwise we estimate it
by fitting to the HST photometry.

Table~1 also lists the sources for our kinematical data.  We restrict
ourselves to reasonable quality CCD-based spectroscopy and do not use
kinematical data beyond about two-thirds of the maximum radius for
which photometry is available.  Whenever there are many sources for a
given slit position of a galaxy, we generally choose those with the
best seeing.  If an estimate for the seeing is unavailable we simply
assume a FWHM of 2'': the MDO masses yielded by our models are fairly
insensitive to the precise value used, as long as it lies between 1''
and 3''.  

The observations yield line-of-sight rotation speeds and velocity
dispersions, convolved with seeing and averaged over spatial ``bins''
determined by the slit width and pixel size.  We combine the measured
rotation speed $v_j$ and the velocity dispersion $\sigma_j$ in each
bin $j$ to obtain an estimate of the second-order moment
$\mu_j^2=v_j^2+\sigma_j^2$, which is the input used by our models.
Strictly speaking, the $v_j$ and $\sigma_j$ quoted by
observers do not individually have any direct connection with the
moments of the line-of-sight velocity profiles (VPs), since they are
usually obtained by fitting Gaussians to the VPs (van der Marel \&
Franx 1993).  However, tests with flattened isotropic toy galaxies
(Dehnen \& Gerhard 1994; Magorrian \& Binney 1994) show that there is
typically an almost-constant difference of about 10\% between the
combination $v_j^2+\sigma_j^2$ and the true second-order moments, with
the sign of the difference changing from the major to the minor axis.
Figure~\ref{figvsigtest} shows a typical example.  Since these
differences are almost constant, they do not affect the MDO masses
fitted by our models.  They could, however, have a small effect on the
fitted mass-to-light ratios.

Deriving the observational uncertainty $\Delta\mu_j$ is a vexing
problem, because analysis methods used by different observers yield a
range of error estimates $\Delta v_j$ and $\Delta\sigma_j$, which
usually do not take systematic effects, such as template mismatch,
into account.  We have tried the following three methods of dealing
with this problem:
\begingroup\parskip=0pt\parindent=2em
\item{(i)} Simply take all quoted errors at face value;

\item{(ii)} Replace $\Delta v_j$ with $\max(\Delta v_j,5\kms)$ and
similarly for $\Delta\sigma_j$;

\item{(iii)} Scale the errors for each exposure along each slit
position such that they are consistent with axisymmetry.  More
precisely, suppose there are $n$ measurements $\sigma_j^+$ along one
side of a galaxy, with corresponding measurements $\sigma_j^-$ along
the other side.  We scale the $\Delta\sigma_j$ by a constant factor such that
$$\chi^2_\sigma\equiv\sum_{j=1}^n
{(\sigma_j^+ -\sigma_j^-)^2\over
(\Delta\sigma_j^+)^2 + (\Delta\sigma_j^-)^2}
=n,
$$
and similarly for the $\Delta v_j$ (Davies \& Birkinshaw 1988).

\endgroup
Notice that the last method implicitly assumes that the errors are
Gaussian.  This is almost certainly wrong, but it is the best we can
do given the heterogeneous nature of our data.  Given the ``improved''
observational errors, the error in $\mu_j$ is
$\Delta\mu_j=\left(v_j^2(\Delta
v_j)^2+\sigma_j^2(\Delta\sigma_j)^2\right)^{1/2}/\mu_j$ to first
order.  We use (iii) above wherever possible, but the results of our
models are usually not significantly affected by which procedure we
employ.

\section Modelling Procedure

\nameplace{\secmodel} We assume that each galaxy is axisymmetric with some 
unknown inclination angle~$i$, and work in cylindrical coordinates
$(R,\phi,z)$ where the $z$-axis is the symmetry axis of the galaxy.  A
lower bound on~$i$ comes from requiring that all isophotes have an
intrinsic axis ratio no less than~$0.3$ (i.e. no flatter than
E7). Each galaxy can have a central MDO of arbitrary mass, but
otherwise the mass-to-light ratio $\Upsilon$ is assumed to be
independent of position.  The distribution function of the stars is
assumed to be a function only of two integrals of motion, the energy
and the $z$-component of angular momentum.  The advantage of these
assumptions is that the (even part of the) kinematics follows uniquely
from the three-dimensional luminosity distribution~$\nu(R,z)$ (e.g.,
Lynden-Bell 1962; Dejonghe 1986).  The disadvantage is that there is
no reason why real galaxies should obey our assumptions.  In
particular, bright, core galaxies are usually non-rotating, and many
studies of them (e.g., van der Marel 1991) show some evidence for
radial anisotropy.  Our two-integral models of flattened, non-rotating
galaxies are {\it tangentially} anisotropic.

We use the modelling procedure introduced by Binney, Davies \&
Illingworth (1990) to predict the kinematics of each galaxy for any
assumed inclination angle~$i$ and MDO mass~$M_\bullet$.  It predicts
the second-order moments, convolved with seeing and averaged over the
same $j=1,\ldots,n$ spatial bins used in the observations.  The
procedure is as follows:

\begingroup\parindent=2em\parskip=0pt
\item{1.} Use a scheme based on maximum penalized
likelihood to find a smooth luminosity density~$\nu(R,z)$ that
projects to an acceptable fit to the observed surface brightness
(Magorrian 1997). The density $\nu$ is not uniquely determined by the
surface brightness unless the galaxy is edge-on (Rybicki 1987).
Romanowsky \& Kochanek (1997) demonstrate that even for quite high
inclinations there can be a large range in $\nu$ consistent with a
given surface brightness; however, they find that the range of
projected second-order moments associated with this uncertainty is
quite small.  We have carried out some experiments that confirm that the
allowable MDO masses are not strongly affected by the indeterminacy in
$\nu(R,z)$;

\item{2.} Calculate the gravitational potential and forces using an assumed
stellar mass-to-light ratio $\Upsilon_0$ and MDO mass $M_\bullet$;

\item{3.} Use the Jeans equations
to calculate the second-order moments $\nu\={v_\phi^2}$
and $\nu\={v_R^2}=\nu\={v_z^2}$;

\item{4.} Project the luminosity-weighted zeroth- and second-order moments
of the line-of-sight velocity along the line of sight; convolve with
seeing; and average over the same spatial bins used in the
observations.

\endgroup

\noindent Dividing the binned, seeing-convolved second-order moment
by the corresponding zeroth-order one yields the model's
predictions~$\hat\mu_j^2(i,\Upsilon_0,M_\bullet)$ in each bin.  These
predictions scale trivially with mass-to-light ratio $\Upsilon$
through
$$\hat\mu_j^2(i,\Upsilon,M_\bullet)={\Upsilon\over\Upsilon_0}
\hat\mu^2_j\left(i,\Upsilon_0,\Upsilon M_\bullet/\Upsilon_0\right).
$$

\subsection Estimation of $M_\bullet$ and $\Upsilon$.

We assume that the measurement errors in the $\mu_j$ are Gaussian and
uncorrelated.  Then the likelihood of the photometric and
kinematic data $D$ given the model parameters $(i,\Upsilon,M_\bullet)$
is
$$\pr(D\mid i,\Upsilon,M_\bullet)\propto
\exp\left(-{\textstyle{1\over2}}\chi^2\right),
\eqname\mdoprob
$$
where
$$\chi^2(i,\Upsilon,M_\bullet)\equiv\sum_{j=1}^n \left(
\mu_j-\hat\mu_j\over\Delta\mu_j \right)^2.
\eqname\eqchisq
$$

We obtain the best-fitting values $M_\bullet$ and $\Upsilon$ and their
confidence intervals as follows.
By Bayes' theorem, the posterior distribution of~$i$, $\Upsilon$ and
$M_\bullet$  given the data $D$ is
$$\pr(i,\Upsilon,M_\bullet\mid D)\propto\pr(D\mid i,\Upsilon,M_\bullet)
\pr(i\mid q')\pr(\Upsilon)\pr(M_\bullet),$$
where we have made the assumption that $i$, $\Upsilon$ and $M_\bullet$
are {\it a priori} independent.  Our priors $\pr(M_\bullet)$ and
$\pr(\Upsilon)$ are flat in $M_\bullet$ and $\log\Upsilon$
respectively.  We make the reasonable assumption that the prior
for~$i$ depends only on the observed axis ratio $q'$ of the galaxy.
Then $\pr(i\mid q')$ can be related to $N(q)\,\d q$, the probability
that a randomly chosen galaxy will have an intrinsic axis ratio lying
between $q$ and $q+\d q$, by a further application of Bayes' theorem:
$$\eqalign{ \pr(i\mid q') & = {\pr(i)\pr(q'\mid i)\over\pr(q')}\cr &
  \propto \pr(i)\int\pr(q'\mid i,q) N(q)\,\d q\cr & \propto
  {1\over\sqrt{q'^2-\cos^2i}}
  N\left((q'^2-\cos^2i)^{1/2}\over\sin i\right),\cr }
$$
where we have made the natural assumption that $\pr(i)=\sin i$ and
have used the relation $\pr(q'\mid i,q) \propto
q'\delta(q^2\sin^2i+\cos^2i-q'^2)$.  We approximate the $N(q)$ obtained
by Tremblay \& Merritt (1995) by a Gaussian centred on $q=0.7$ with
standard deviation $0.1$.  Our results are only very weakly dependent
on this form.

We are interested mainly in $M_\bullet$ and
$\Upsilon$, not in~$i$.  Marginalizing the latter, we get the joint
posterior distribution of $\Upsilon$ and~$M_\bullet$ as
$$\pr(\Upsilon,M_\bullet\mid D)=\pr(D\mid \Upsilon,M_\bullet)
\pr(\Upsilon)\pr(M_\bullet),\eqname\probmups$$
where
$$\pr(D\mid \Upsilon,M_\bullet)\equiv \int
\pr(D\mid i,\Upsilon,M_\bullet) \pr(i\mid q')\,\d i.\eqname\likemups$$
The posterior distributions $\pr(M_\bullet\mid D)$ and
$\pr(\Upsilon\mid D)$ follow by marginalizing~\ref{probmups} again.

\section Results for individual galaxies

We have made models of each of our 36 galaxies for a range
of~$M_\bullet\ge 0$, $\Upsilon$ and $i$.  The models do not provide
adequate descriptions of the kinematics of four of the galaxies
(NGC~1700, NGC~4365, NGC~4494 and NGC~4589).  For these four,
Figure~\ref{figbad}(a) shows how $\chi^2$ of equation~\ref{eqchisq}
varies with MDO mass~$M_\bullet$ and inclination angle~$i$.
Figure~\ref{figbad}(b) plots the kinematics of the models with the
best-fitting values of $M_\bullet$ against the observations.  All four
galaxies are known to have kinematically distinct cores (e.g., Forbes
et al.~1996), so it is perhaps not surprising that our axisymmetric
models do not work for them.  We omit these four in the demographical
analysis in the next section.  For comparison, only two (NGC~3608 and
NGC~4278) of the 32 galaxies that our models do fit are known to have
kinematically distinct cores.

The models describe the kinematics of all of the remaining 32 galaxies
reasonably well for some value of $M_\bullet$.
Figure~\ref{figgood}(a) shows the posterior distribution
$\pr(\Upsilon,M_\bullet\mid D)$ in each case.
Figure~\ref{figgood}(b) shows the kinematics of the best-fitting
models along each slit position fitted.
Six of these galaxies have independent determinations of the mass of a
central MDO.  The comparison between the best-fit masses as determined
here and the mass estimates in the literature for these six is
presented in Figure~\ref{figmdotest}.  For all but one galaxy we
obtain MDO masses that are in good agreement with those from earlier
work.  This gives us some degree of confidence in the assumptions that
go into our models.  The one exception is NGC~3115 for which Kormendy
et al.\ (1996a) claim an MDO mass of about $2\times10^9M_\odot$, some
four times larger than our present mass estimate.

Our models imply that only three of the 32 galaxies (NGC~2778,
NGC~4467, and NGC~7332) are consistent (at the 68\% confidence level)
with $M_\bullet=0$.  However, Figures~\ref{figgood}(a) and (b) show
that the available data for each of these three are also consistent
with a reasonably large value of $M_\bullet$.  NGC~7332 has the
strongest upper limit on $M_\bullet$.  In fact, the central dip in its
dispersion profile is suggestive of either a mass-to-light ratio that
decreases close to the centre, or else strong tangential anisotropy.
Kormendy (1993b) has suggested that its formation history may be
different from the other galaxies in the sample.  We do not, however,
omit it in the analysis below.  There are a further three galaxies
(NGC~4168, NGC~4473 and NGC~4636) consistent with $M_\bullet=0$ at the
95\% confidence level.  All the rest have $M_\bullet>0$.

Seven of the 32 galaxies show evidence for nuclear activity or strong
dust obscuration.  For these we make two types of models: one under
the na{\"\i}ve assumption that all the observed light near the galaxy
centre comes from stars, the other that only uses the photometry
beyond a radius $R_{\rm min}$, where $R_{\rm min}$ is given in
Table~1.  We find that the MDO masses predicted by the two types of
models generally agree quite well.  This is unsurprising given the
relatively poor spatial resolution of the kinematical data.  In what
follows we use only the MDO masses obtained by omitting photometry
within $R_{\rm min}$.

Table~2 lists the 68\% confidence bounds that our models place on
$M_\bullet$ and $\Upsilon$ for the 32 galaxies.  The correlations
between $\Upsilon$ and~$L$ and between $M_\bullet$ and $M_{\rm bulge}$
are plotted on Figure~\ref{figcorrel}.  Ignoring the error bars on
$\Upsilon$ the formal best-fit straight line to $(\log
L,\log\Upsilon)$ is
$$
\log(\Upsilon_{\rm fit}/\Upsilon_\odot) = (-1.12\pm0.33) +
(0.18\pm0.03)\log(L/L_\odot)\eqname\eqfita
$$
with an RMS deviation between $\log\Upsilon_{\rm fit}$ and
$\log\Upsilon$ of $0.12$.
Our crude fit is broadly consistent with
the fundamental-plane correlation $\Upsilon\propto L^{0.2}$ predicted
using the virial theorem
(e.g., Faber et al.~1987; Bender,
Burstein \& Faber~1992).  Similarly, the correlation between
$M_\bullet$ and $M_{\rm bulge}$ for those galaxies with $M_\bullet>0$
can be described by
$$\log (M_{\bullet,\rm fit}/M_\odot) = (-1.82\pm1.36) + (0.96\pm0.12)
\log (M_{\rm bulge}/M_\odot),\eqname\eqfitb
$$
with an RMS $(\log M_{\bullet,\rm fit}-\log M_\bullet)$ of $0.49$.
This result is consistent with the proportionality $M_\bullet\propto
M_{\rm bulge}$ that was first pointed out by Kormendy (1993a) and KR95.
This apparent correlation is the subject of the next section.

Finally, we check whether there is any correlation between the
residuals $x_i\equiv\log M_{\bullet,i}-\log M_{\bullet,\rm fit,i}$ and
$y_i\equiv\log\Upsilon_i-\log\Upsilon_{\rm fit,i}$.  One might expect
a negative correlation if our models were fitting spuriously high MDO
masses for some galaxies, thus depressing the fitted value of
$\Upsilon$.  Figure~\ref{figcorrel}(c) shows that there is no such
correlation.  The correlation coefficient $r_{xy}=0.014$,
which is not significant.

\section MDO mass distribution

What do these new results tell us about the distribution of MDOs
among galaxies?  Let us assume initially that the MDO mass
distribution of our sample depends only on $x\equiv M_\bullet/M_{\rm
bulge}$ and is characterized by some other parameters~$\omega$; that
is, that there is some function $\pr(x\mid\omega)\,\d x$ which is the
probability that a galaxy has an MDO with mass in the range $[x,x+\d
x]$.  

\topinsert
\begingroup
\offinterlineskip
\halign{\quad \bf \hfil#&\vrule#& \quad#\quad & \quad$#$\quad & \quad#\hfil\cr
          &&\strut\hfil$\omega$\hfil  & \hfil \pr_+(x\mid\omega)\hfil  \cr
&height3pt&&\cr
\noalign{\hrule}
&height3pt&&\cr
Power Law 1 -- $P_{\rm PL1}$ && $(f,\log x_0,\alpha)$ &
Nx^\alpha \hbox{\ if $x<x_0$; zero otherwise}\quad(\alpha>-1)\hfil\cr
&height3pt&&\cr
Power Law 2 -- $P_{\rm PL2}$ && $(f,\log x_0,\alpha)$ &
Nx^\alpha \hbox{\ if $x>x_0$; zero otherwise}\quad(\alpha<-1)\hfil\cr
&height3pt&&\cr
Schechter -- $P_{\rm S}$  && $(f,\log x_0,\alpha)$ & N\left(x/x_0\right)^\alpha
\exp\left(-x/x_0\right)\quad(\alpha>-1)\hfil\cr
&height3pt&&&\cr Gaussian -- $P_{\rm G}$ && $(f,\log x_0,\log \Delta)$
& N\exp\left[-{1\over2} (x-x_0)^2/\Delta^2\right] \cr &height3pt&&&\cr
Log Gaussian -- $P_{\rm LG}$ && $(f,\log x_0,\log\Delta)$ &
N\exp\left[-{1\over2} (\log x-\log x_0)^2/\Delta^2\right] \cr }
\endgroup
\eightpoint
{\bf Table~3.} The five parameterizations for $\pr(x\mid\omega)$
considered here.  The variable $x\equiv M_\bullet/M_{\rm bulge}$ where
$M_\bullet$ is the mass of the MDO and $M_{\rm bulge}$ is the mass of
the hot stellar component of the galaxy.  For a given set of
parameters $\omega$, the probability that a galaxy has an MDO with
mass in the range $[x,x+\d x]$ is $\pr(x\mid\omega,P)\,\d
x=(1-f)\delta(x)\,\d x+f\pr\!{}_+(x\mid\omega,P)\,\d x$, where $f$ is
the fraction of galaxies with $M_\bullet>0$ and the $N(\omega)$ in
$\pr\!{}_+(x\mid\omega,P)$ is a normalizing factor
(equation~\ref{eqnorm}).  The prior probability $\pr(\omega\mid P)$ is
assumed to be flat in the parameters $\omega$.
\endinsert

We experiment with several parameterizations $P$ for~$\pr(x\mid\omega)$,
as shown in Table 3.  In each case, one of the parameters, $f$, is the
fraction of galaxies with $M_\bullet>0$, so that
$\pr(x\mid\omega)$ is of the form
$$\pr(x\mid\omega,P)=(1-f)\delta(x)+f\pr\!{}_+(x\mid\omega,P),$$ where
$\pr_+(x\mid\omega,P)$ describes the distribution of MDOs with
$M_\bullet>0$.  The $N(\omega)$ in $\pr\!{}_+(x\mid\omega,P)$ is a
normalizing factor chosen such that
$$\int_0^\infty\pr\!{}_+(x\mid\omega,P)\,\d x=1.\eqname\eqnorm
$$
The parameterizations $P_{\rm PL2}$ and $P_{\rm LG}$ assume that there
is a genuine ridge line in $\pr(x)$ at $x=x_0$, whereas the other
three also test whether KR95's apparent ridge at $x\simeq0.005$
is just the upper envelope of some ridgeless $\pr(x)$.

For each parameterization $P=(P_{\rm PL1},P_{\rm PL2},P_{\rm S},P_{\rm
G},P_{\rm LG})$, we first seek the most likely set of
parameters~$\omega$ given our data~$D$.  By Bayes' theorem, the
posterior distribution of~$\omega$ and mass-to-light ratios
$\b\Upsilon\equiv(\Upsilon_1,\cdots,\Upsilon_N)$ of the 32 galaxies is
$$\eqalign{
\pr(\omega \b\Upsilon\mid D,P) & \propto
 \pr(\omega\mid P)\pr(\b\Upsilon)
  \pr(D\mid \omega\b\Upsilon,P)\cr
&\propto \pr(\omega\mid P)\pr(\b\Upsilon)
 \int \pr(D\mid \b\Upsilon\b x)
 \pr(\b x\mid \omega,P)\,\d\b x,\cr
}$$
where $\pr(D\mid\b\Upsilon\b x)$ is a product of factors of the form
of equation~\ref{likemups}.  The prior $\pr(\omega\mid P)$ is assumed
flat in the parameters $\omega$ given in Table~3.
We are interested only in the parameters~$\omega$, not in~$\b\Upsilon$.
Marginalizing the latter yields
$$\eqname\eqposterior\eqalign{
\pr(\omega\mid D,P) & = \int \pr(\omega\b\Upsilon\mid D,P)\,
\d\b\Upsilon\cr
& \propto \pr(\omega\mid P) 
\int \pr(D\mid\b x)
\pr(\b x\mid \omega,P)\,\d\b x,\cr
& \propto \pr(\omega\mid P)
\prod_{j=1}^N \int \pr(D_j\mid x_j)\pr(x_j\mid \omega,P)\,\d x_j,\cr
}$$
where we have defined
$$
\pr(D\mid\b x) \equiv \int \pr(D\mid \b\Upsilon\b x)
 \pr(\b\Upsilon)\, \d\b\Upsilon,
$$
and similarly for $\pr(D_j\mid x_j)$.  

The posterior distributions $\pr(\omega\mid D,P)$ for each
parameterization are plotted on Figure~\ref{figpost}.  Table~4 lists
the best-fitting parameters with their 68\% confidence intervals and
Figure~\ref{figdist} plots $\pr(x\mid\omega,P)$ for the best-fitting
parameters $\omega$ in each case.  According to all parameterizations
except $P_{\rm PL1}$, nearly all galaxies have MDOs ($f\simeq0.96$)
with means $\langle x\rangle\simeq0.01$ and $\langle\log
x\rangle\simeq-2.25$, consistent with the KR95 interpretation.
However, the best-fitting parameters from both $P_{\rm PL2}$ and
$P_{\rm LG}$ imply that there is a genuine ridge in $\pr(x)$ at this
mean $x$, whereas both $P_{\rm PL1}$ and $P_{\rm S}$ say there is {\it
no} ridge, since they prefer $\alpha<0$.  The Gaussian
parameterization $P_{\rm G}$ is inconclusive: there is not a strong
lower limit on~$x_0$ in this case, since the most likely value of the
other parameter~$\Delta$ is comparable in size to~$x_0$.

Which of the five parameterizations gives the better description of
the real $\pr(x)$?  Using Bayes' theorem again, the plausibility of the
parameterization~$P$ given the available data~$D$ is
$$\eqalign{
\pr(P\mid D) & = {\pr(P)\pr(D\mid P)\over \pr(D)}\cr
       & = {\pr(P)\over \pr(D)}\int \pr(D\mid \omega,P)\pr(\omega\mid
P)\,\d\omega.\cr }\eqname\bookies
$$
If we assume that all of the parameterizations are {\it a priori}
equally likely, i.e., $\pr(P_{\rm PL1})=\pr(P_{\rm PL2})=\pr(P_{\rm
S})=\pr(P_{\rm G})=\pr(P_{\rm LG})$, then we find that $\pr(P_{\rm
PL2}\mid D)=4.0\pr(P_{\rm LG}\mid D)\simeq2\times10^5\pr(P_{\rm S}\mid
D)\simeq10^{10}\pr(P_{\rm PL1}\mid D)
\simeq7\times10^{10}\pr(P_{\rm G}\mid D)$: $P_{\rm PL2}$ and
$P_{\rm LG}$ provide by far the best description of the five.  This
result suggests that there really is a ridge in $\pr(x)$ at
$\log x\simeq-2.2$.

It is also instructive to try to obtain a ``non-parametric'' estimate
of $\pr(x)$.  We take $n$ parameters $\omega_1\ldots\omega_n$ with
$n=50$.  We define $\omega_i$ as the probability that a randomly
chosen galaxy has an MDO whose mass lies between $x_{i-1}$ and $x_i$,
where $x_i$ runs logarithmically from $x_1=10^{-5}$ to $x_{50}=1$,
and $x_0=0$.  A reasonable prior guess for $\pr(x)$ (and therefore the
$\omega_i$) is a power law.  So we choose
$$\log\pr(\omega)=-{\lambda\over n}\sum_{i=2}^{n-1} 
\left(\log\omega_{i+1}-2\log\omega_i+\log\omega_{i-1}\over(\Delta\log
x)\right)^2,
$$
where the free parameter $\lambda$ controls how smooth (i.e., how far
from a pure power law) we think an acceptable $\pr(x)$ ought to be.
We use $10^6$ iterations of the Metropolis algorithm (Metropolis et
al. 1953; see also Saha \& Williams 1994) to obtain the posterior
distribution $\pr(\omega\mid D)$ for each $\omega_i$ for a range of
$\lambda$.  The results for $\lambda=5$ are plotted on
Figure~\ref{figdist}(b).  The ``non-parametric'' distributions
$\pr(x)$ calculated in this way are broadly the same as the those
obtained from the best parameterizations $P_{\rm PL2}$ and $P_{\rm
LG}$.

Thus far we have assumed that the MDO mass distribution depends only
on $x\equiv M_\bullet/M_{\rm bulge}$, but there is no good reason for
assuming that $M_\bullet$ should be correlated with the mass rather
than, say, the luminosity of the bulge.  Consider a more general
form,
$$x'\equiv\left(M_\bullet\over M_\odot\right) \left(L_\odot\over
L\right) \left(\Upsilon_\odot\over\Upsilon\right)^a,\eqname\aparam
$$
where $a$ is a free parameter.  The analysis above can be carried out
with $x$ replaced by $x'$.  Setting $a=1$ tests the correlation of
$M_\bullet$ with $M_{\rm bulge}$ (the case we have just considered),
whereas setting $a=0$ tests its correlation with $L$.  The results of
our calculations of $\pr(P\mid D,a)$ for a range of $a$ are plotted on
Figure~\ref{figcompa}.  Clearly, the case $a=1$ is the most plausible --
$M_\bullet$ is much more strongly correlated with the mass of the bulge
than the luminosity.

Finally, a simple confirmation of the proportionality between
$M_\bullet$ and $M_{\rm bulge}$ can be obtained by splitting the
sample of 32 galaxies into the most luminous half and the least
luminous half, and then calculating $\pr(\omega\mid D,P)$ for each
subsample for each parameterization.  With the exception of the poorly
fitting parameterization $P_{\rm PL1}$, we find that the best-fitting
parameters $\omega$ calculated using each subsample lie within the
95\% confidence region of the parameters calculated using the full
sample.

\section Conclusions

We have examined a sample of 36 galaxy bulges and found that the
kinematics of~32 of them are described well by two-integral
axisymmetric models.  Among these~32, a substantial MDO is required in
all but four in order for our models to reproduce the observed
kinematics.  We have considered a range of models for the demography
of these MDOs.  In the best-fitting models about 96\% of galaxies have
an MDO.  The mass of this MDO is strongly correlated with the bulge
mass, with an MDO-to-bulge mass ratio of around 0.005.  Possible
explanations for this correlation have already been discussed by Faber
et al.\ (1997).  The galaxies without MDOs perhaps have a different
formation history; one possible scenario has been put forward by
Kormendy (1993b).

The mass-to-light ratios $\Upsilon$ fit by our models scale with
luminosity $L$ as $\Upsilon\propto L^{0.2}$, which is just the usual
fundamental plane correlation.  Since our models take full account of
the shape of the light distribution of each galaxy, they rule out any
attempts to explain the slope of the fundamental plane by a
``non-homology'' of the light profiles (e.g., Graham \& Colless 1997,
and references therein).  Our models do not, however, consider the
possibility of a systematic change in orbital anisotropy with
luminosity (Ciotti et al.\ 1996).

These results are based on an ``assembly-line'' approach to building
galaxy models, which is necessarily less accurate than building models
for each galaxy by hand. In particular:
\begingroup\parindent=2em\parskip=0pt
\item{(i)} Some or all of the galaxies may not be axisymmetric.

\item{(ii)} Even if the galaxies are axisymmetric, our two-integral models 
are not the most general possible. For some or all of the galaxies,
there may exist more general three-integral models that can reproduce
the observed kinematics (and, indeed, the full line-of-sight velocity
profiles) without needing to invoke MDOs -- see Kormendy et
al. (1997a) for an example.

\item{(iii)} The selection criteria used to
derive this sample are heterogeneous and impossible to quantify, although any
biases introduced by properties such as luminosity, core size, and surface
brightness are accounted for by the analysis procedure in \S 3.

\item{(iv)} The assumption that the mass-to-light ratio is independent of
position outside the centre may not be correct. 

\endgroup
Of the points above, the most important is perhaps (ii) -- our
conclusions are most uncertain due to our assumption of a two-integral
distribution function.  We should know soon whether more general
three-integral models (e.g., Rix et al.\ 1997, Gebhardt et al.\ 1997,
Richstone et al.\ 1997) will relax the need for MDOs in at least some
of the galaxies in our sample.  However, it is not yet clear what
mechanism could effect just the right degree of radial anisotropy in
each galaxy to cause the apparent correlation $M_\bullet\propto M_{\rm
bulge}$ as seen by our two-integral models.

{\bf Acknowledgements:} We thank Jes\'us Gonz\'alez for providing
kinematical data from his thesis.  Our collaboration was supported by
HST data analysis funds through GO grants GO-2600.01.87A and
GO-06099.01-94A, by NASA grant NAS-5-1661 to the WFPC1 IDT, and by
grants from NSERC.  ST acknowledges support from an Imasco fellowship.
We thank the Fields Institute for Research in Mathematical Sciences at
the University of Toronto and OCIW for their hospitality during part
of this work.

\vfill\eject
\def\apj #1 #2{ApJ, #1, #2}
\def\apjs #1 #2{ApJS, #1, #2}
\def\aj #1 #2{AJ, #1, #2}
\def\mn #1 #2{MNRAS, #1, #2}
\def\aa #1 #2{A\&A, #1, #2}
\def\araa #1 #2{ARA\&A, #1, #2}

\biblio\overfullrule=0pt
\bibitem Bender R., D\"obereiner S., M\"ollenhoff C., 1987 \aa 177 {53 (BDM)}
\bibitem Bender R., Nieto J.-L., 1990, \aa 239 {97 (BN)}
\bibitem Bender R., Burstein D., Faber S.M., 1992, \apj 399 462
\bibitem Bender R., Saglia R.P., Gerhard O.E., 1994, \mn 269 {785 (BSG)}
\bibitem Bender R., Kormendy J., Dehnen W., 1996, \apj 464 {L119 (BKD)}
\bibitem Binney J.J., Davies R.L., Illingworth G.D., 1990, \apj 361 {78 (BDI)}
\bibitem Byun Y.I., et al., 1996, \aj 111 1889
\bibitem Ciotti L., Lanzoni B., Renzini A., 1996, \mn 282 1
\bibitem Davies R., Birkinshaw M., 1988, \apjs 68 {409 (DB)}
\bibitem Dehnen W., Gerhard O.E., 1994, \mn 268 1019
\bibitem Dejonghe H., 1986, Phys. Rep., 133, 218
\bibitem Faber S.M., Dressler A., Davies R.L., Burstein D.,
Lynden-Bell D., Terlevich R., Wegner G., 1987, in ``Nearly Normal
Galaxies, From the Planck Time to the Present'', ed. S.M. Faber (NY:
Springer), 317
\bibitem Faber S.M., et al., 1997, preprint
\bibitem Fisher D., Illingworth G., Franx M., 1994, \aj 107 {160 (FIF94)}
\bibitem Fisher D., Illingworth G., Franx M., 1995, \apj 438 {539 (FIF95)}
\bibitem Forbes D. A., Franx M., \& Illingworth G. D. 1995, \aj 109 1988
\bibitem Forbes D.A., Franx M., Illingworth D., Carollo C.M., 1996, \apj 467 126
\bibitem Franx M., Illingworth G., Heckman T., 1989, \apj 344 {613 (FIH)}
\bibitem Gebhardt K., Richstone D., et al., 1997, (G97) (n3379)
\bibitem Gonz\'alez J., 1993, Ph.D. thesis, UC Santa Cruz (G93)
\bibitem Graham A., Colless M.M., submitted to MNRAS (astro-ph/9701020)
\bibitem Grillmair C.J., et al. 1994, \aj 108 102
\bibitem Harms R.J., et al., 1994, \apj 435 L35
\bibitem Jaffe W., 1983, \mn 202 995
\bibitem Jaffe W., et al., 1994, \aj 108 1567
\bibitem Jedrzejewski R., Schechter P.L., 1989, \aj 98 147 (JS)
\bibitem Kent S.M., 1987, \aj 94 {306 (Ke87)}
\bibitem Kormendy J., 1988, \apj 335 {40 (K88) (n4594)}
\bibitem Kormendy J., 1993a, in ``The Nearest Active Galaxies'',
ed. J.~Beckman, L.~Colina \& H.~Netzer (Madrid: Consejo Superior de
Investigaciones Cientificas), p.~197
\bibitem Kormendy J., 1993b, in IAU Symposium 153, ``Galactic
Bulges'', eds. H.~Habing and H.~Dejonghe, Kluwer: Dordrecht
\bibitem Kormendy J., Richstone D., 1992, \apj 393 {559 (KR92)}
\bibitem Kormendy J., Richstone D., 1995, \araa 33 {581 (KR95)}
\bibitem Kormendy J., et al., 1996a, \apj 459 {L57 (K96a) (n3115)}
\bibitem Kormendy J., et al., 1996b, \apj 473 {L91 (K96b) (n4594)}
\bibitem Kormendy J., et al., 1997a, \apj 482 {L139 (K97a) (n4486b)}
\bibitem Kormendy J., et al., 1997b, preprint (K97b) (n3377).
\bibitem Kormendy J., Bender R., 1997, preprint (KB97)
\bibitem Lauer T.R., et al., 1992a, \aj 103 703 (L92a) (M87)
\bibitem Lauer T.R., et al., 1992b, \aj 104 552 (L92b) (M32)
\bibitem Lauer T.R., et al., 1993, \aj 106 1436
\bibitem Lauer T.R., et al. 1995, \aj 110 2622 
\bibitem Lynden-Bell D., 1962, \mn 123 447
\bibitem Magorrian S.J., 1997, in preparation
\bibitem Magorrian S.J., Binney J.J., 1994, \mn 271 949
\bibitem Metropolis N., et al., 1953, J. Chem. Phys., 21, 1087
\bibitem Peletier R.F., Davies R.L., Illingworth G.D., Davis L.E., Cawson M.,
1990, \aj 100 {1091 (PDIDC)}
\bibitem Richstone D., Bower G., Dressler A., 1990, \apj 353 118
\bibitem Richstone D., et al., 1997, in preparation
\bibitem Rix H.W., de Zeeuw P.T., Carollo C.M., Cretton N., van der
Marel R.P., 1997, preprint
\bibitem Romanowsky A. J., \& Kochanek C. S., 1997, \mn 287 35
\bibitem Rybicki G., 1987, in Structure and Dynamics of Elliptical
Galaxies, ed. T. de Zeeuw (Dordrecht: Kluwer), 397.
\bibitem Saha P., Williams T.B., 1994, \aj 107 1295
\bibitem Scorza C., Bender R., 1995, \aj 293 {20 (SB)}
\bibitem Tremaine S., 1995, \aj 110 628
\bibitem Tremblay B., Merritt D., 1995, \aj 110 1039
\bibitem van der Marel R.P., 1991, \mn 253 710
\bibitem van der Marel R.P., 1994, \mn 270 {271 (vdM94a) (M87)}
\bibitem van der Marel R.P., Franx M., 1993, \apj 407 525
\bibitem van der Marel R.P., Rix H.-W., Carter D.,Franx M., White S.D.M., de
 Zeeuw T., 1994, \mn 268 {521 (vdM94b) (M31,M32,n3115,n4594)}
\bibitem van der Marel R.P., de Zeeuw P., Rix H.-W., Quinlan G.D., 1997, Nature
385, 610
\bibitem Young P., Westphal J.A., Kristian J., Wilson C.P., Landauer
F.P., 1978, \apj 221 721
\endbiblio
\vfill\eject

\appendices
\section Appendix: Notes on individual galaxies

{\bf M31:} This galaxy has a double nucleus (e.g., Lauer et al.\ 1993),
but axisymmetric models should still provide a reasonable description
of the gross features of its kinematics.  We use kinematical data from
van der Marel et al.\ (1994) and Kormendy \& Bender (1997).  The former
appear to measure major-axis radii from the photometric centre of the
galaxy, rather than the kinematic centre, which we assume to be
coincident with the fainter nucleus (e.g., Tremaine 1995).  Thus we
add 0.3 arcsec to van der Marel et al.'s quoted major-axis positions.
We do not fit to kinematical data beyond 10 arcsec because the outer
photometry we use (Kent 1987) consists only of a major- and a
minor-axis profile with no additional isophote shape information.

{\bf NGC 1600:} Both Jedrzejewski \& Schechter (1989) and Bender,
Saglia \& Gerhard (1994) give major- and minor-axis kinematics for
this galaxy.  There are many more outlier points in the latter data,
so we reject it.

{\bf NGC~2778:} Fisher, Illingworth \& Franx (1995; FIF95) and
Gonz\'alez (1993; G93) give major-axis profiles.  Both
G93 and Jedrzejewski \& Schechter (1989) give minor axis profiles.
FIF95's central dispersion is inconsistent with the others, so we
reject their data for this galaxy.

{\bf NGC 3379:} The best kinematical data comes from Gebhardt et
al. (1997).  We restrict our model fits to their ground-based data
within 12 arcsec of the centre.

{\bf NGC 4486:} We use the blue G-band kinematics from van der Marel
(1994), and reject his infrared kinematics which are probably affected
by template mismatch.  In the same paper, van der Marel presents
evidence that this galaxy is radially anisotropic in its outer parts.
Therefore we restrict our model fits to the kinematics within the
innermost 5 arcsec.

{\bf NGC 4594:} The ground-based outer photometry (Kormendy~1988)
consists only of a major- and a minor-axis profile.  Because of this,
and because of the problems with dust obscuration, we only use
kinematical data within 8 arcsec.

\vfill\eject

\newfigure\figvsigtest
\figure{
\centerline{\psfig{file=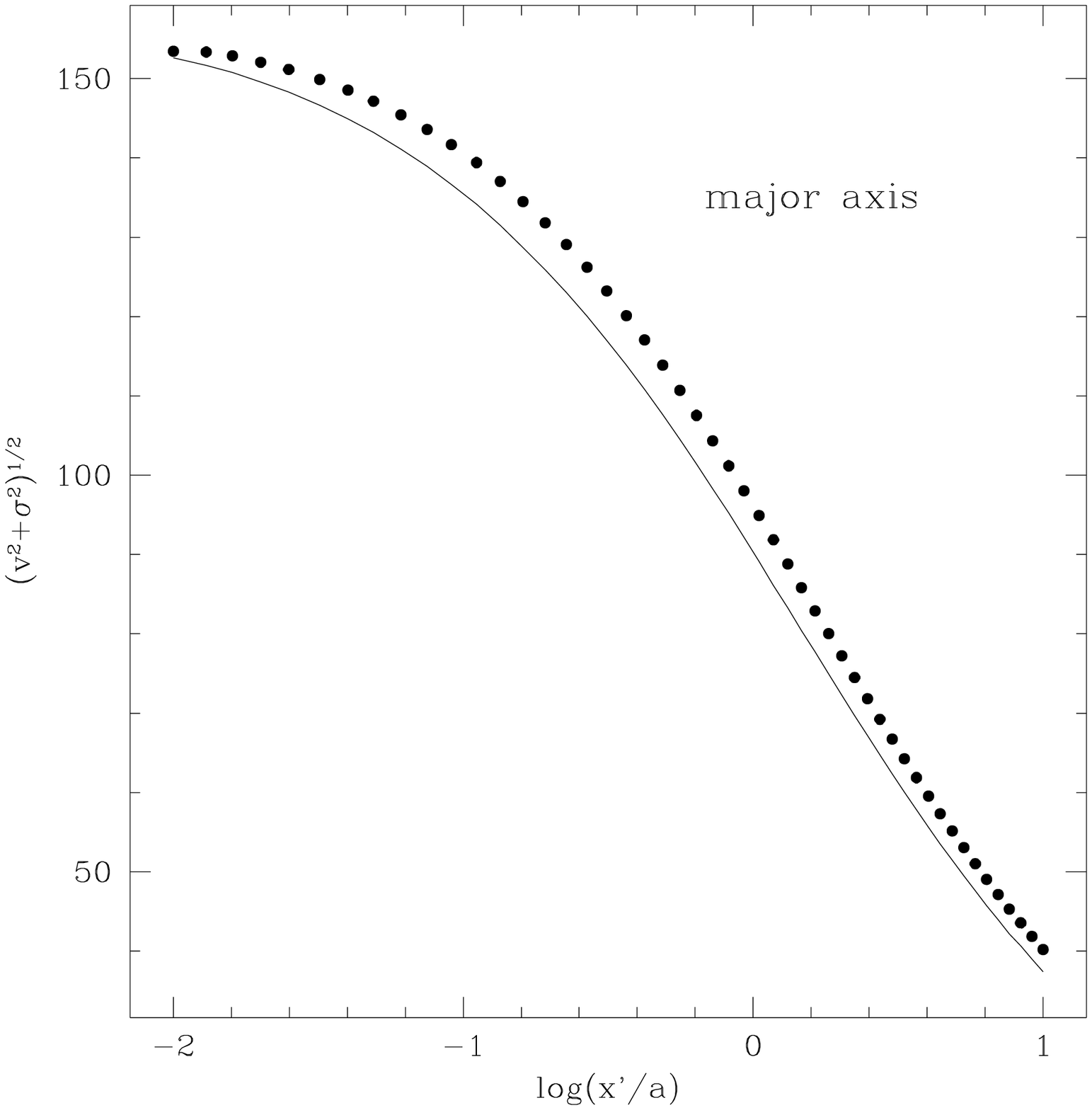,width=\hhsize}
            \psfig{file=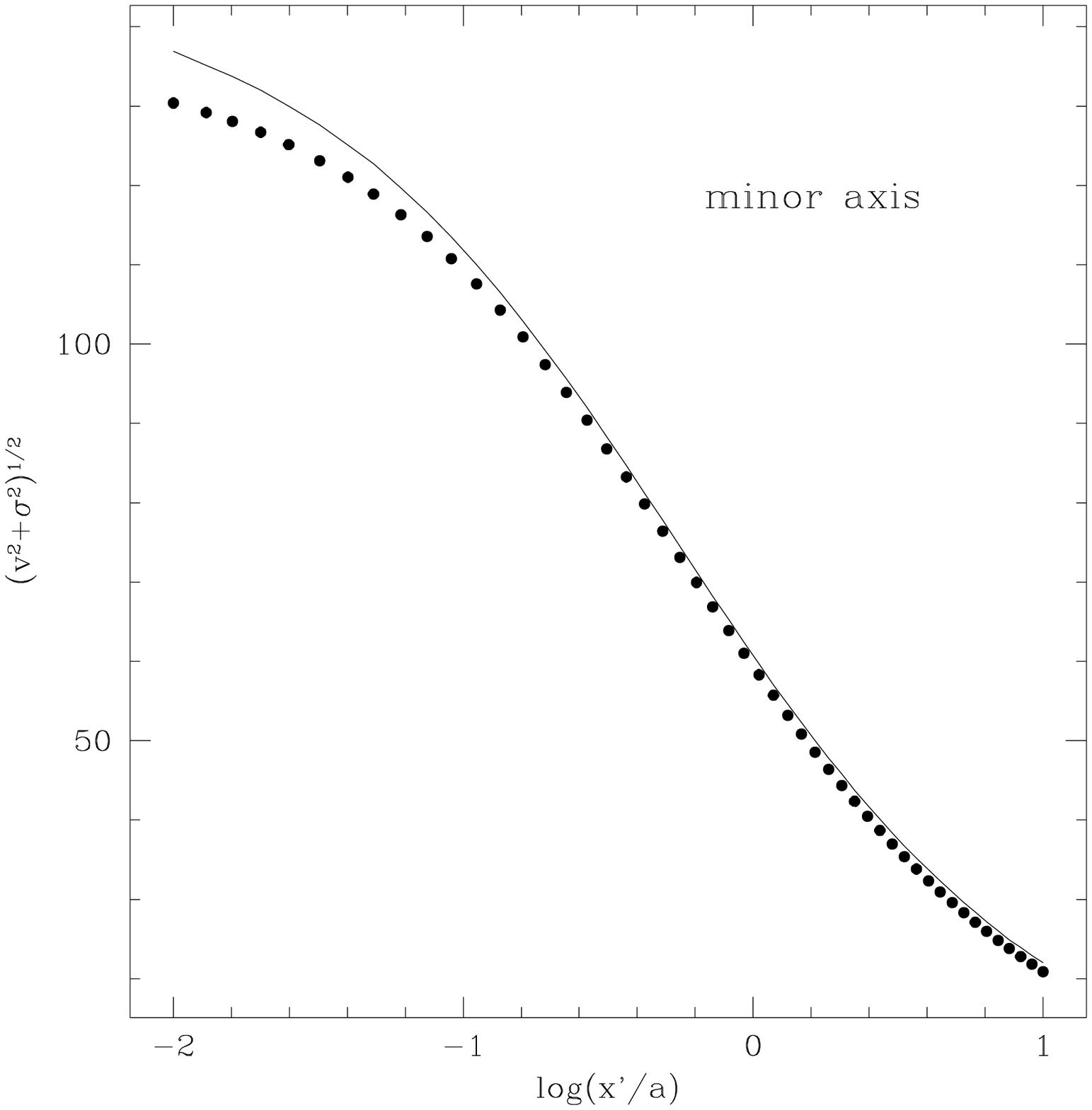,width=\hhsize} }}
{{\bf Figure \ref{figvsigtest}.} Major- and minor-axis projected
second-moment profiles for a flattened (axis ratio $q=0.6$) isotropic
(rotating) Jaffe (1983) model viewed edge on.  The curves show the
(square root of the) classical second-order moments.  The points plot
the approximation $(v^2+\sigma^2)^{1/2}$ where $v$ and~$\sigma$ are
the parameters of the best-fitting Gaussians to the line-of-sight
velocity profiles.}

\newfigure\figbad
\def\caption{%
{{\bf Figure \ref{figbad}(a).} Plots of $\chi^2$
versus $M_\bullet/M_{\rm bulge}$ for the four galaxies that our models
do not describe well.  The different curves on each plot correspond to
different assumed inclinations.  A reasonable fit would have
$\chi^2\approx N_{\rm dof}\pm(2N_{\rm dof})^{1/2}$ where the number of
degrees of freedom, $N_{\rm dof}$, is related to $n$, the number of
kinematical bins used in the fit, by $N_{\rm dof}=n-2$.  The heavy
solid and dashed lines show $\chi^2=N_{\rm dof}$ and $\chi^2=N_{\rm
dof}+(2N_{\rm dof})^{1/2}$ respectively.}  }
\def\captioncont{{\bf Figure \ref{figbad}(a)}...continued.}
\figure{
\centerline{
\psfig{file=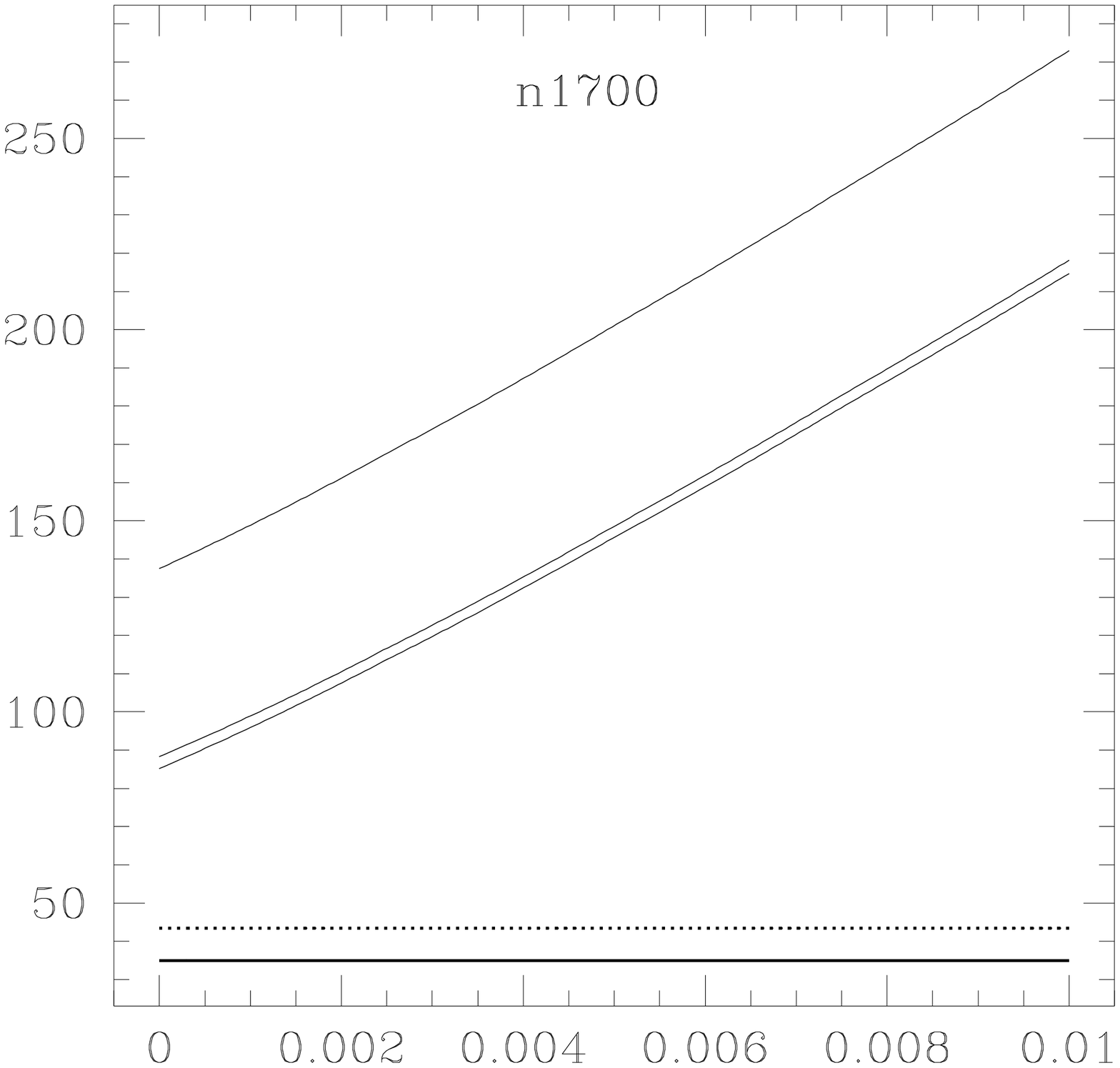,width=0.25\hsize}
\psfig{file=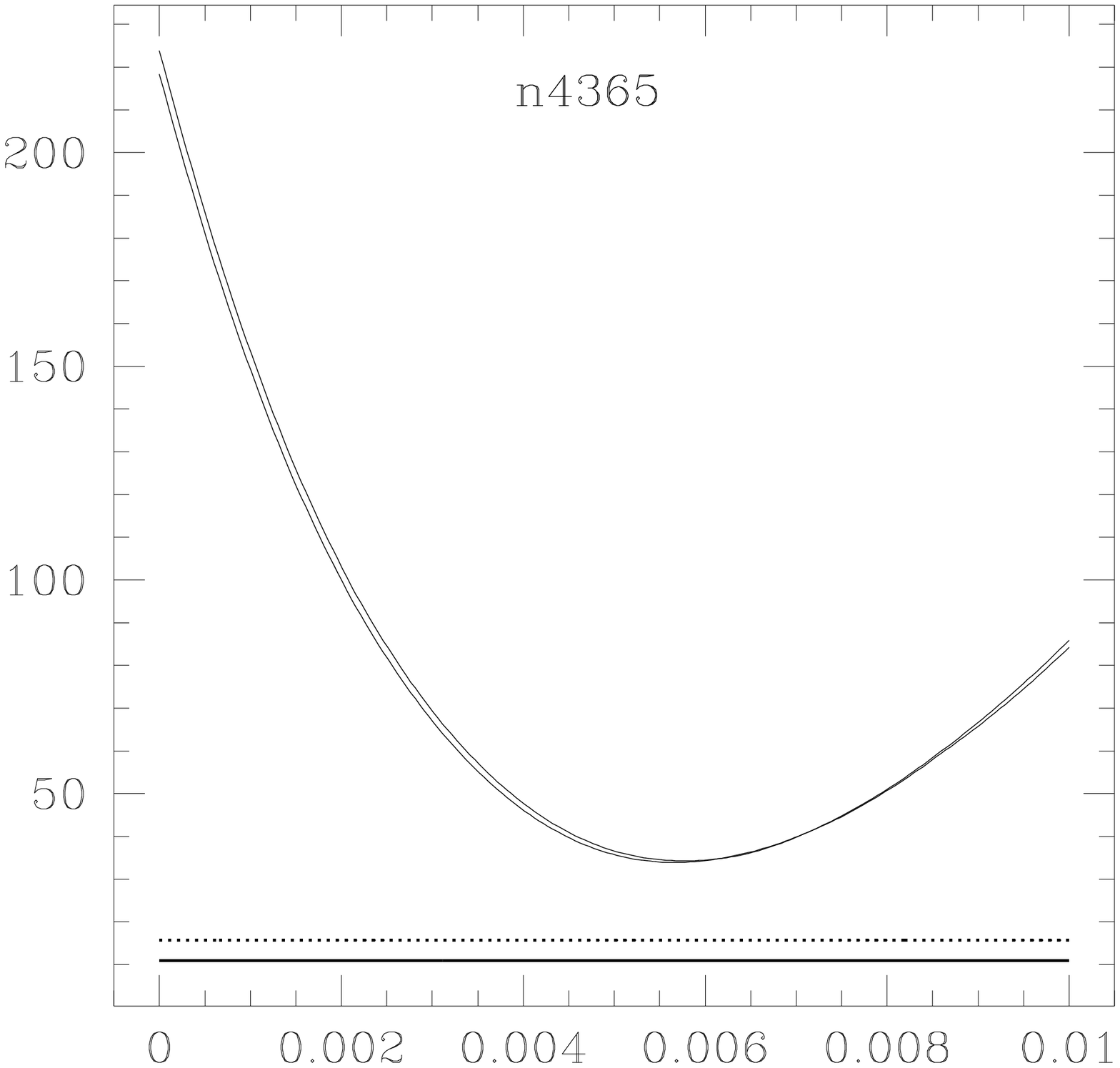,width=0.25\hsize}
\psfig{file=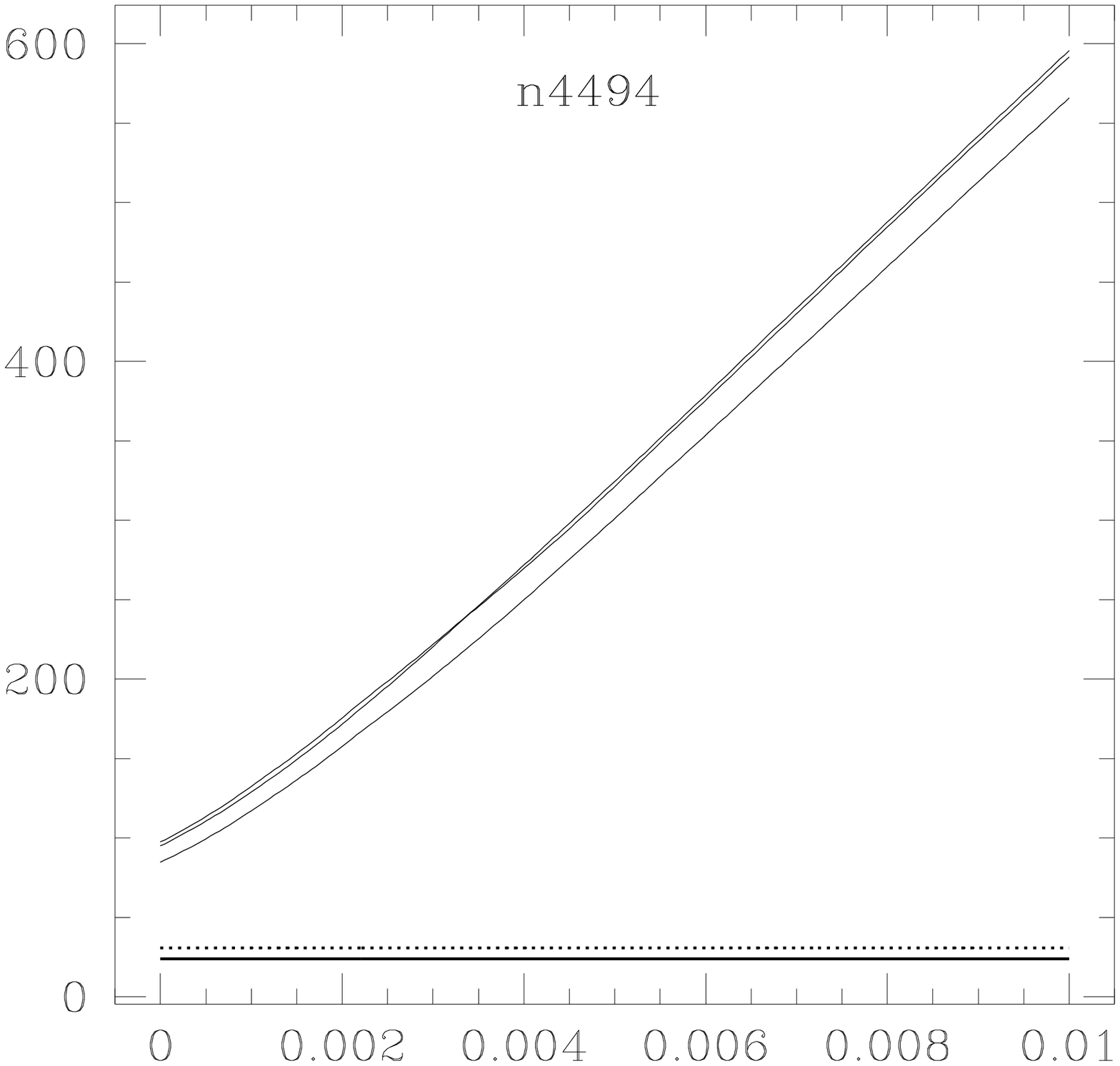,width=0.25\hsize}
\psfig{file=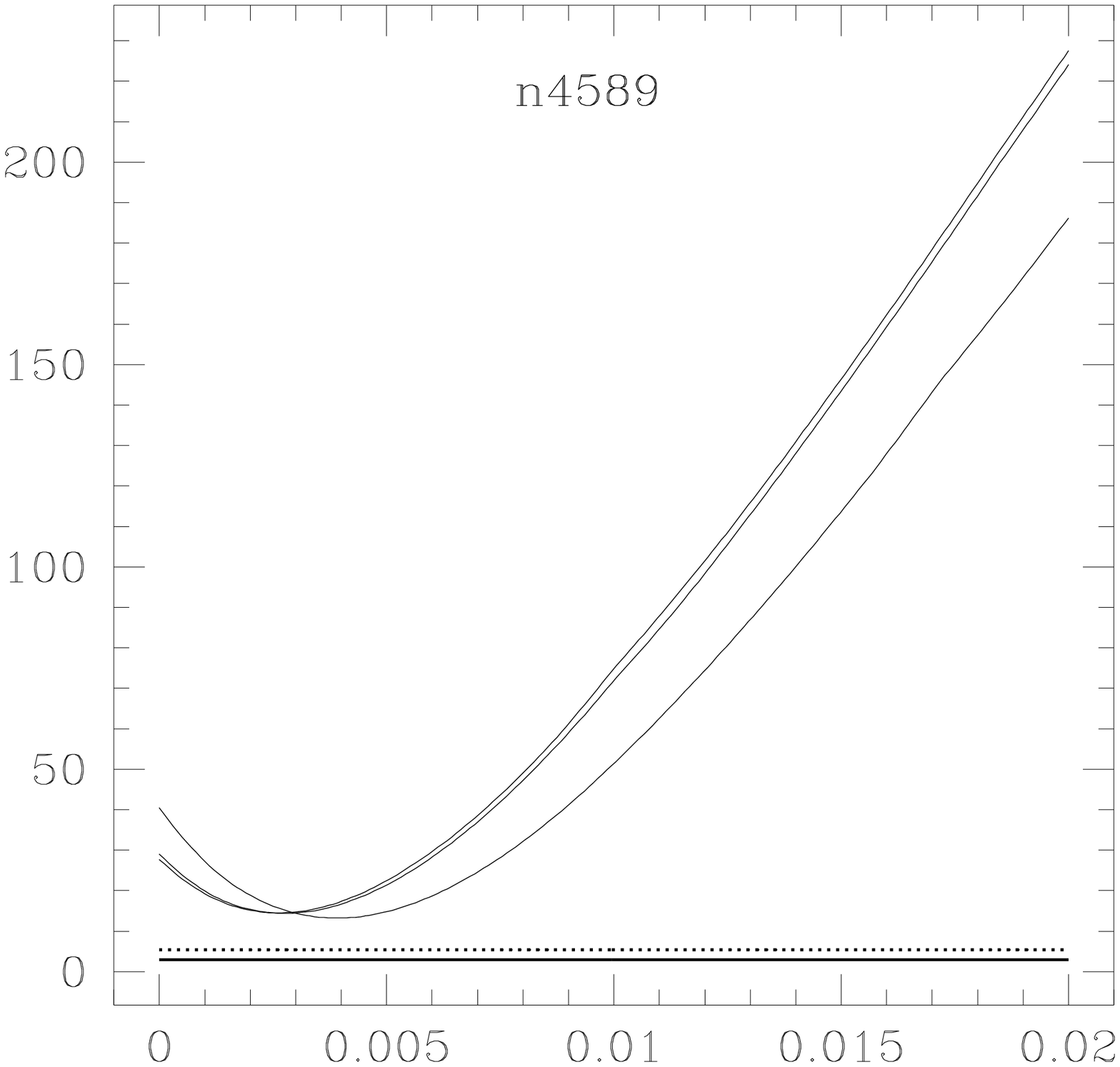,width=0.25\hsize}
}
}{\caption}

\def\caption{%
{{\bf Figure \ref{figbad}(b).} Kinematical profiles of the
best-fitting models along each slit position used for the galaxies in
Figure~\ref{figbad}(a).  The plots show $(v^2+\sigma^2)^{1/2}$ (in units of
$\kms$) versus distance from the centre of the galaxy (in arcsec).
The observed kinematics are plotted as circles, open or closed
depending on which side of the galaxy the observation was made.  The
solid curves show the model predictions (convolved with the same
seeing as the best ground-based observations) for the best-fitting
value of $M_\bullet$ and $\Upsilon$ for each galaxy.  The results for
each assumed inclination angle are plotted as separate curves.  For
comparison the dashed curves plot the model predictions with the same
value of $\Upsilon$ as above but $M_\bullet=0$.  }}
\def\captioncont{{\bf Figure \ref{figbad}(b)}...continued.}
\figure{
\centerline{
\psfig{file=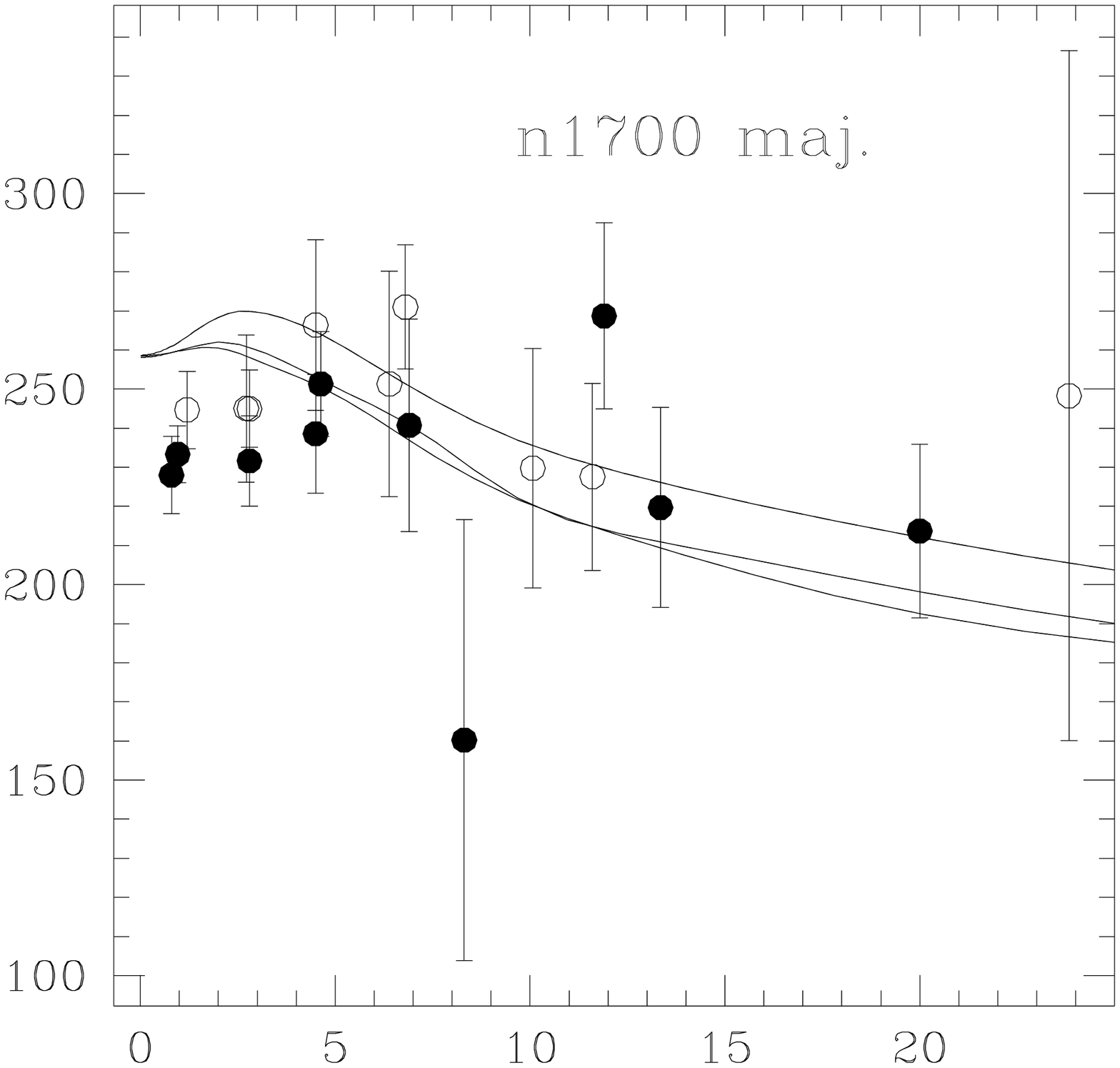,width=0.25\hsize}
\psfig{file=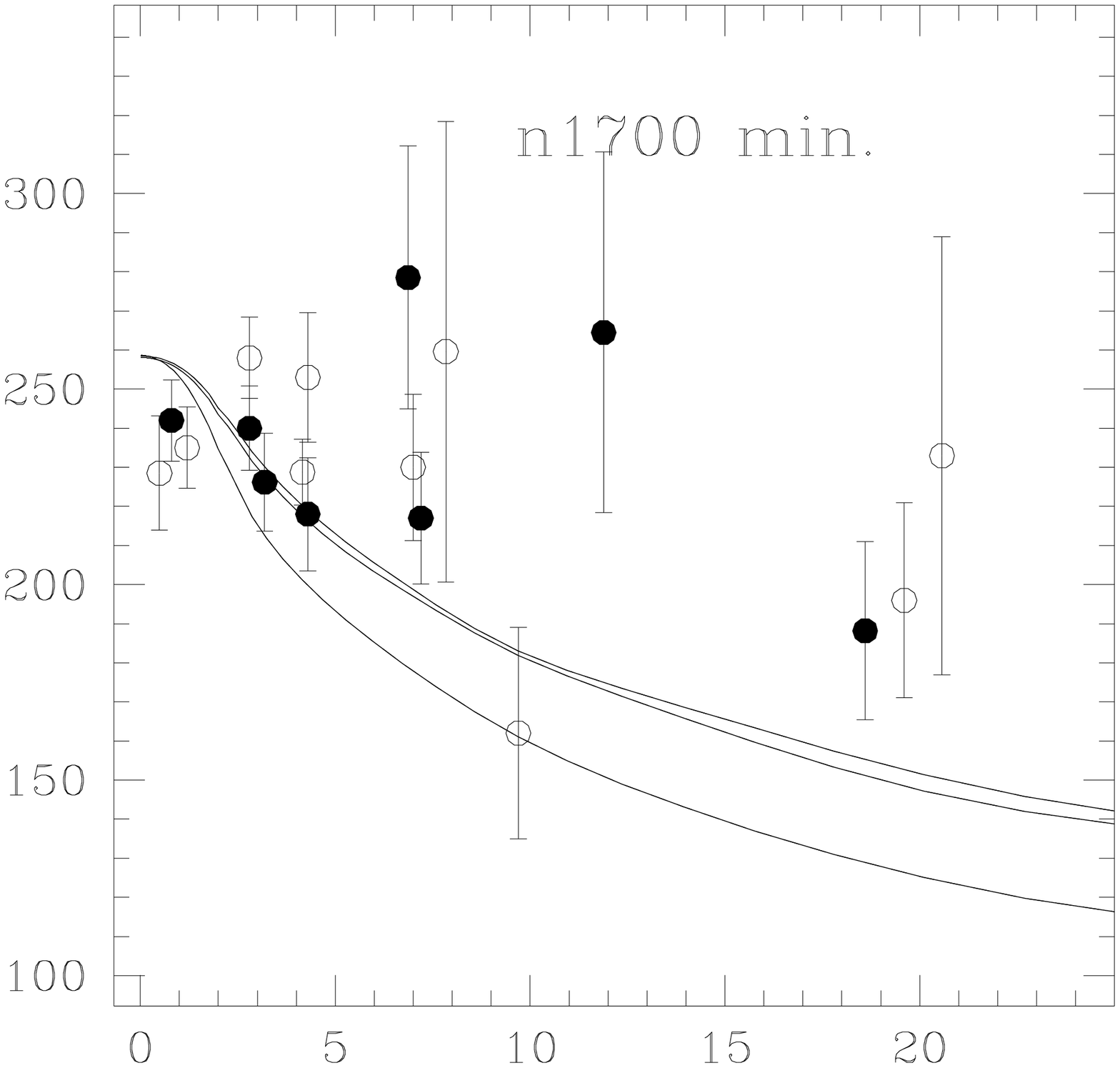,width=0.25\hsize}
\psfig{file=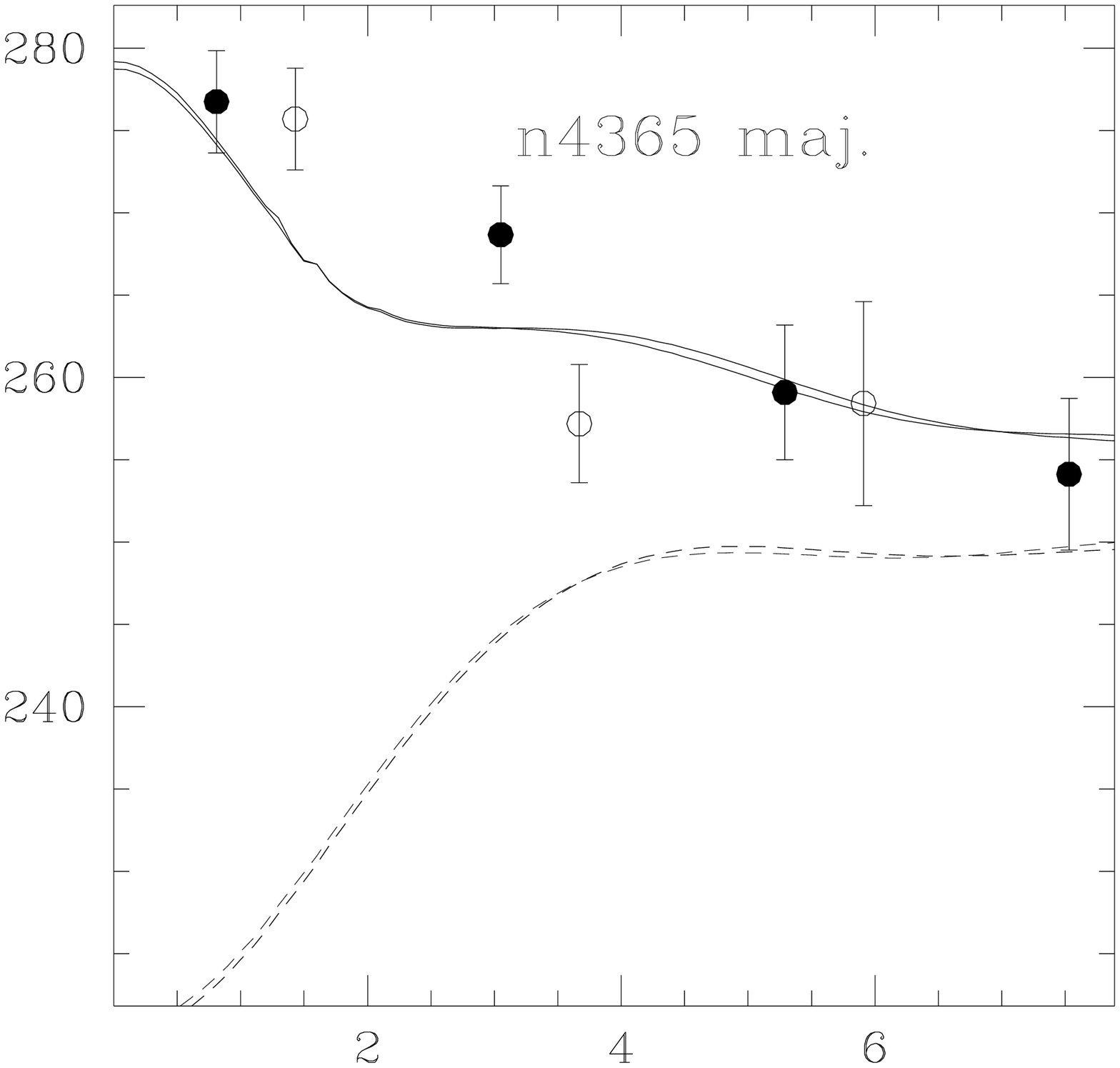,width=0.25\hsize}
\psfig{file=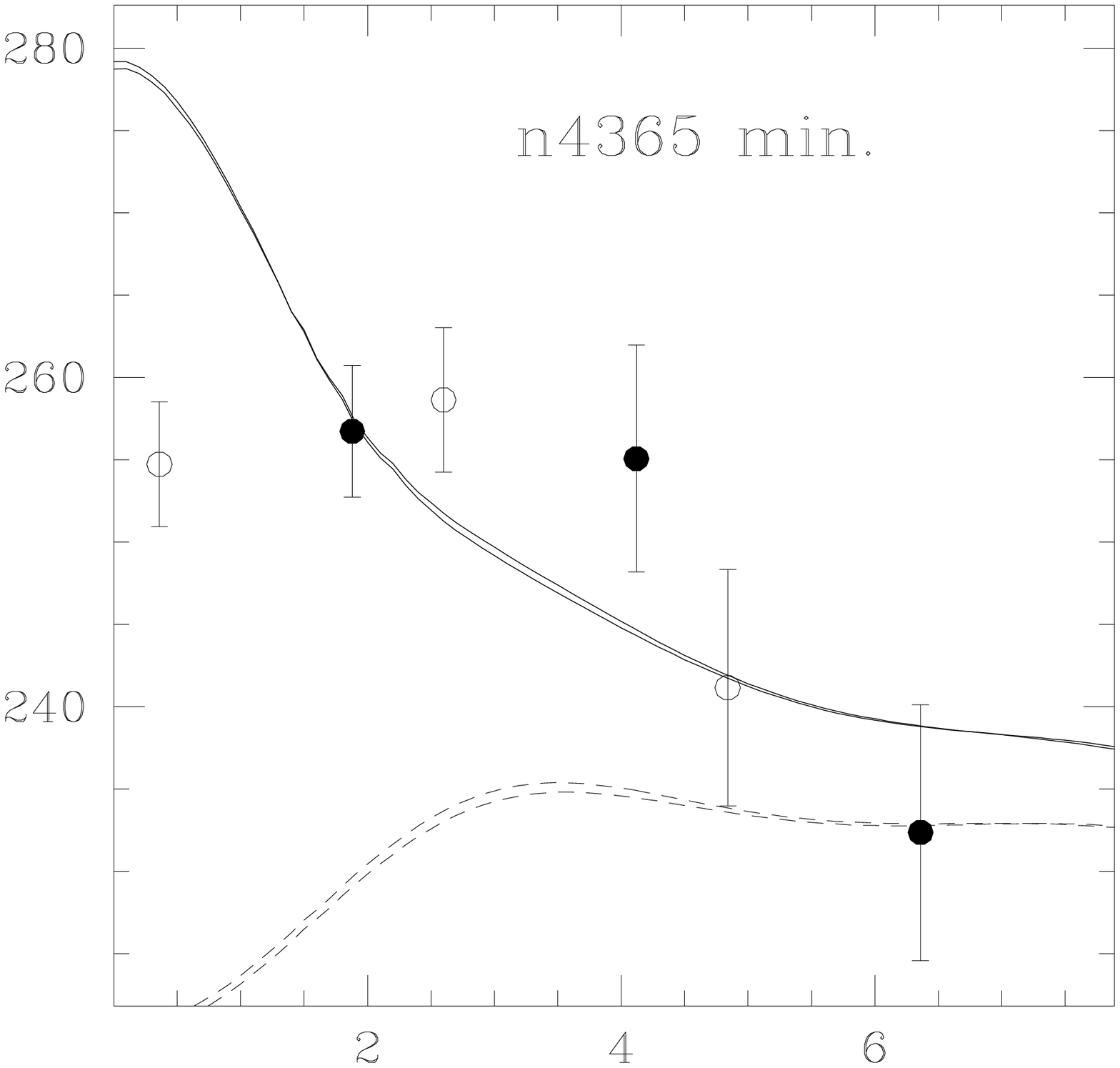,width=0.25\hsize}
}
\centerline{
\psfig{file=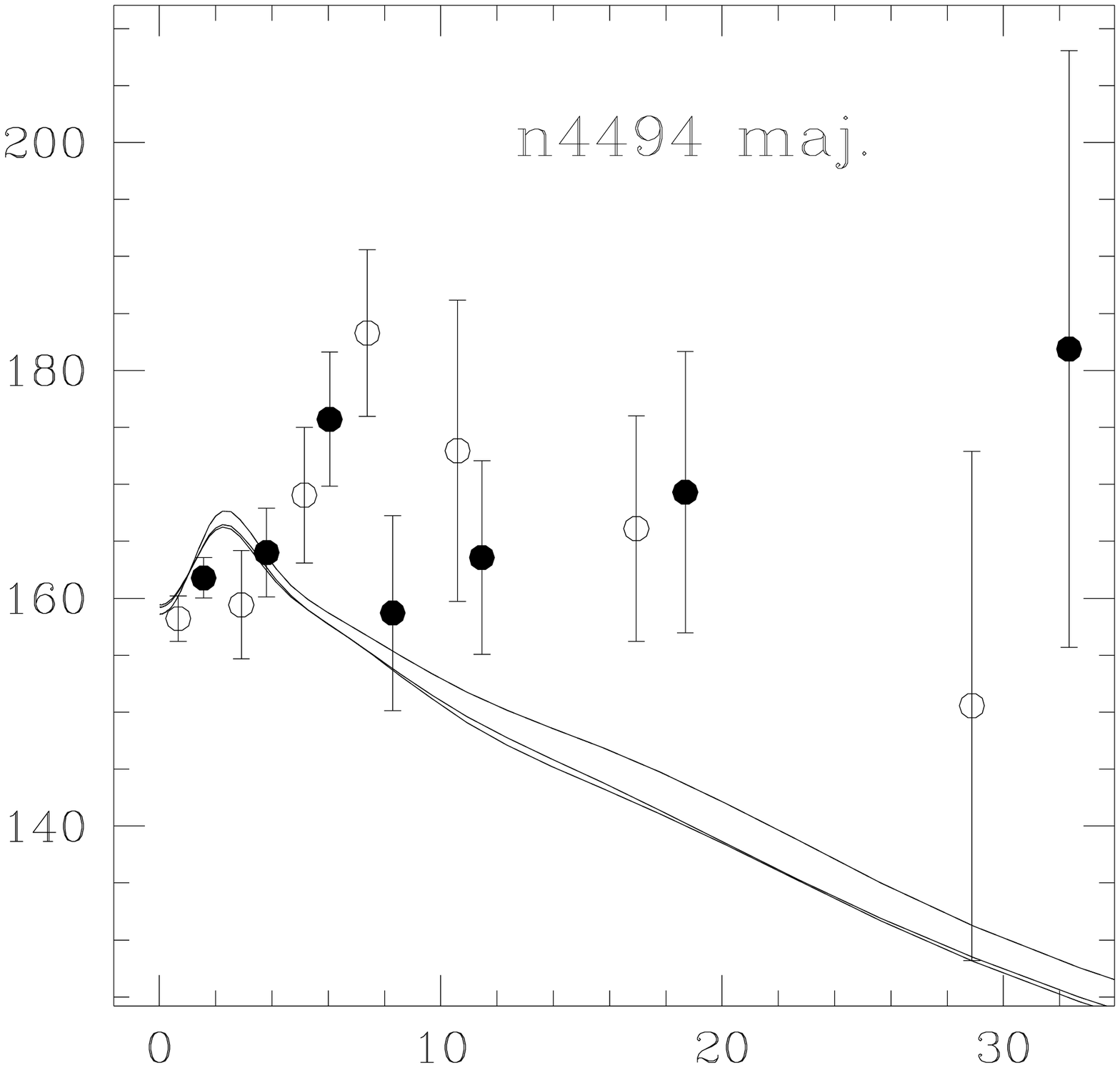,width=0.25\hsize}
\psfig{file=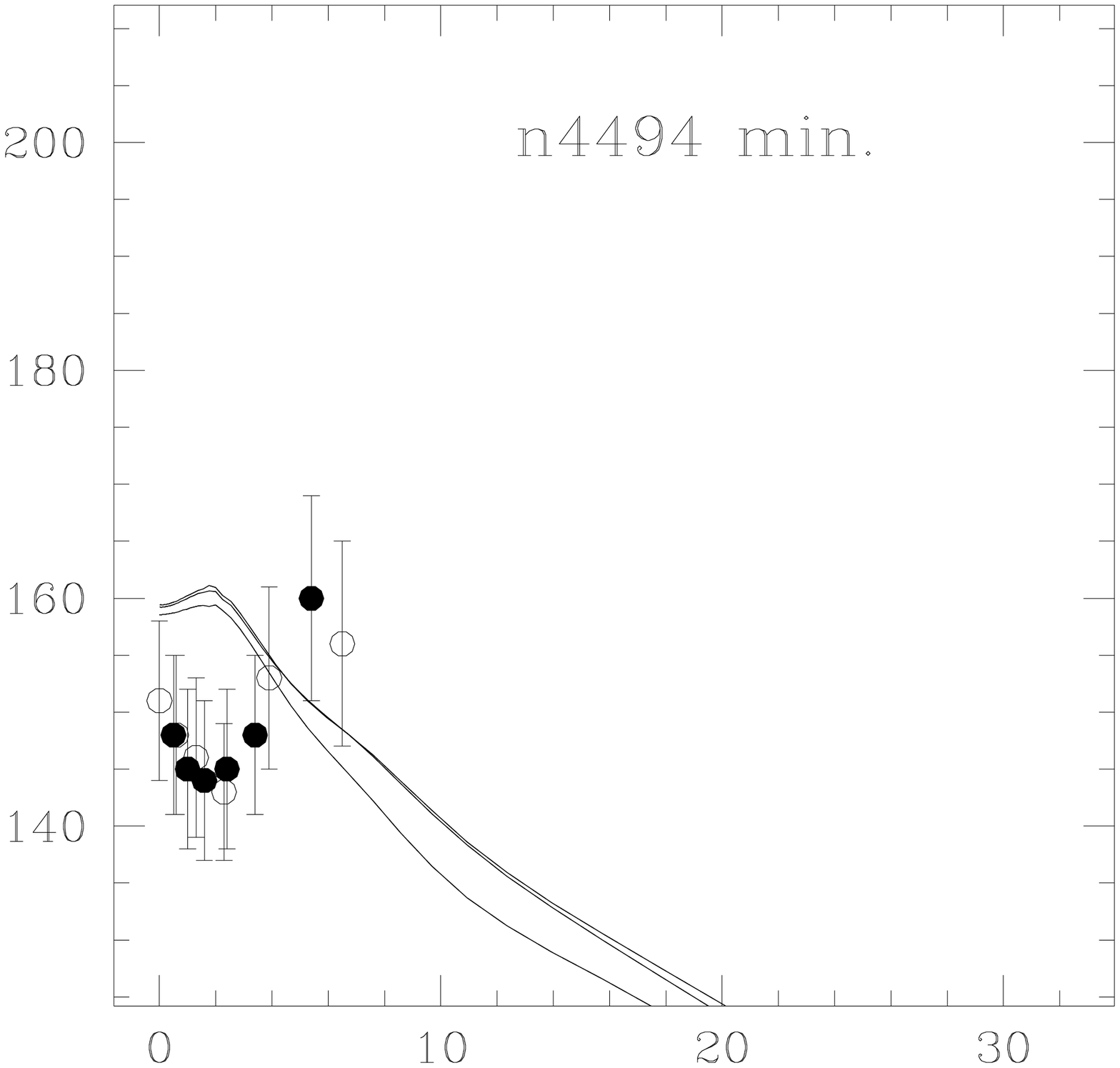,width=0.25\hsize}
\psfig{file=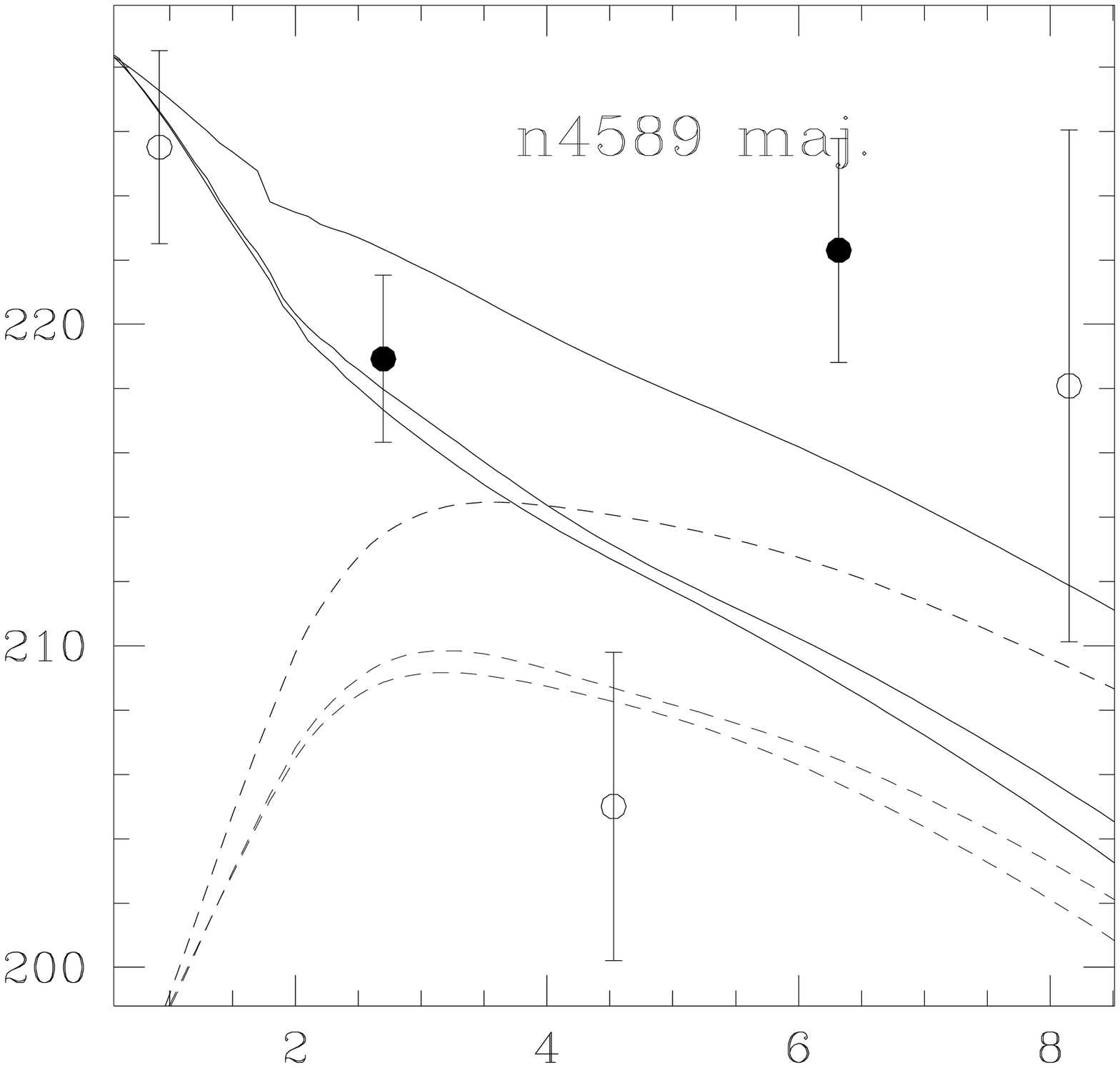,width=0.25\hsize}
\hbox to 0.25\hsize{}
}
}{\caption}

\newfigure\figgood
\def\caption{%
{\bf Figure \ref{figgood}(a).} The posterior distributions
$\pr(\Upsilon,M_\bullet\mid D)$ for all 32 galaxies that our models
describe well.  The vertical and horizontal axes are
$\log(\Upsilon/\Upsilon_\odot)$ and $M_\bullet/M_{\rm bulge}$
respectively.  Successive light contours indicate a factor of ten
change in $\pr(\Upsilon,M_\bullet\mid D)$.  The heavy contours enclose
the 68\% and 95\% confidence regions on $\Upsilon$ and $M_\bullet$.}
\def\captioncont{{\bf Figure \ref{figgood}(a)}...continued.}
\figure{
\centerline{
\psfig{file=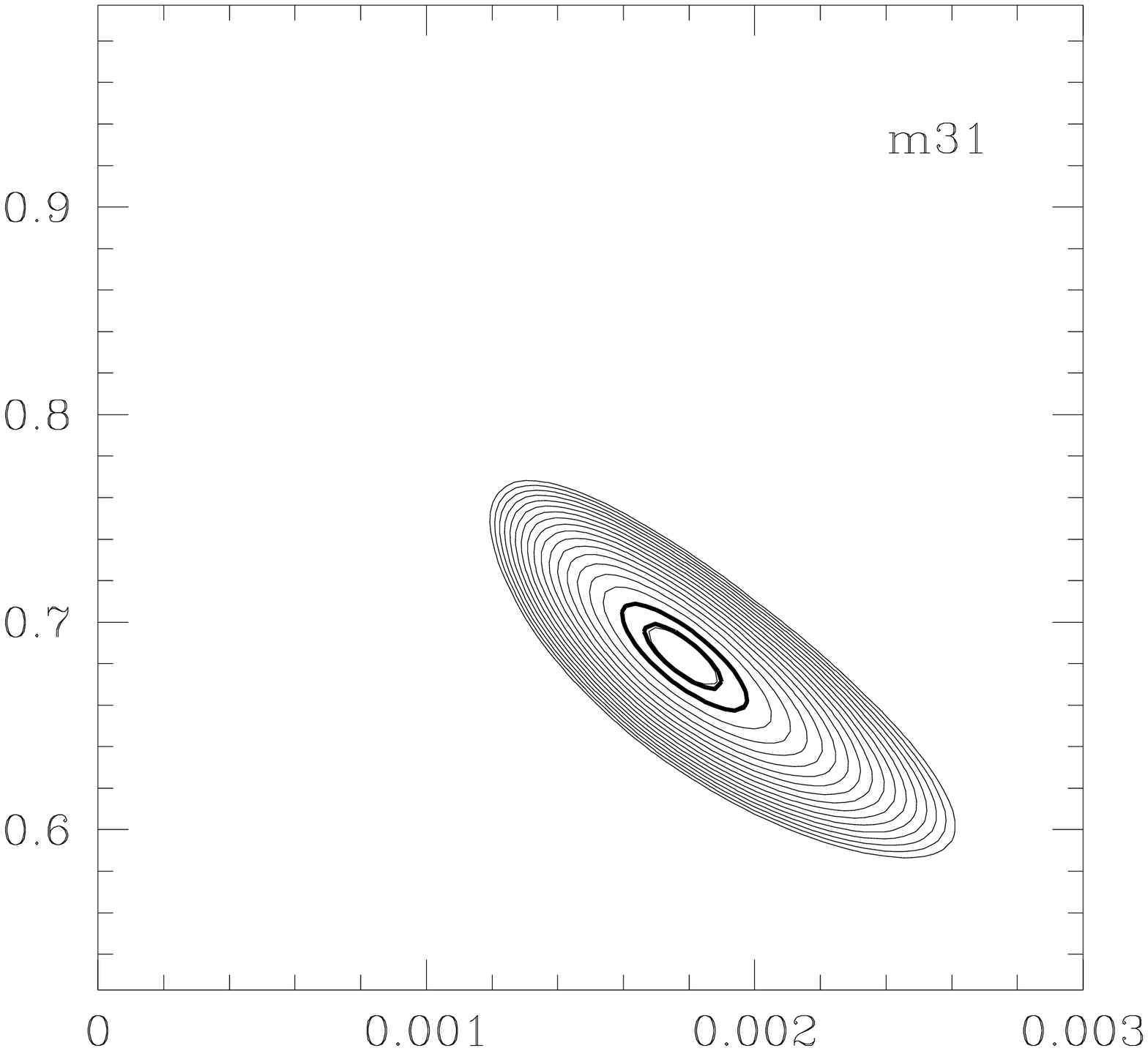,width=0.25\hsize}
\psfig{file=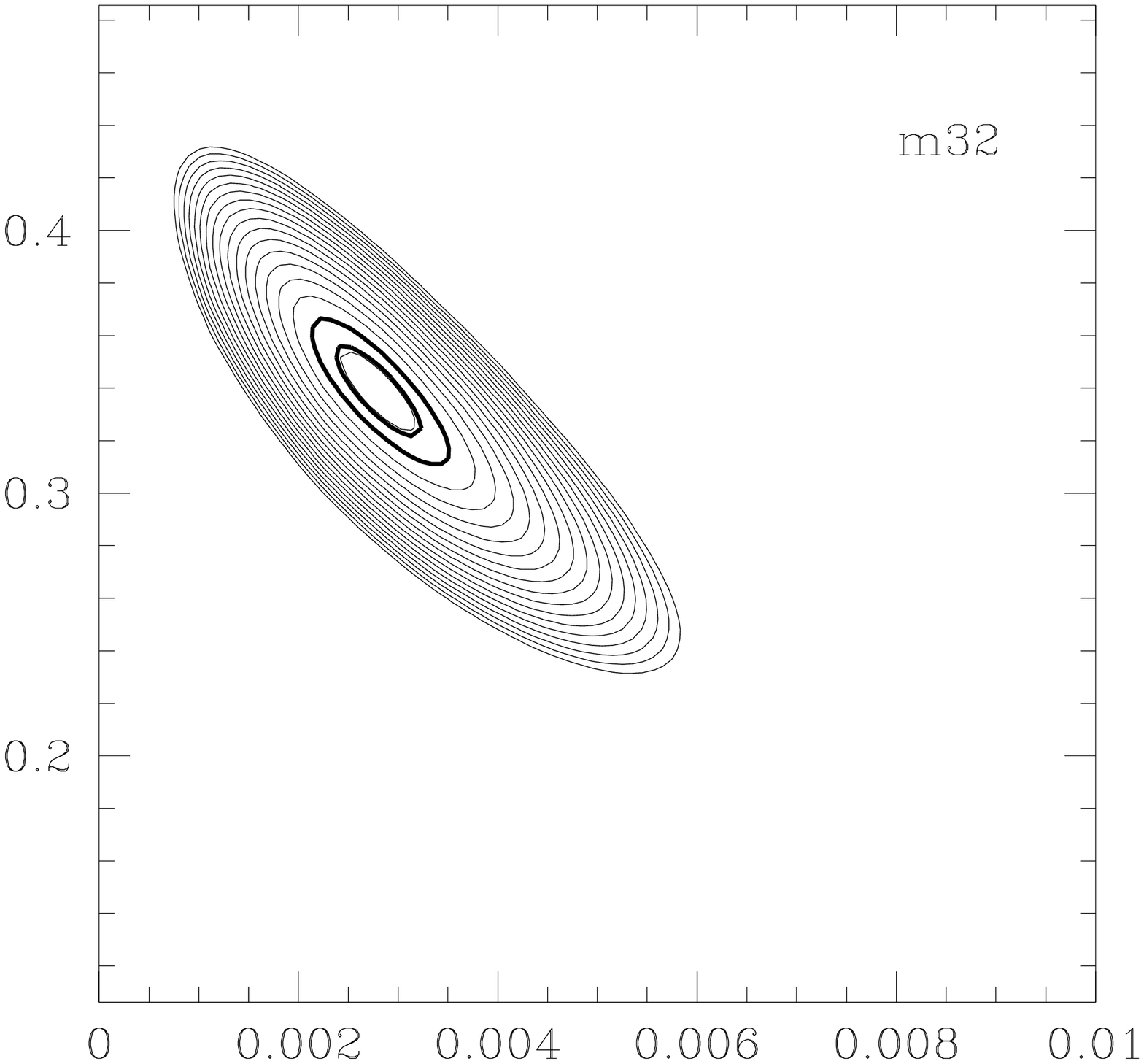,width=0.25\hsize}
\psfig{file=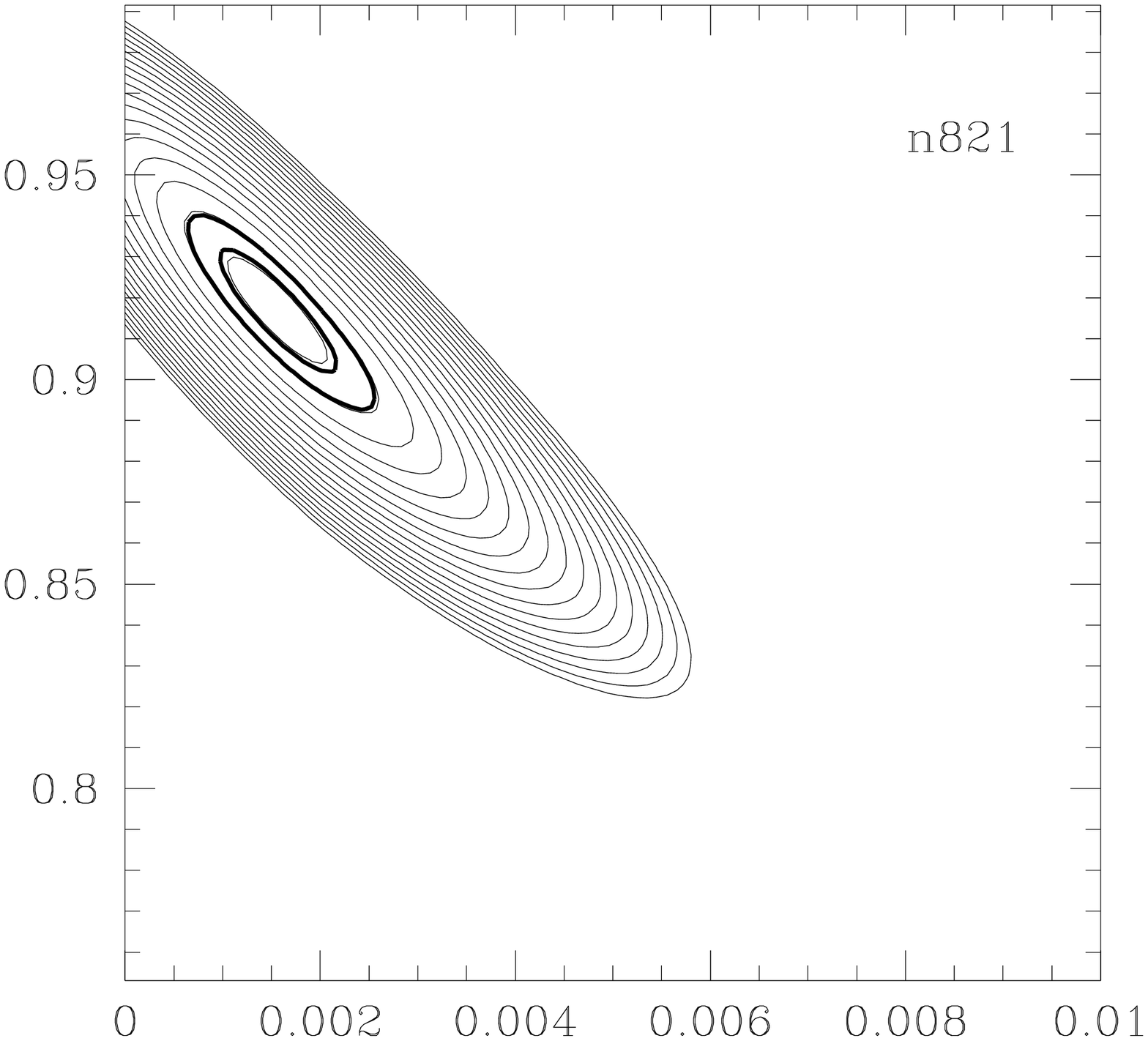,width=0.25\hsize}
\psfig{file=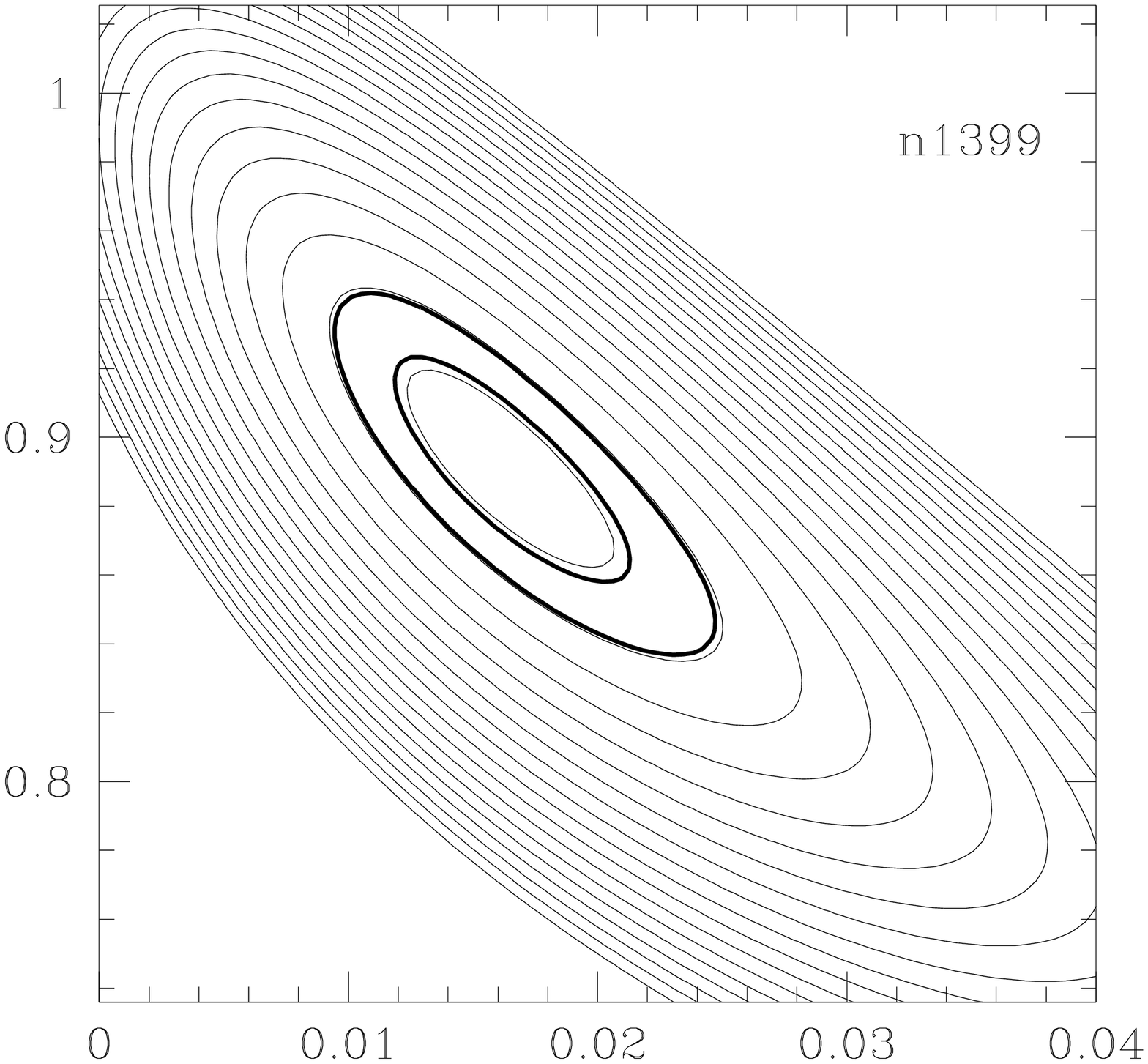,width=0.25\hsize}
}
\centerline{
\psfig{file=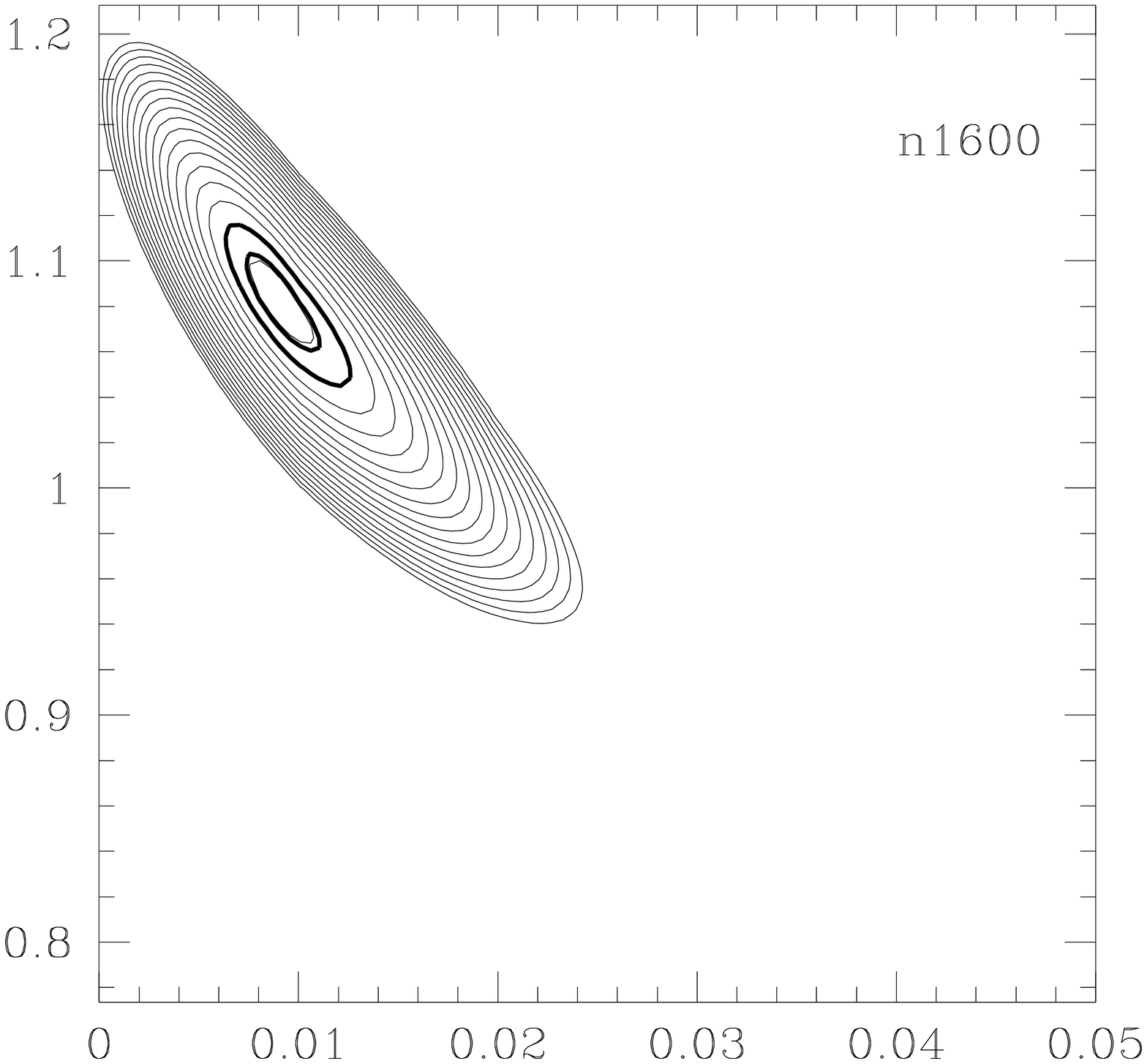,width=0.25\hsize}
\psfig{file=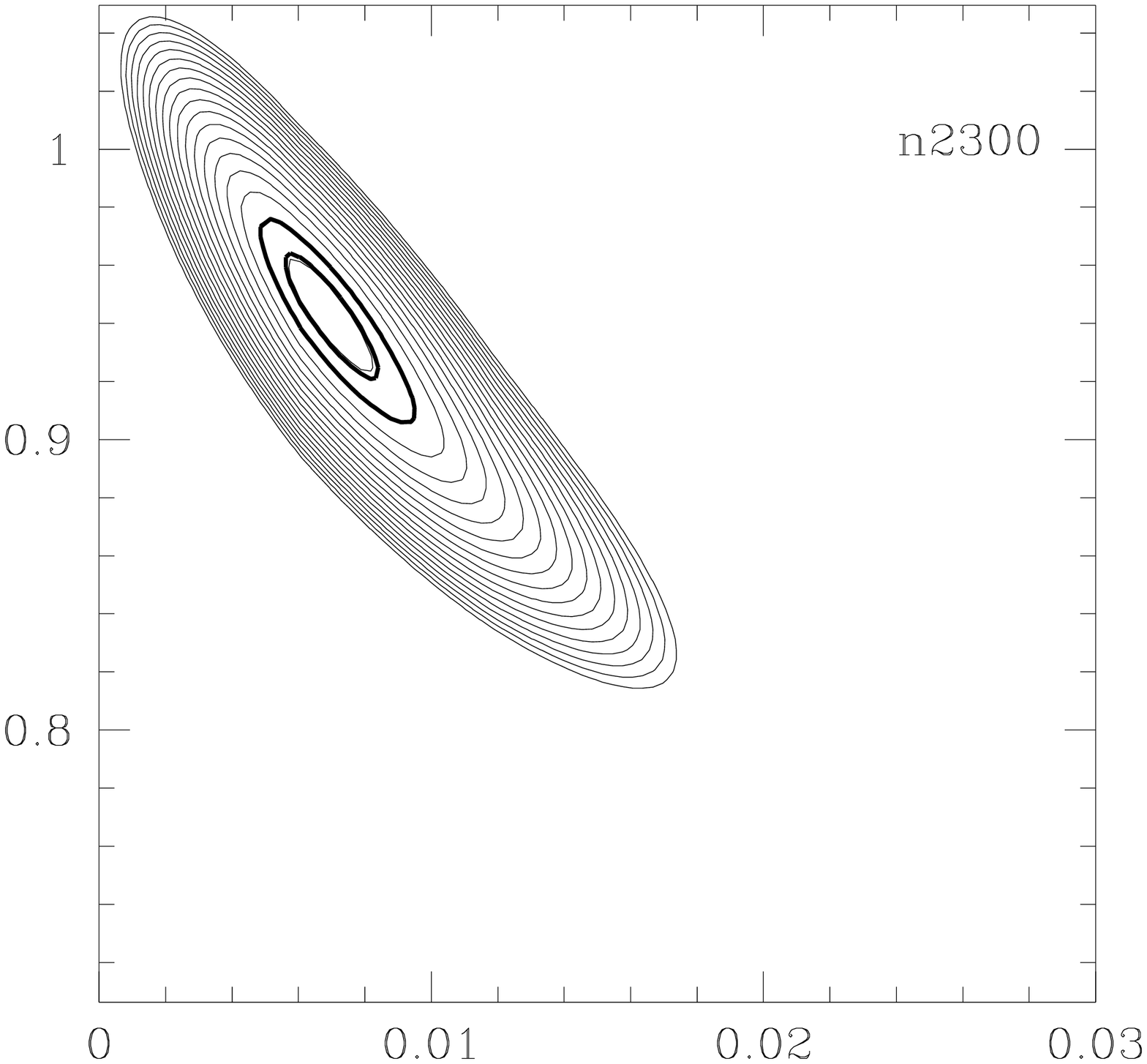,width=0.25\hsize}
\psfig{file=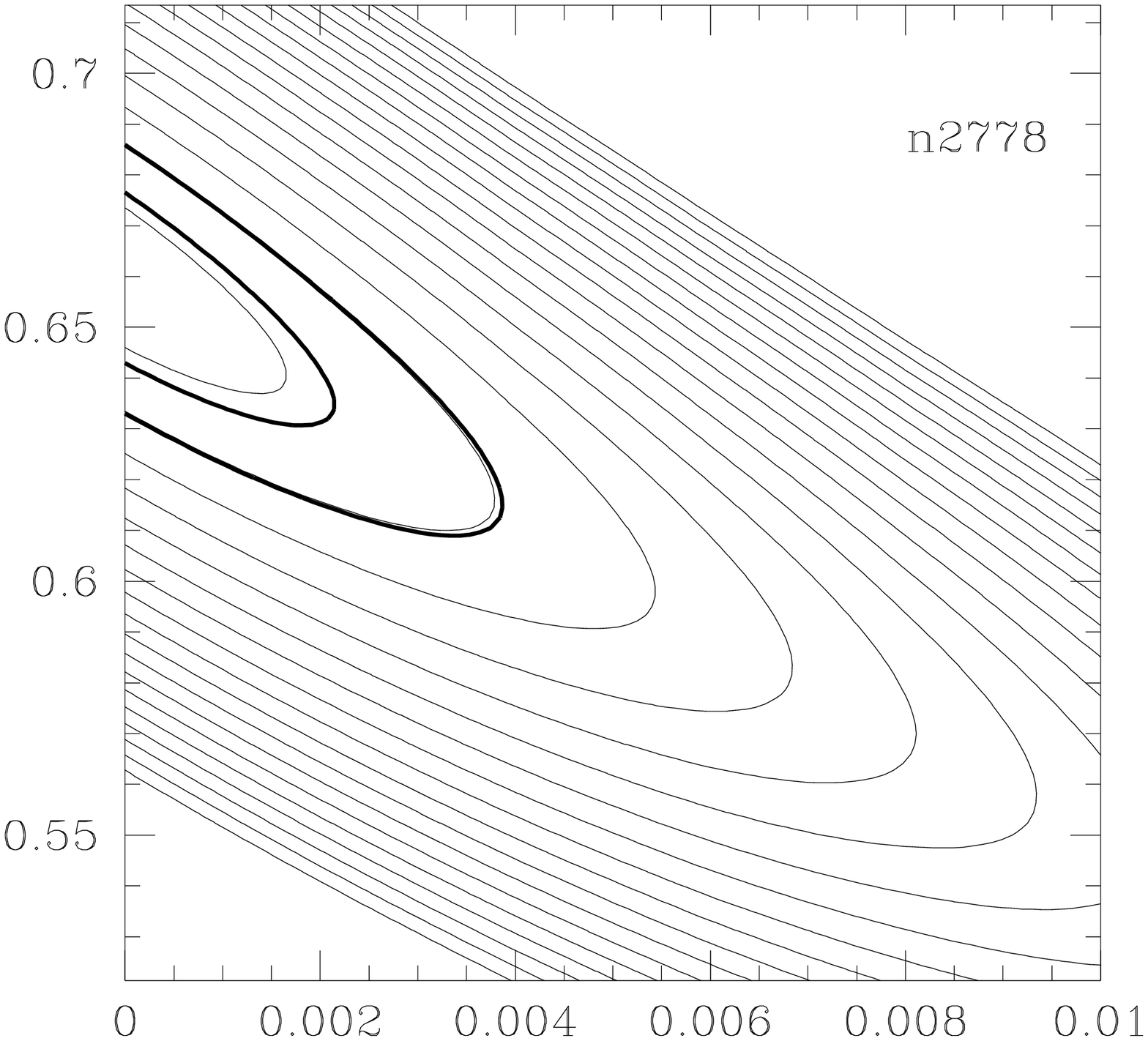,width=0.25\hsize}
\psfig{file=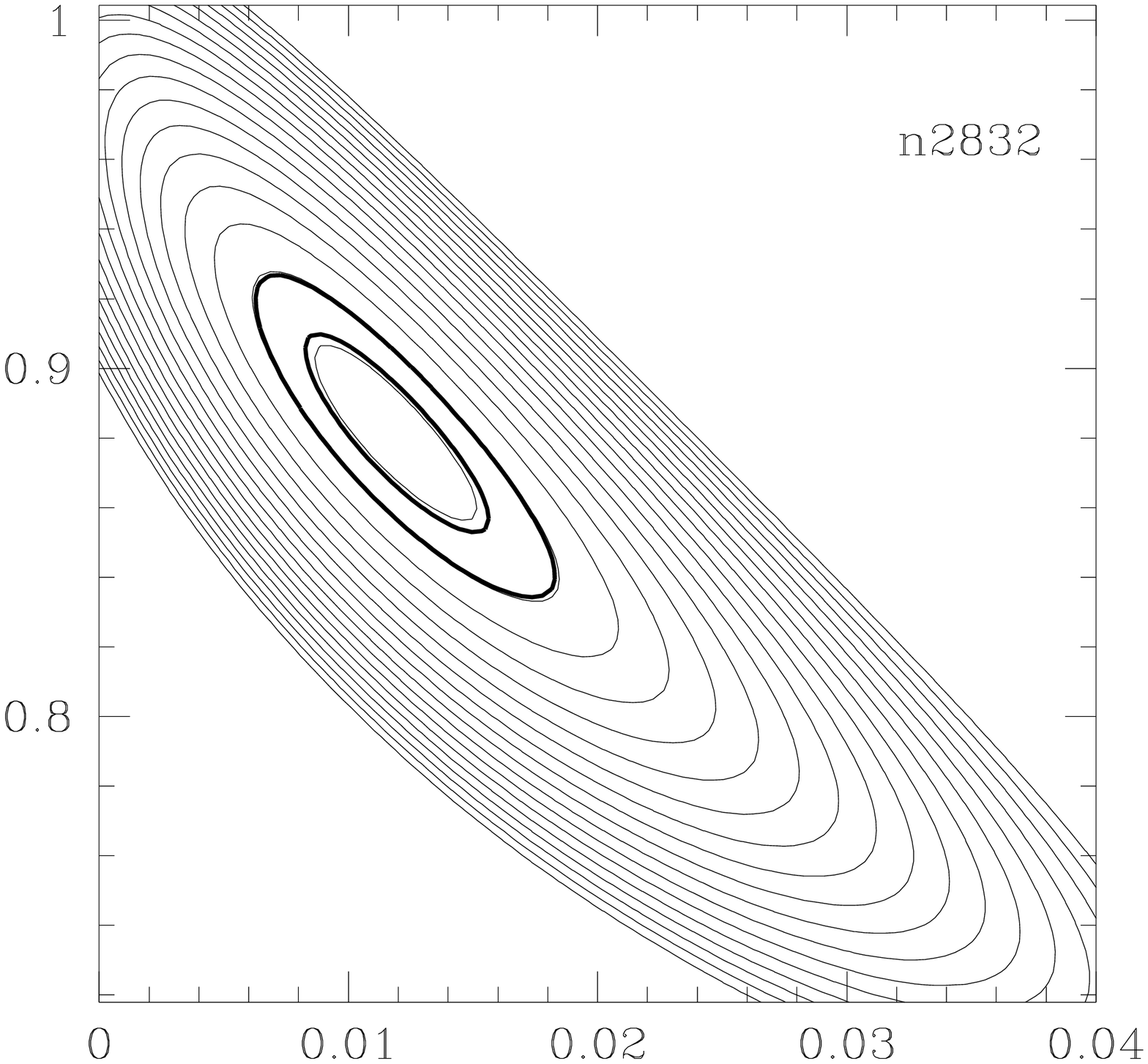,width=0.25\hsize}
}
\centerline{
\psfig{file=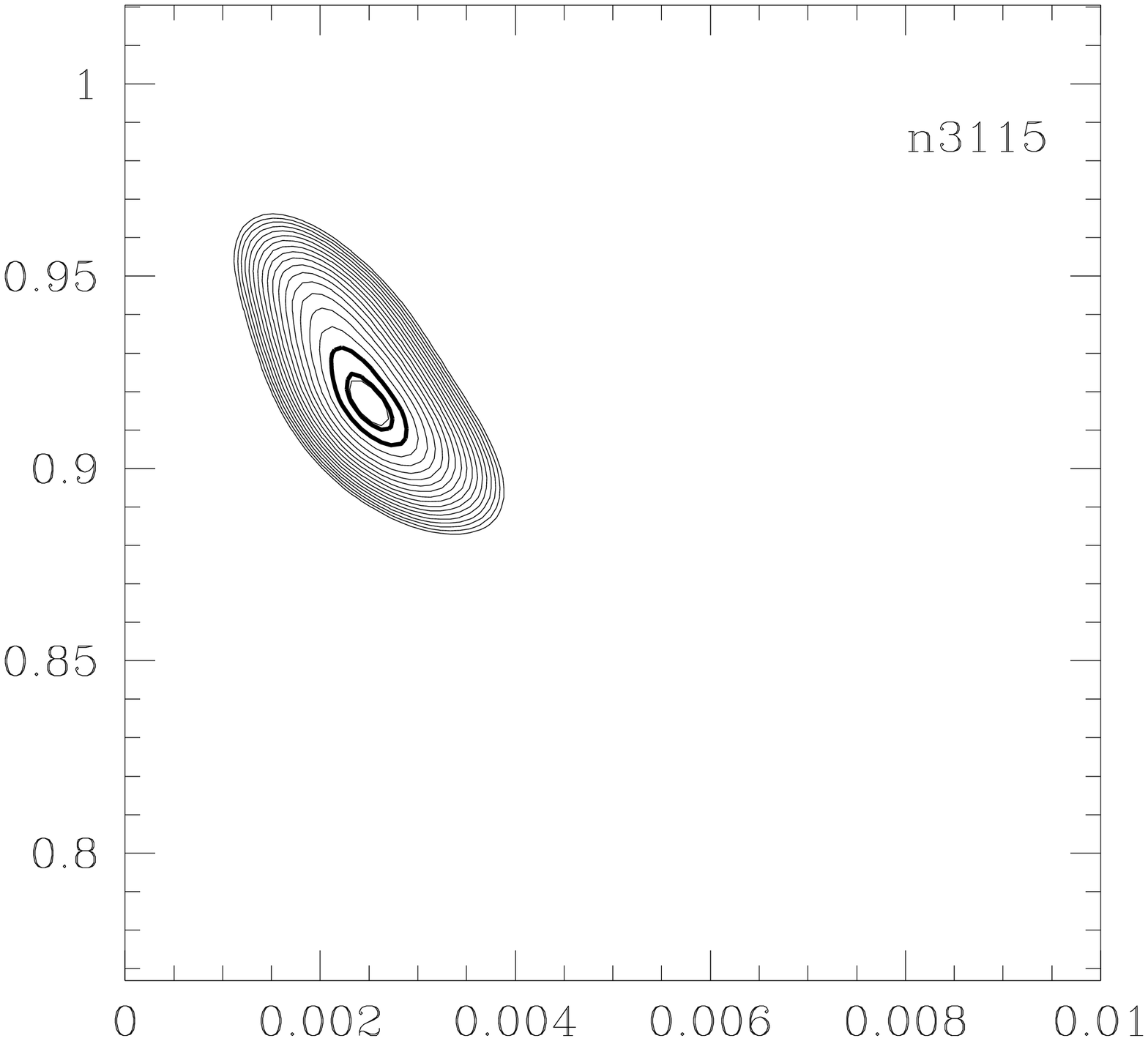,width=0.25\hsize}
\psfig{file=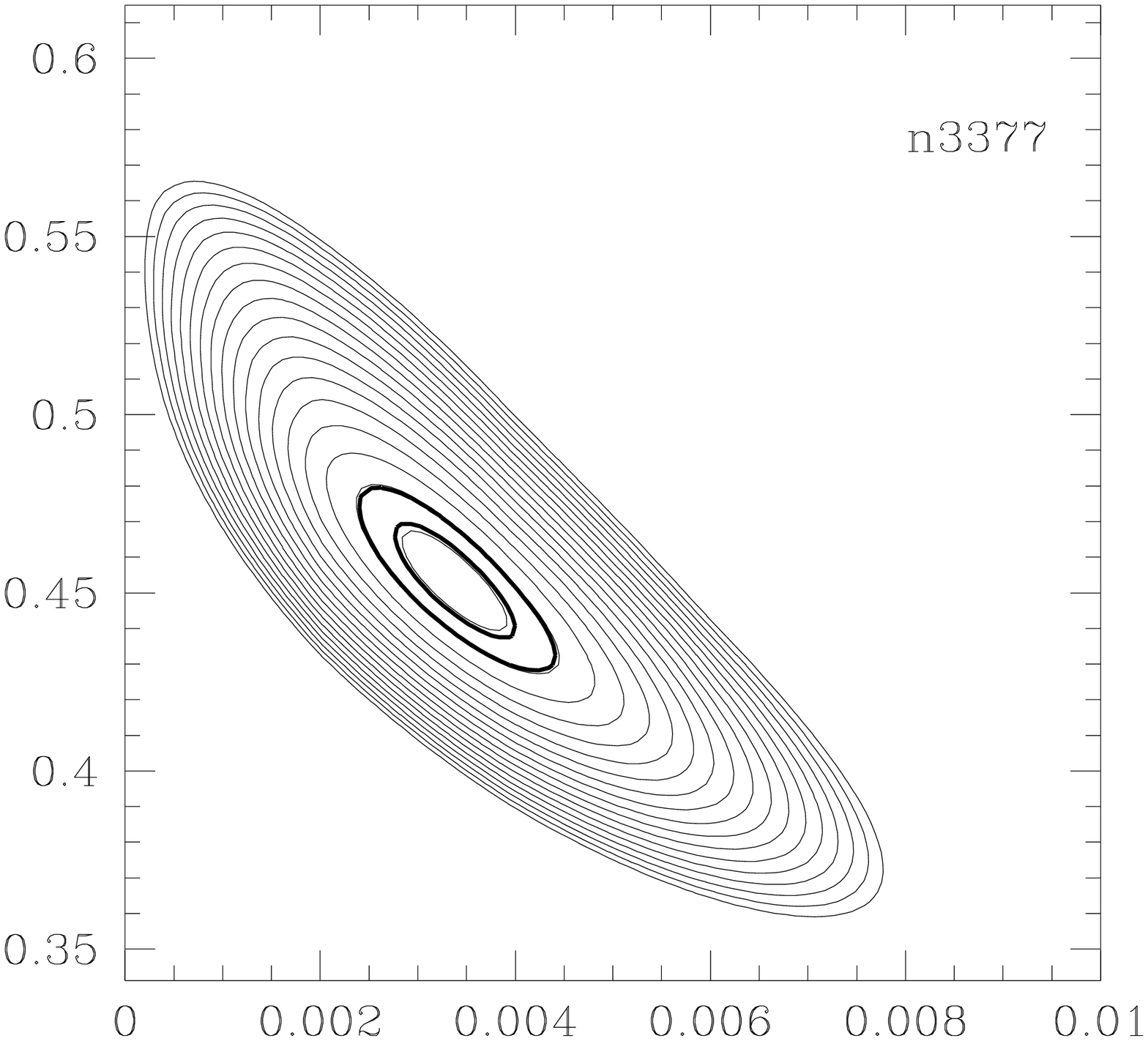,width=0.25\hsize}
\psfig{file=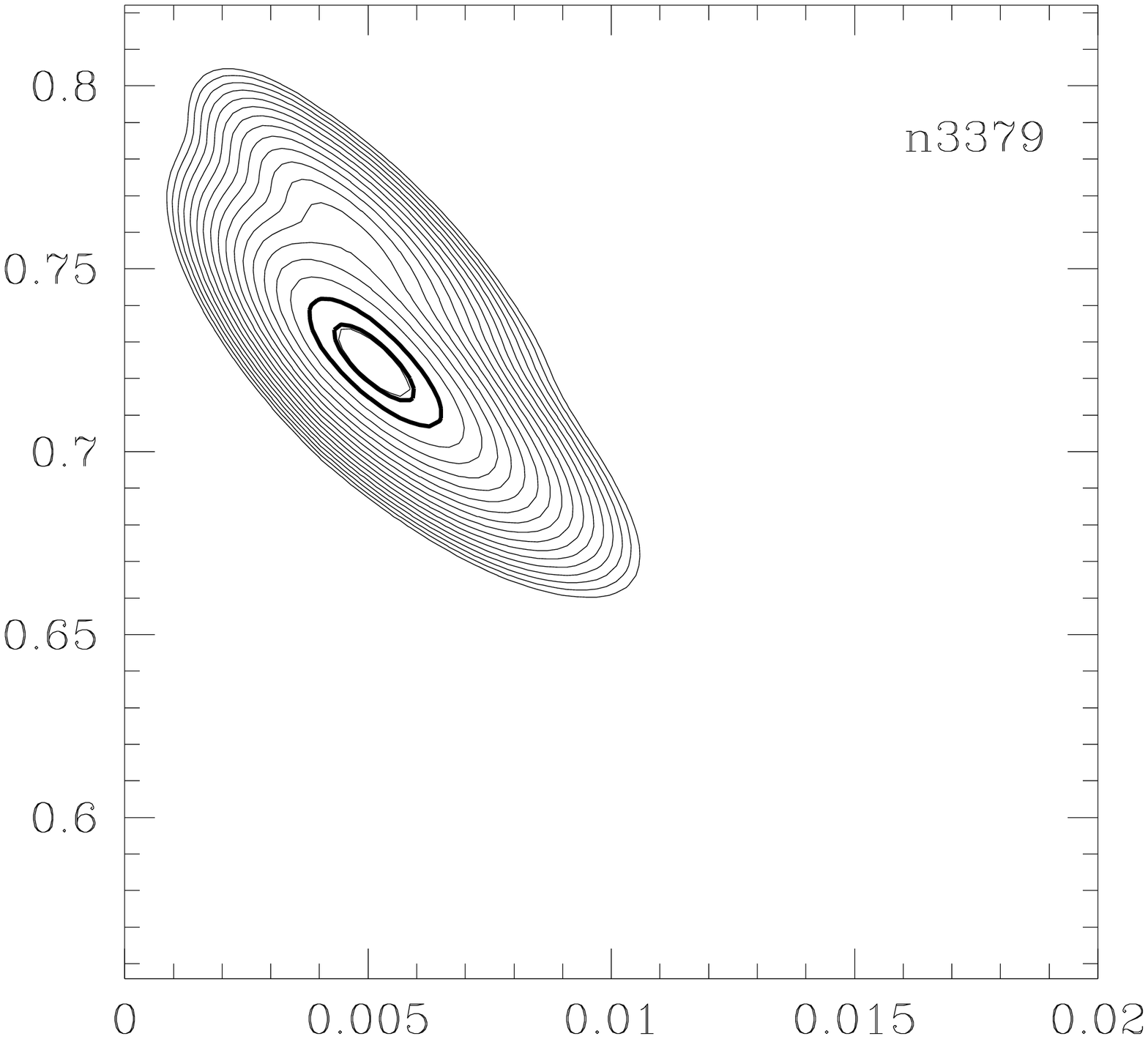,width=0.25\hsize}
\psfig{file=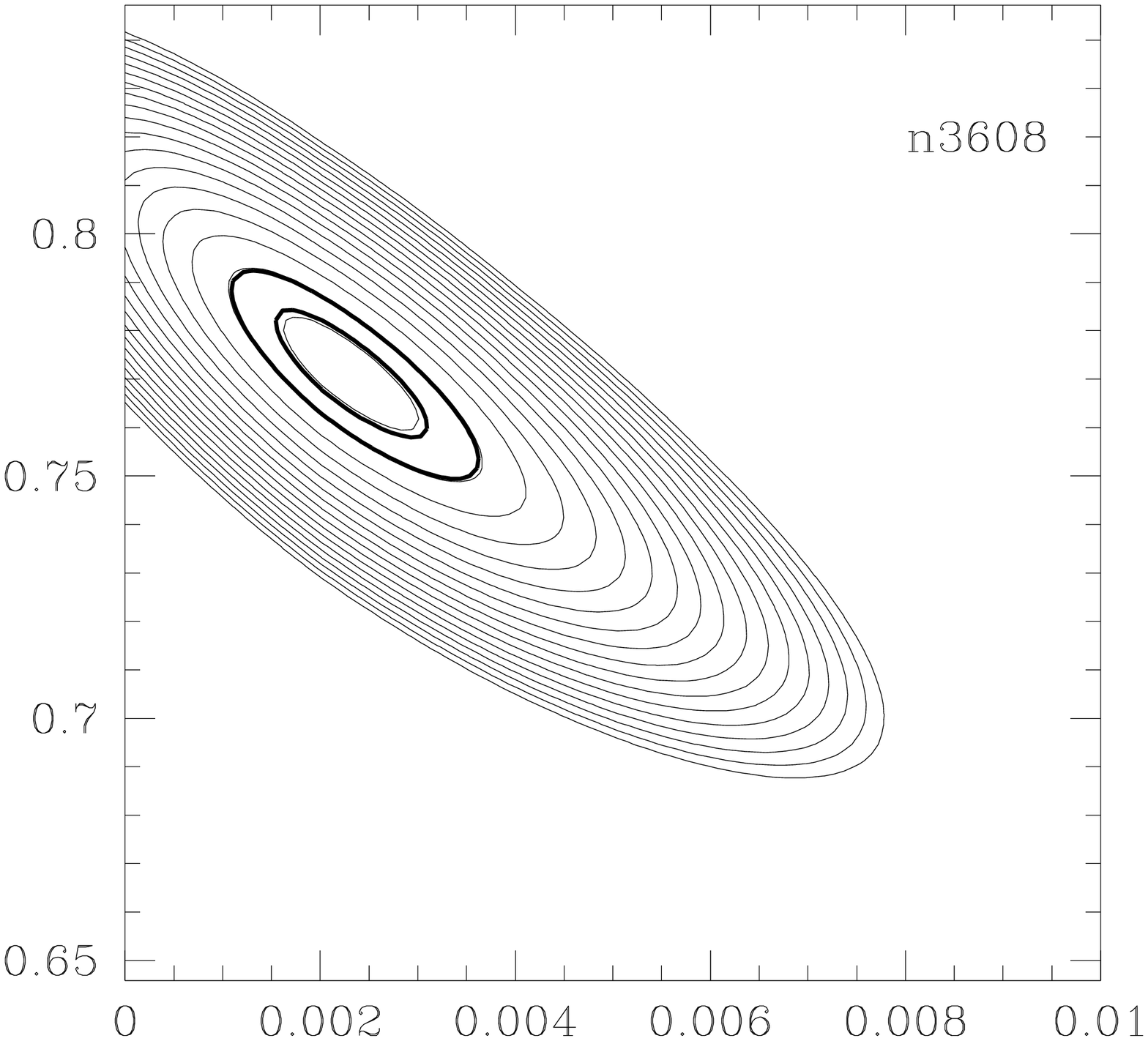,width=0.25\hsize}
}
\centerline{
\psfig{file=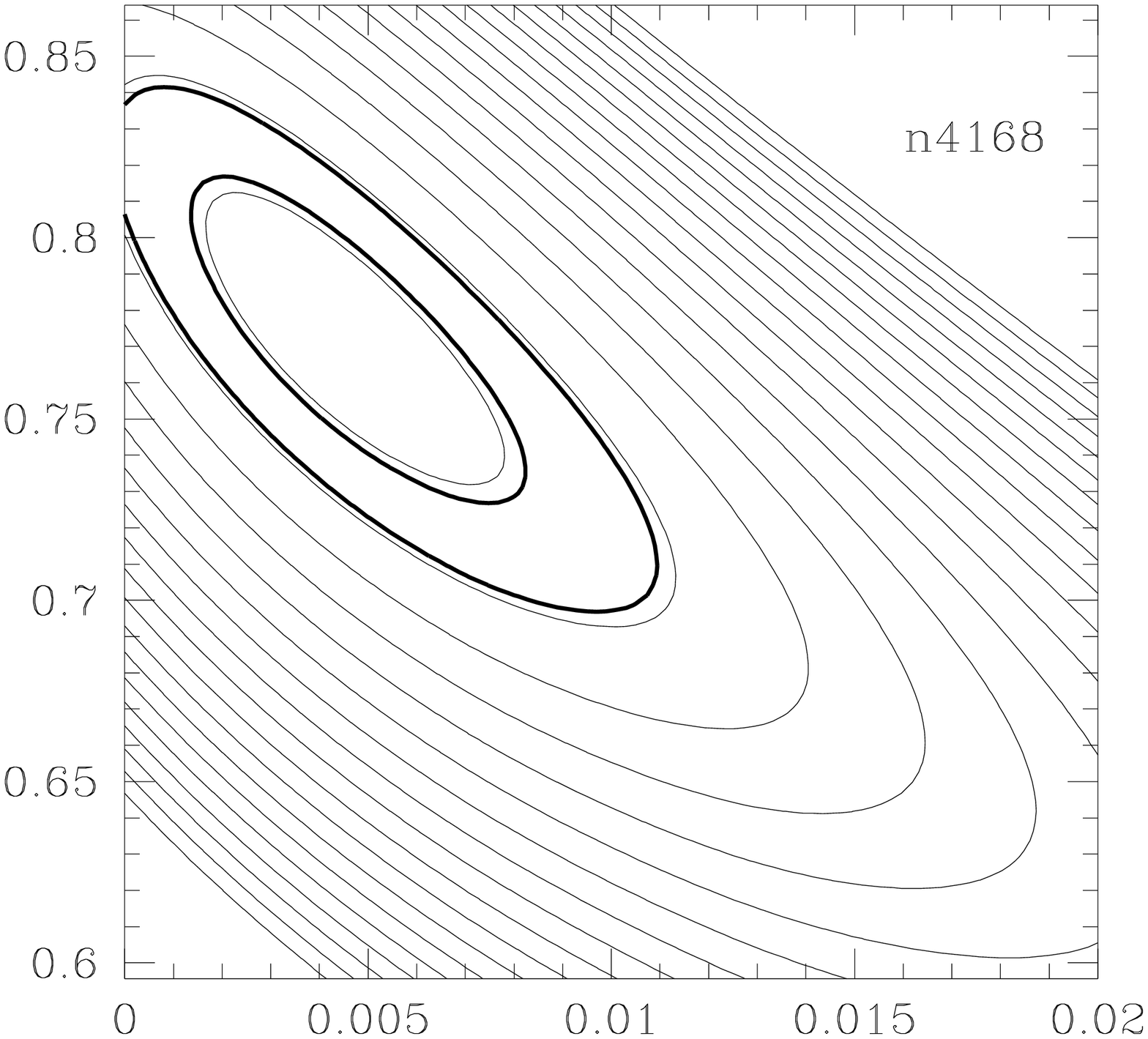,width=0.25\hsize}
\psfig{file=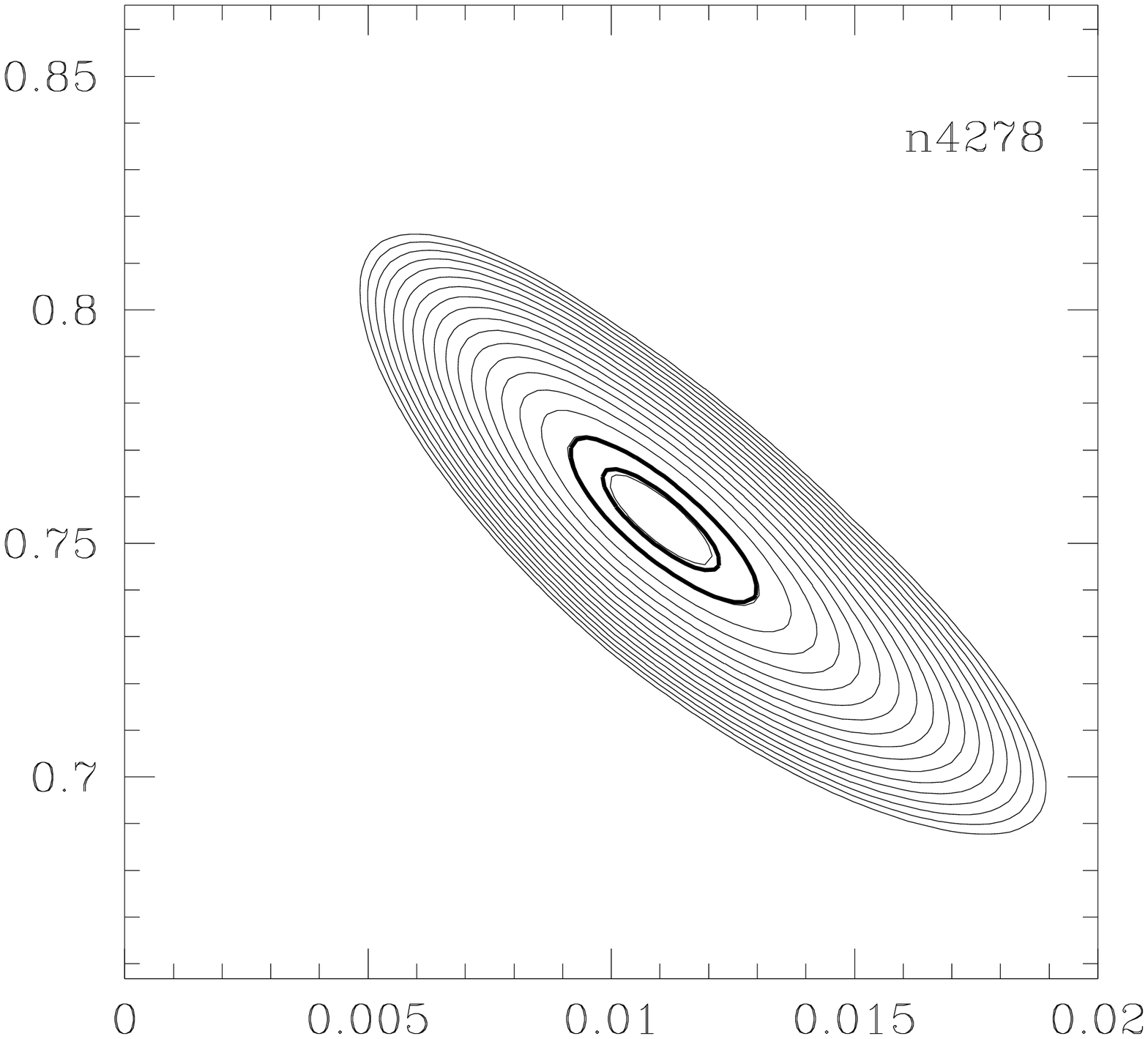,width=0.25\hsize}
\psfig{file=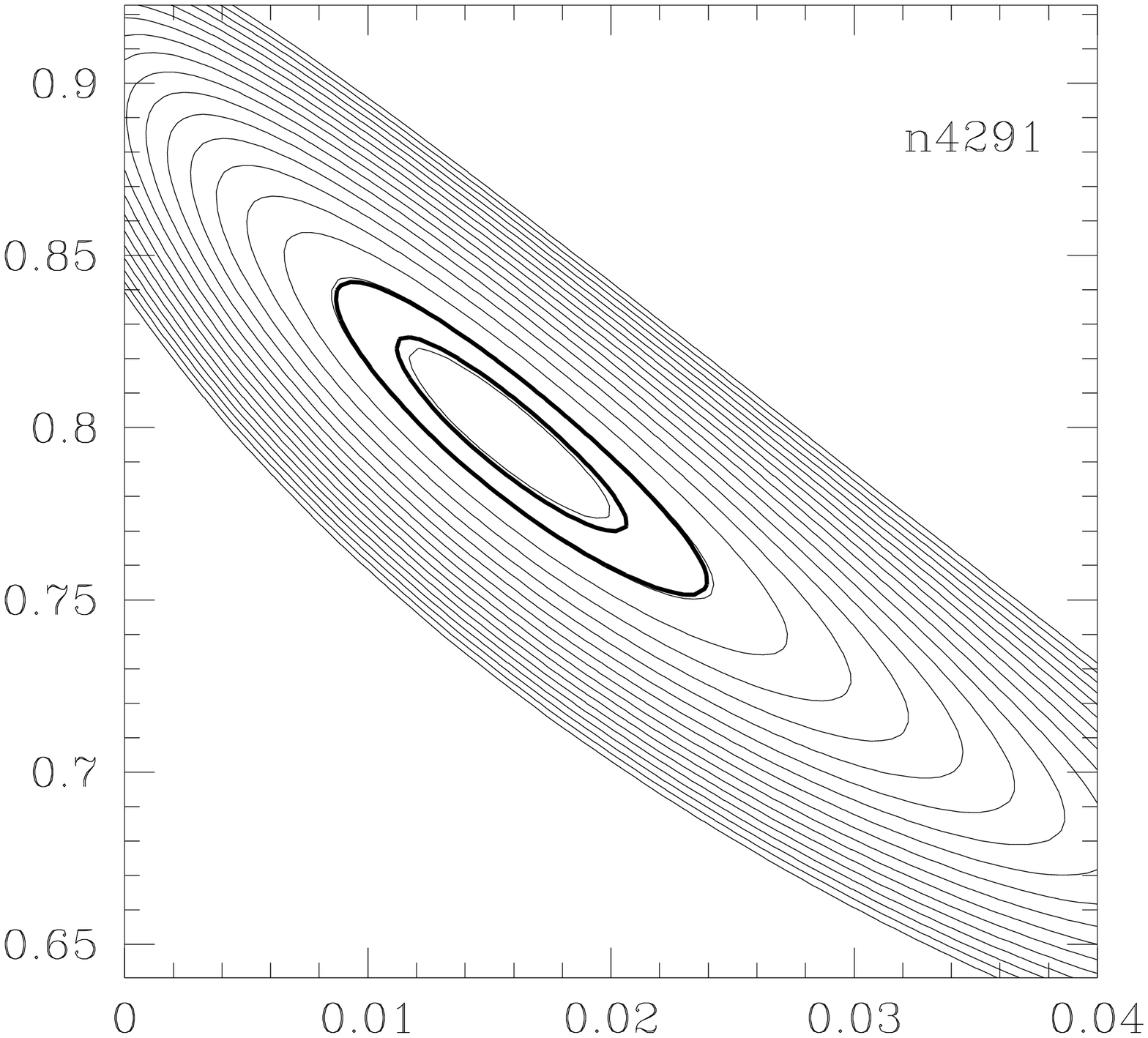,width=0.25\hsize}
\psfig{file=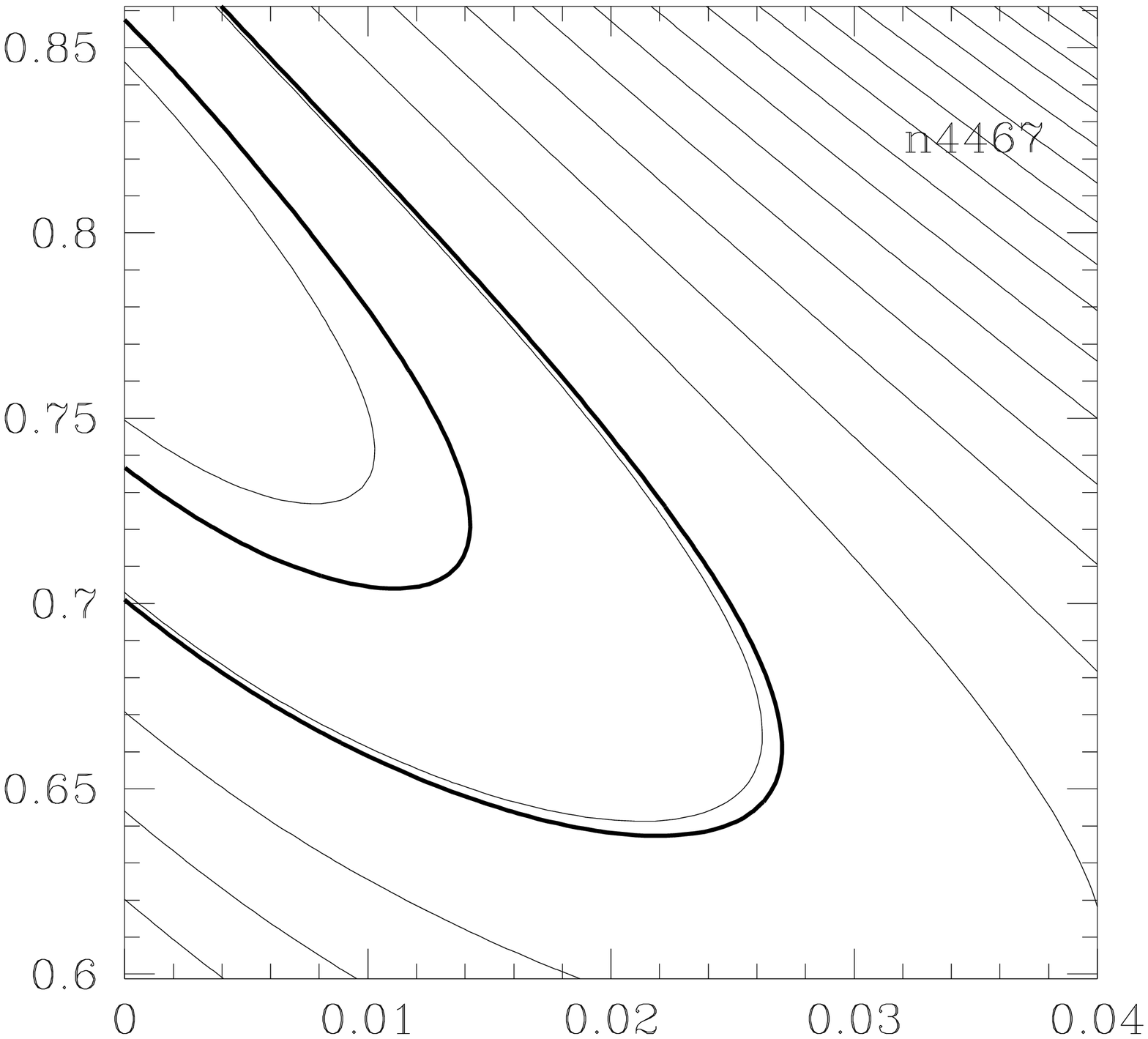,width=0.25\hsize}
}
}{\caption}
\figure{
\centerline{
\psfig{file=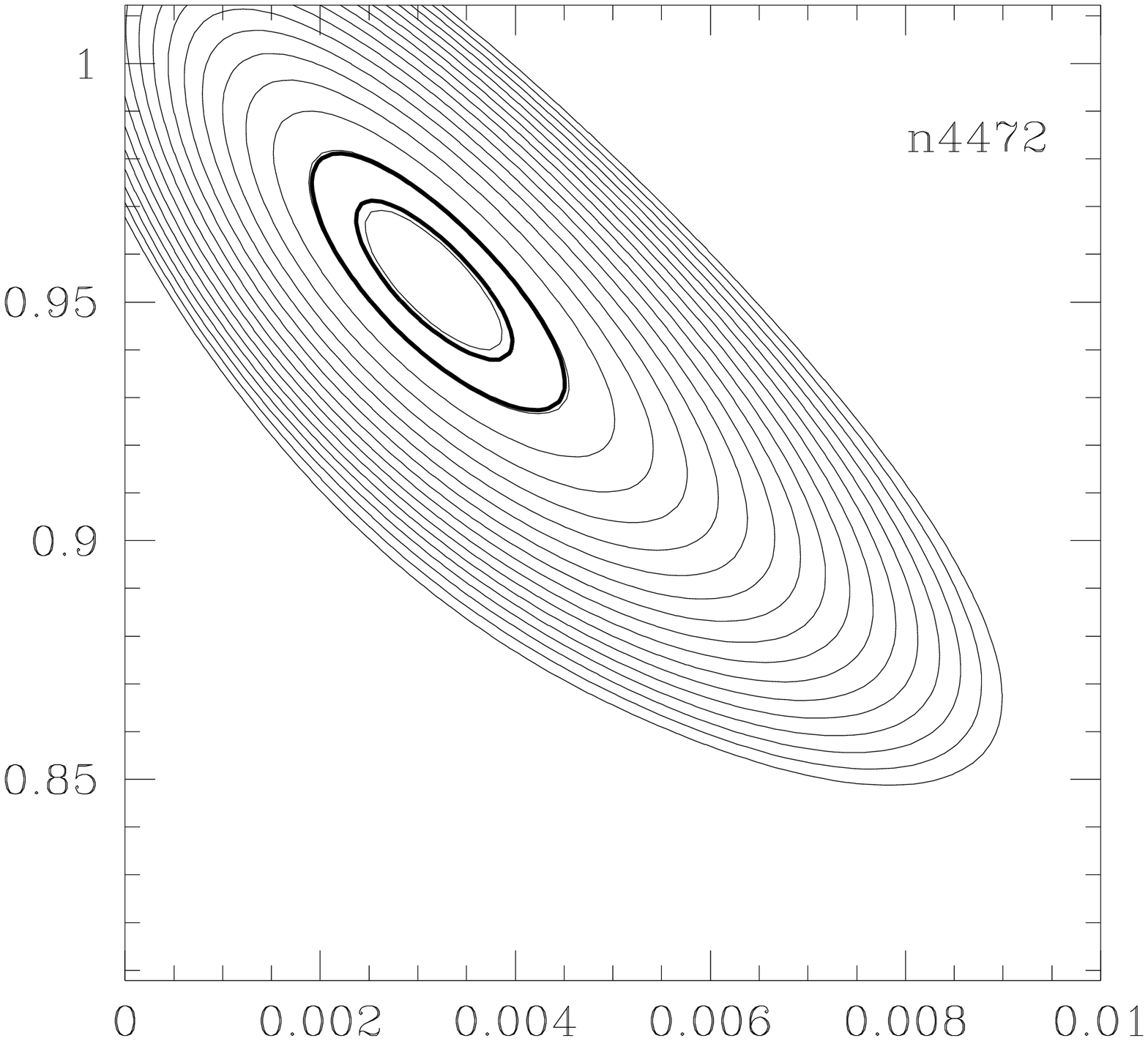,width=0.25\hsize}
\psfig{file=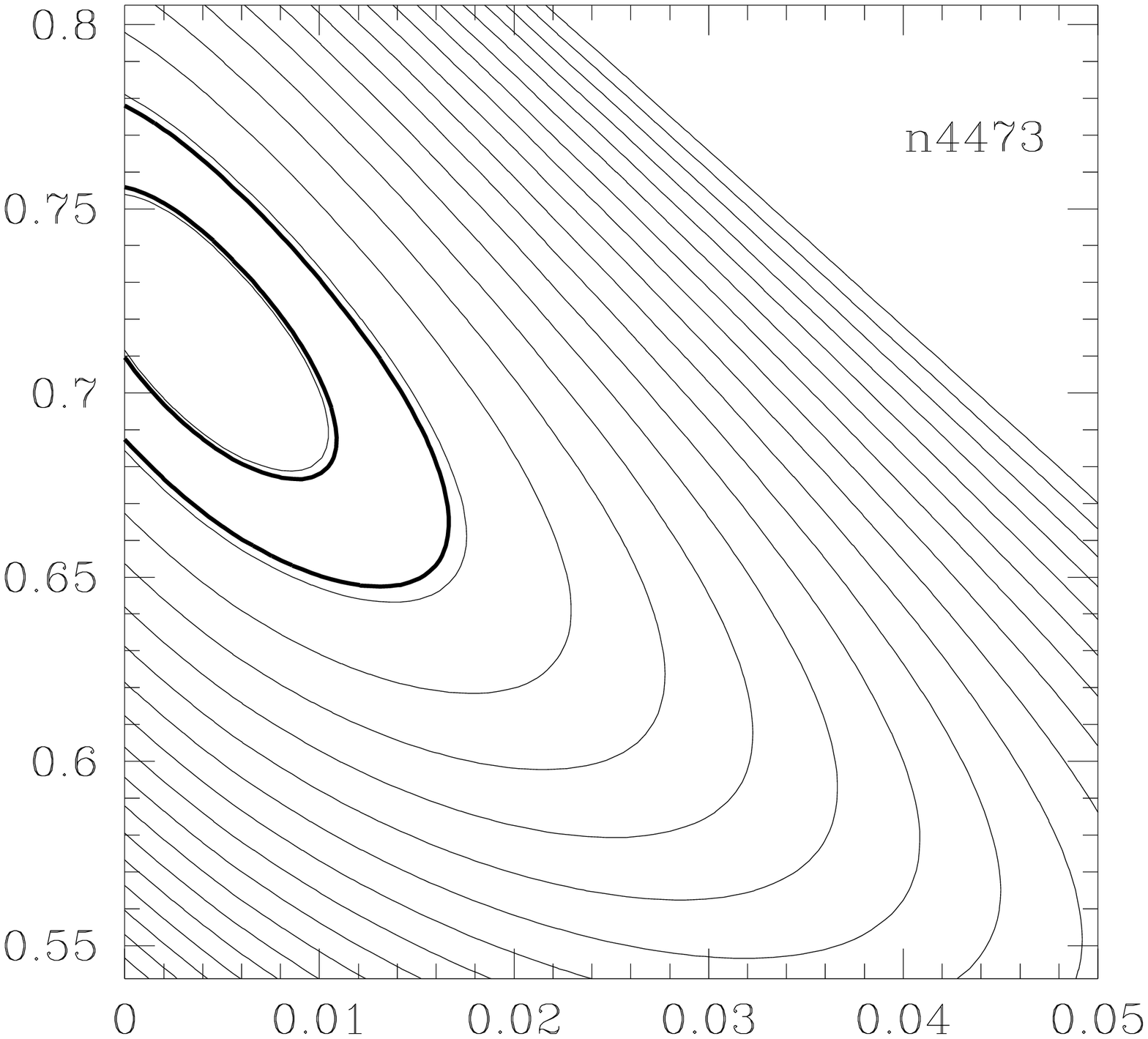,width=0.25\hsize}
\psfig{file=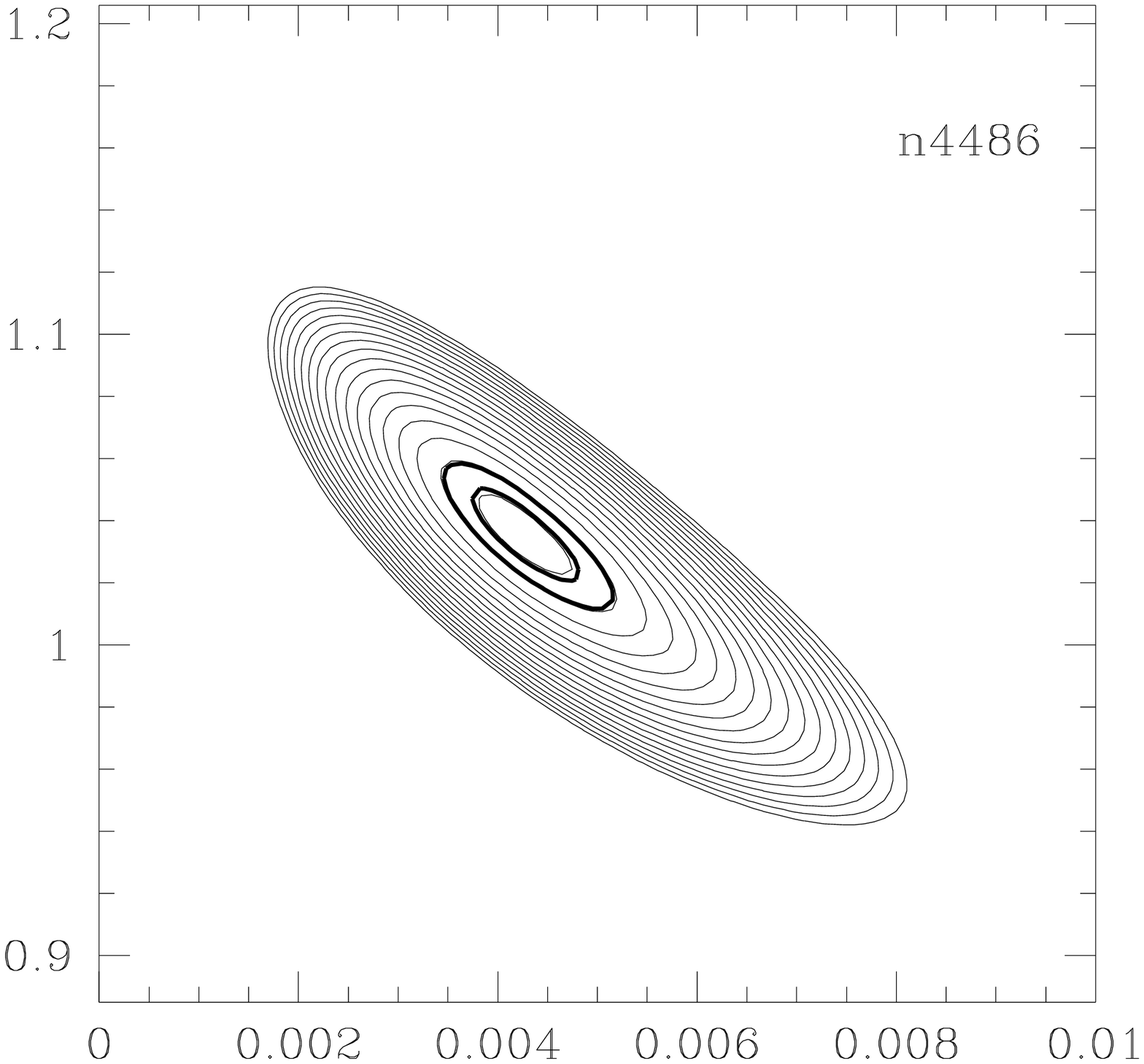,width=0.25\hsize}
\psfig{file=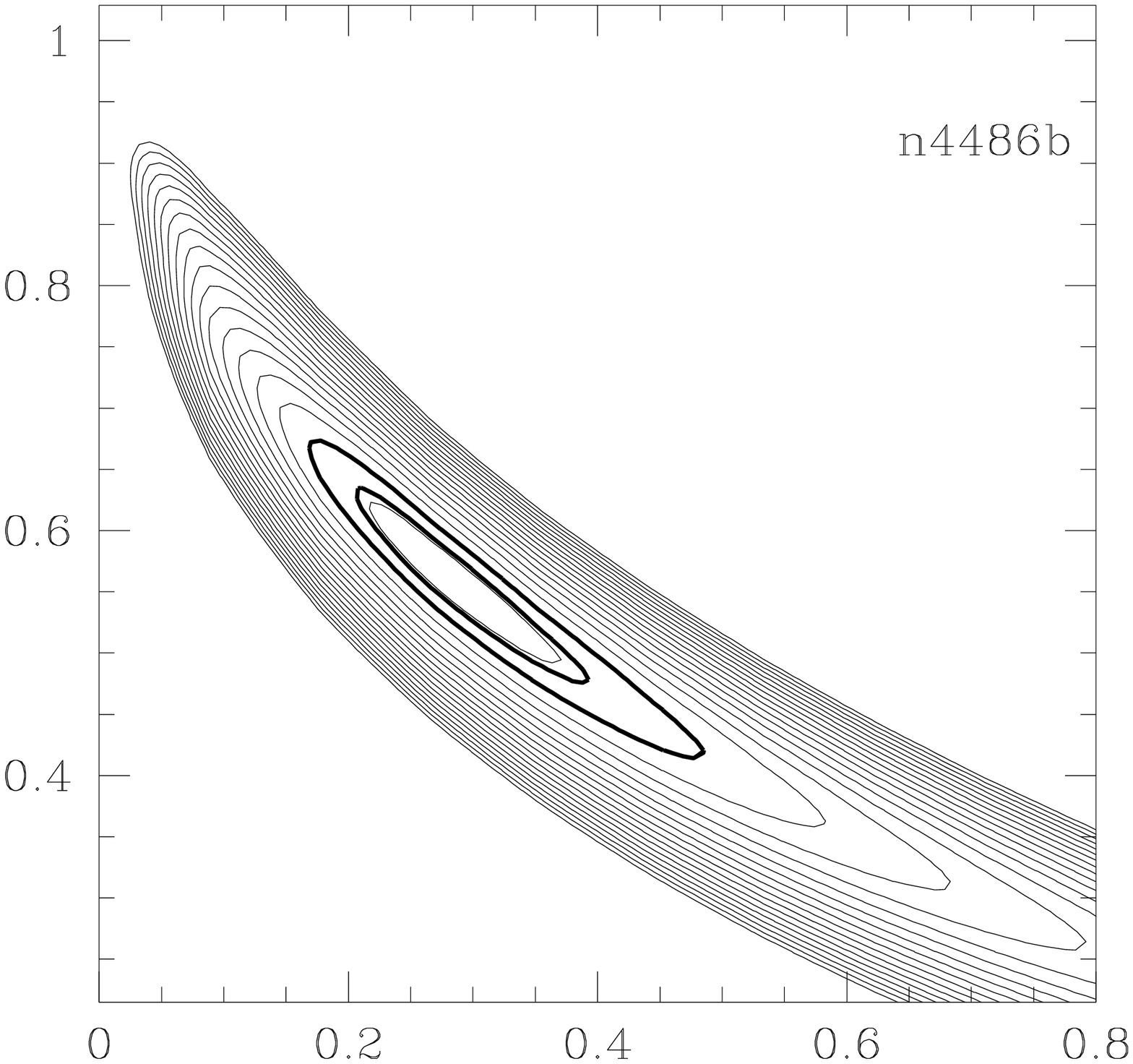,width=0.25\hsize}
}
\centerline{
\psfig{file=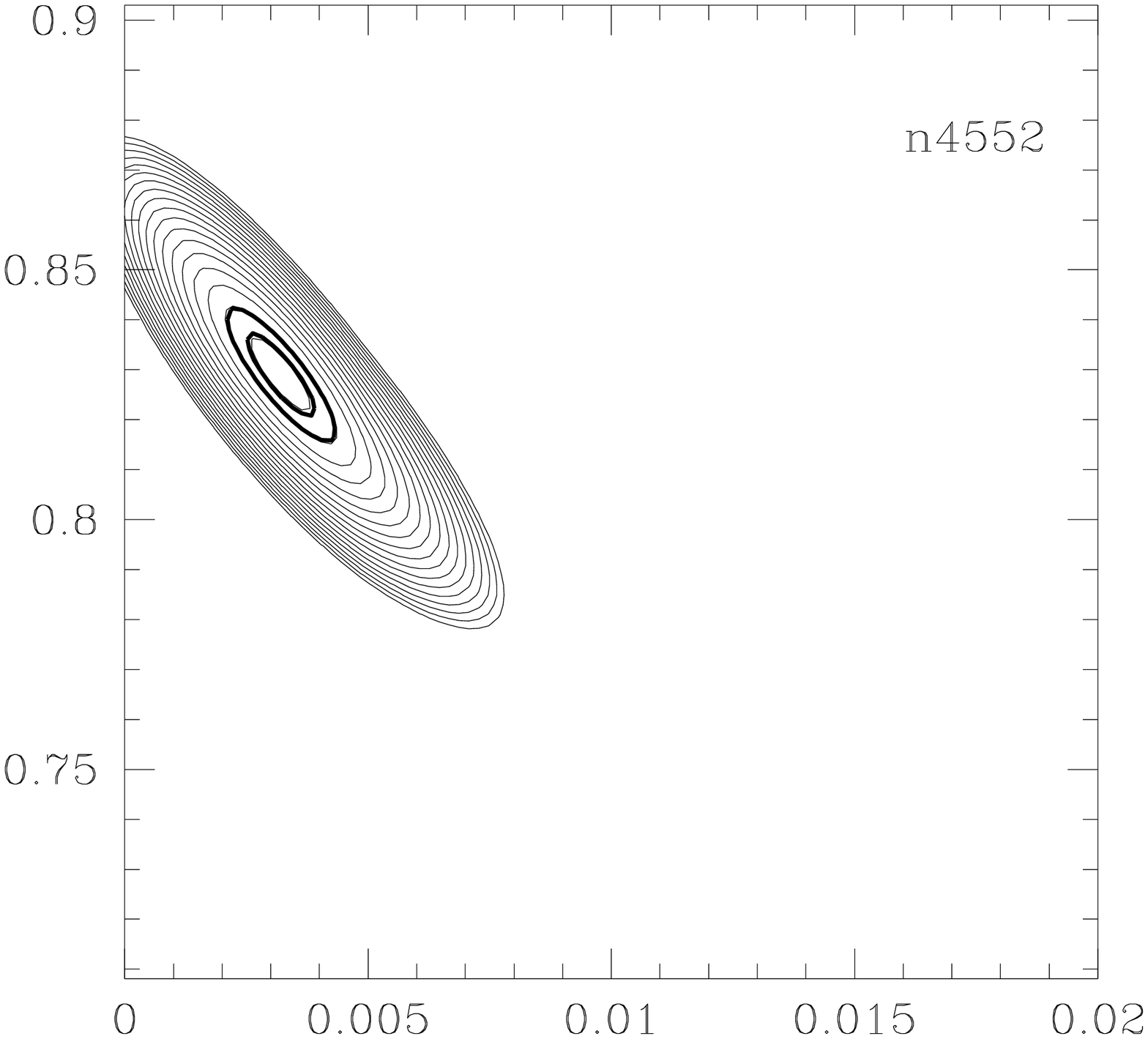,width=0.25\hsize}
\psfig{file=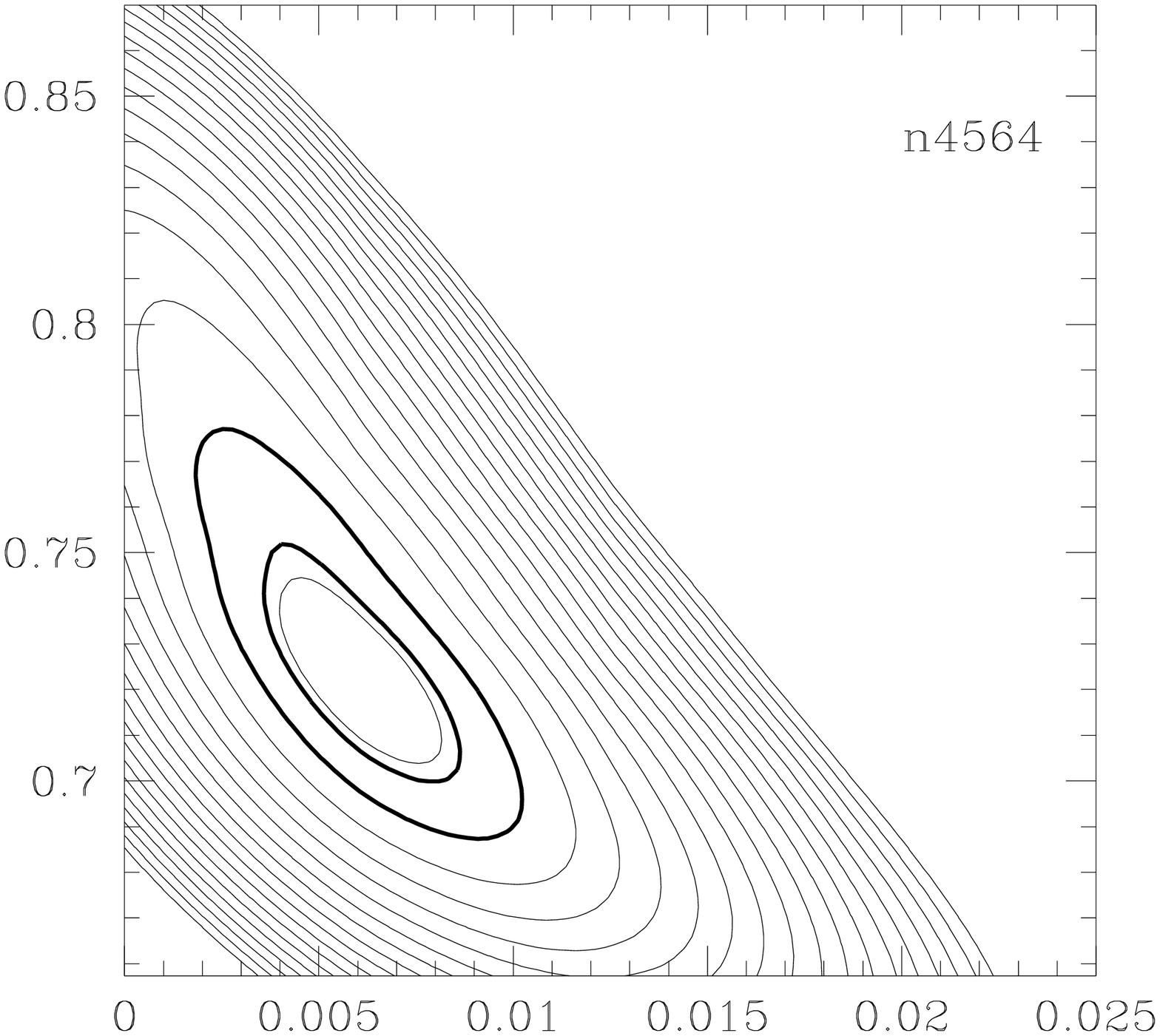,width=0.25\hsize}
\psfig{file=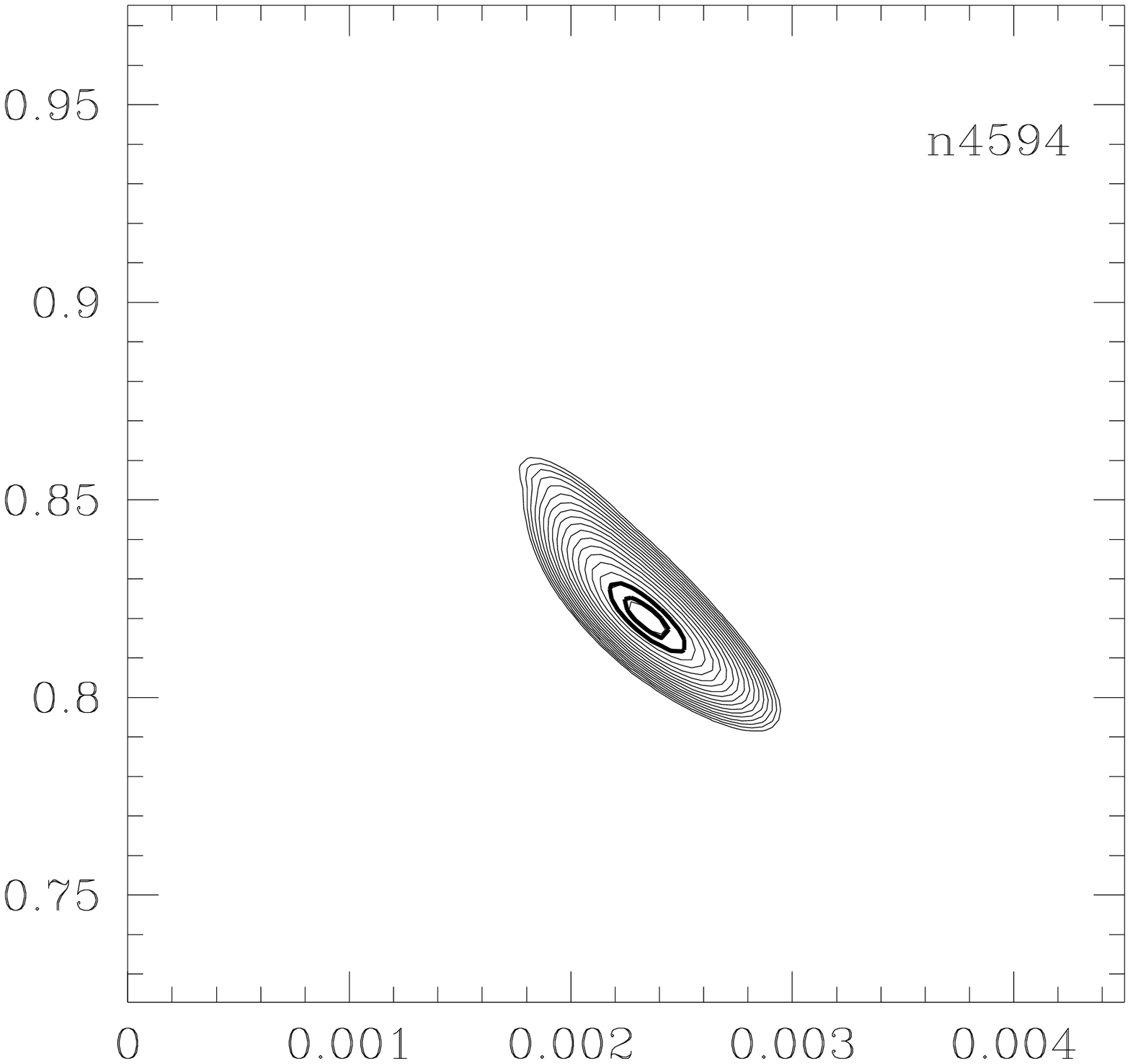,width=0.25\hsize}
\psfig{file=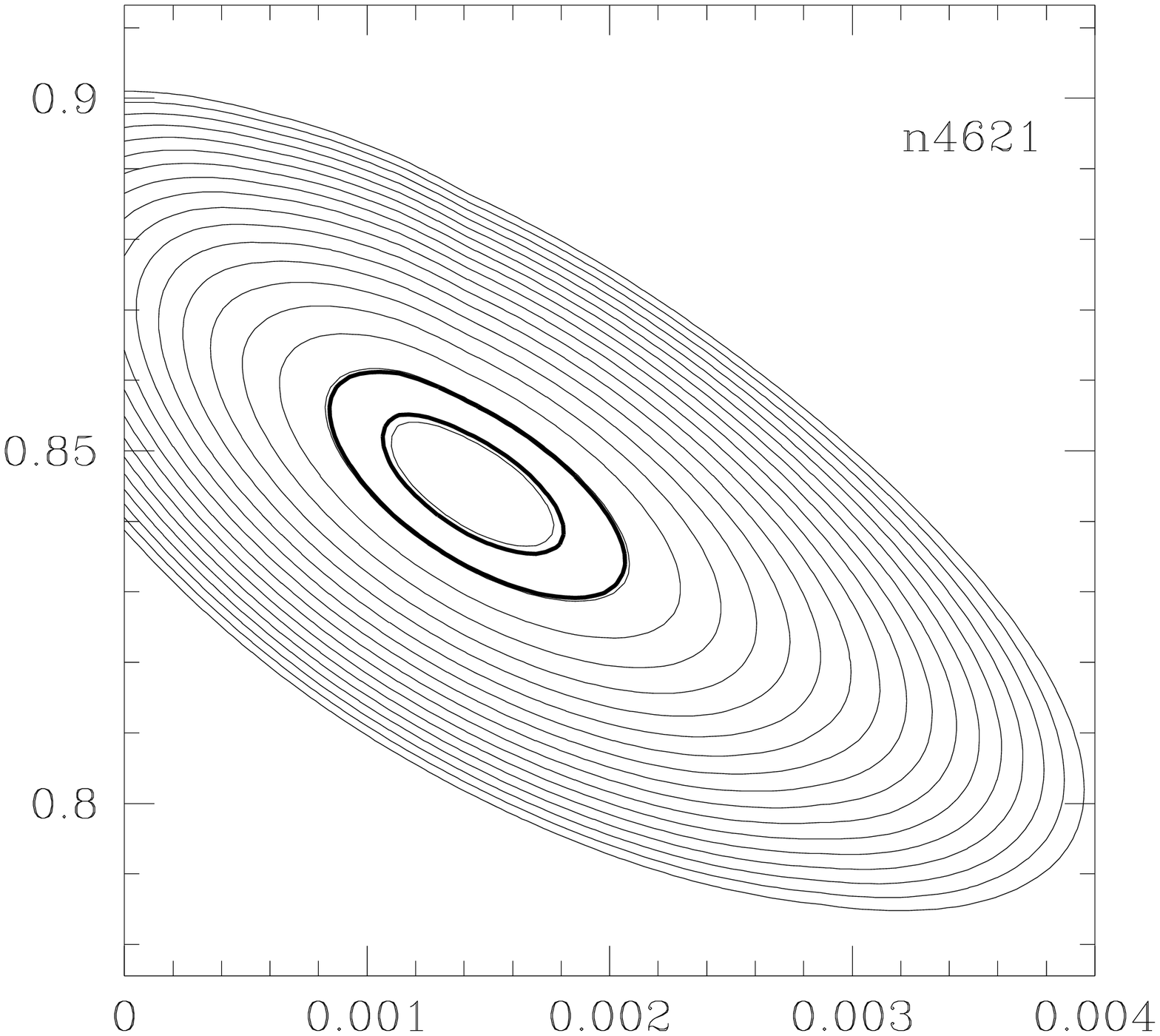,width=0.25\hsize}
}
\centerline{
\psfig{file=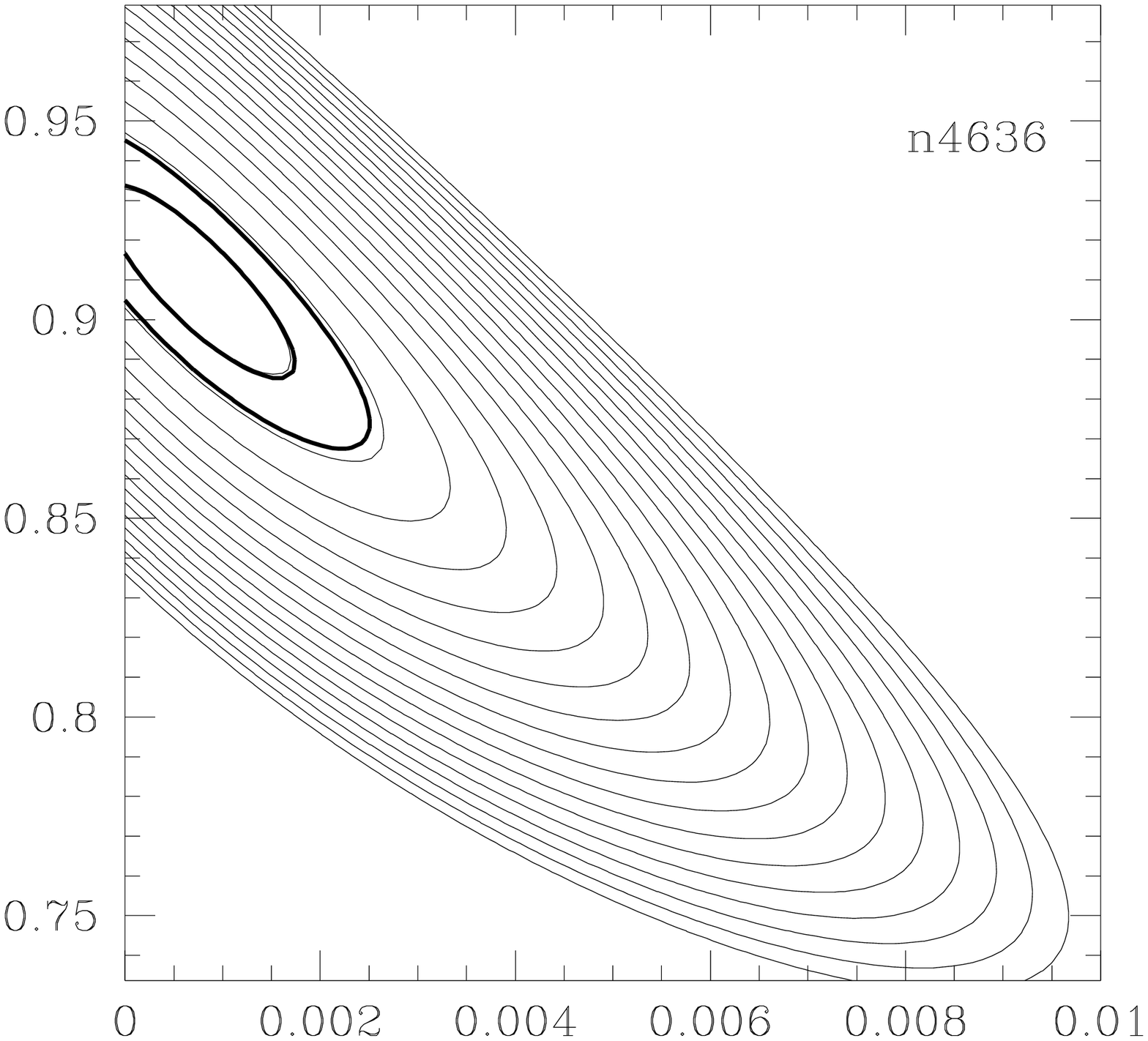,width=0.25\hsize}
\psfig{file=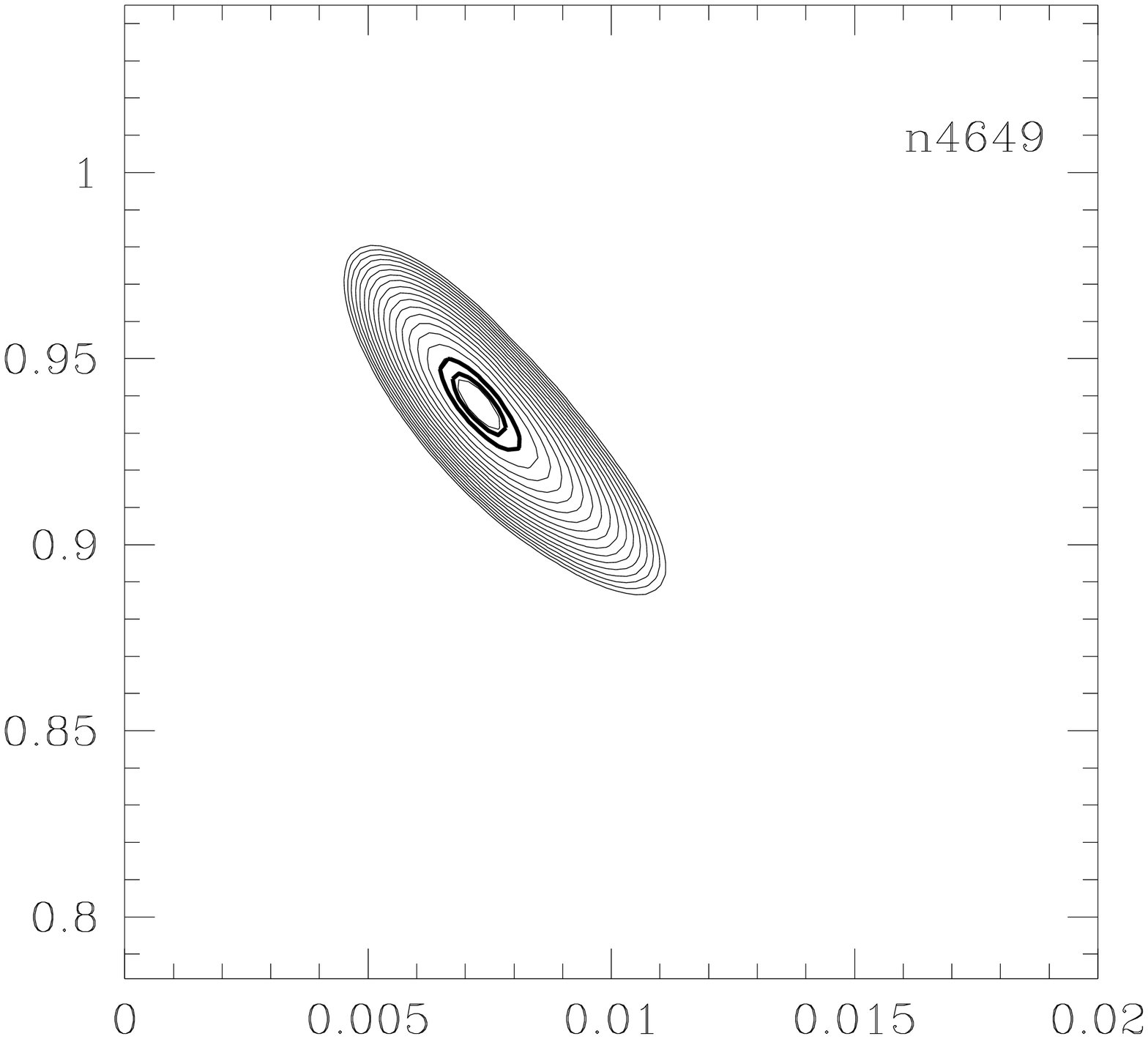,width=0.25\hsize}
\psfig{file=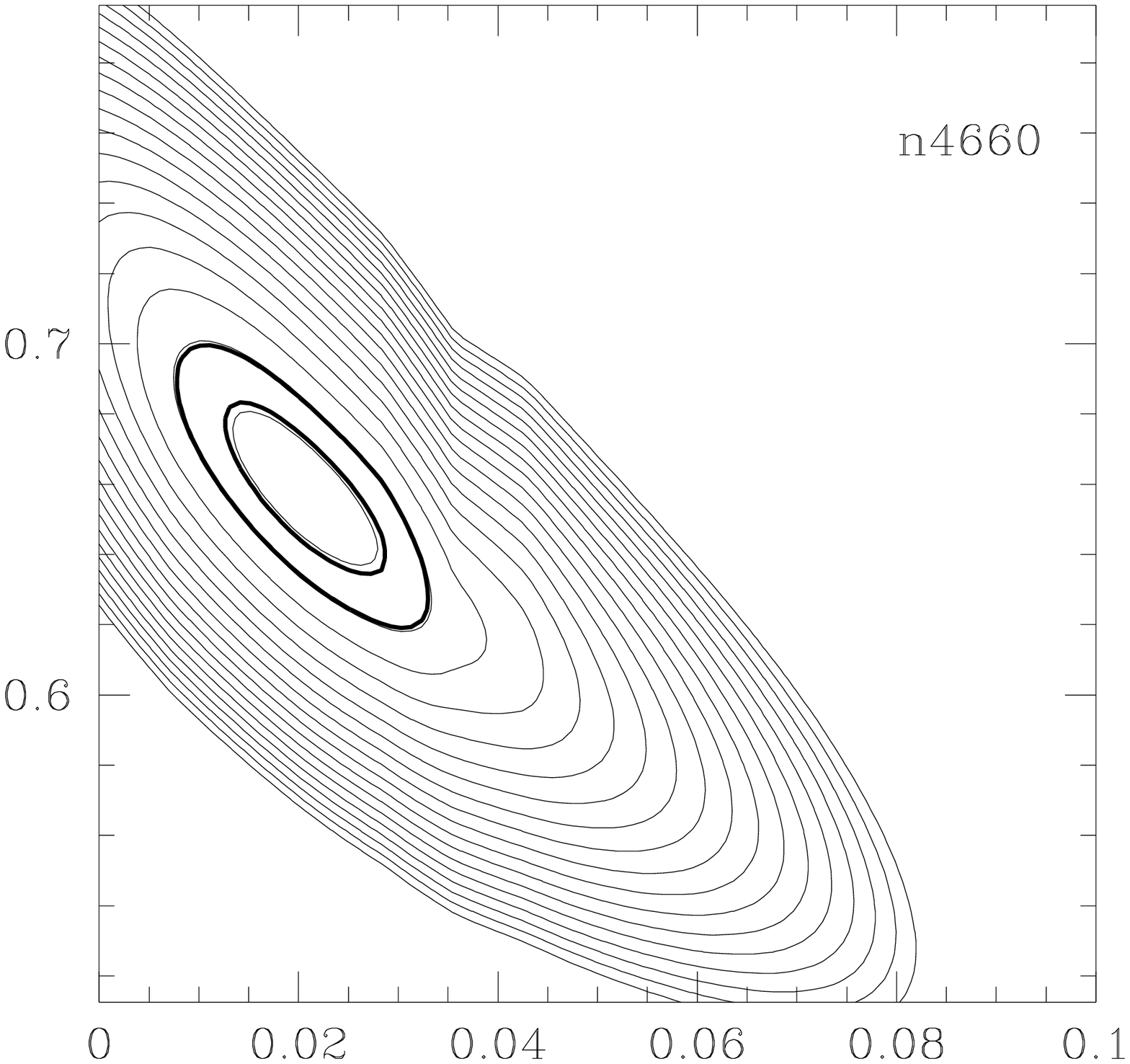,width=0.25\hsize}
\psfig{file=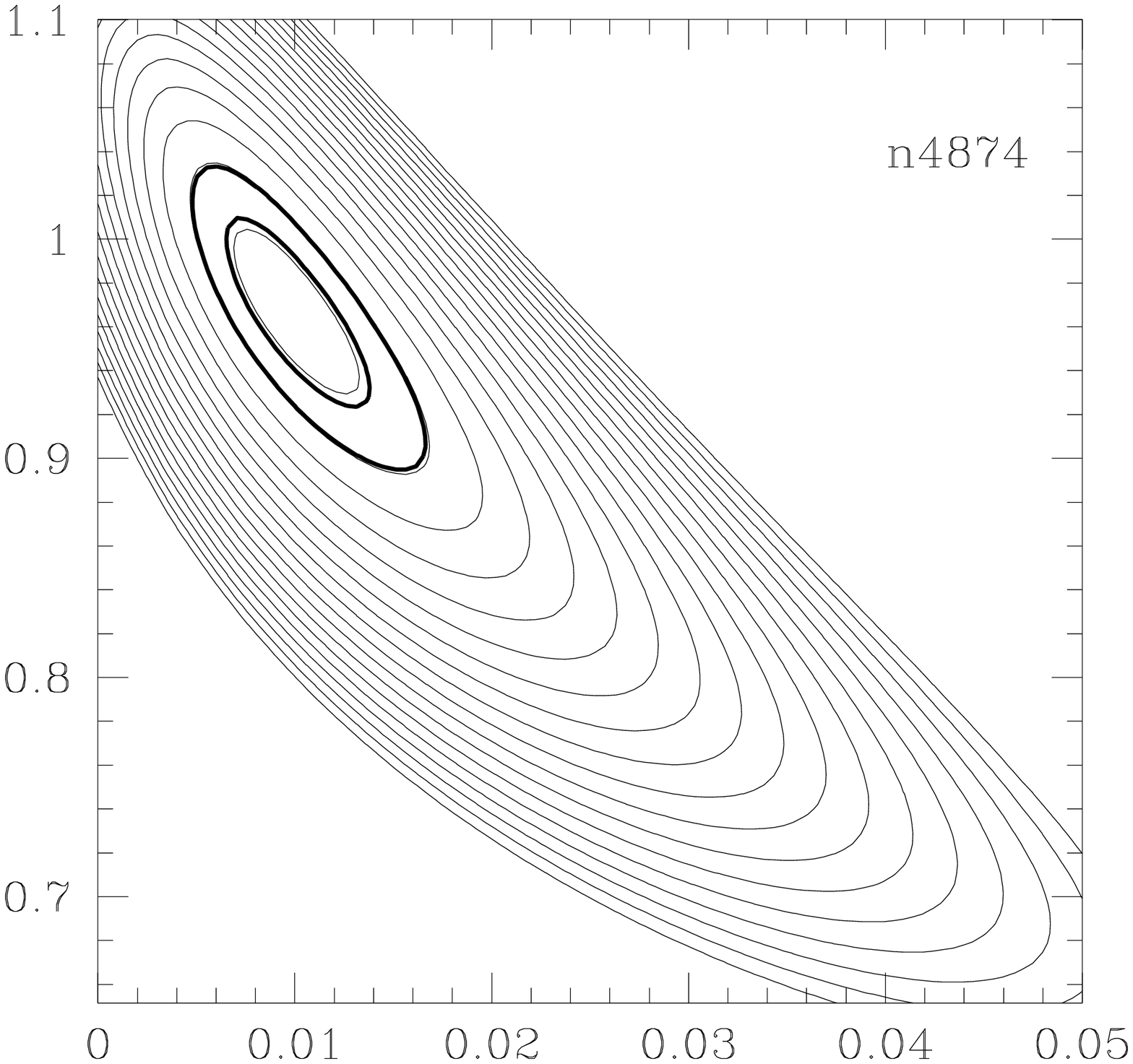,width=0.25\hsize}
}
\centerline{
\psfig{file=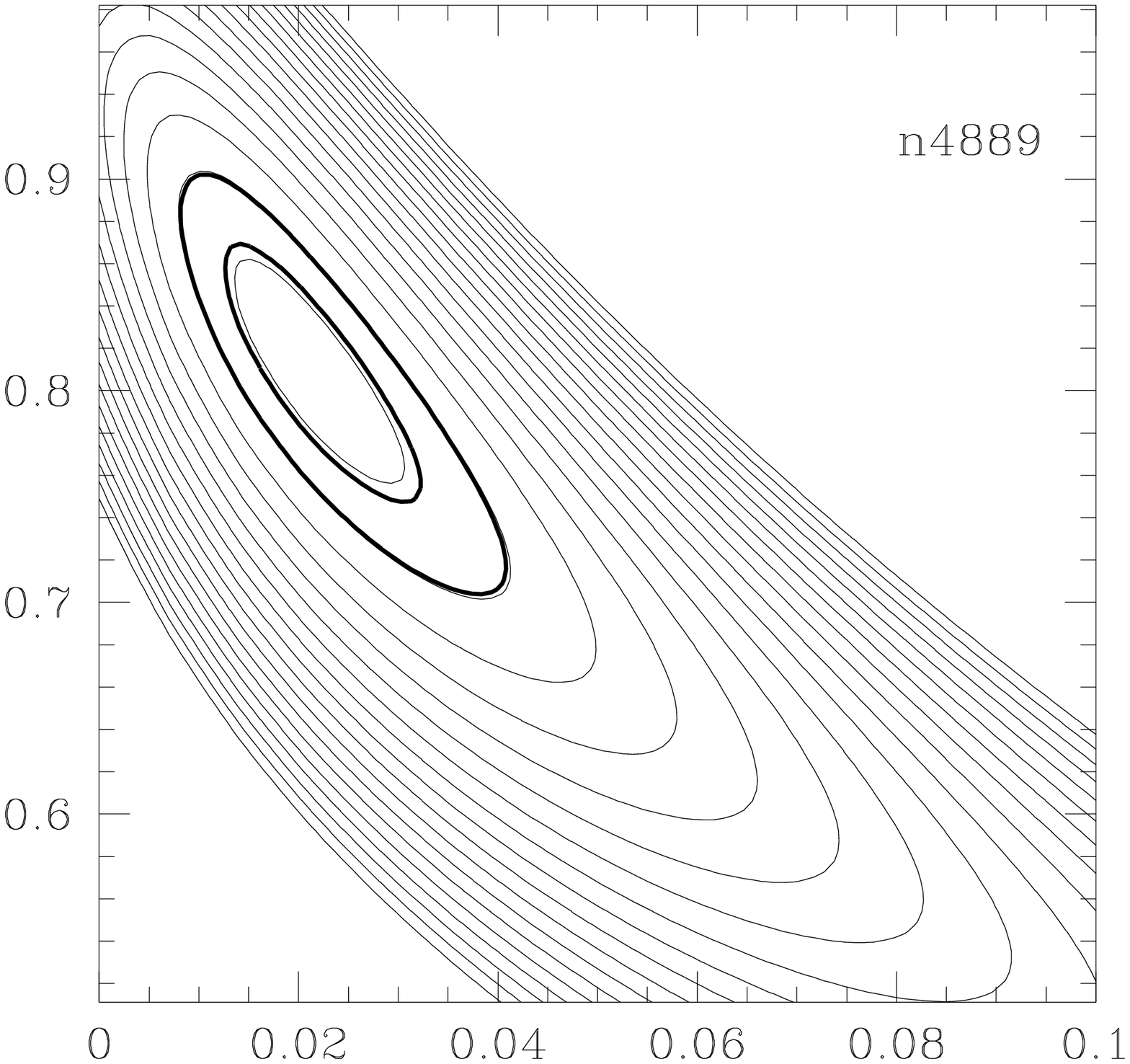,width=0.25\hsize}
\psfig{file=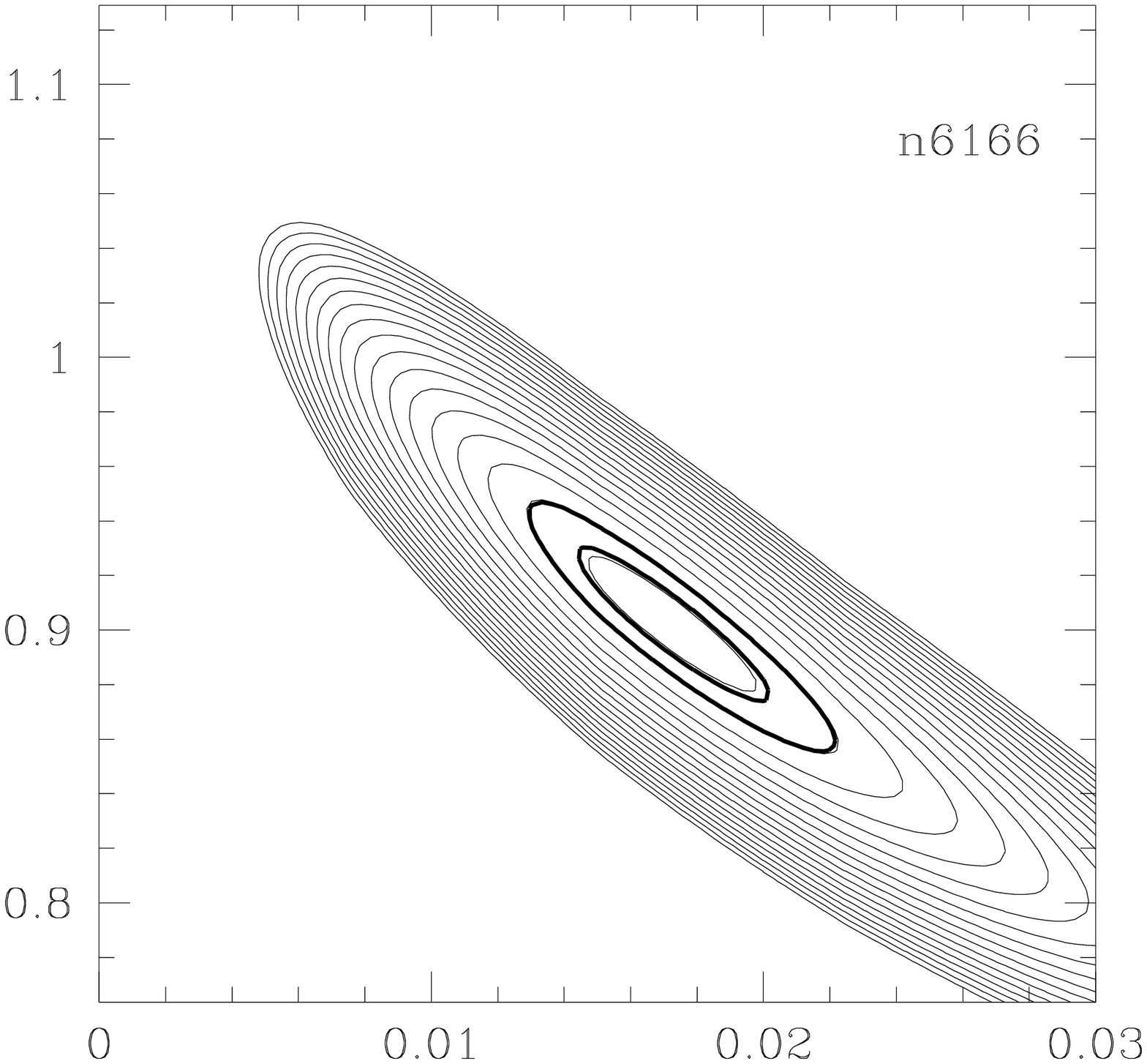,width=0.25\hsize}
\psfig{file=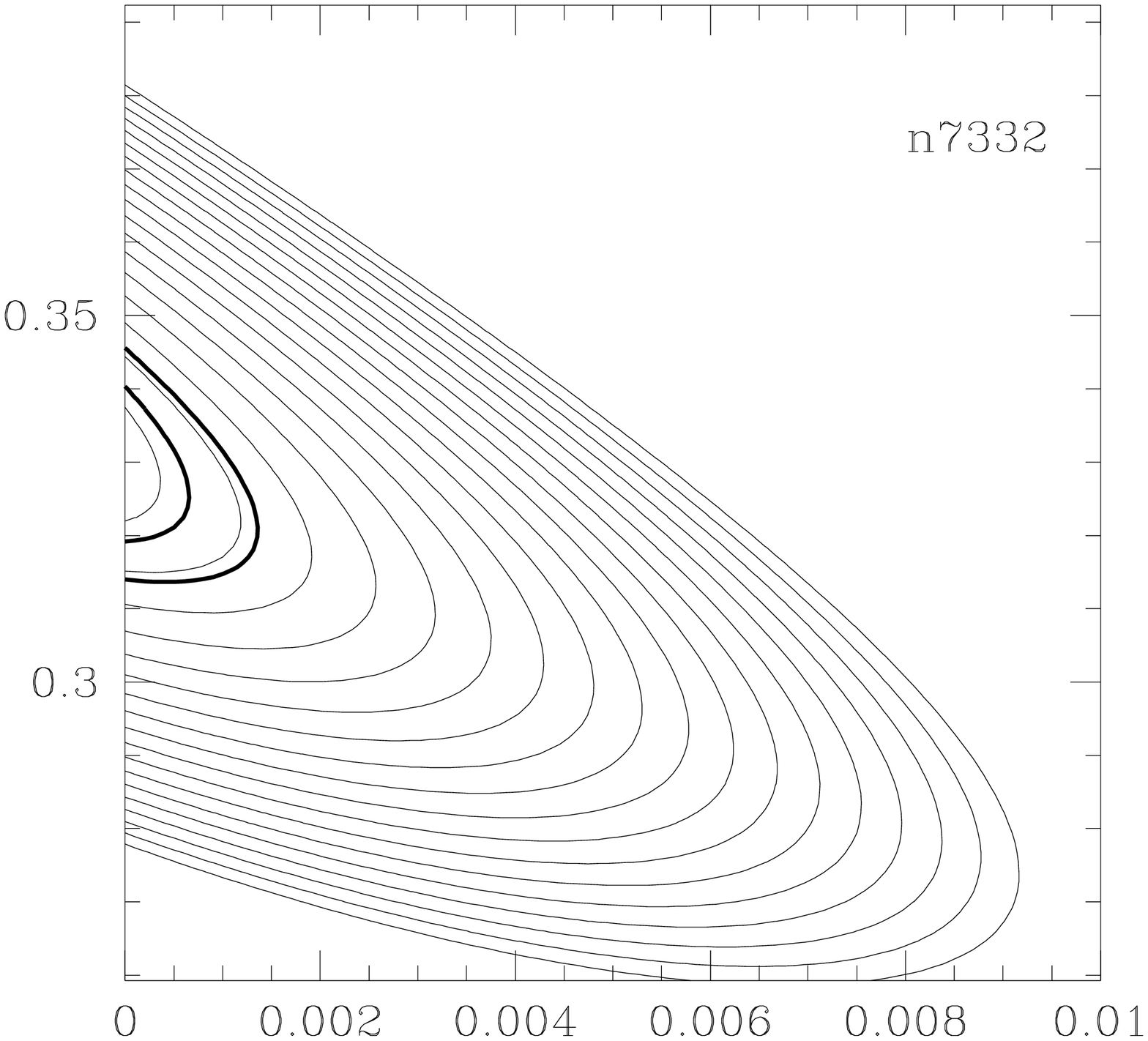,width=0.25\hsize}
\psfig{file=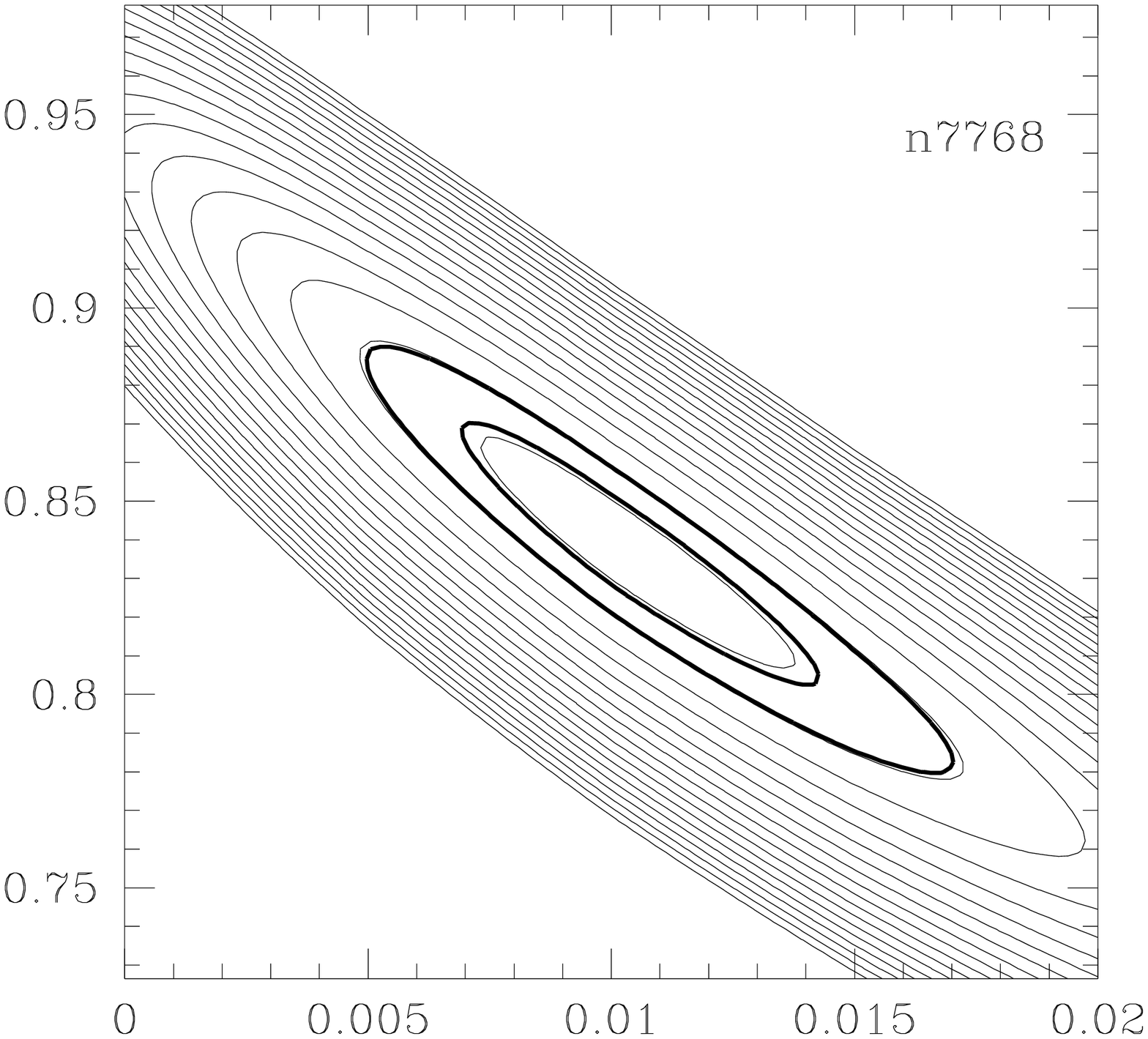,width=0.25\hsize}
}
}{\captioncont}

\def\caption{%
{\bf Figure \ref{figgood}(b).} As for Figure~\ref{figbad}(b), but for
the galaxies that our models describe well.  We also plot crosses to
show the {\it spatially binned}, seeing-convolved model predictions
for those cases where this quantity differs significantly from the
unbinned model predictions (described by the curves).}
\def\captioncont{{\bf Figure \ref{figgood}(b)}...continued.}
\figure{
\centerline{
\psfig{file=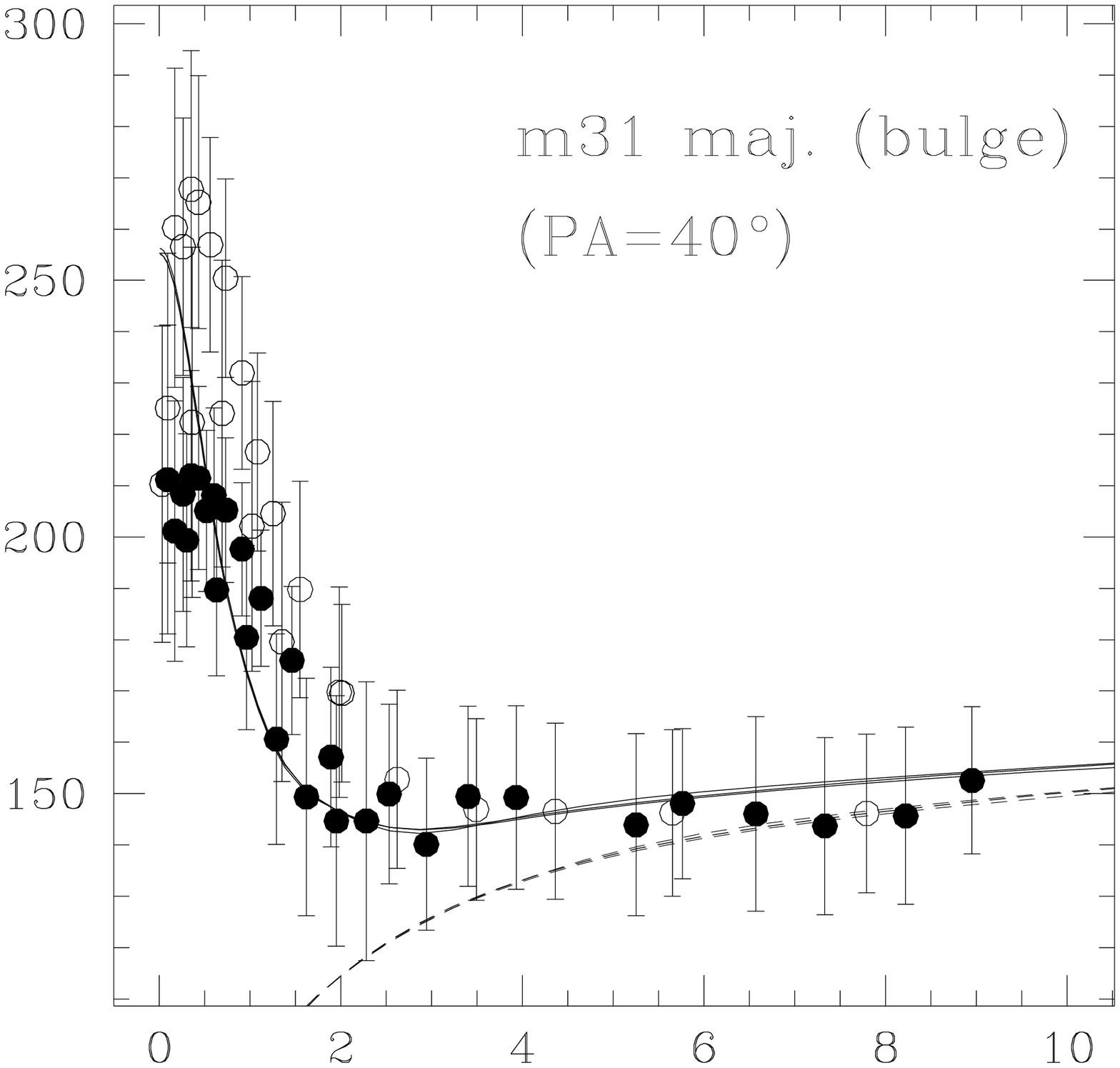,width=0.25\hsize}
\psfig{file=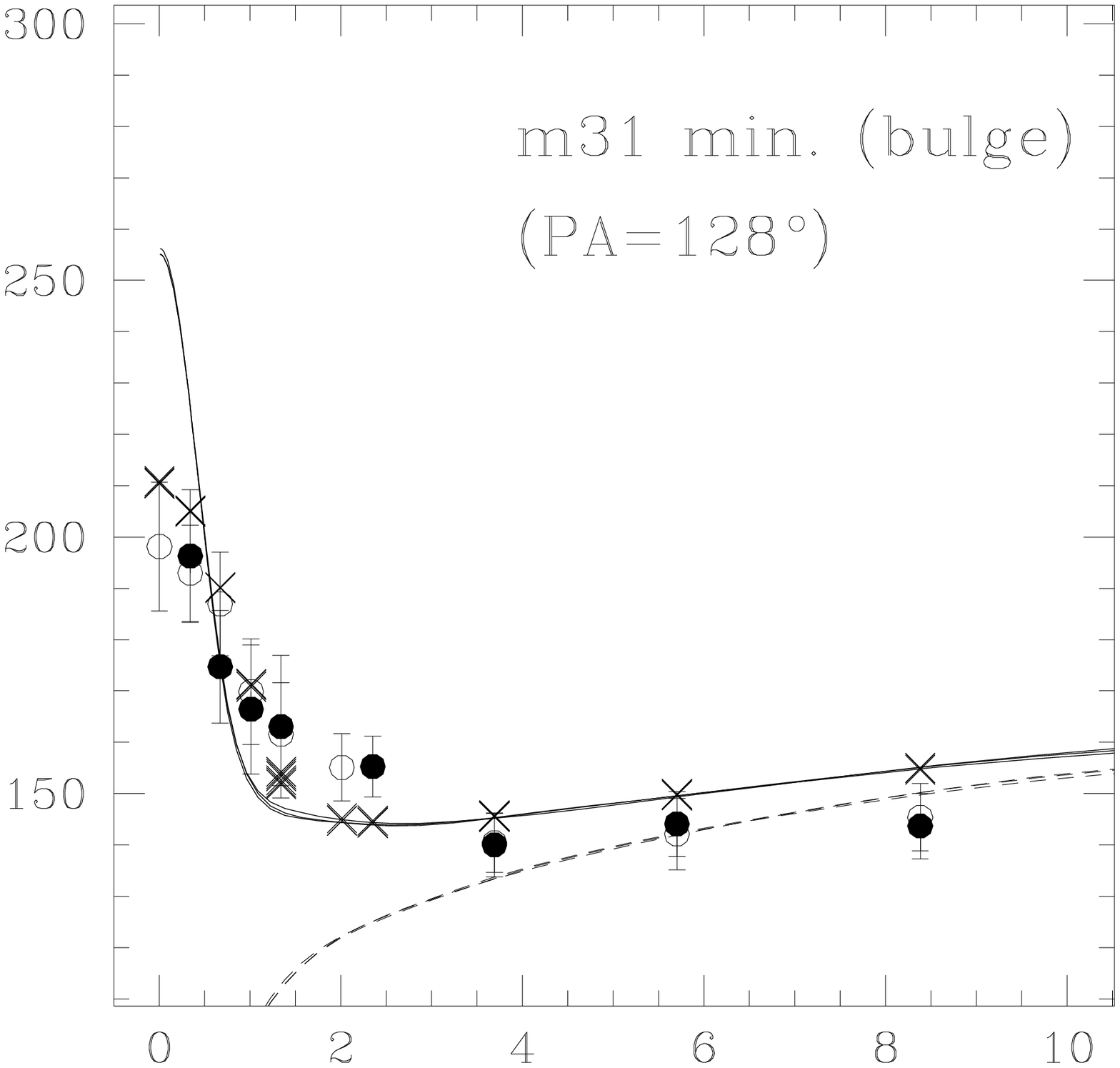,width=0.25\hsize}
\psfig{file=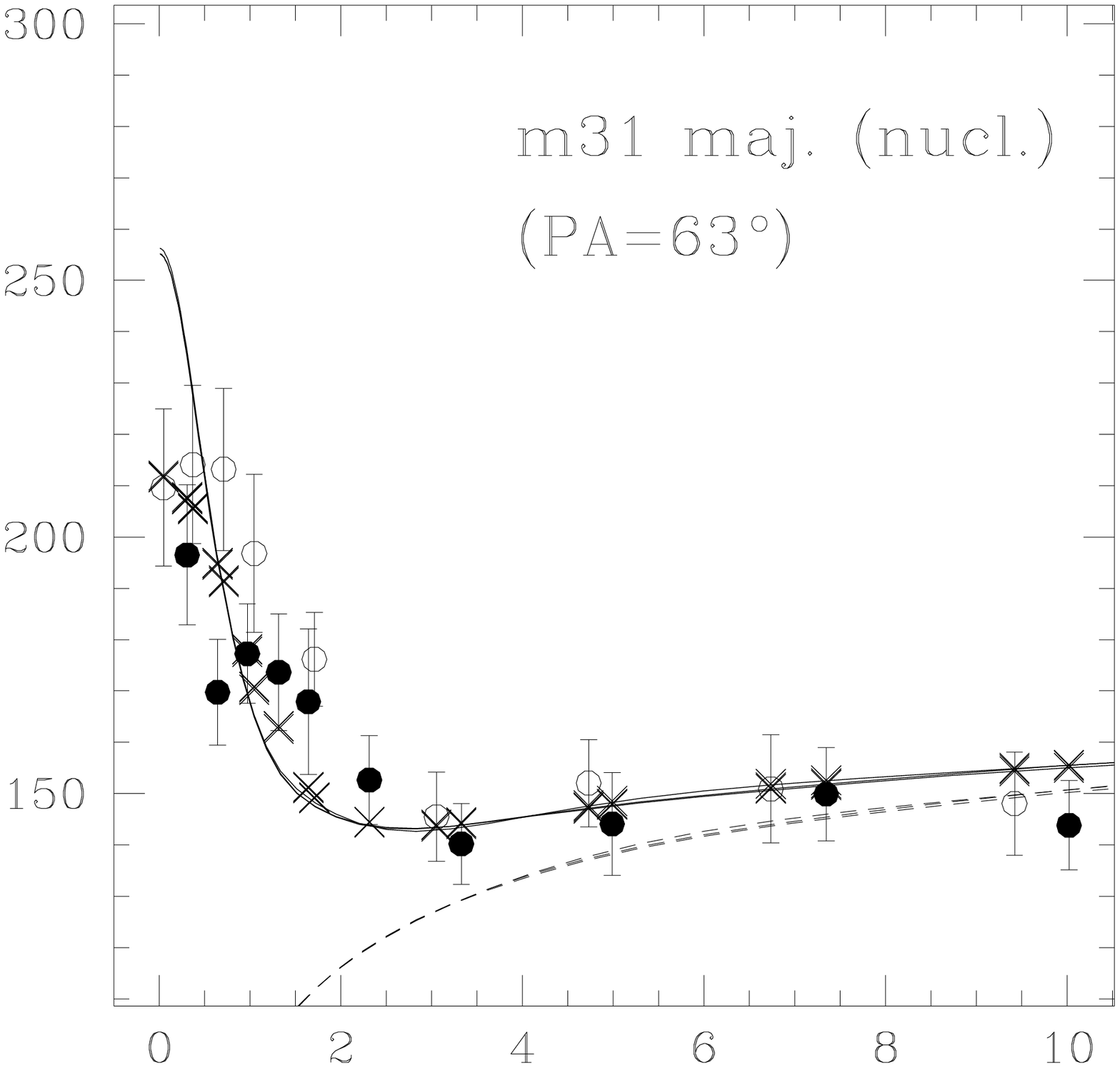,width=0.25\hsize}
\psfig{file=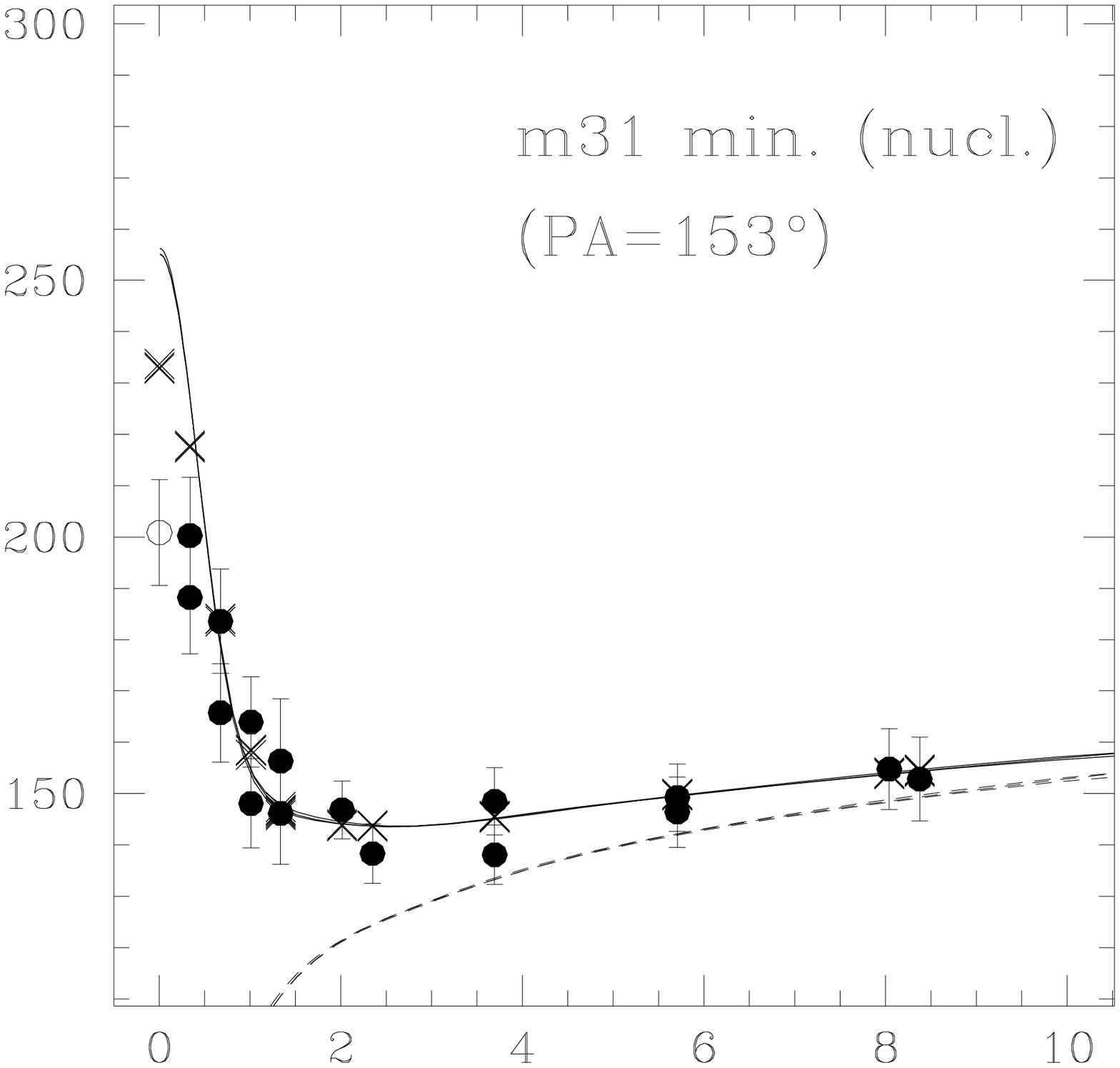,width=0.25\hsize}
}
\centerline{
\psfig{file=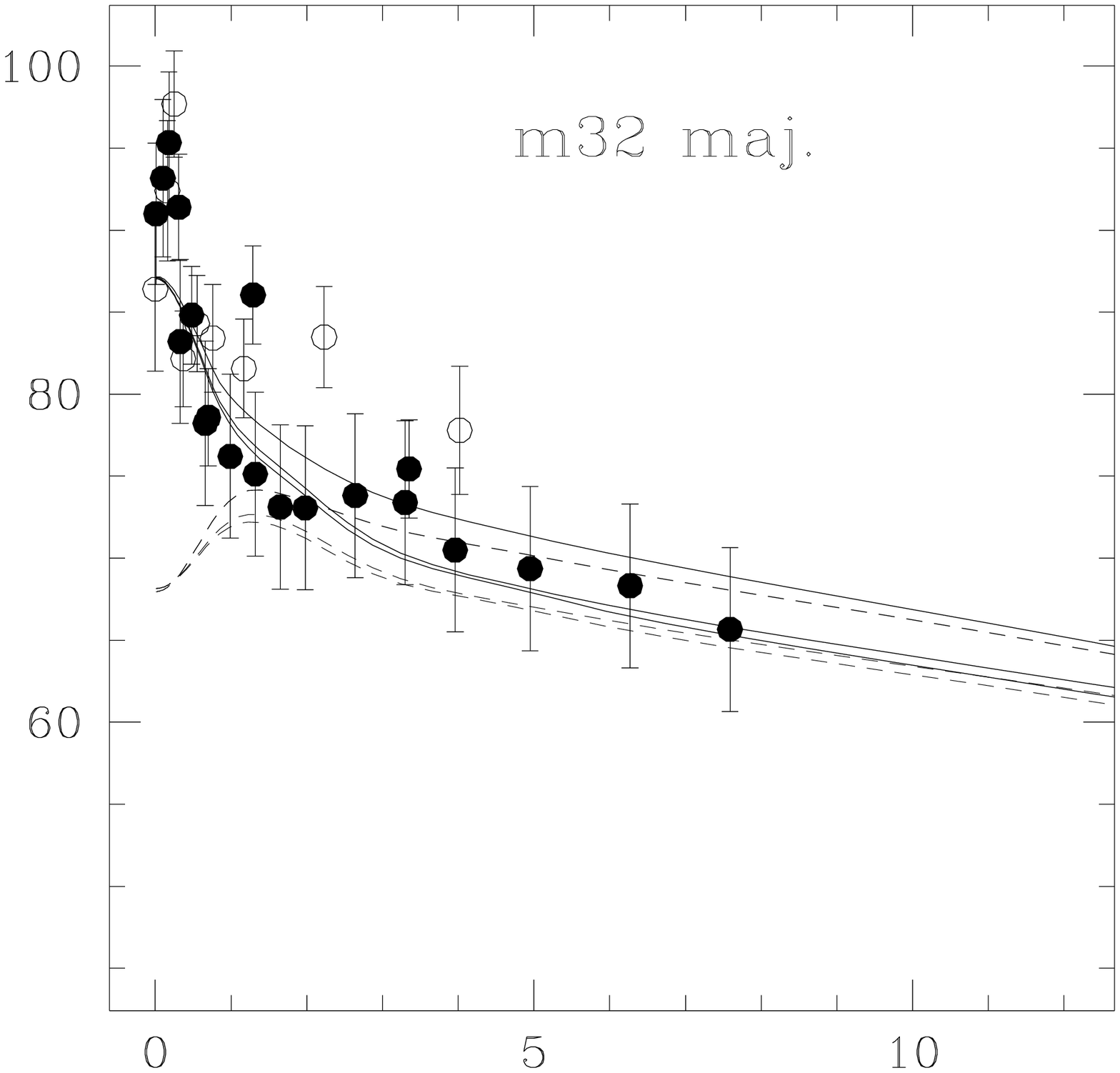,width=0.25\hsize}
\psfig{file=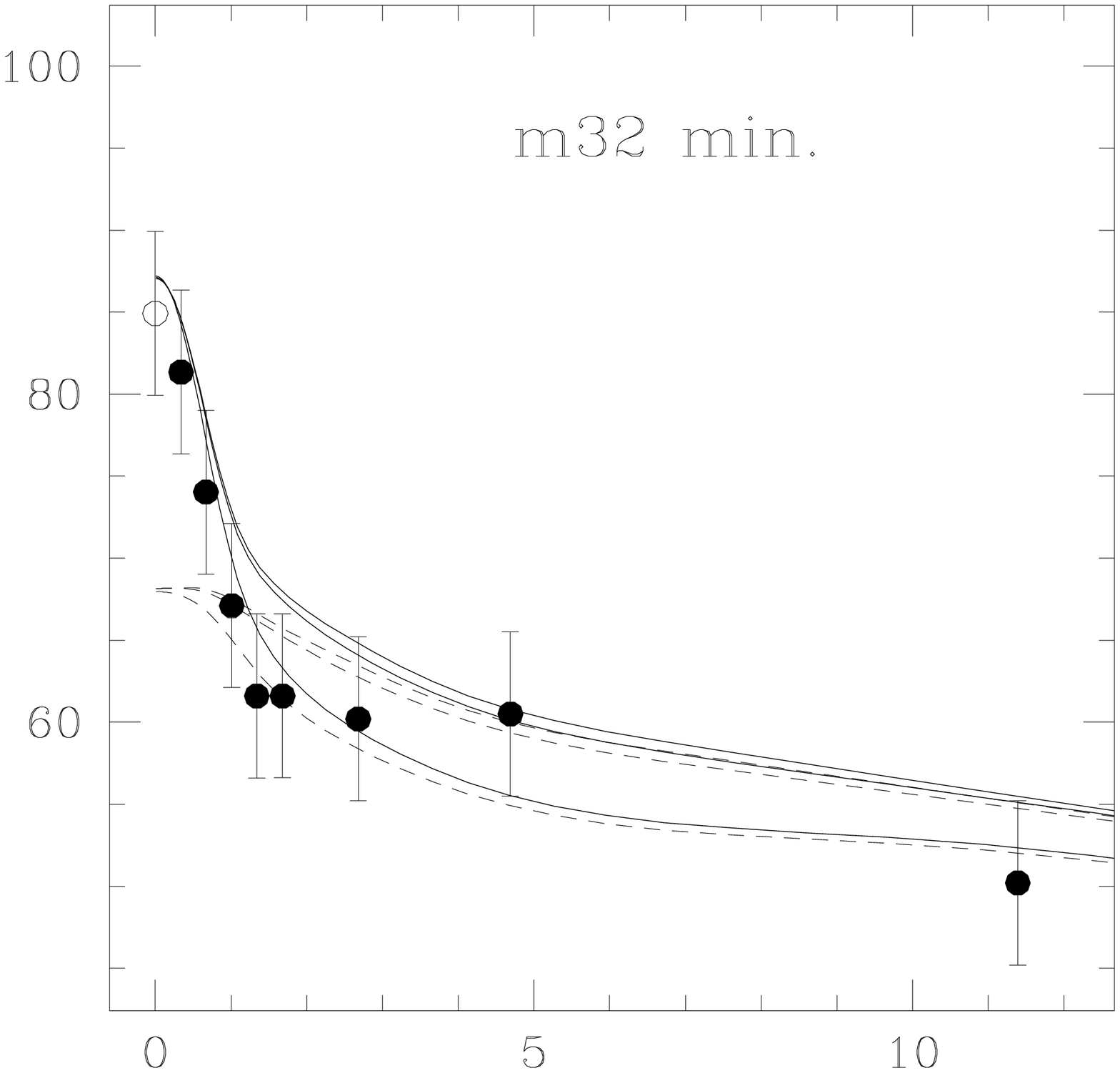,width=0.25\hsize}
\psfig{file=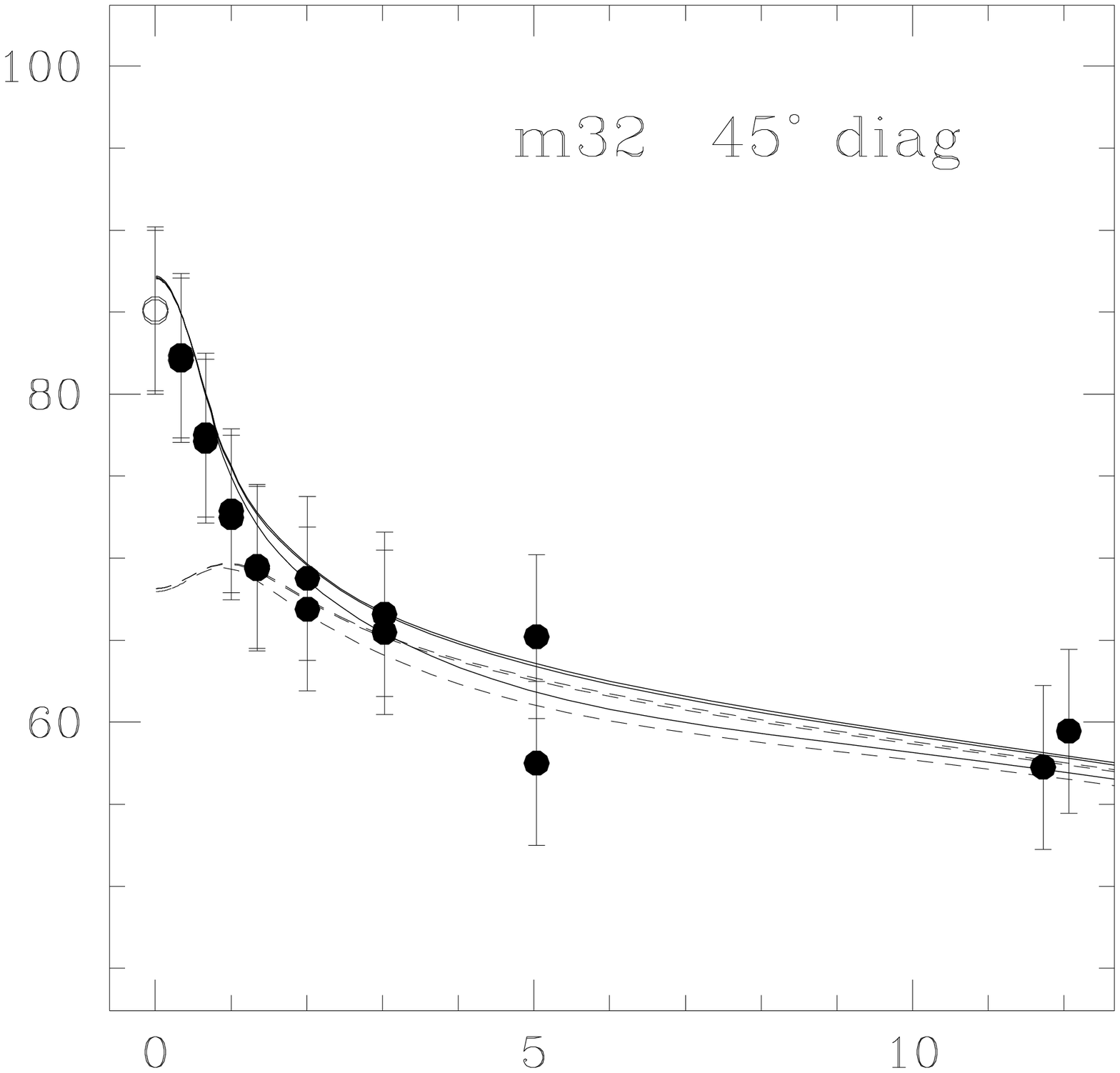,width=0.25\hsize}
\psfig{file=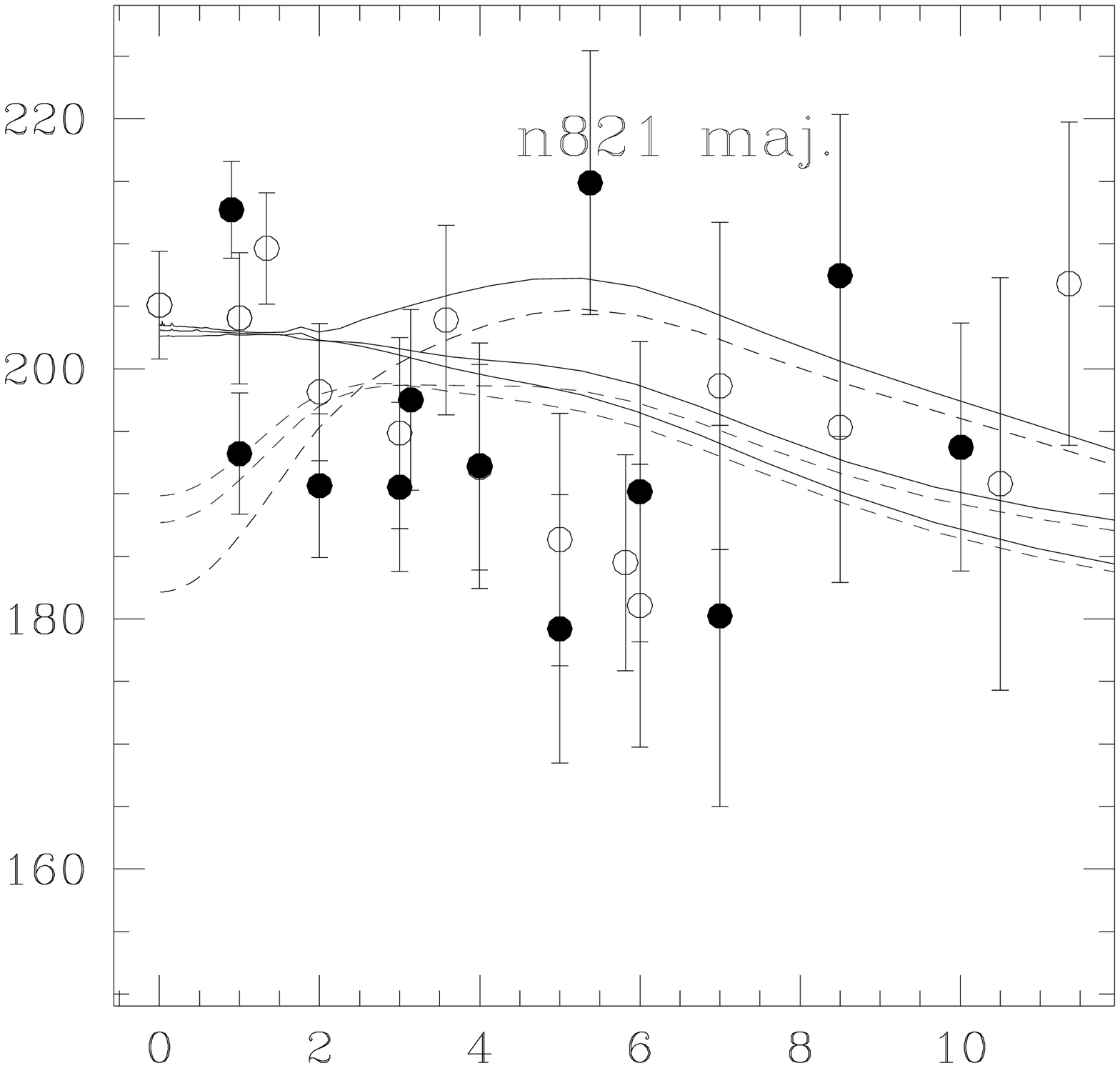,width=0.25\hsize}
}
\centerline{
\psfig{file=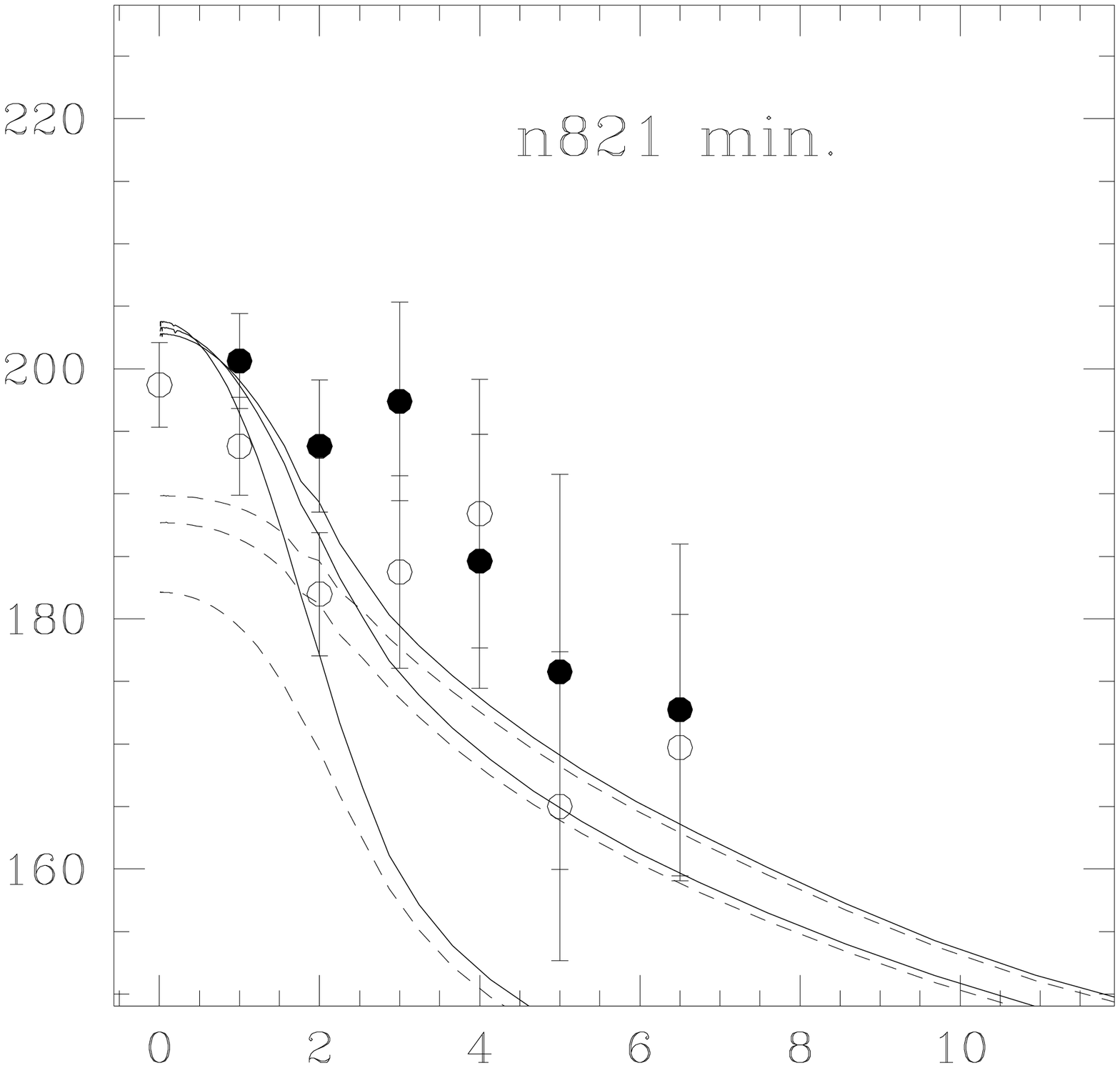,width=0.25\hsize}
\psfig{file=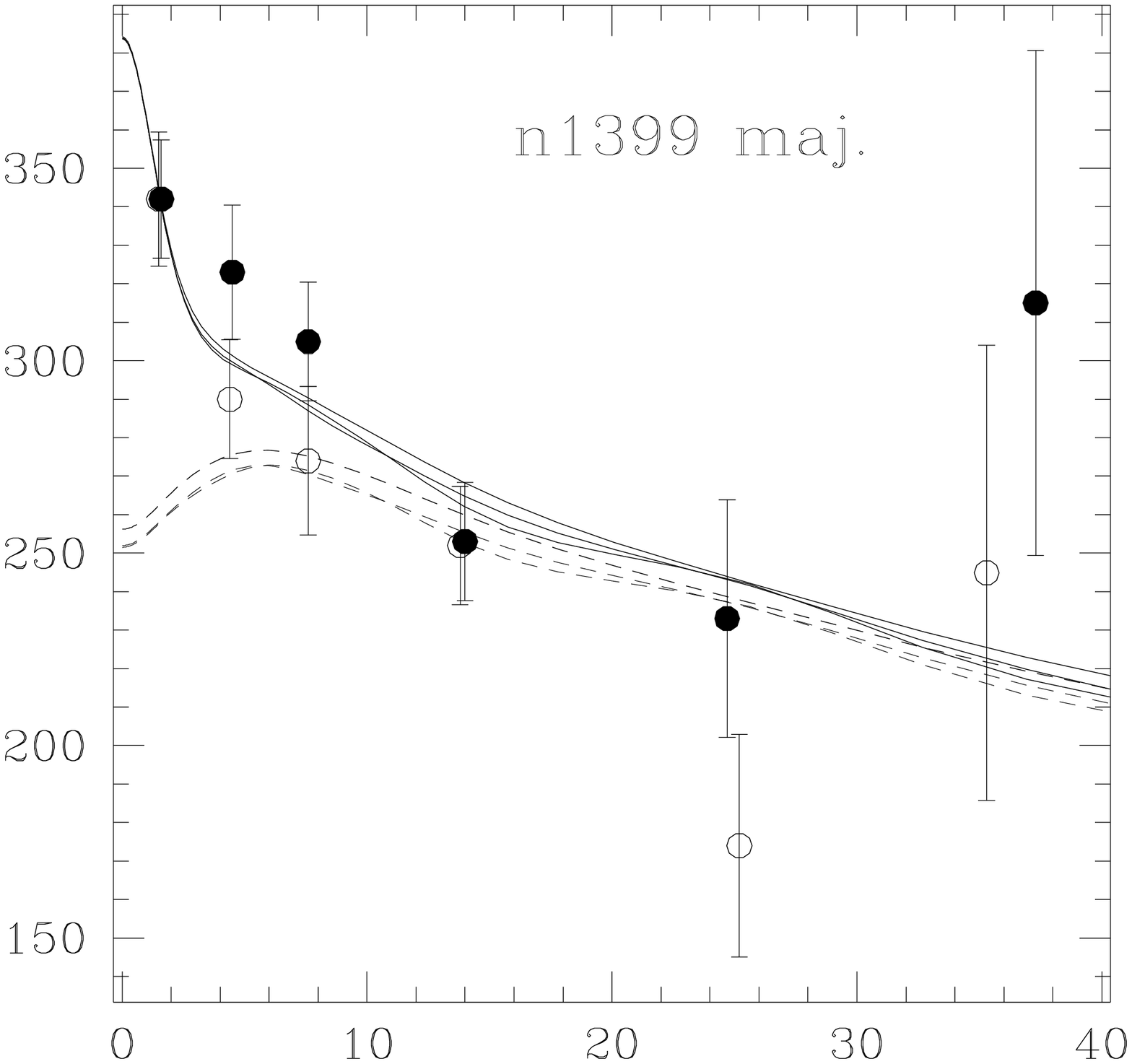,width=0.25\hsize}
\psfig{file=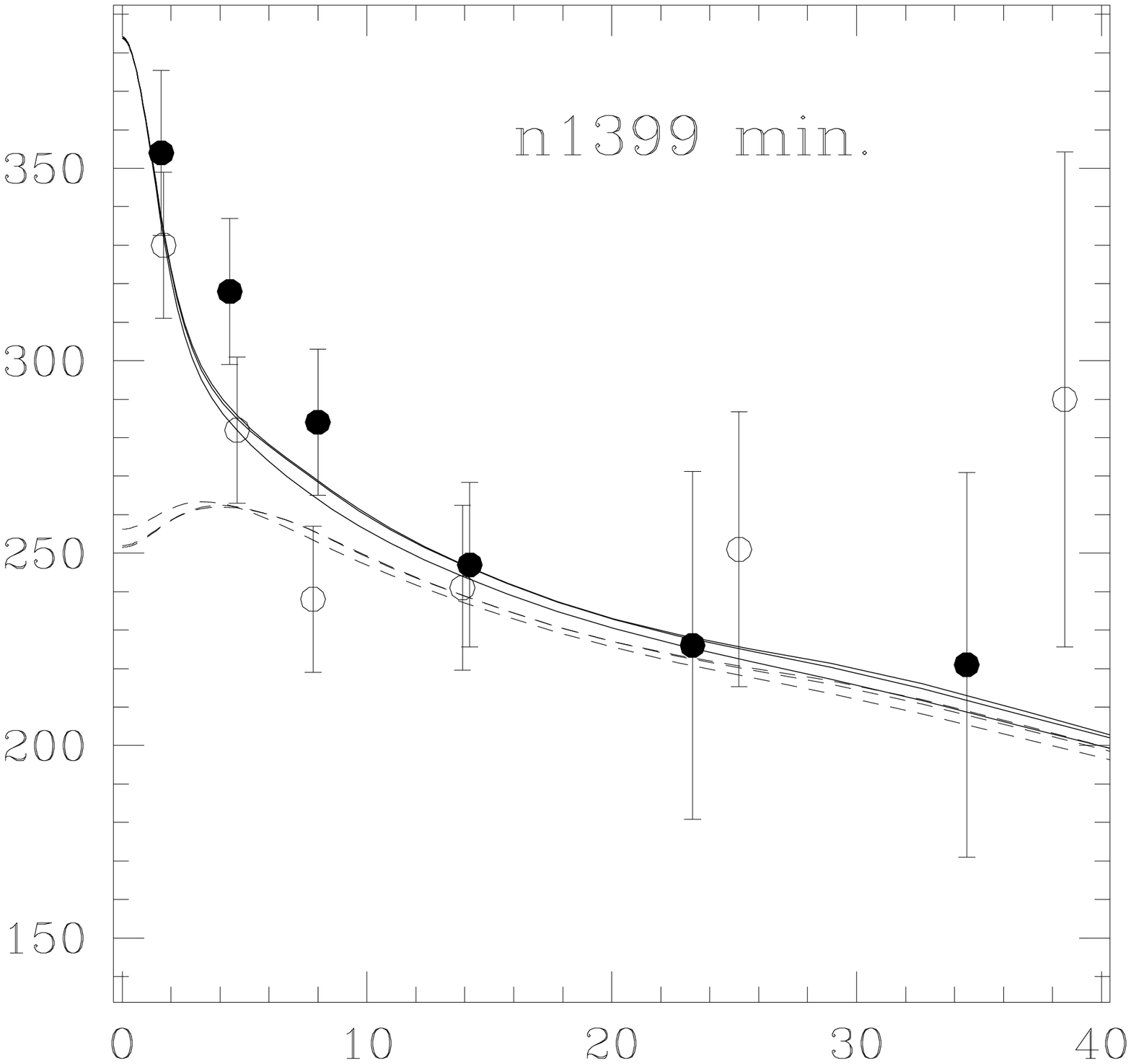,width=0.25\hsize}
\psfig{file=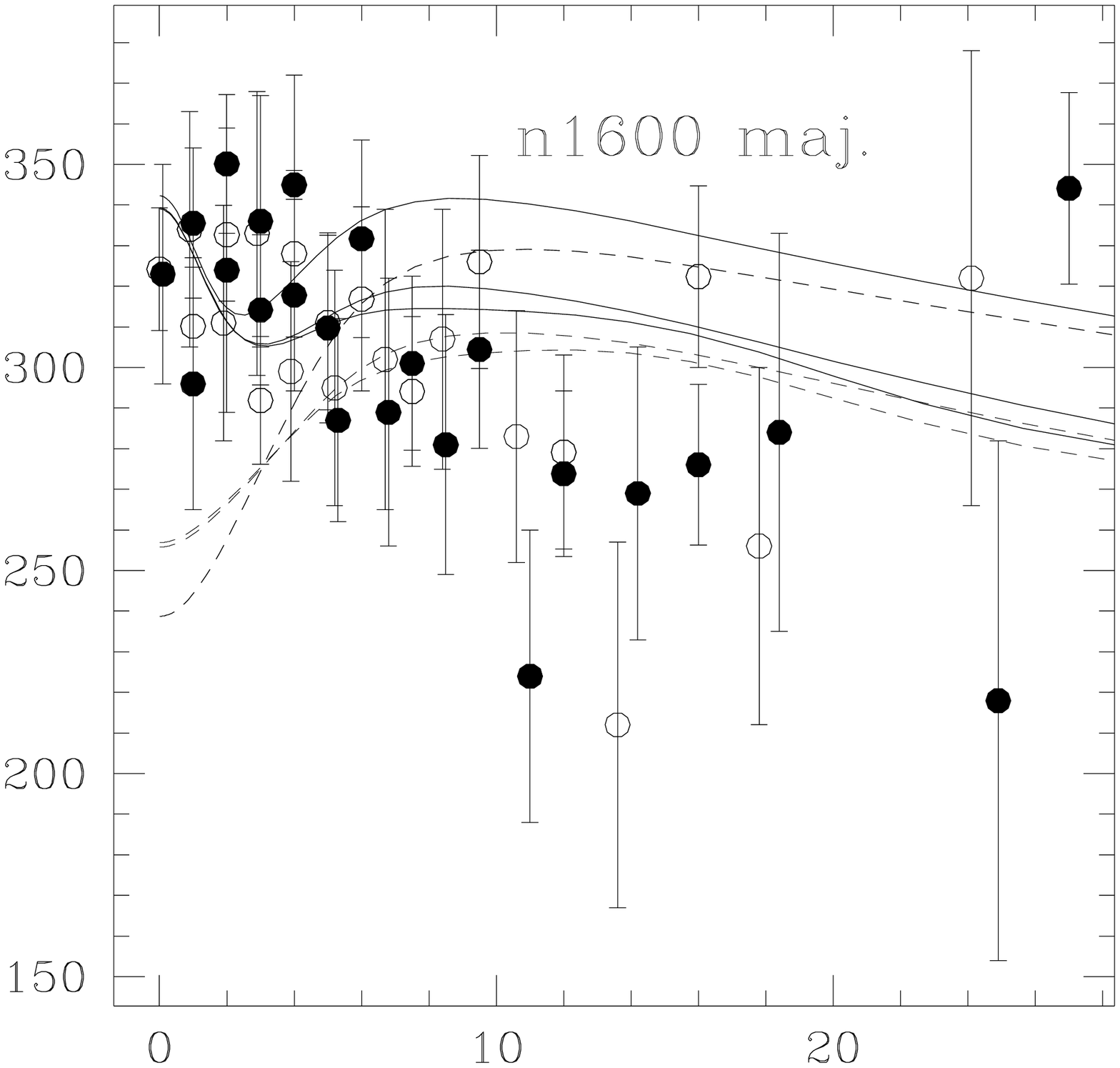,width=0.25\hsize}
}
\centerline{
\psfig{file=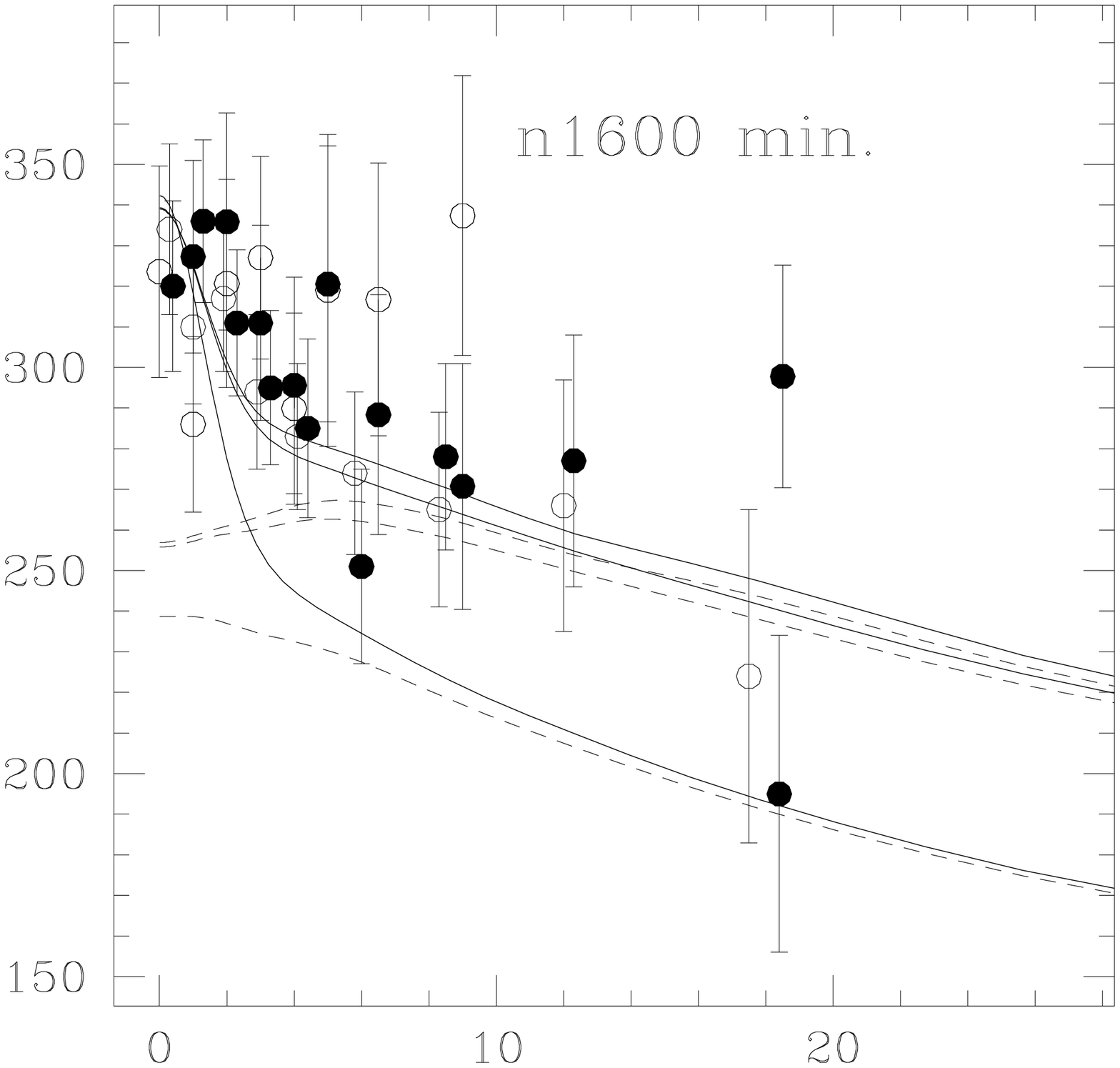,width=0.25\hsize}
\psfig{file=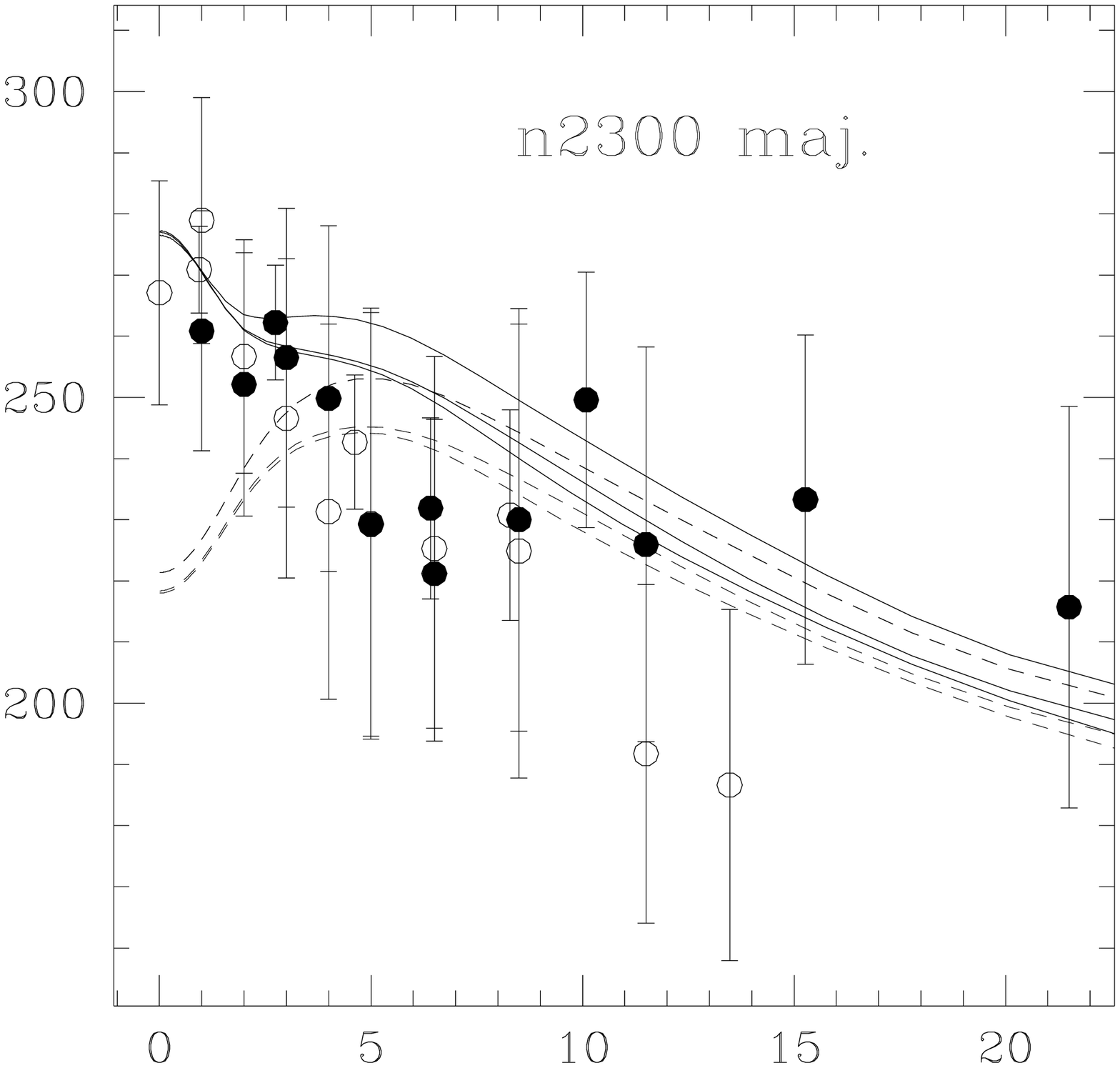,width=0.25\hsize}
\psfig{file=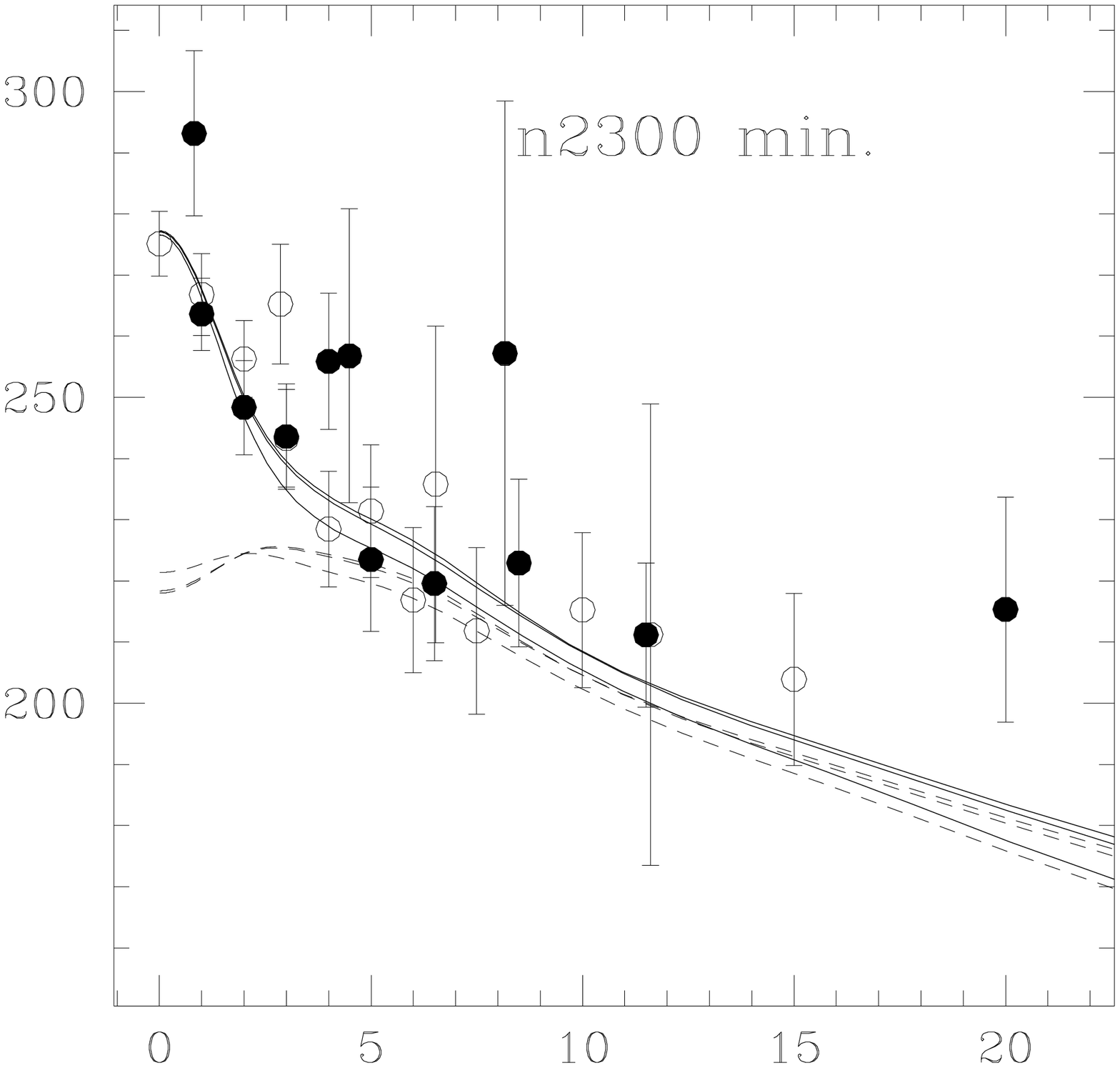,width=0.25\hsize}
\psfig{file=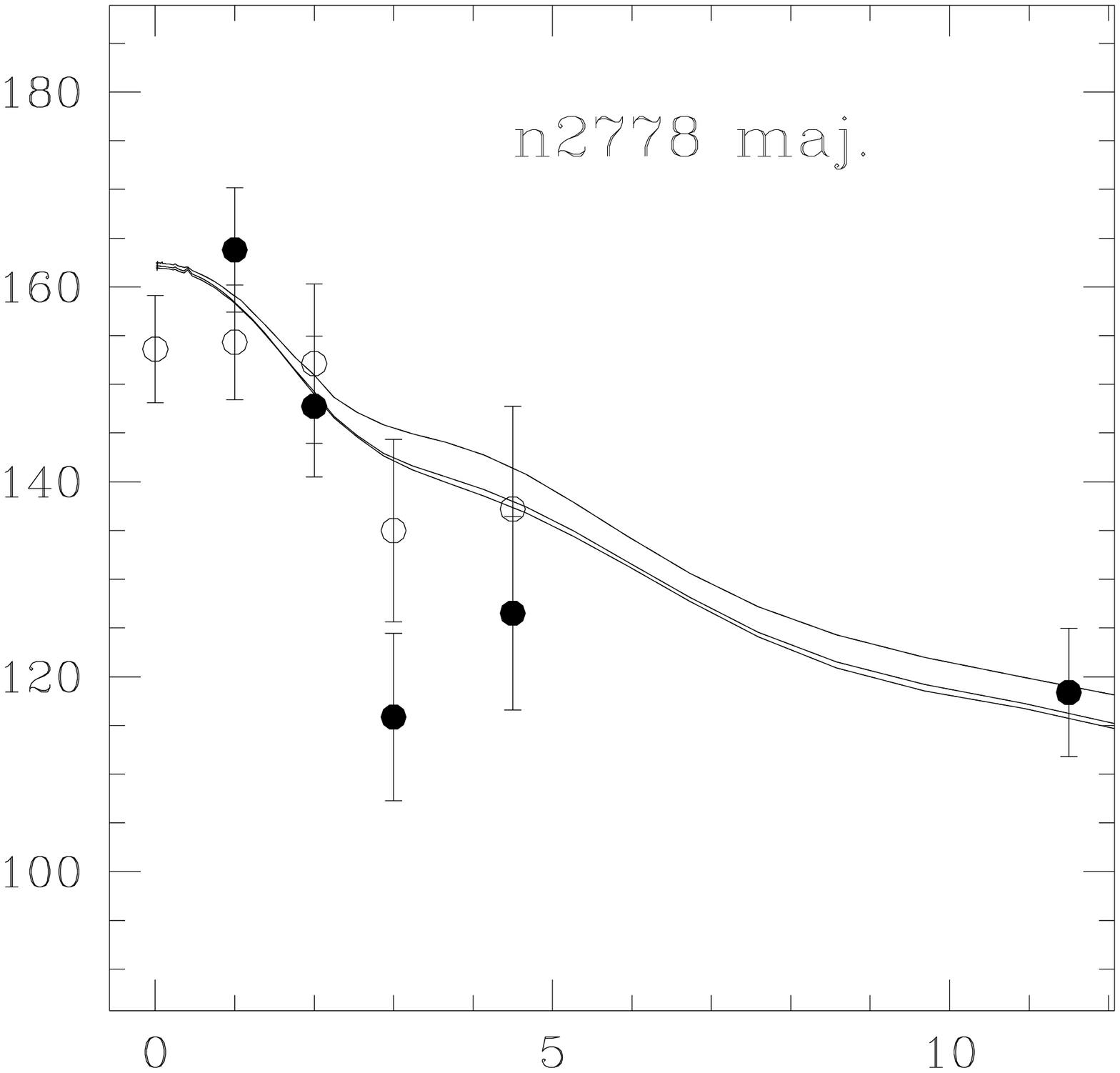,width=0.25\hsize}
}
}{\caption}
\figure{
\centerline{
\psfig{file=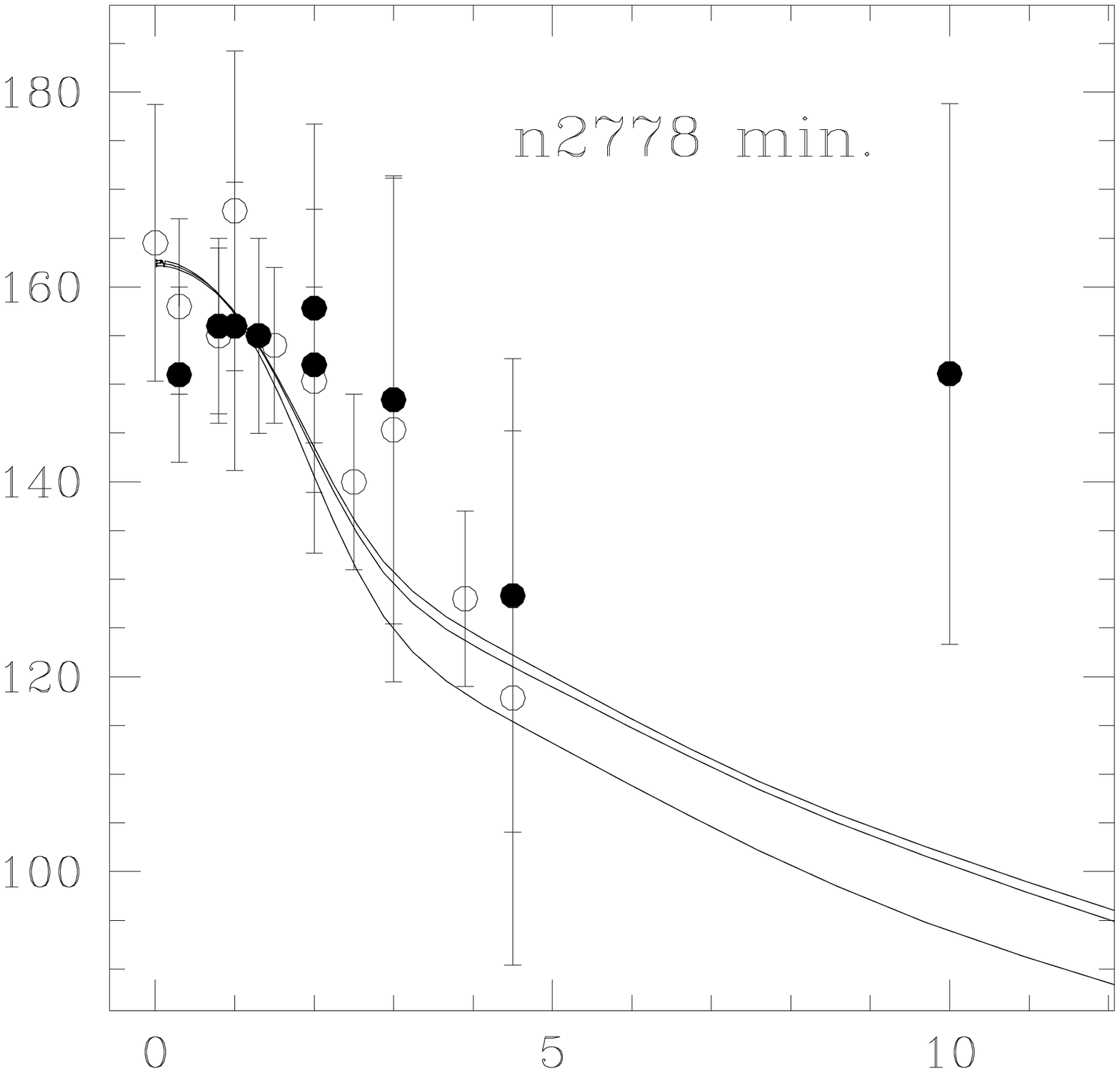,width=0.25\hsize}
\psfig{file=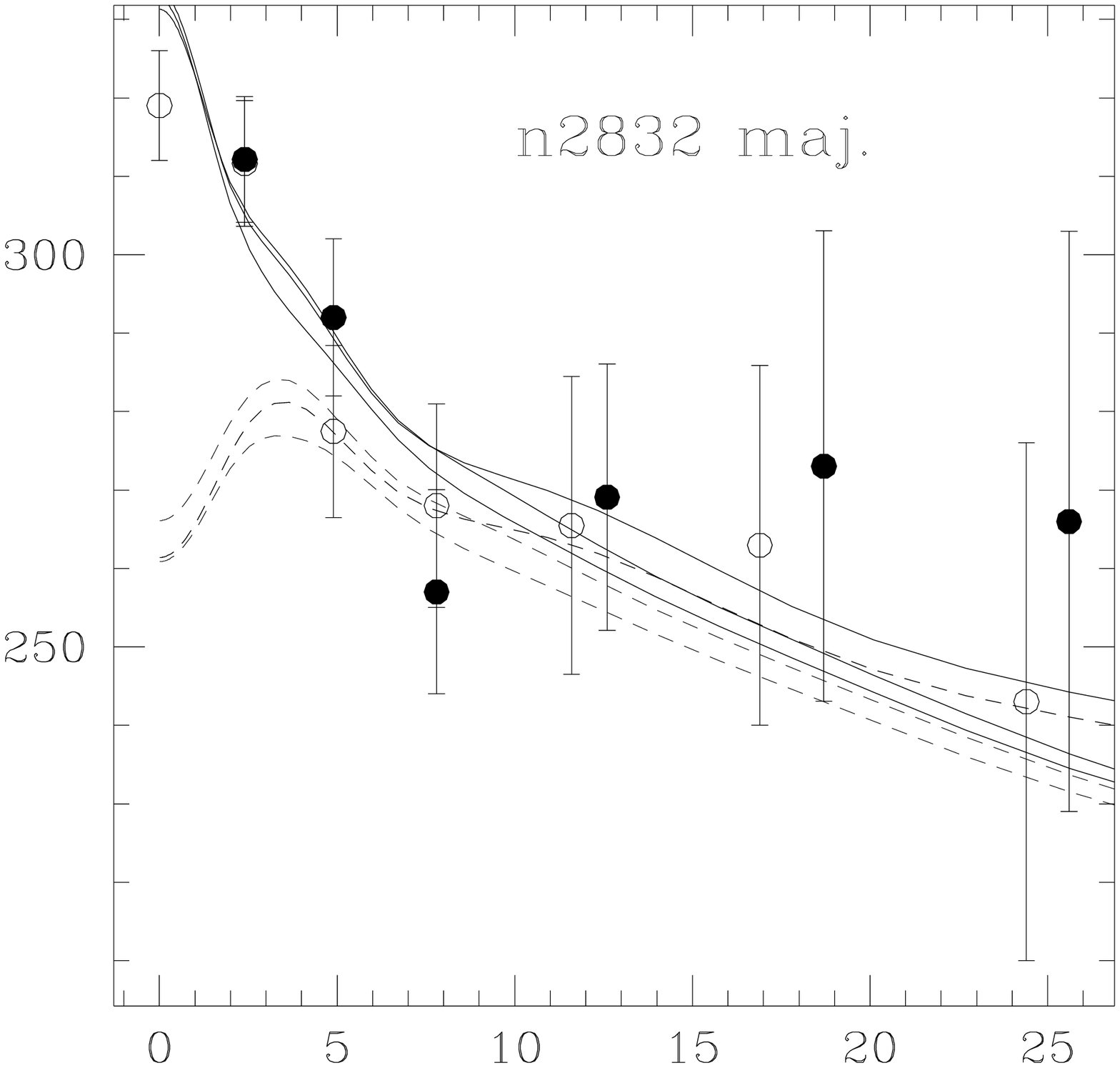,width=0.25\hsize}
\psfig{file=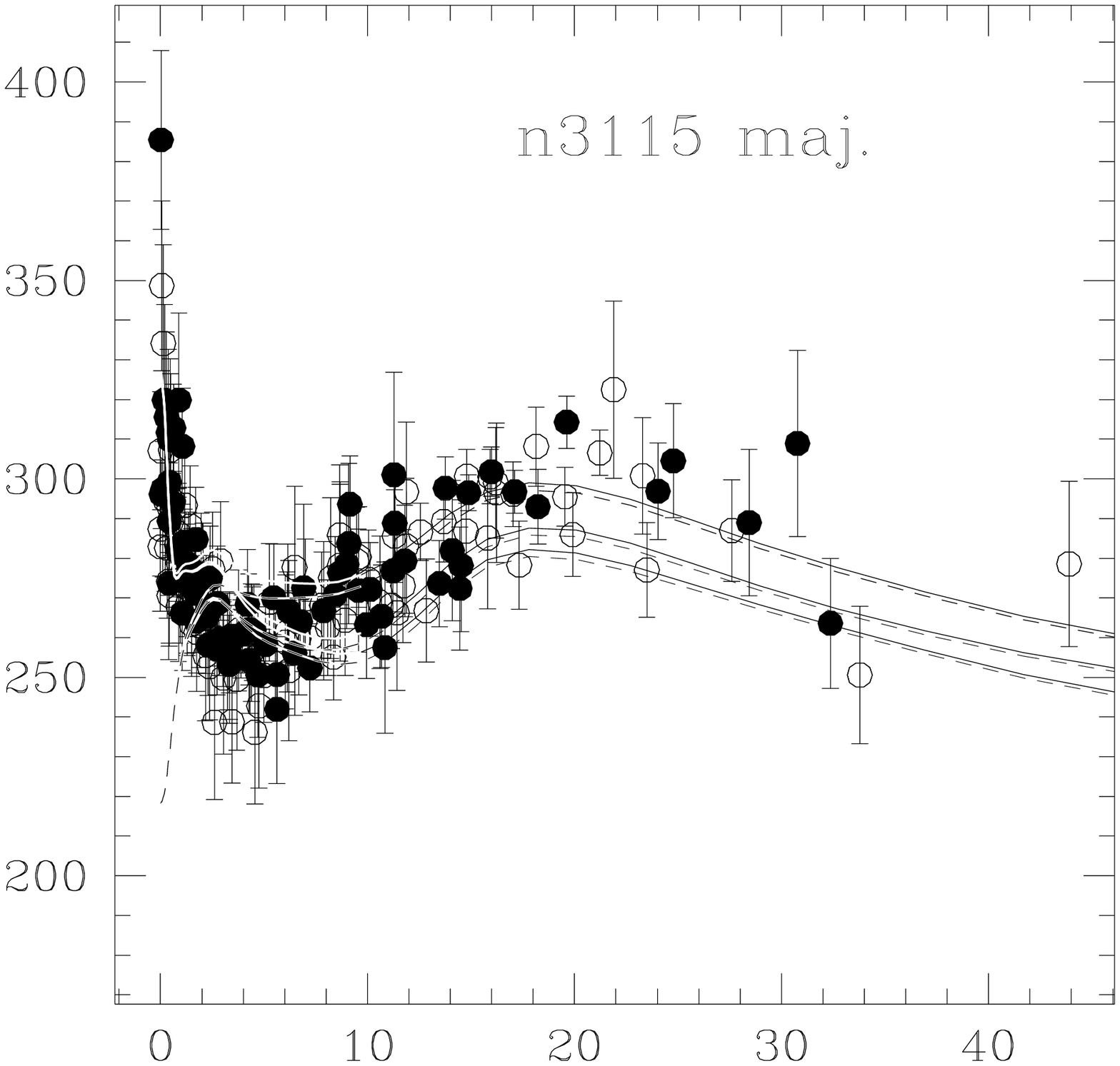,width=0.25\hsize}
\psfig{file=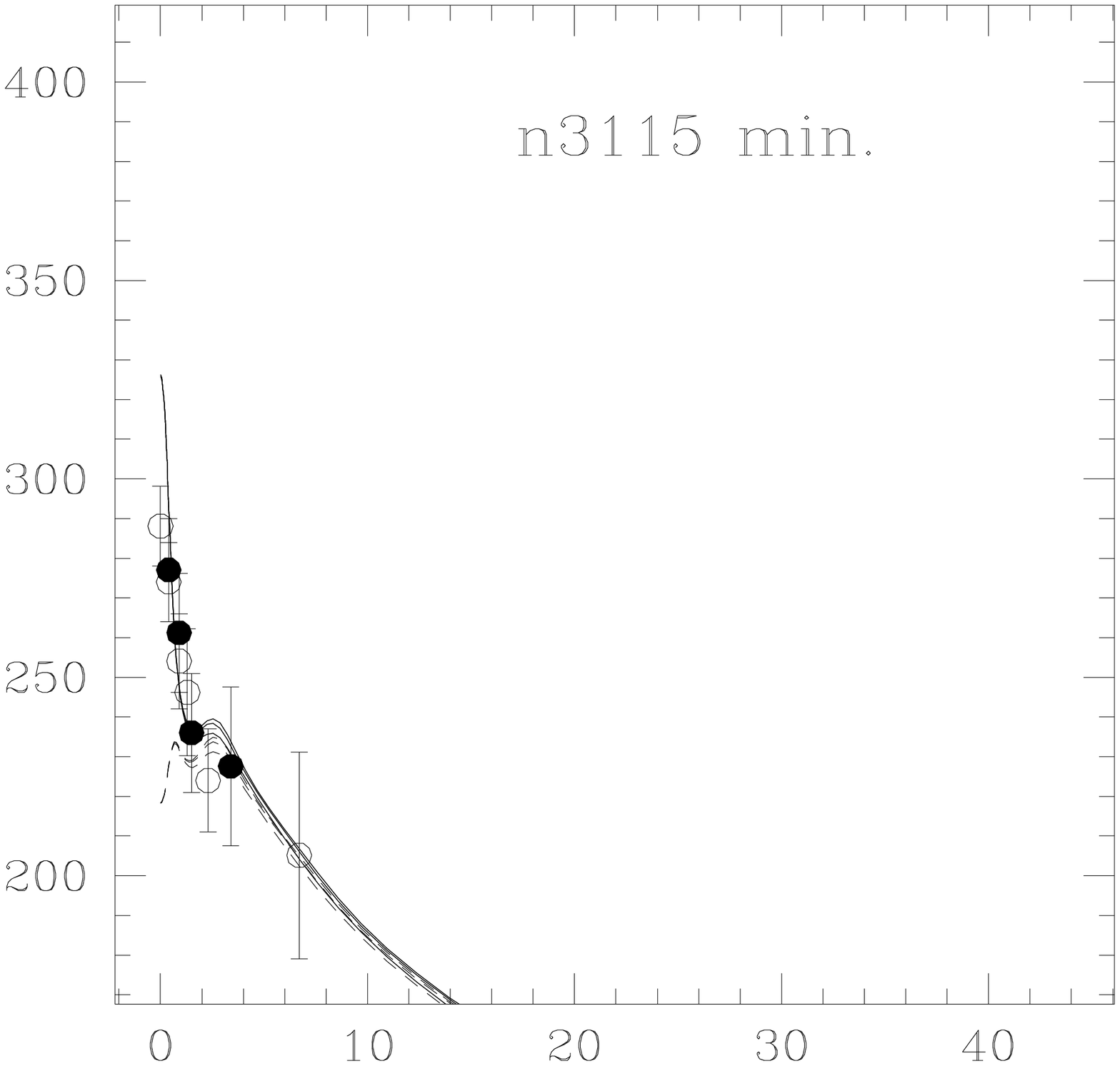,width=0.25\hsize}
}
\centerline{
\psfig{file=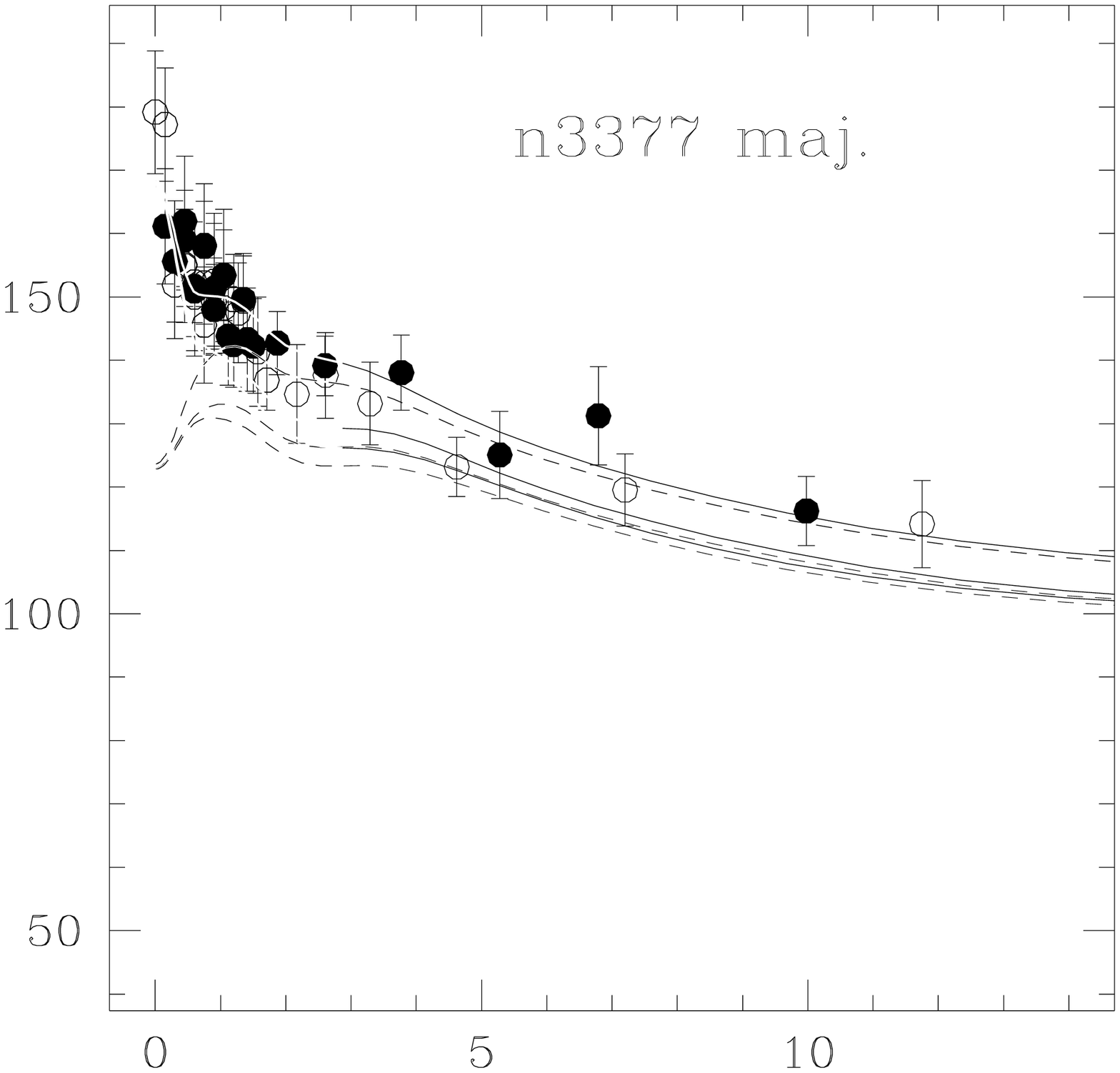,width=0.25\hsize}
\psfig{file=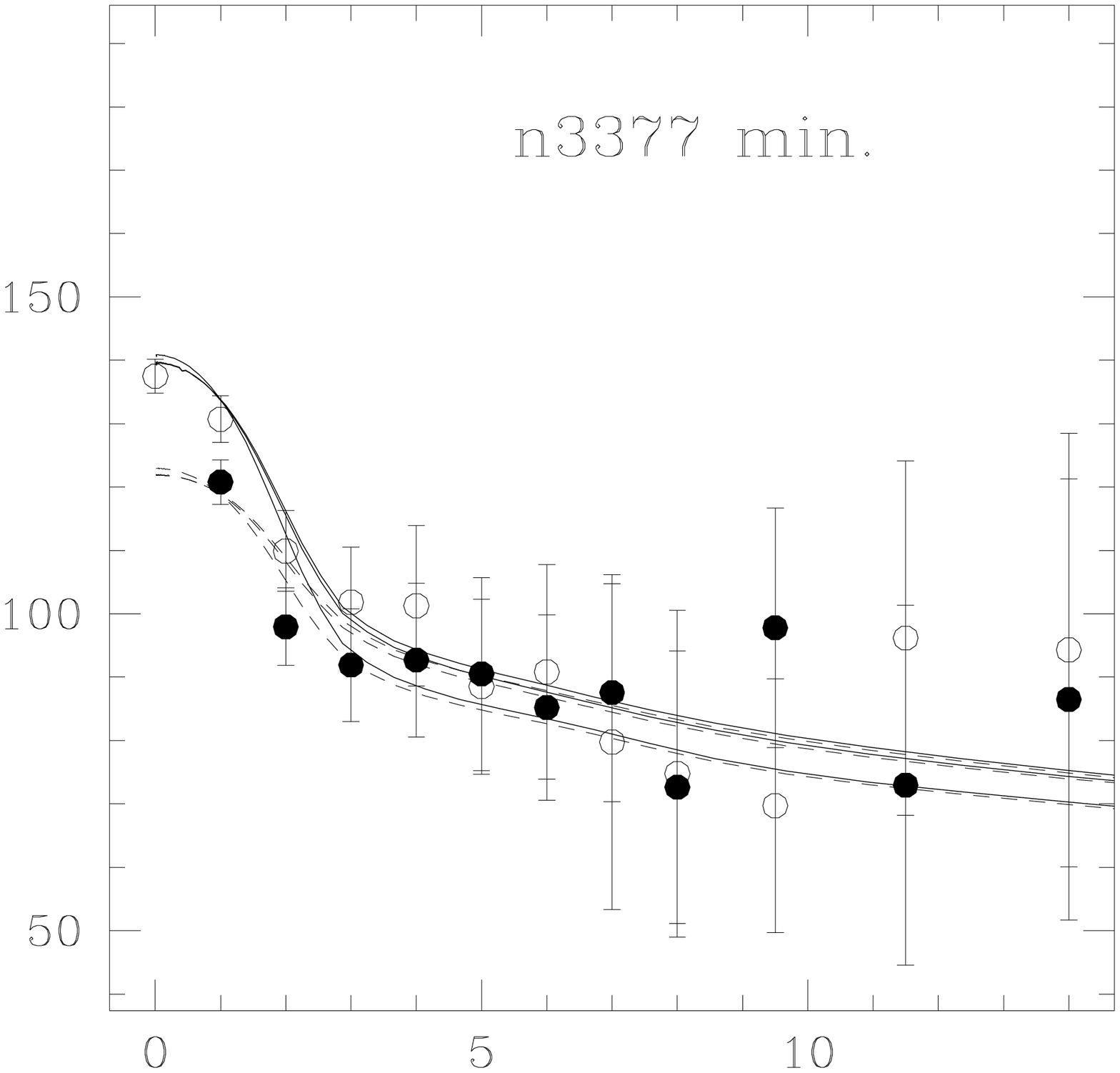,width=0.25\hsize}
\psfig{file=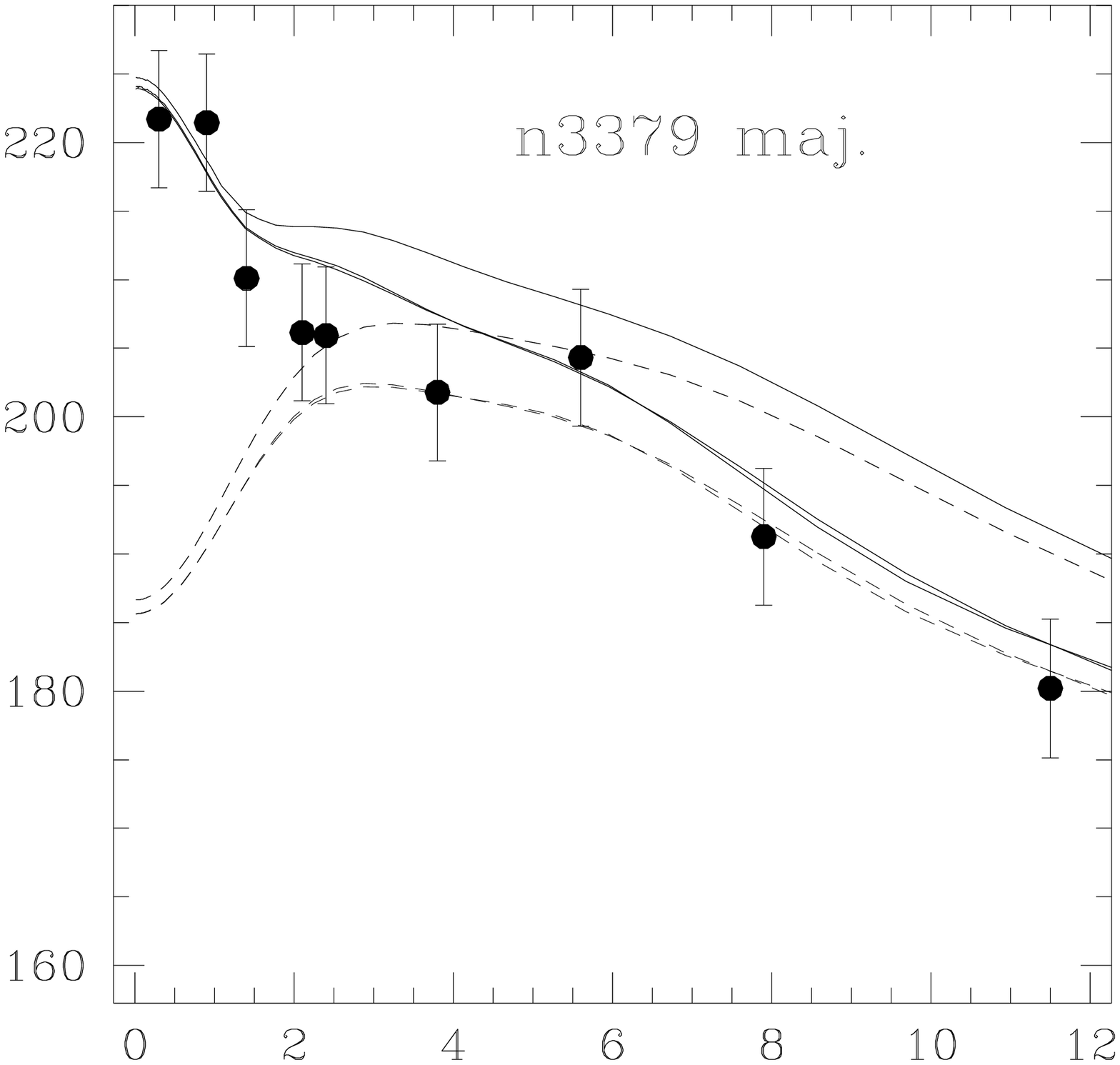,width=0.25\hsize}
\psfig{file=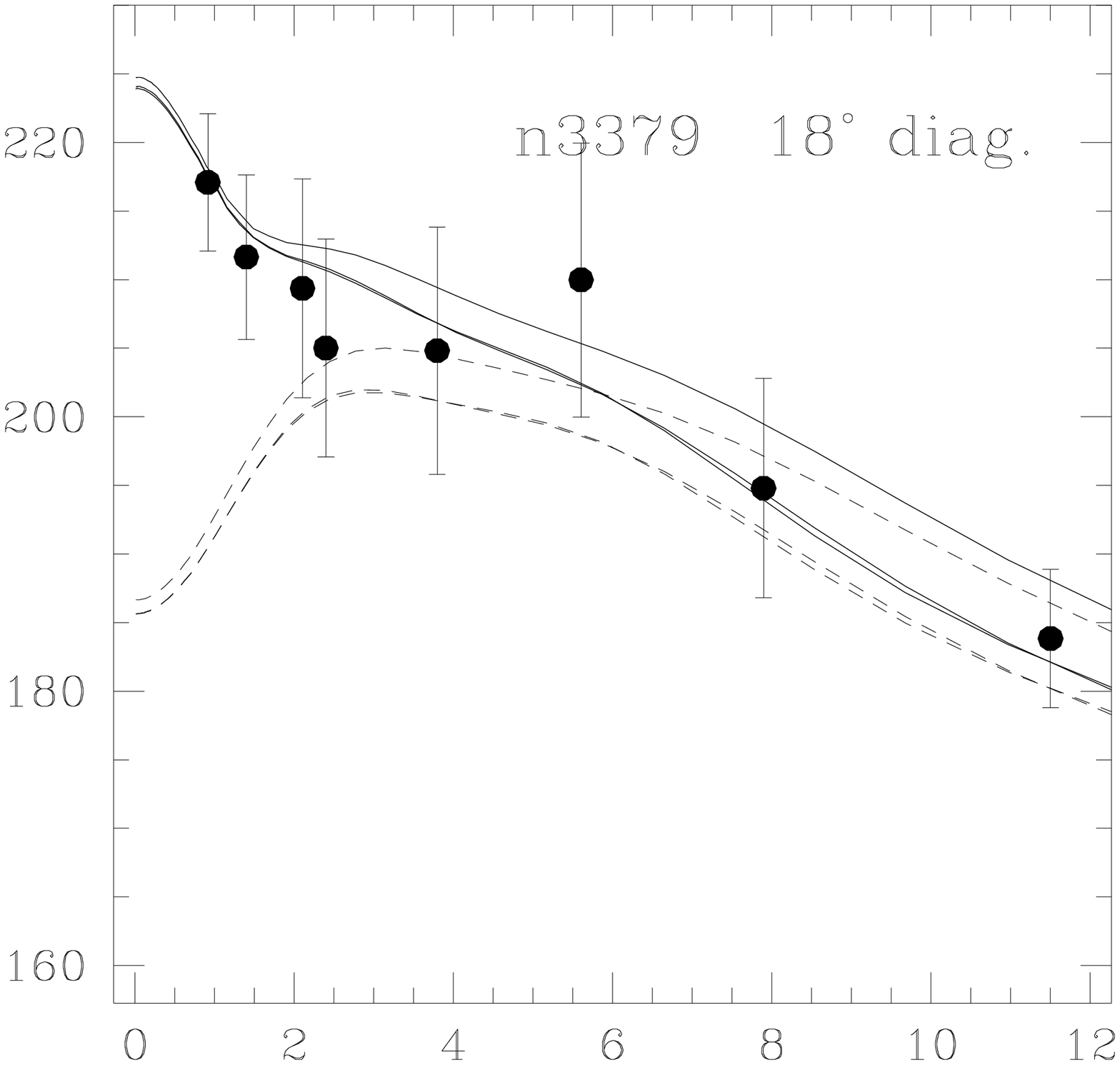,width=0.25\hsize}
}
\centerline{
\psfig{file=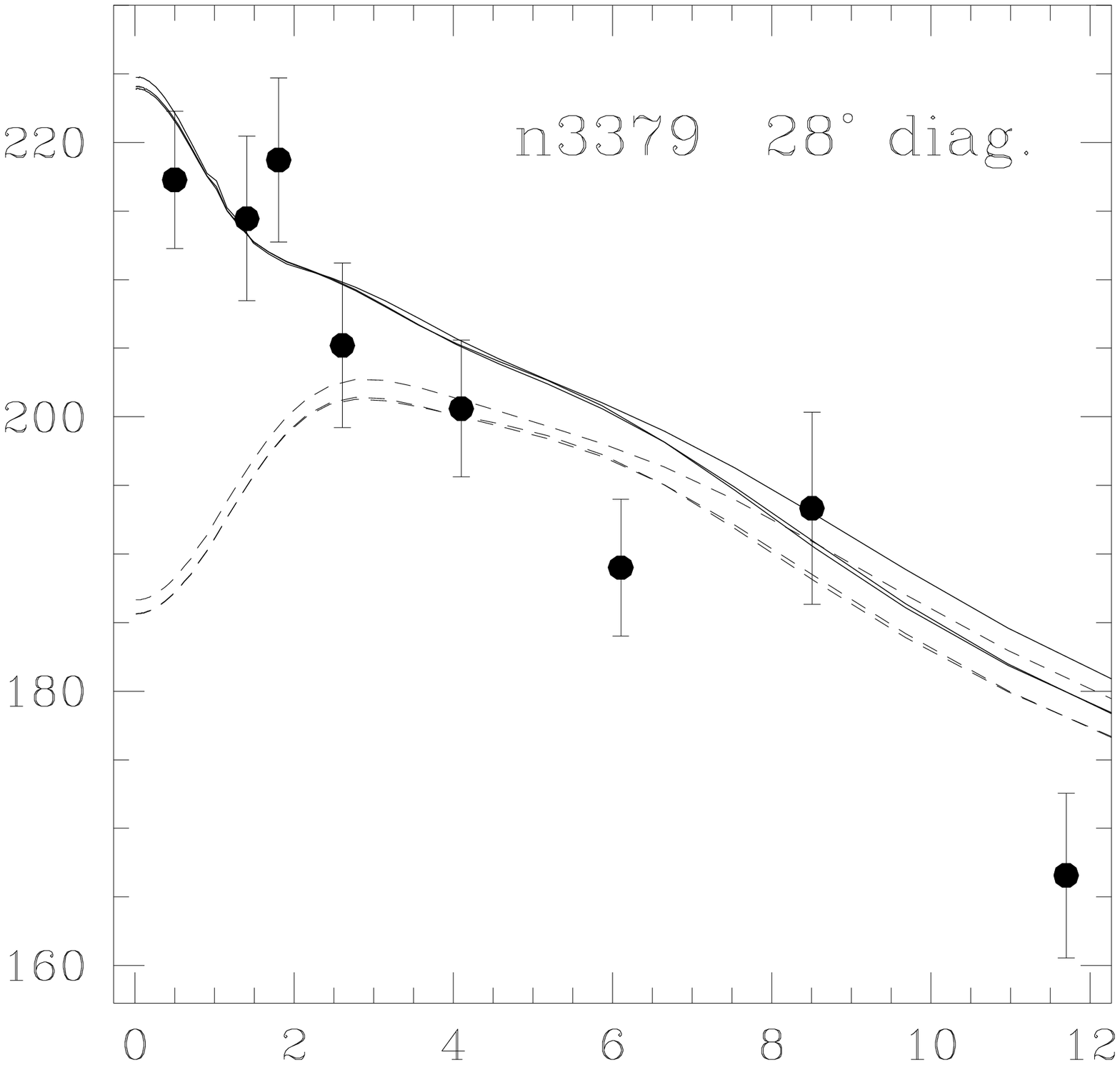,width=0.25\hsize}
\psfig{file=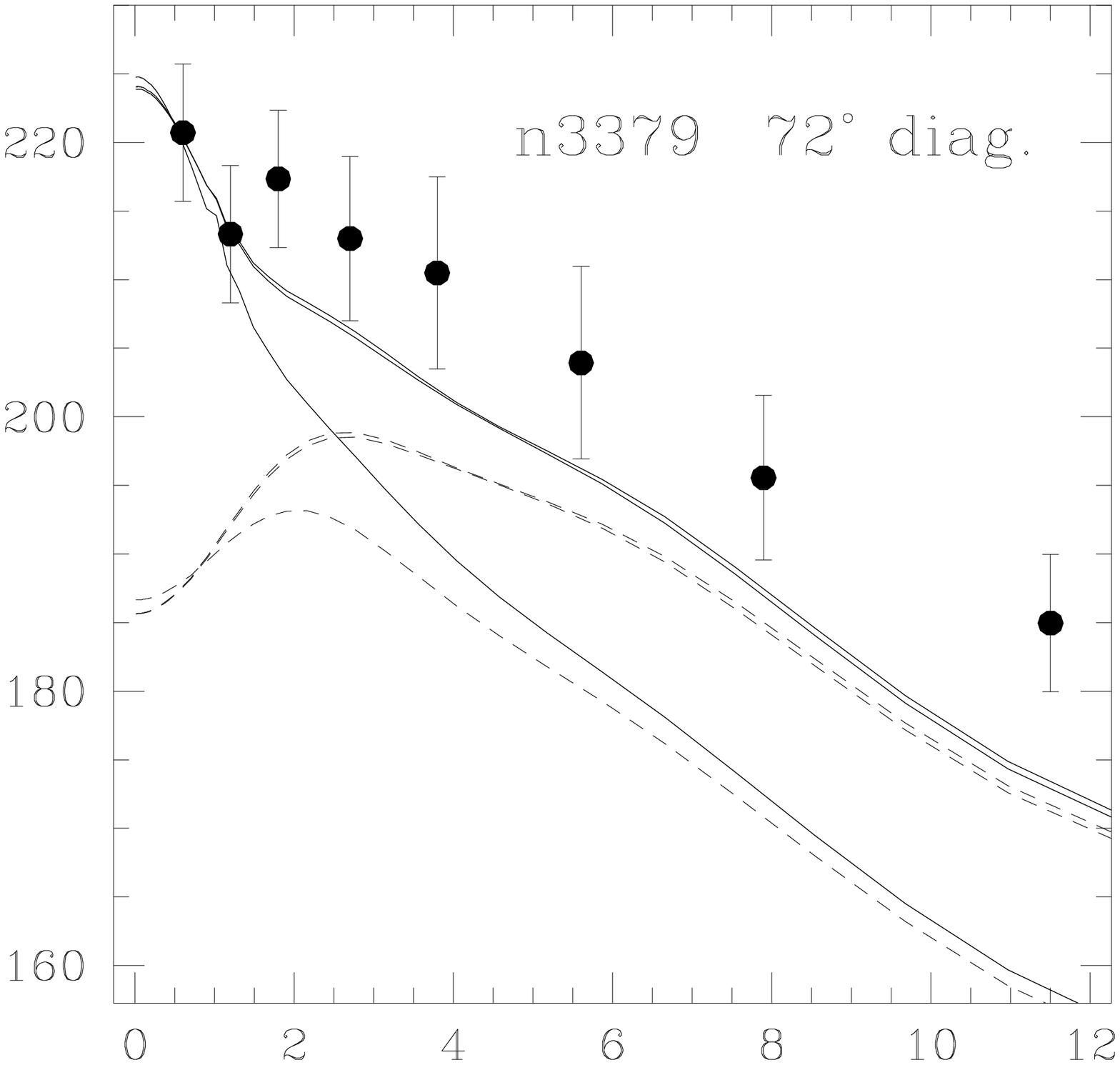,width=0.25\hsize}
\psfig{file=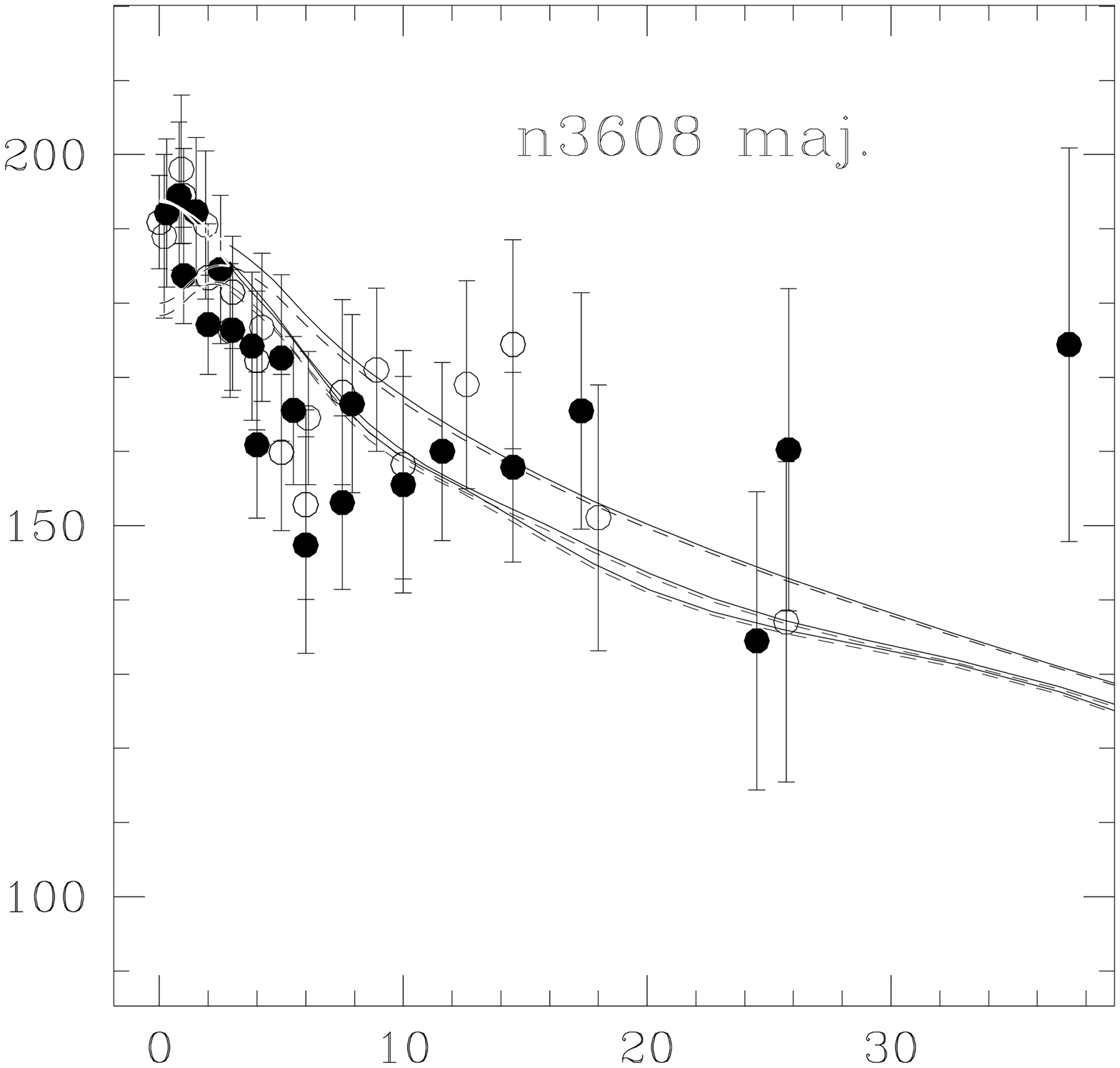,width=0.25\hsize}
\psfig{file=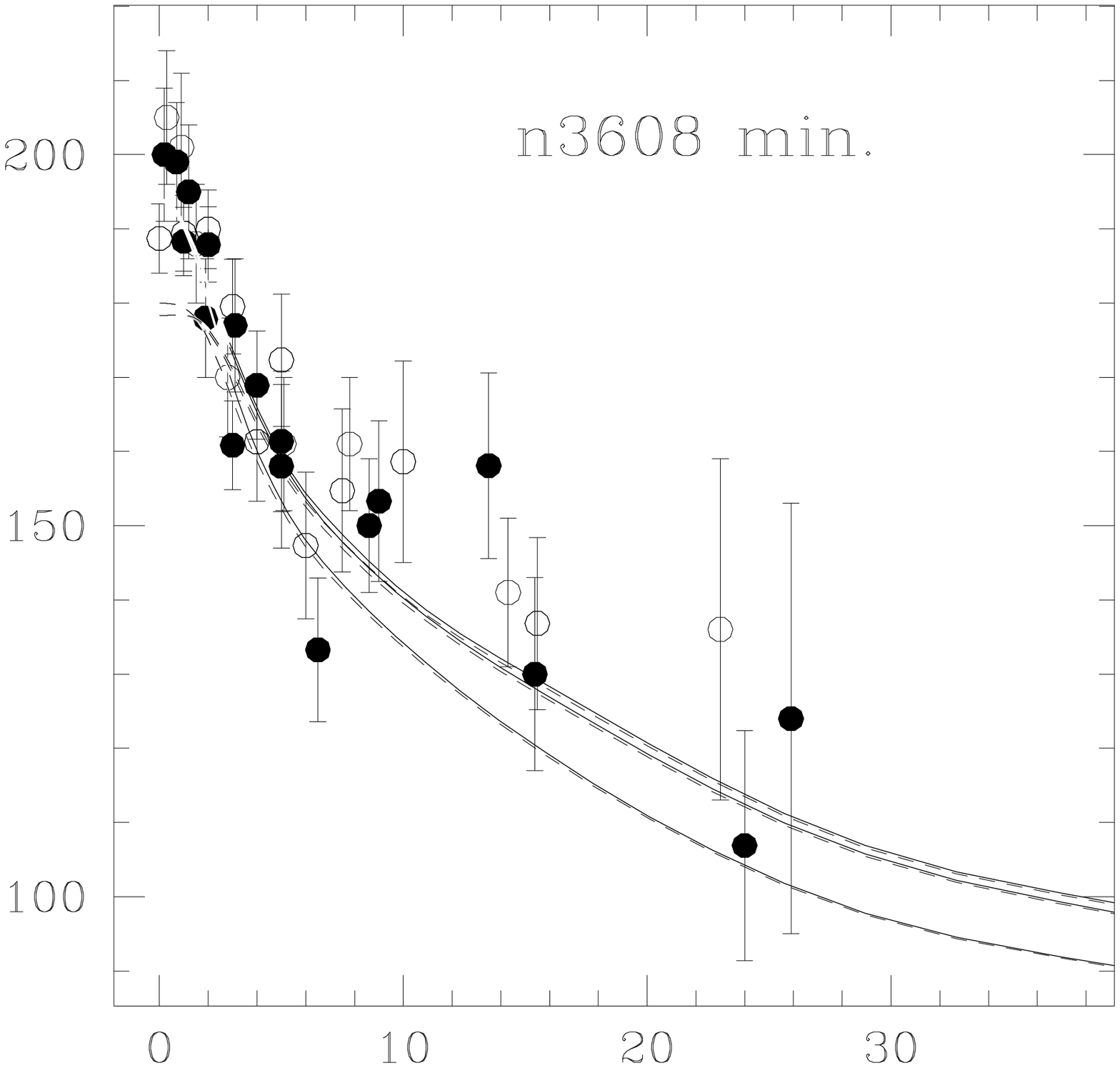,width=0.25\hsize}
}
\centerline{
\psfig{file=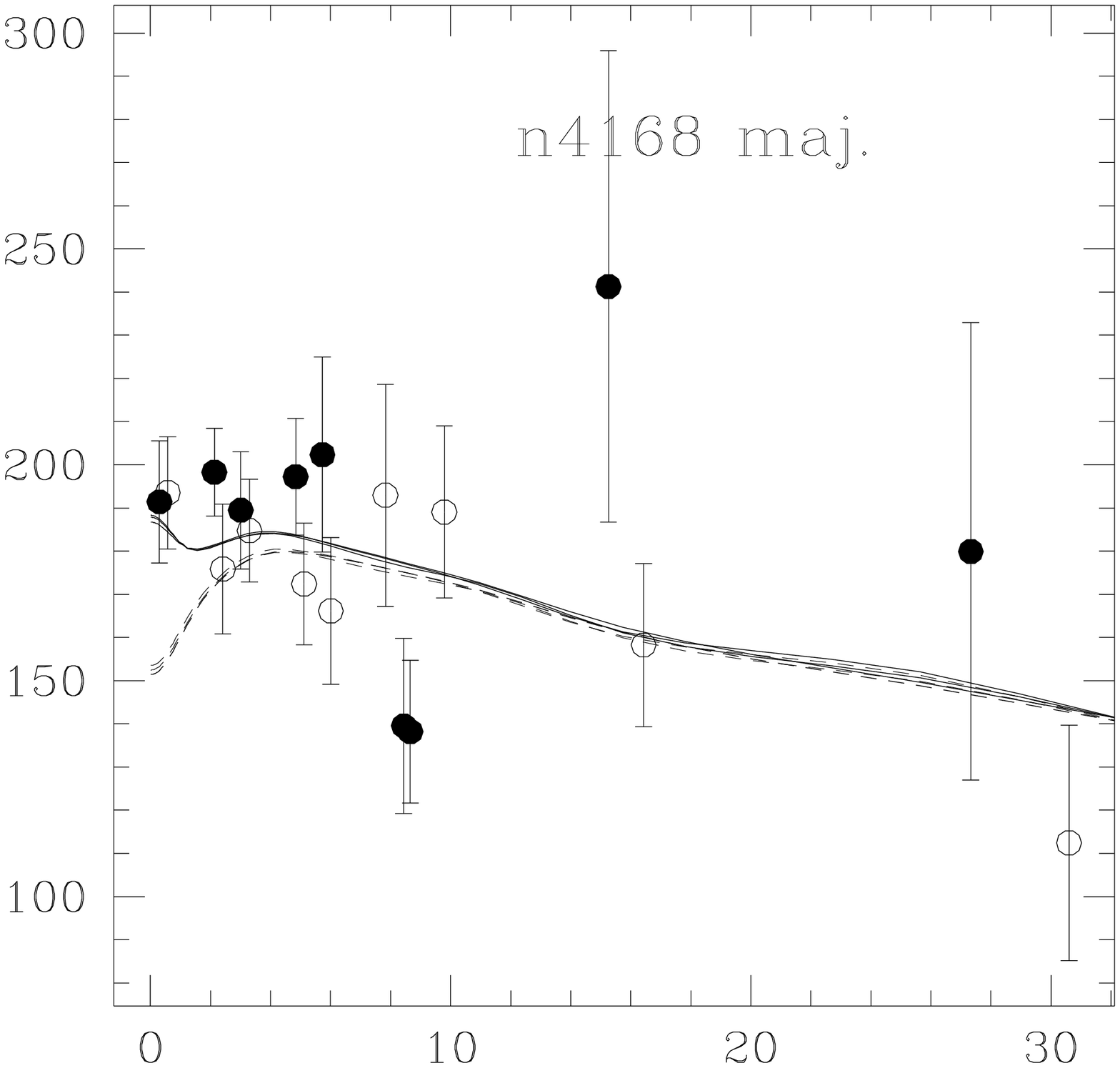,width=0.25\hsize}
\psfig{file=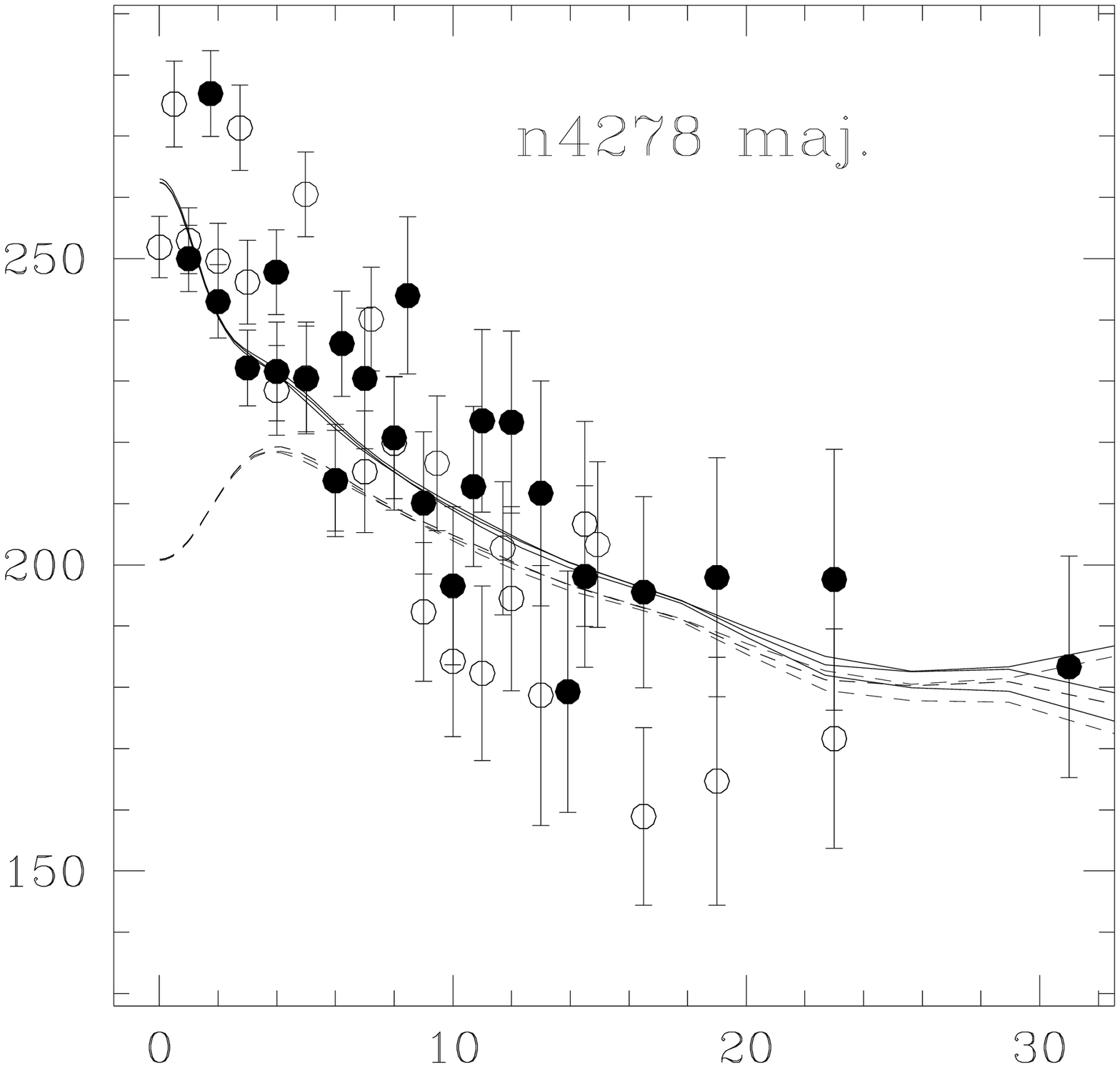,width=0.25\hsize}
\psfig{file=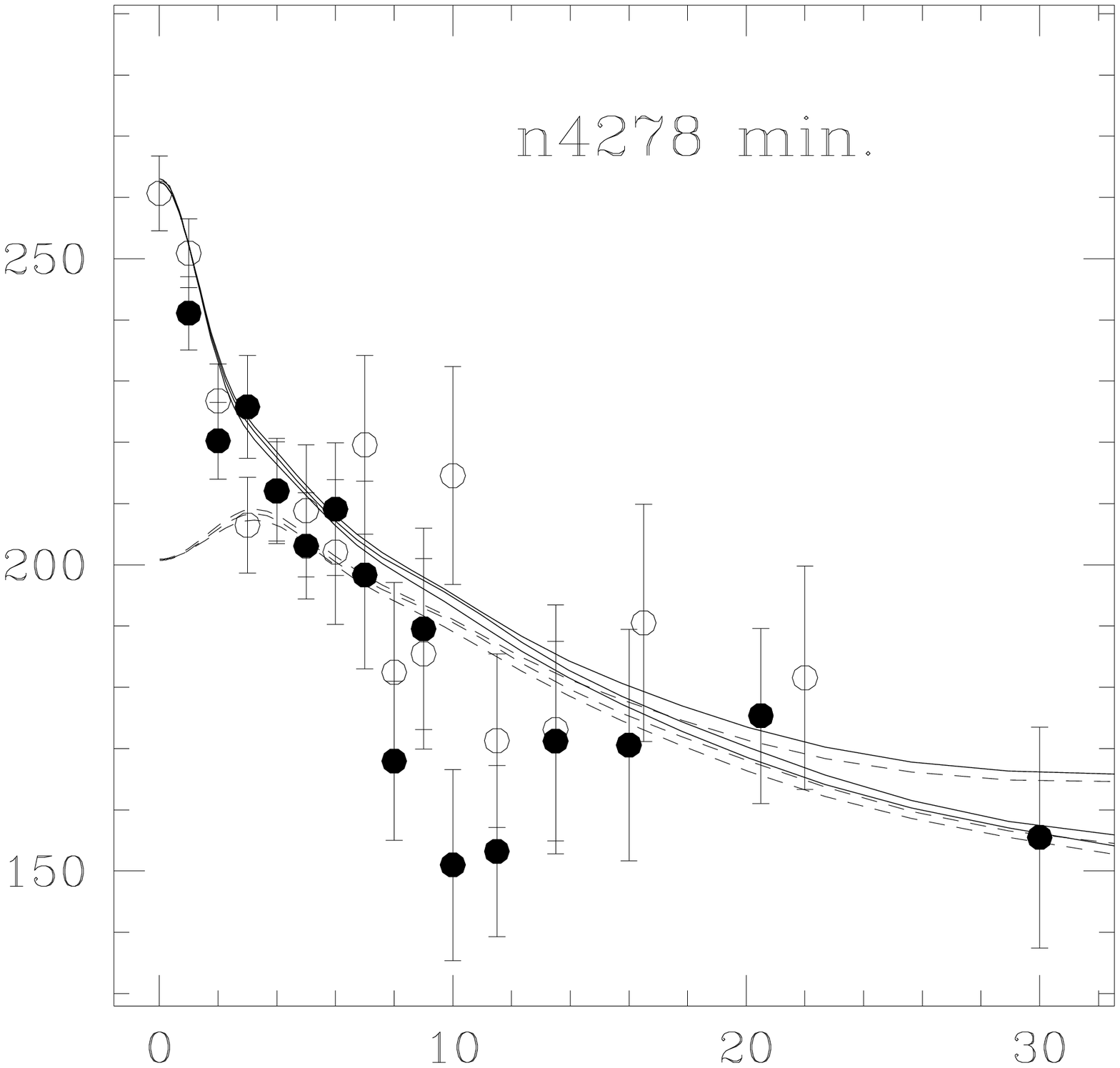,width=0.25\hsize}
\psfig{file=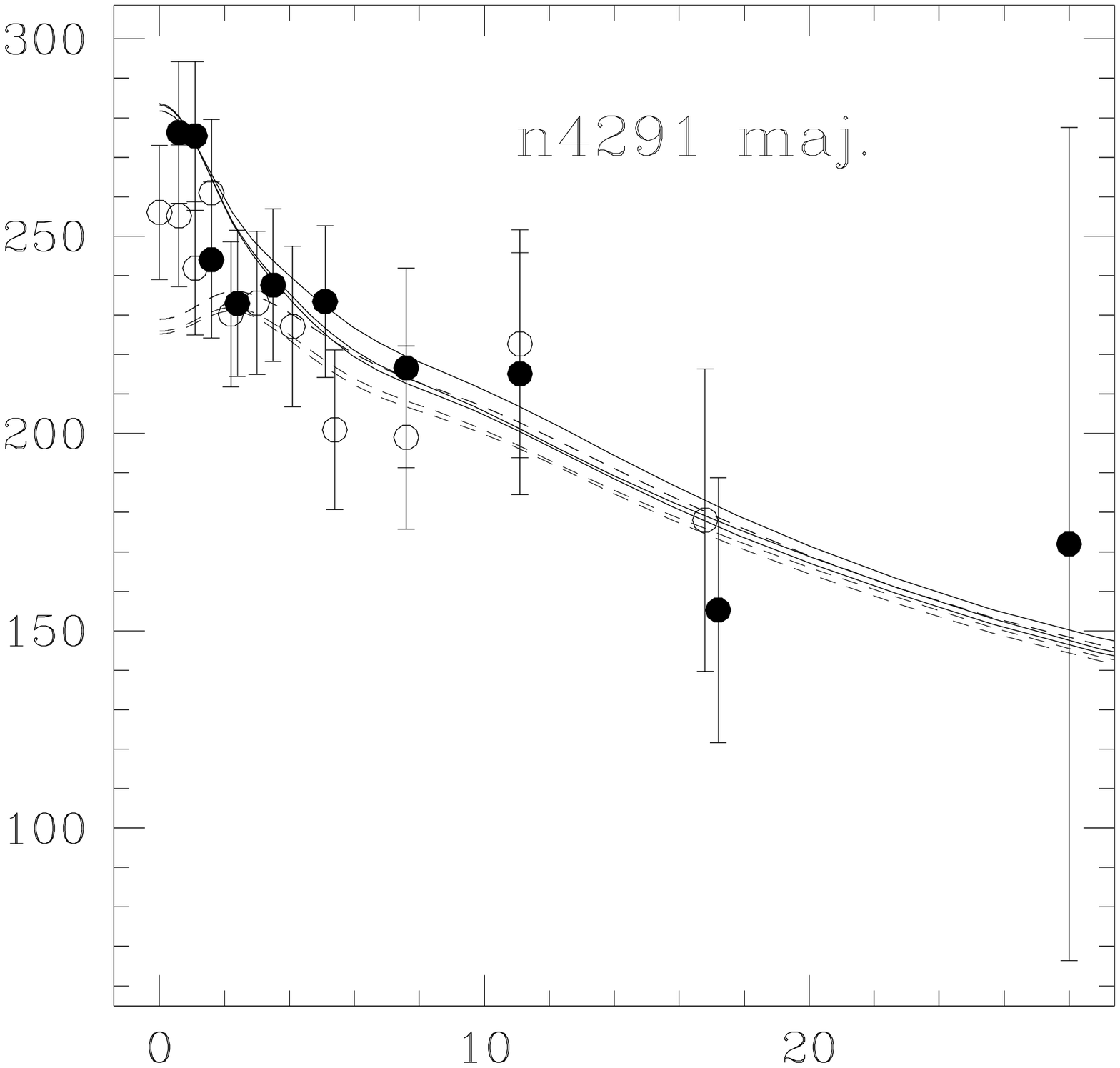,width=0.25\hsize}
}
}{\captioncont}
\figure{
\centerline{
\psfig{file=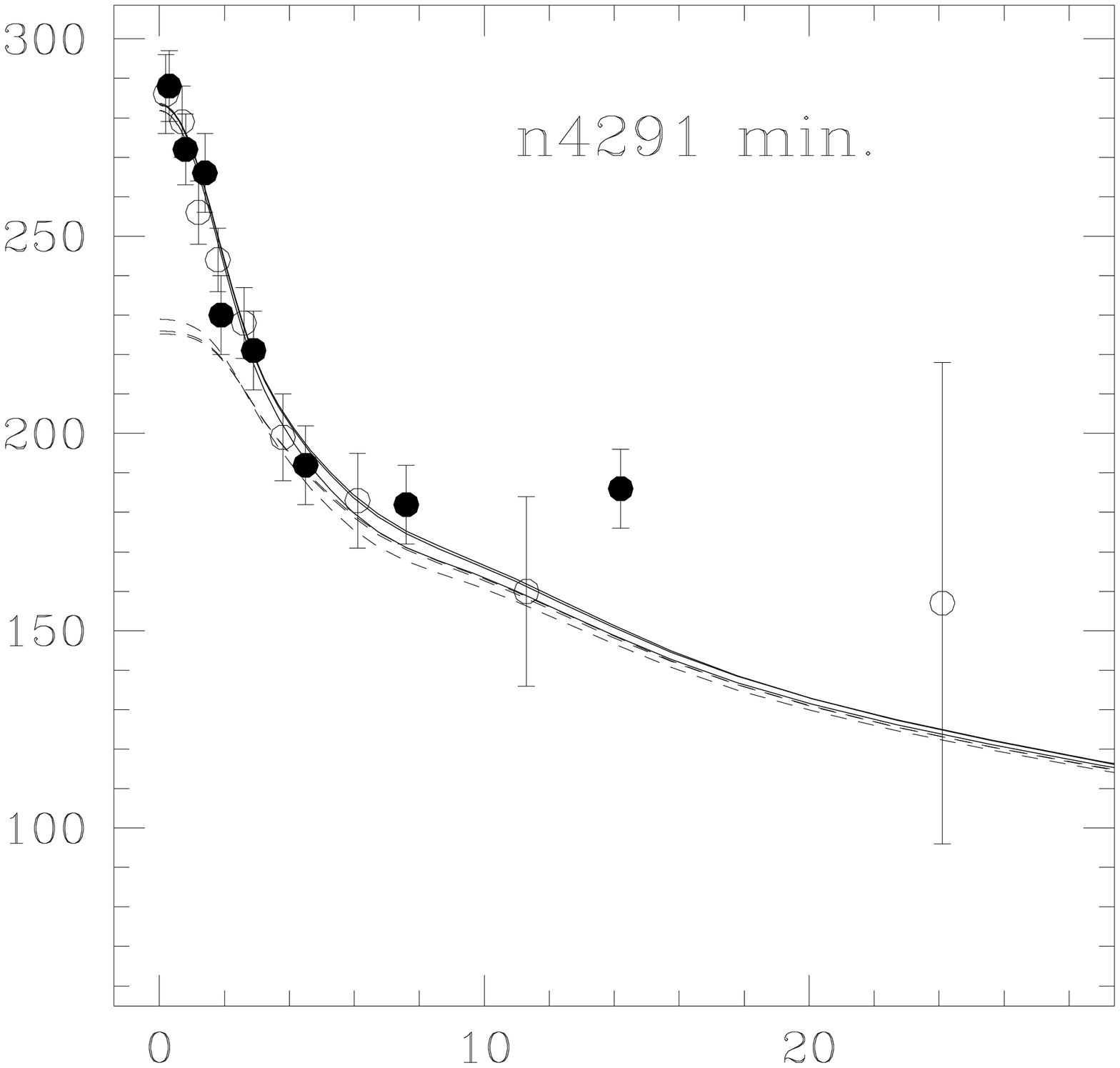,width=0.25\hsize}
\psfig{file=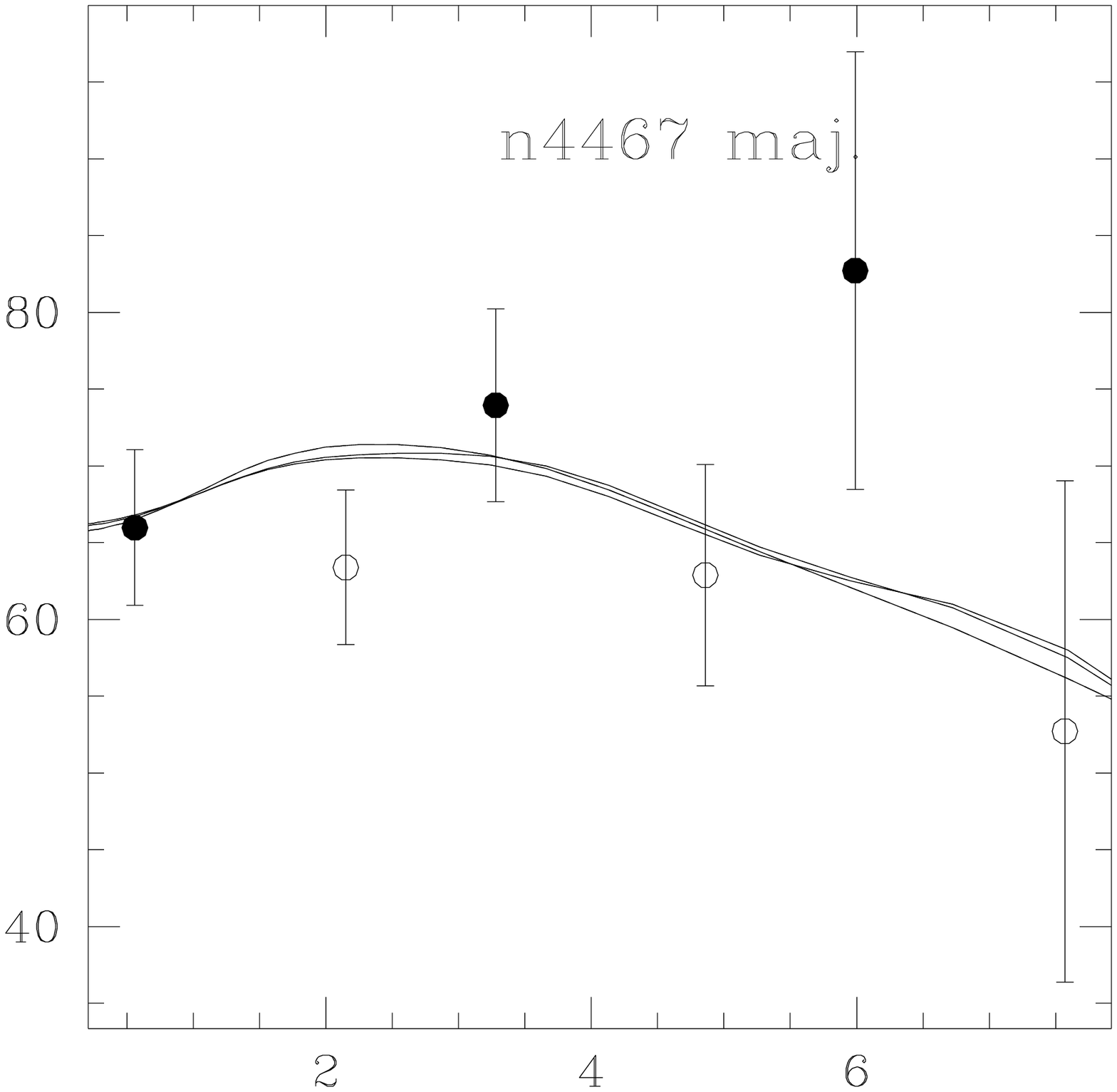,width=0.25\hsize}
\psfig{file=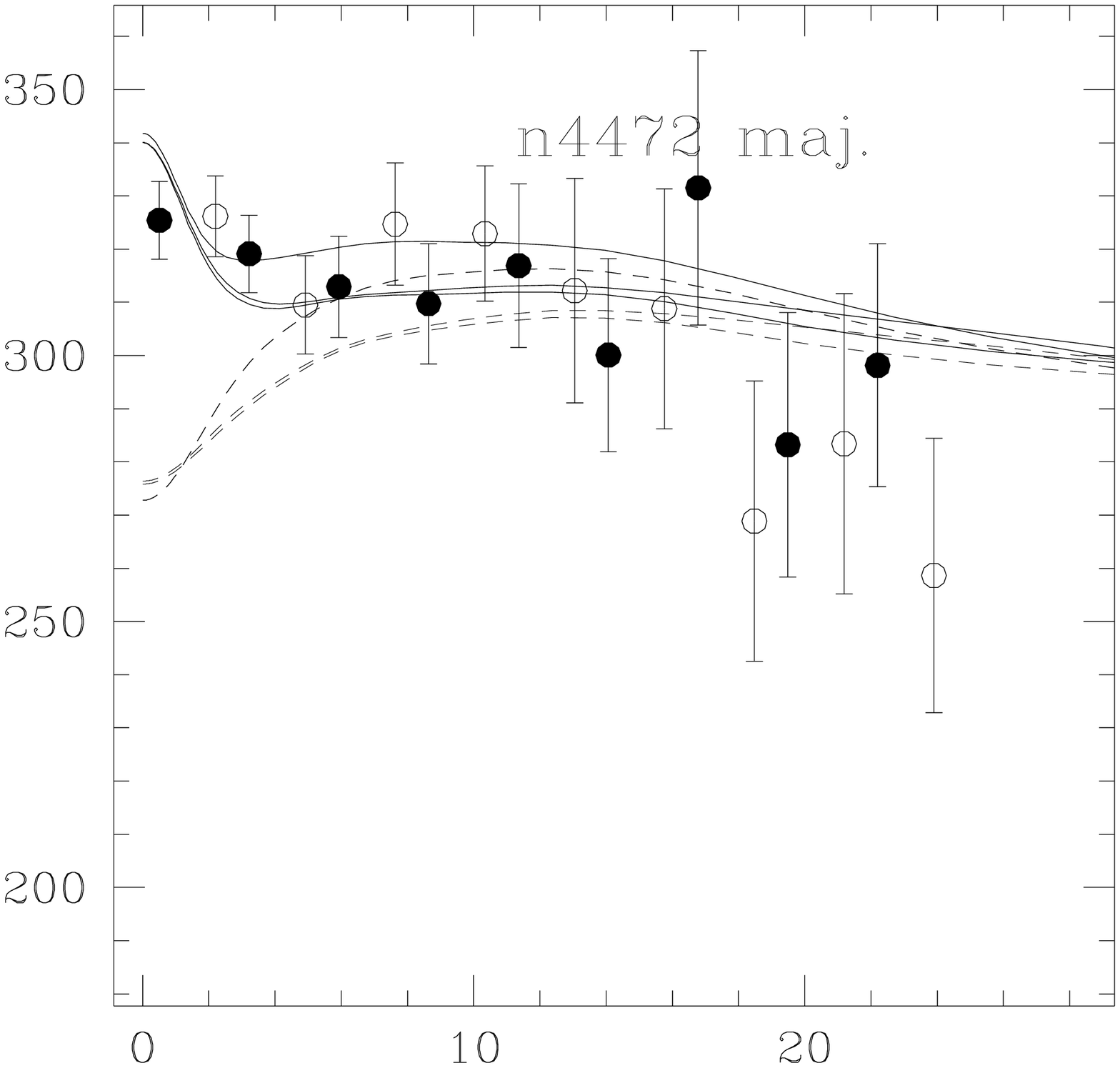,width=0.25\hsize}
\psfig{file=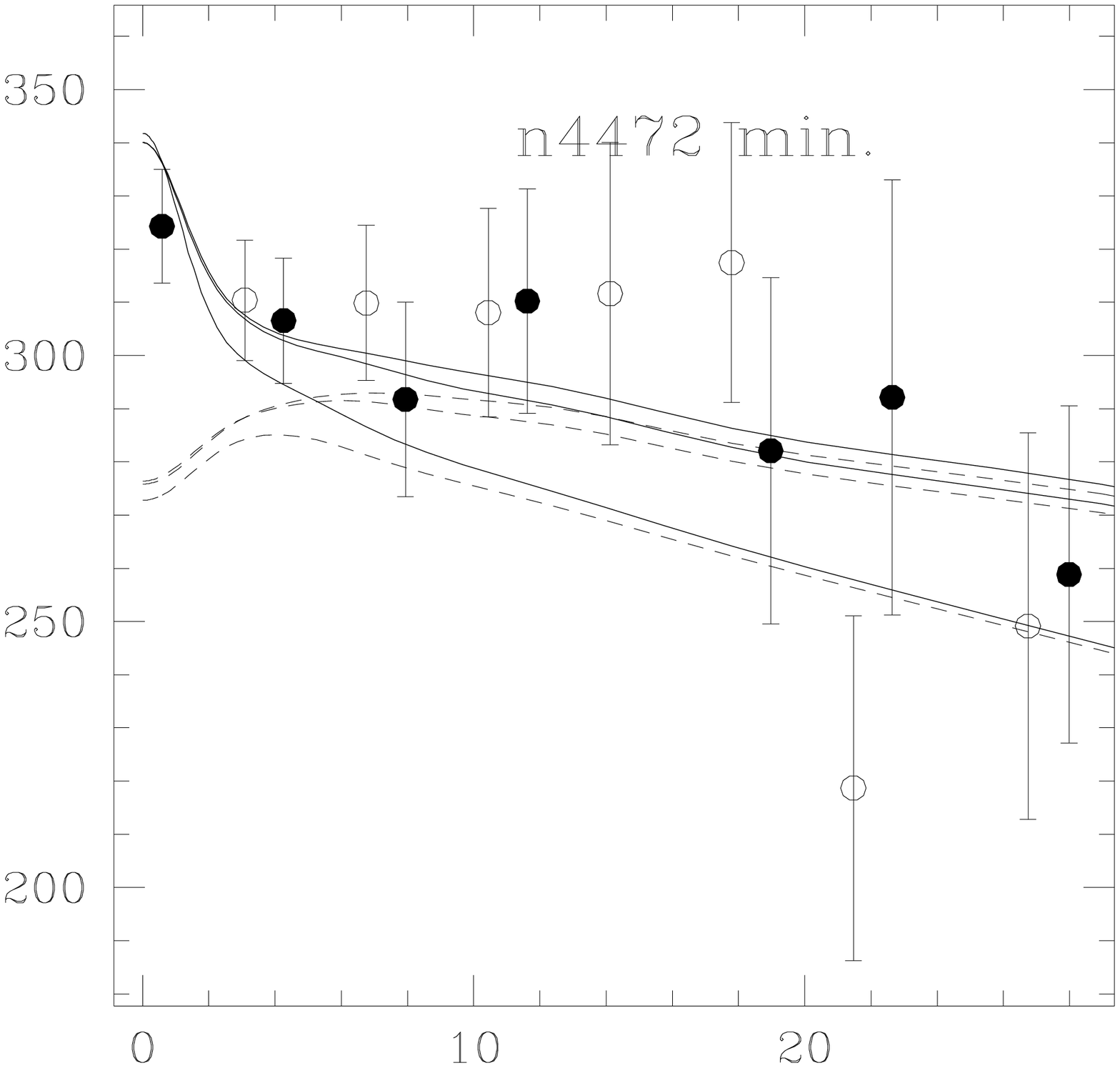,width=0.25\hsize}
}
\centerline{
\psfig{file=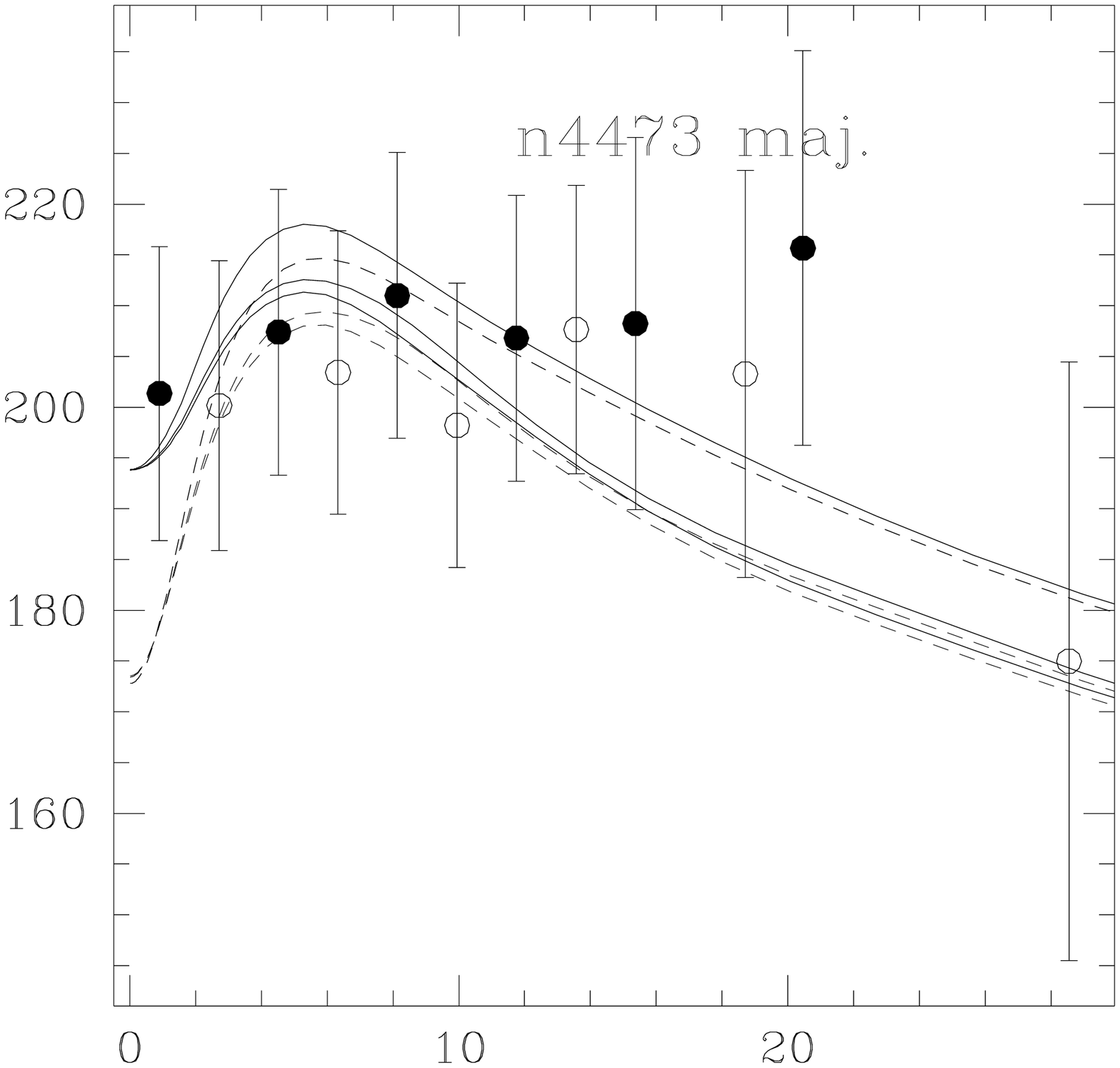,width=0.25\hsize}
\psfig{file=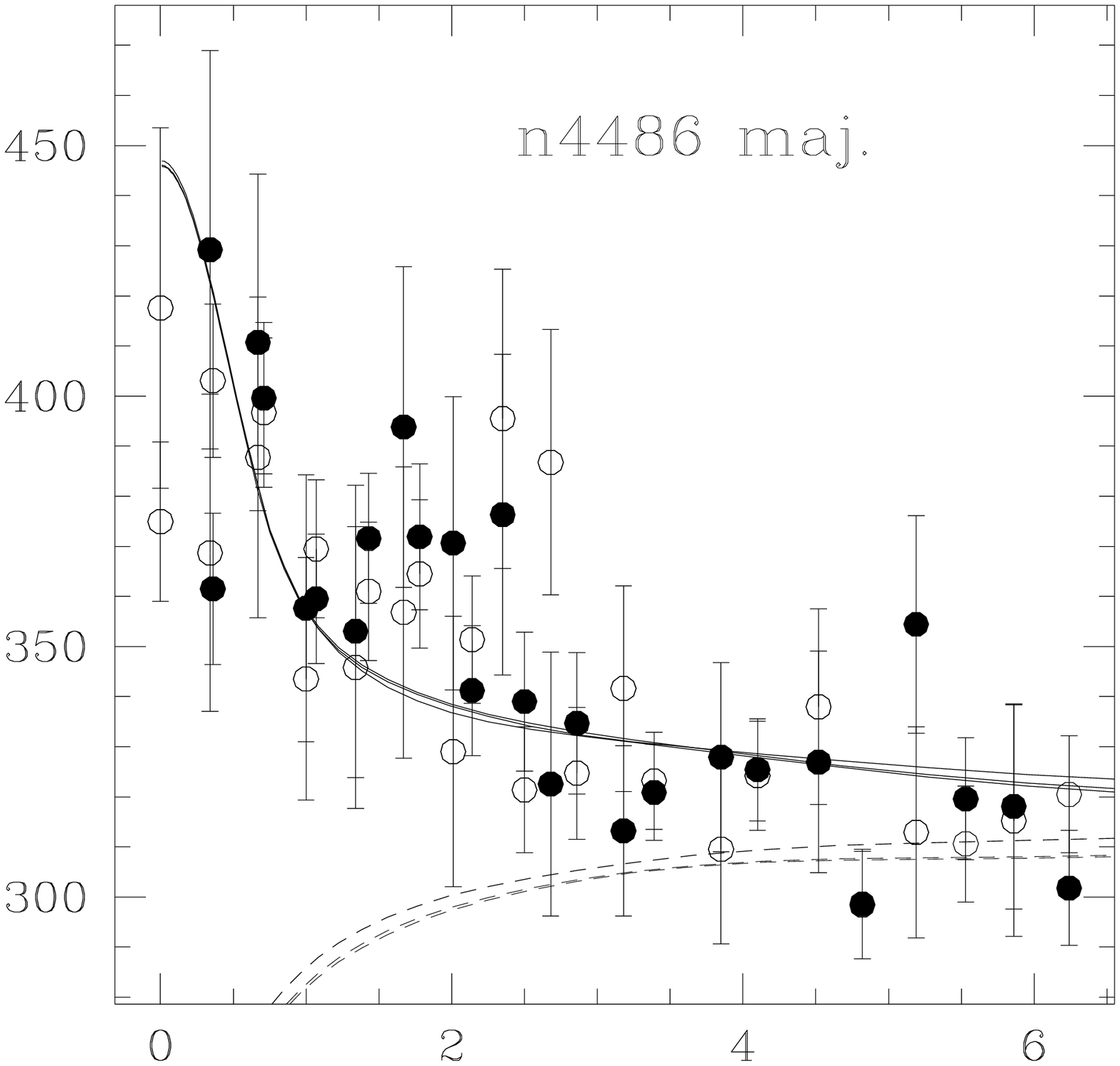,width=0.25\hsize}
\psfig{file=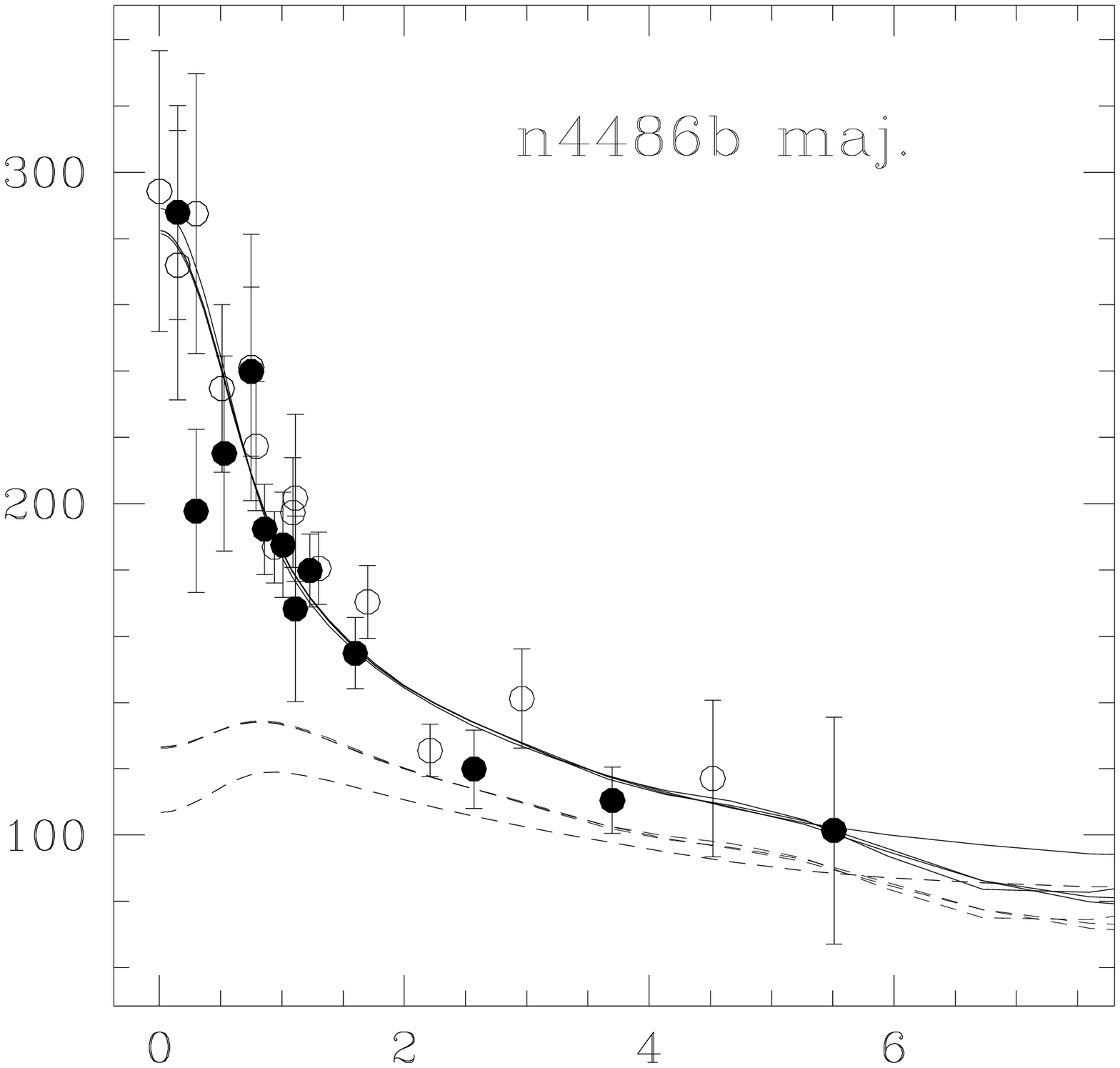,width=0.25\hsize}
\psfig{file=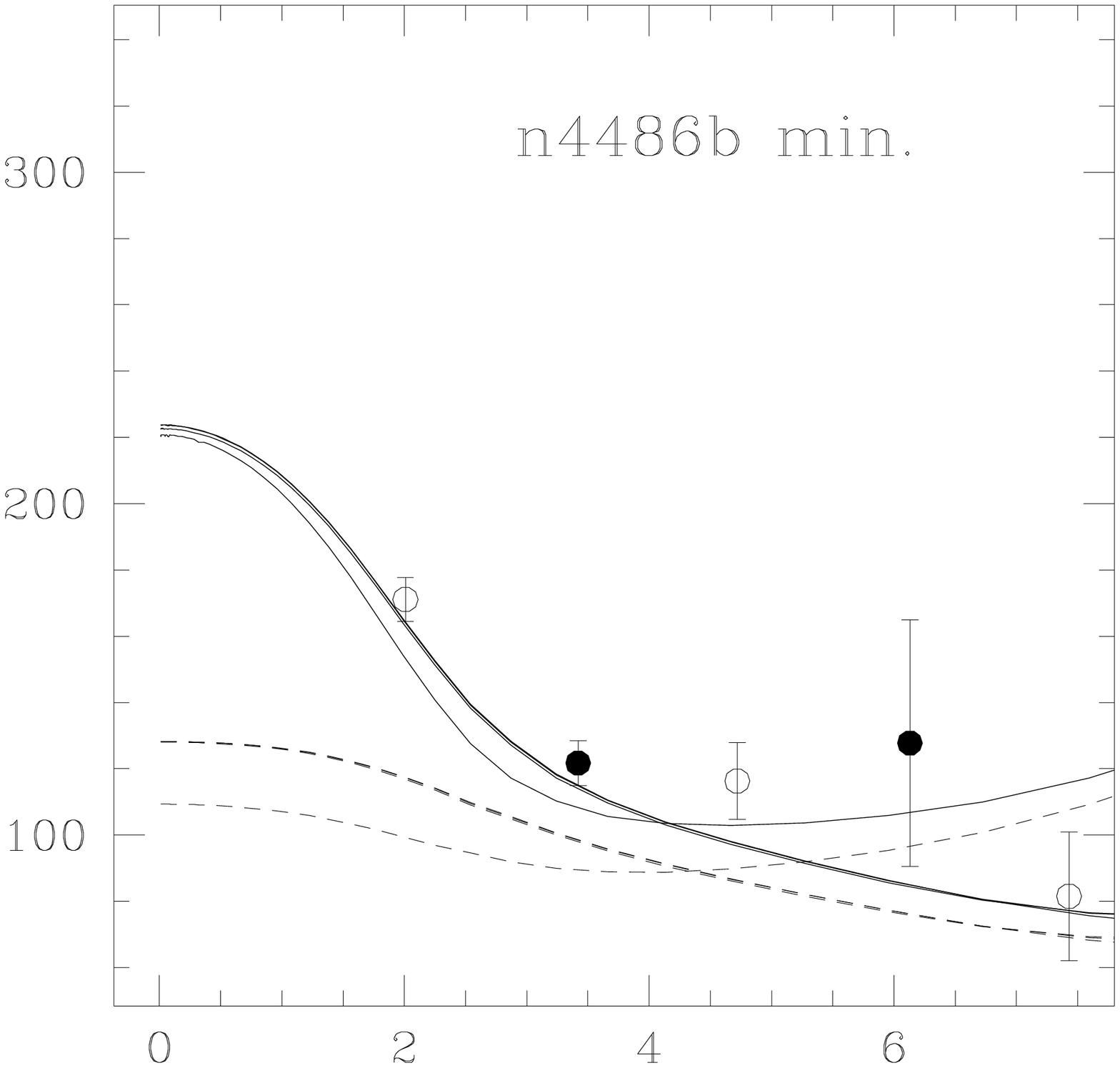,width=0.25\hsize}
}
\centerline{
\psfig{file=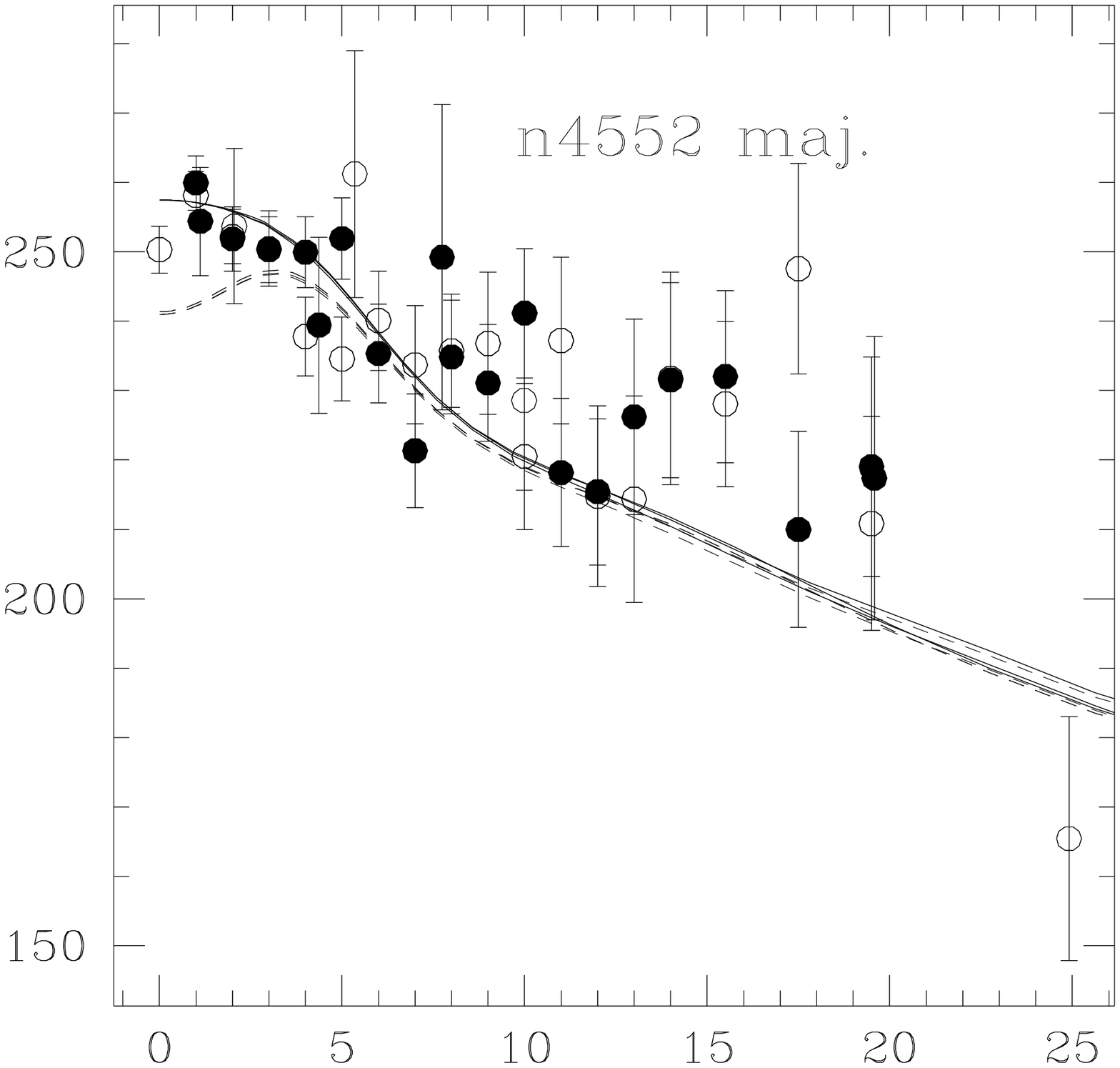,width=0.25\hsize}
\psfig{file=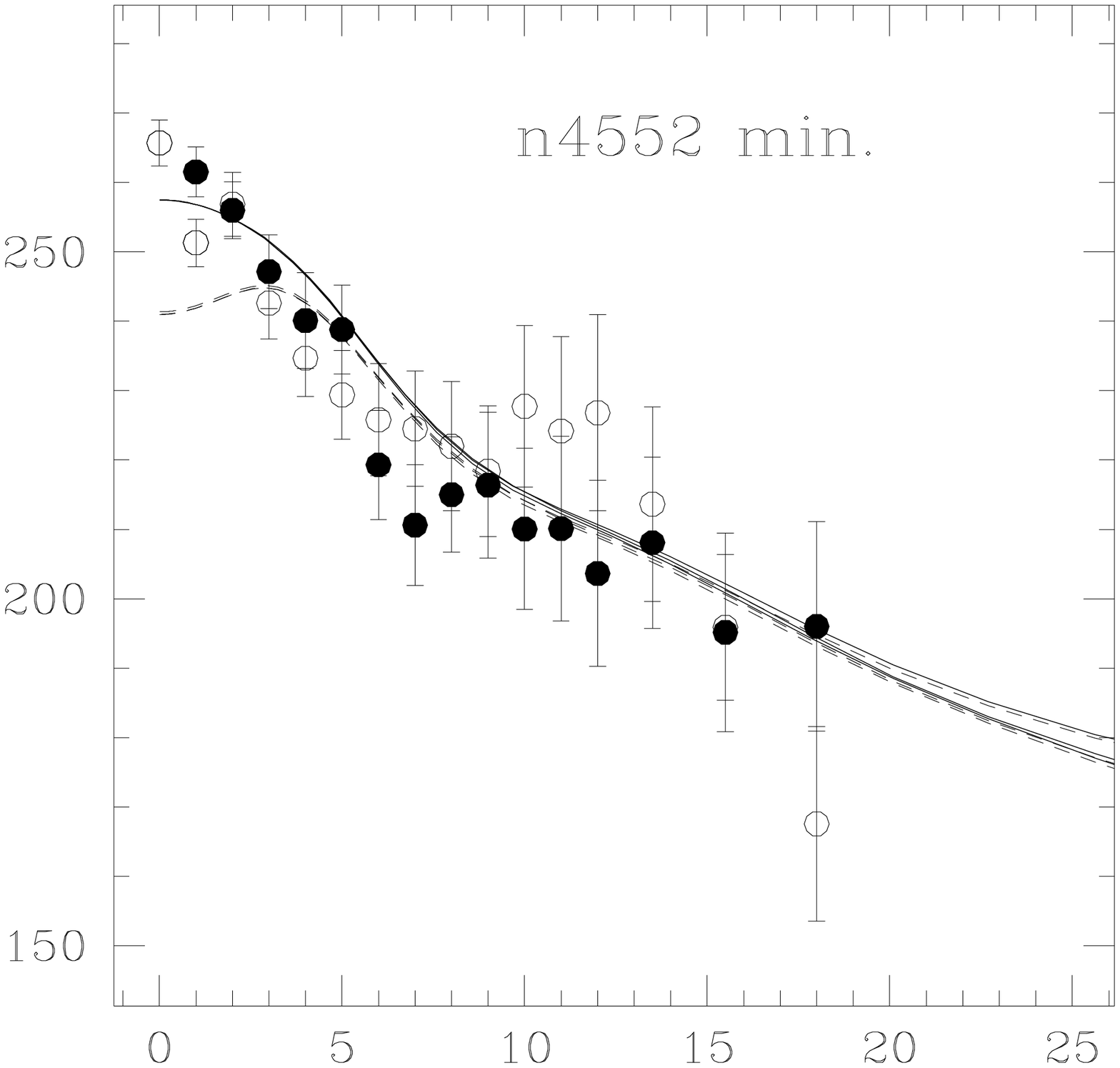,width=0.25\hsize}
\psfig{file=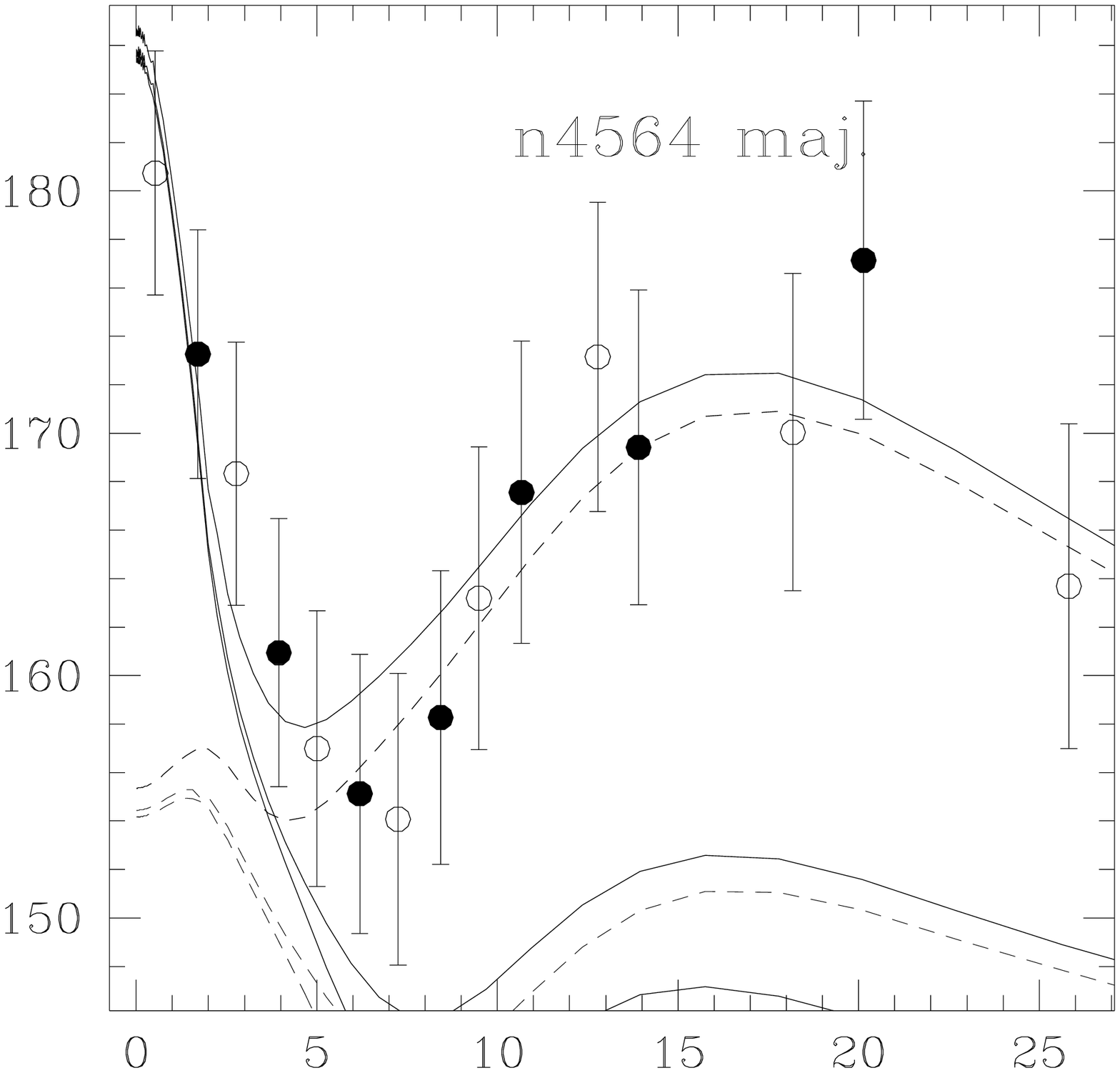,width=0.25\hsize}
\psfig{file=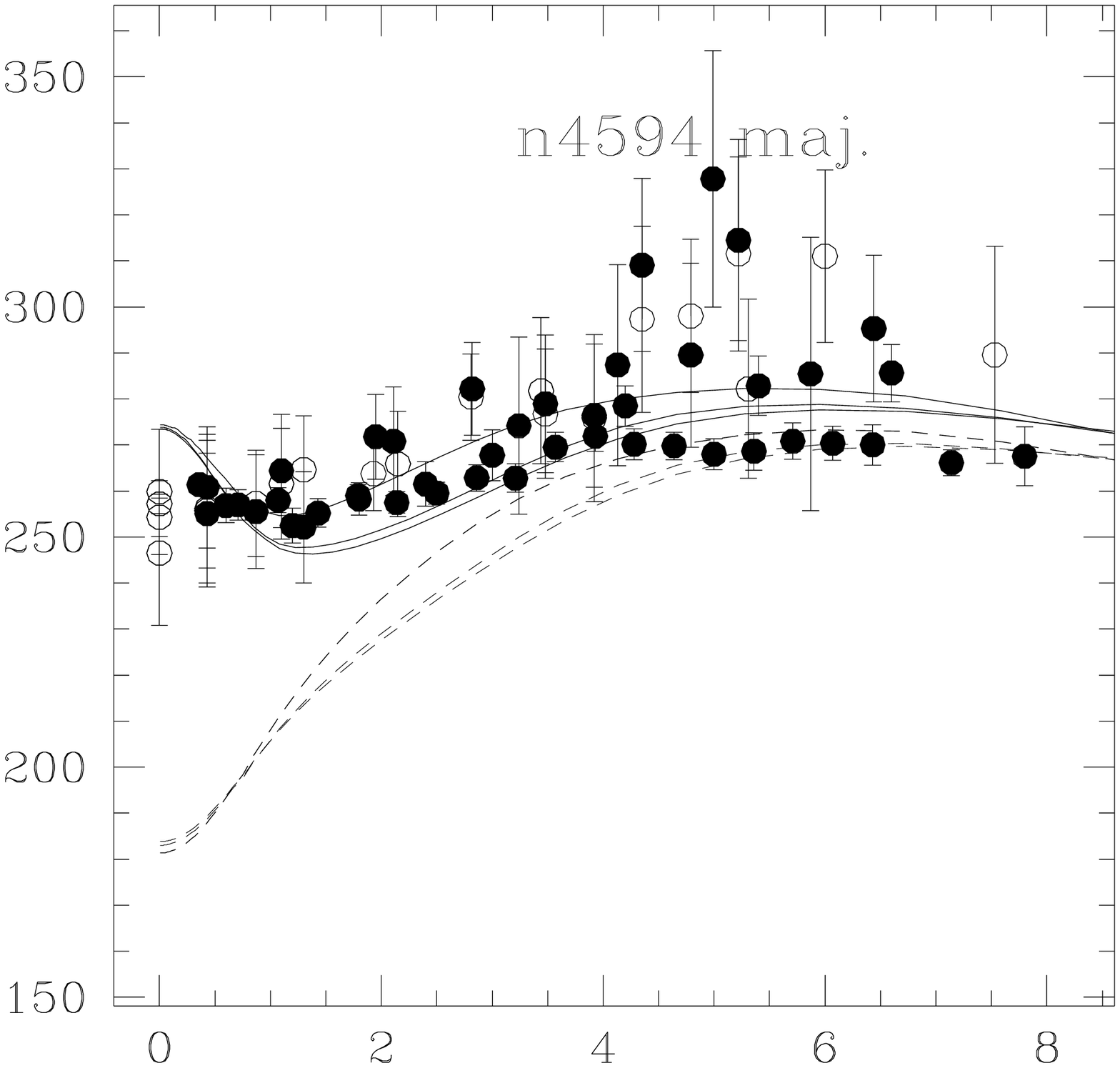,width=0.25\hsize}
}
\centerline{
\psfig{file=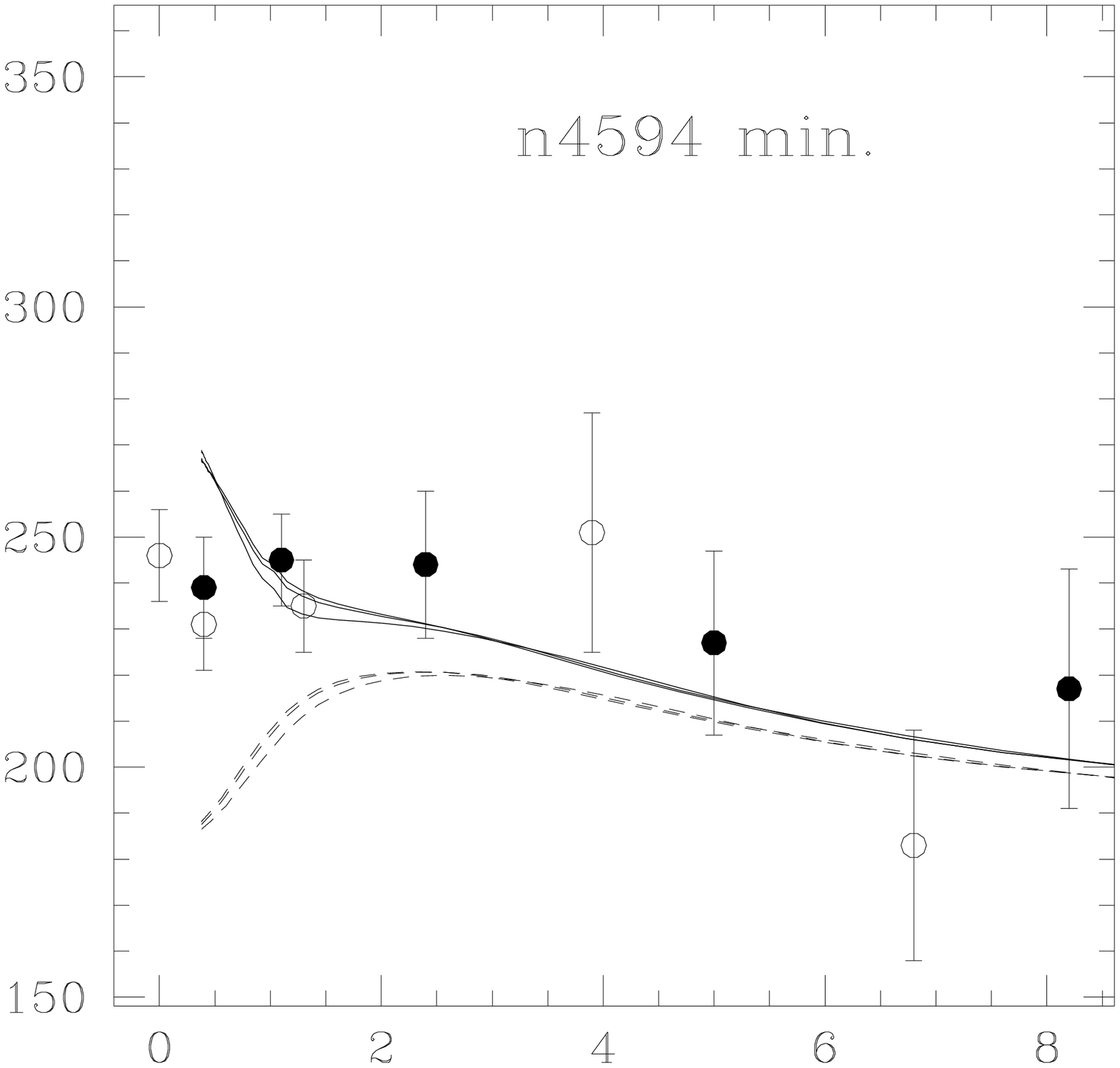,width=0.25\hsize}
\psfig{file=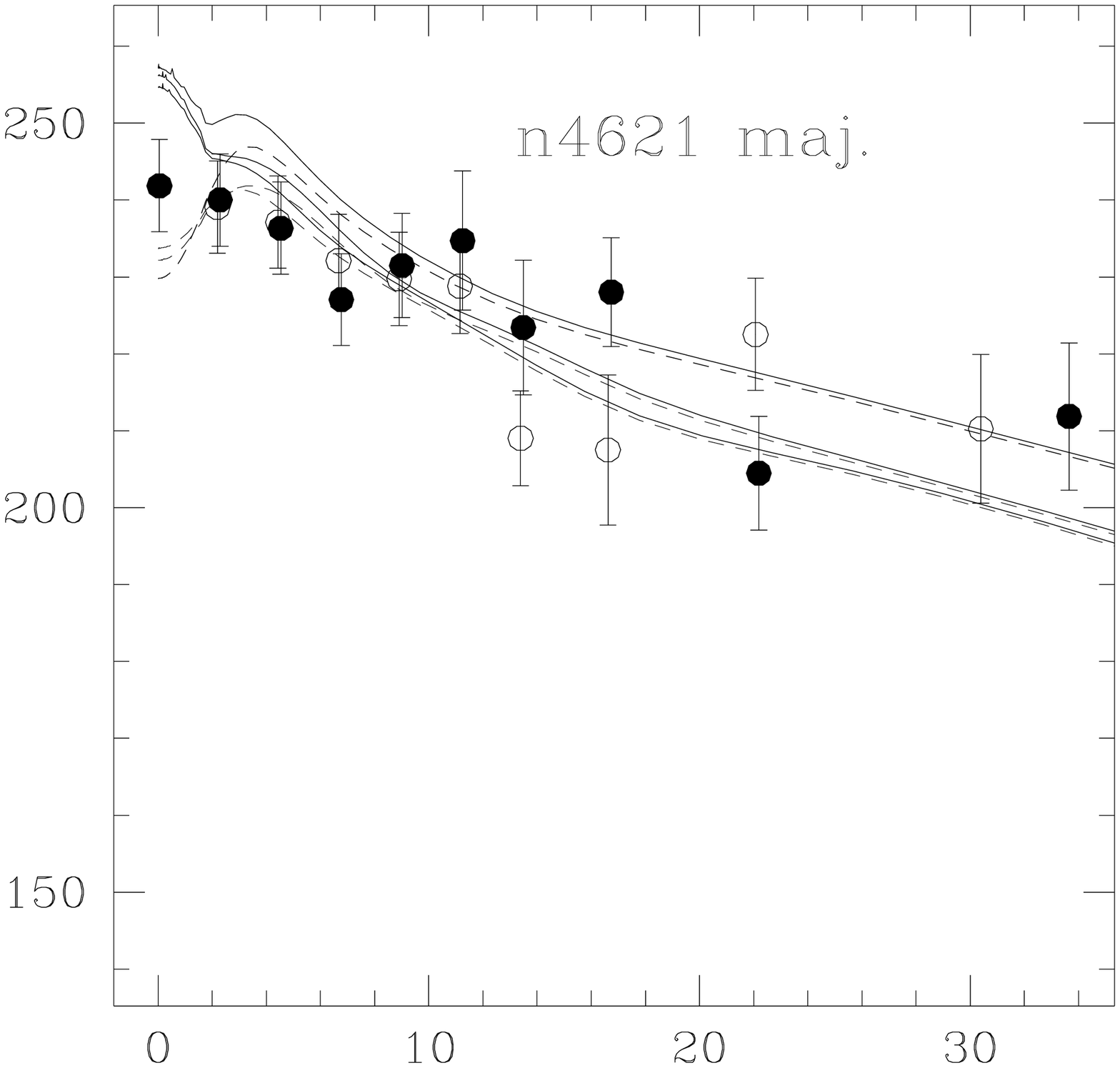,width=0.25\hsize}
\psfig{file=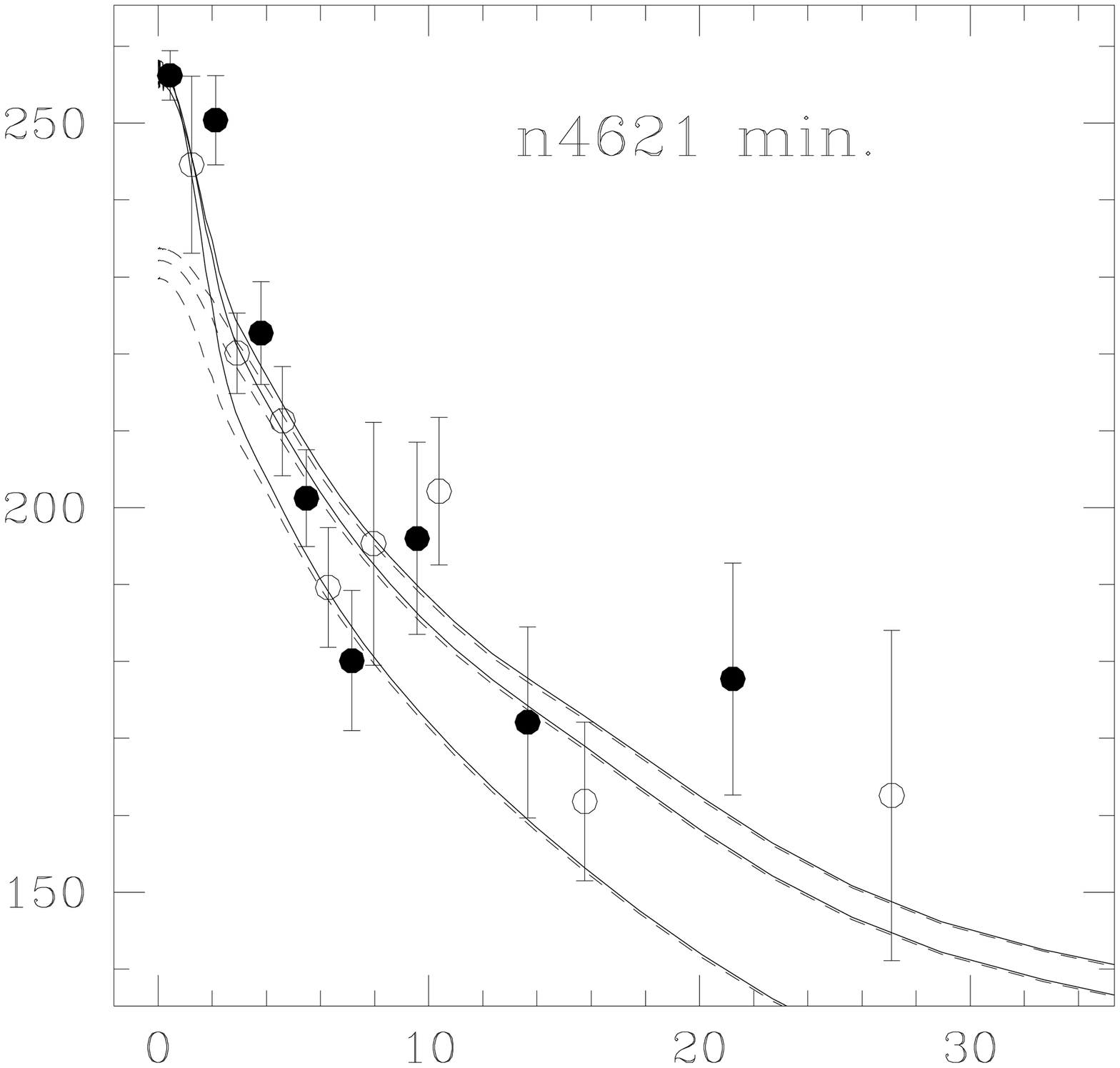,width=0.25\hsize}
\psfig{file=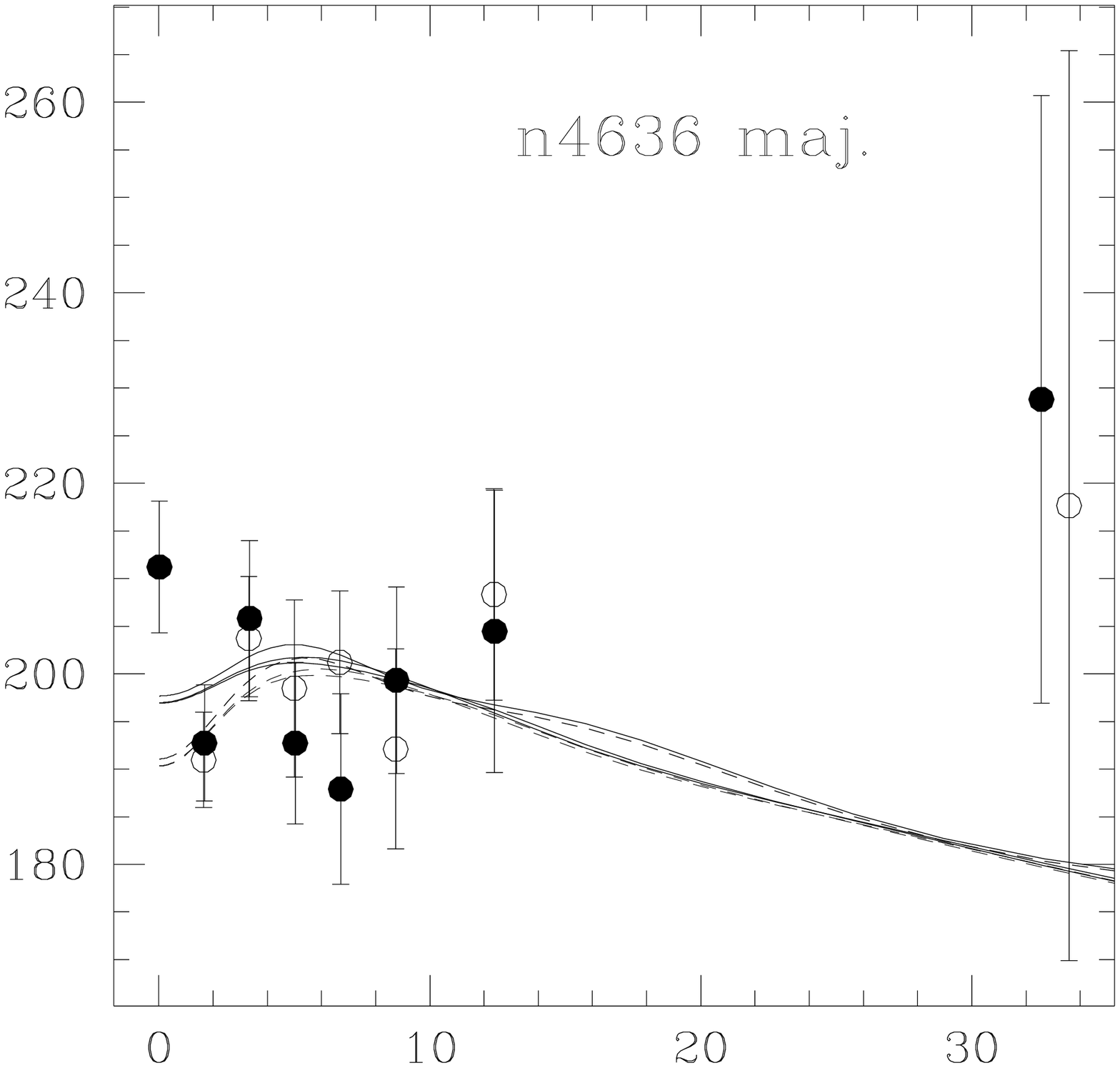,width=0.25\hsize}
}
}{\captioncont}
\figure{
\centerline{
\psfig{file=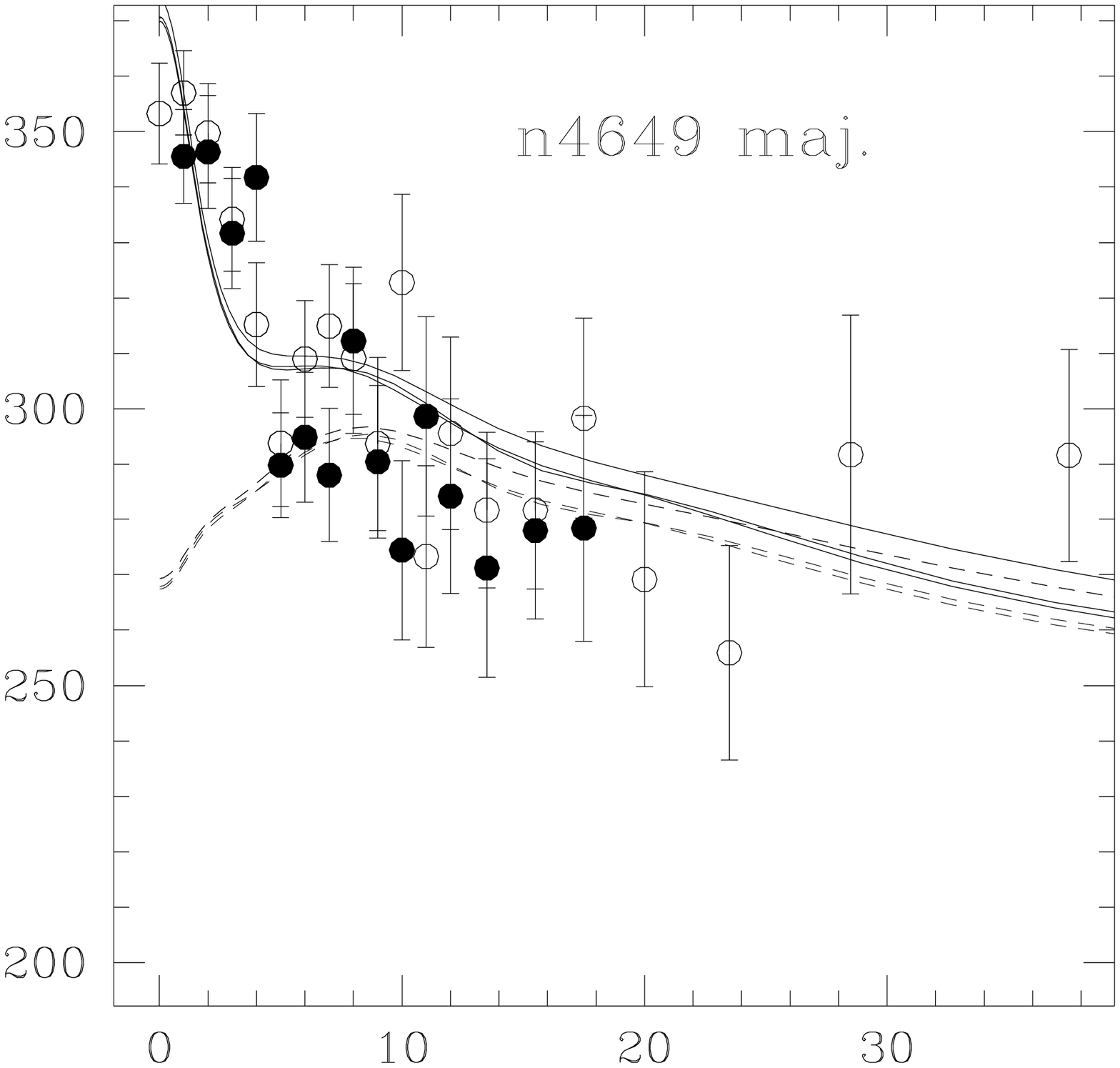,width=0.25\hsize}
\psfig{file=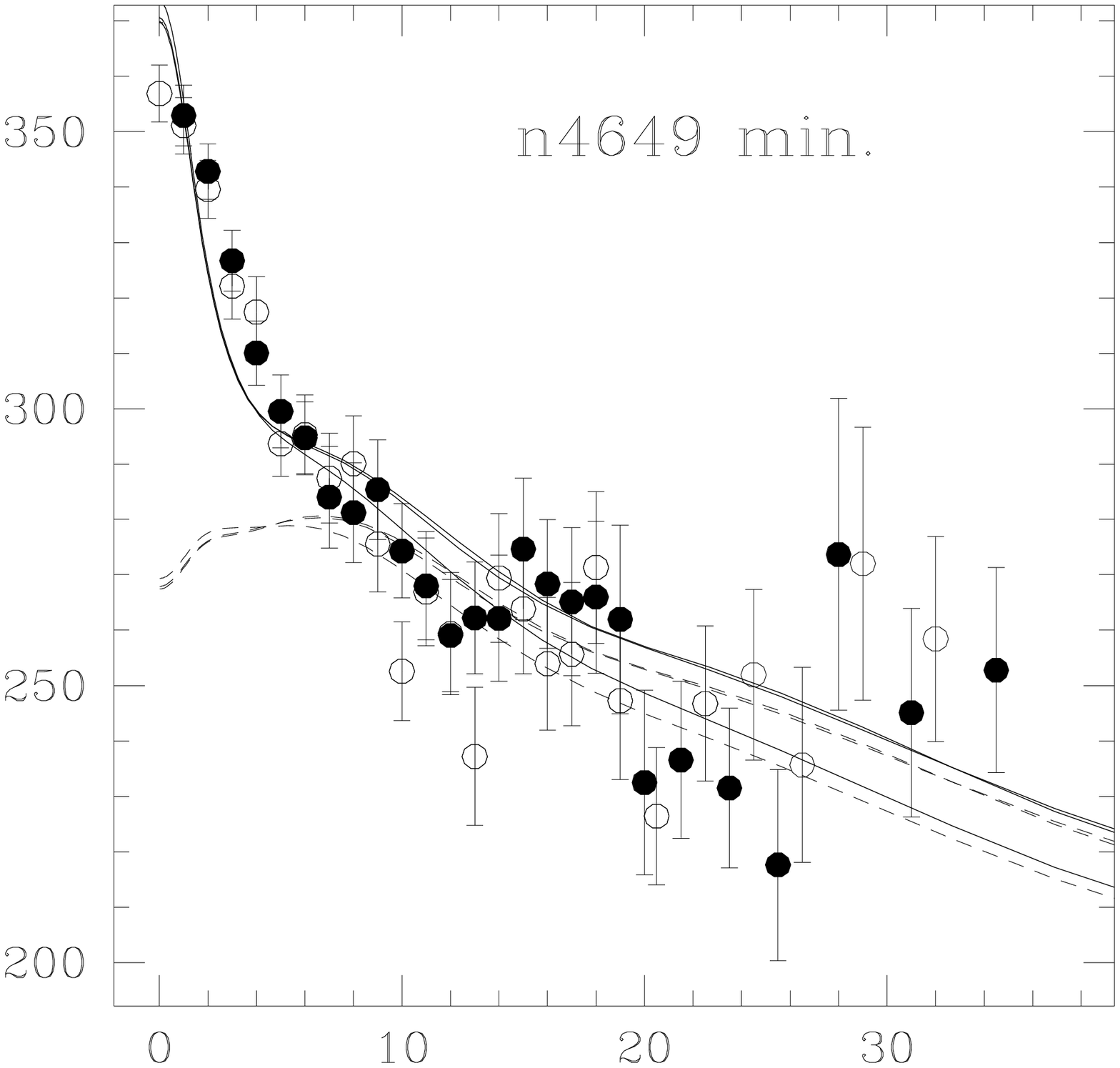,width=0.25\hsize}
\psfig{file=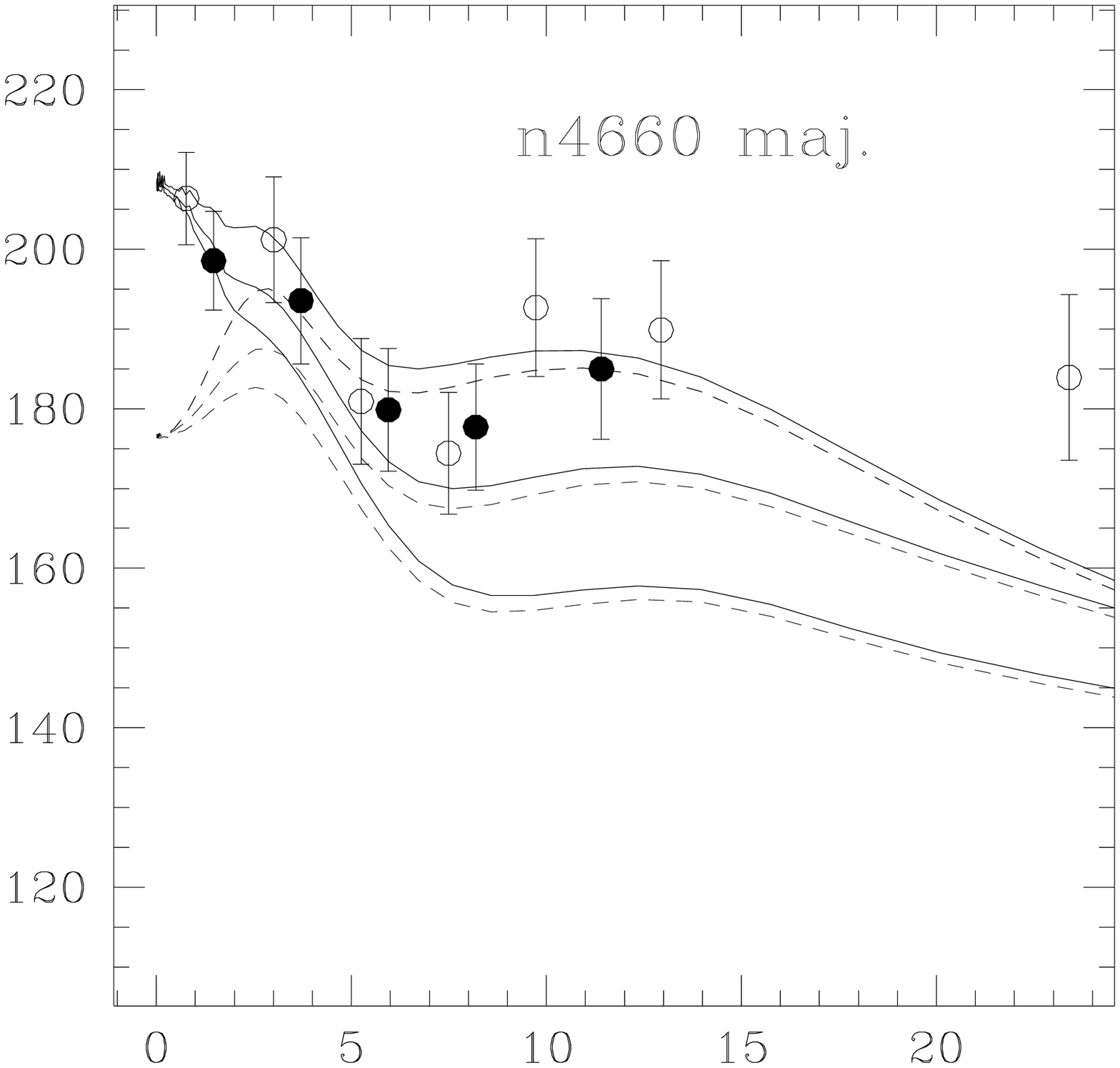,width=0.25\hsize}
\psfig{file=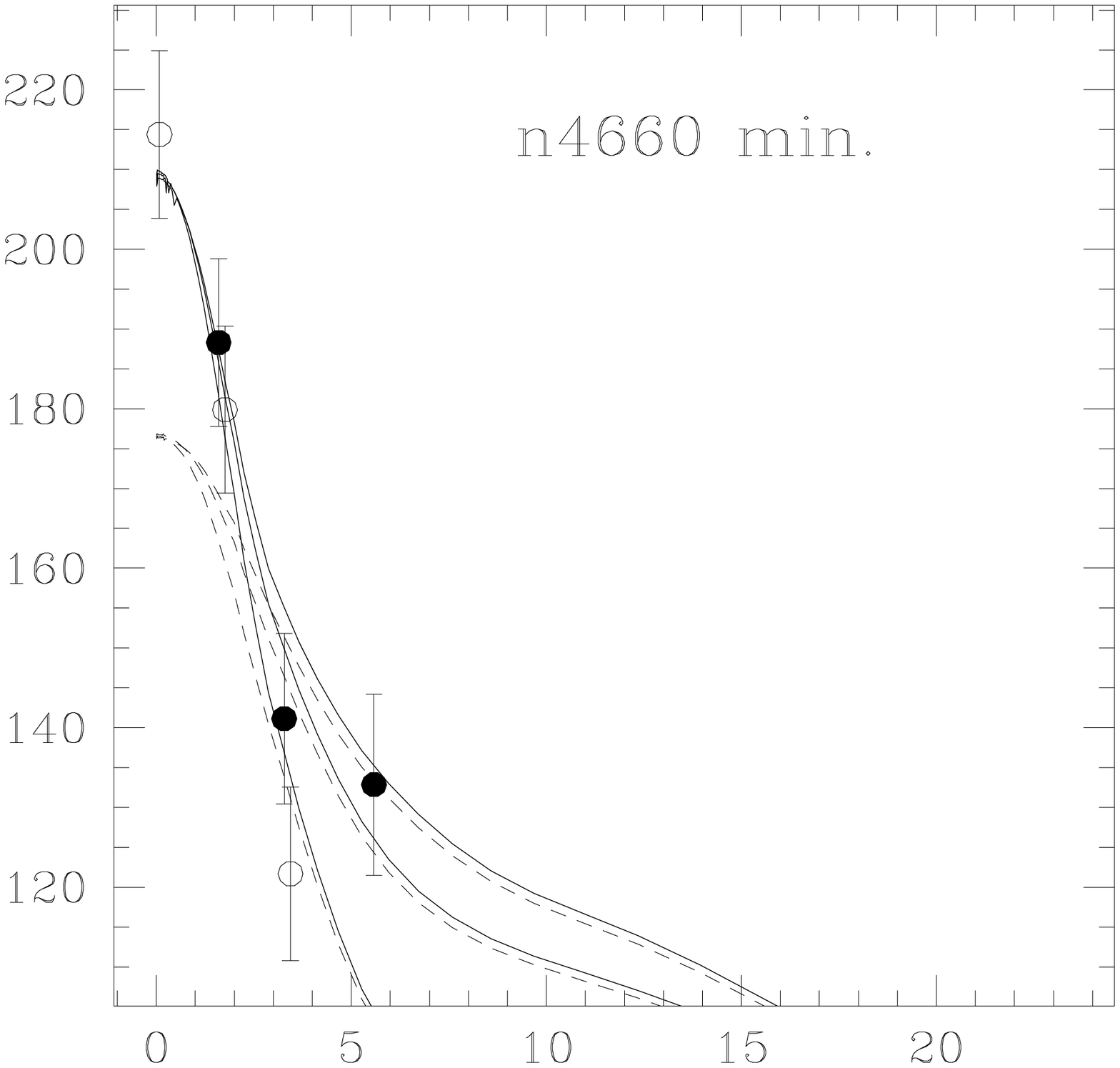,width=0.25\hsize}
}
\centerline{
\psfig{file=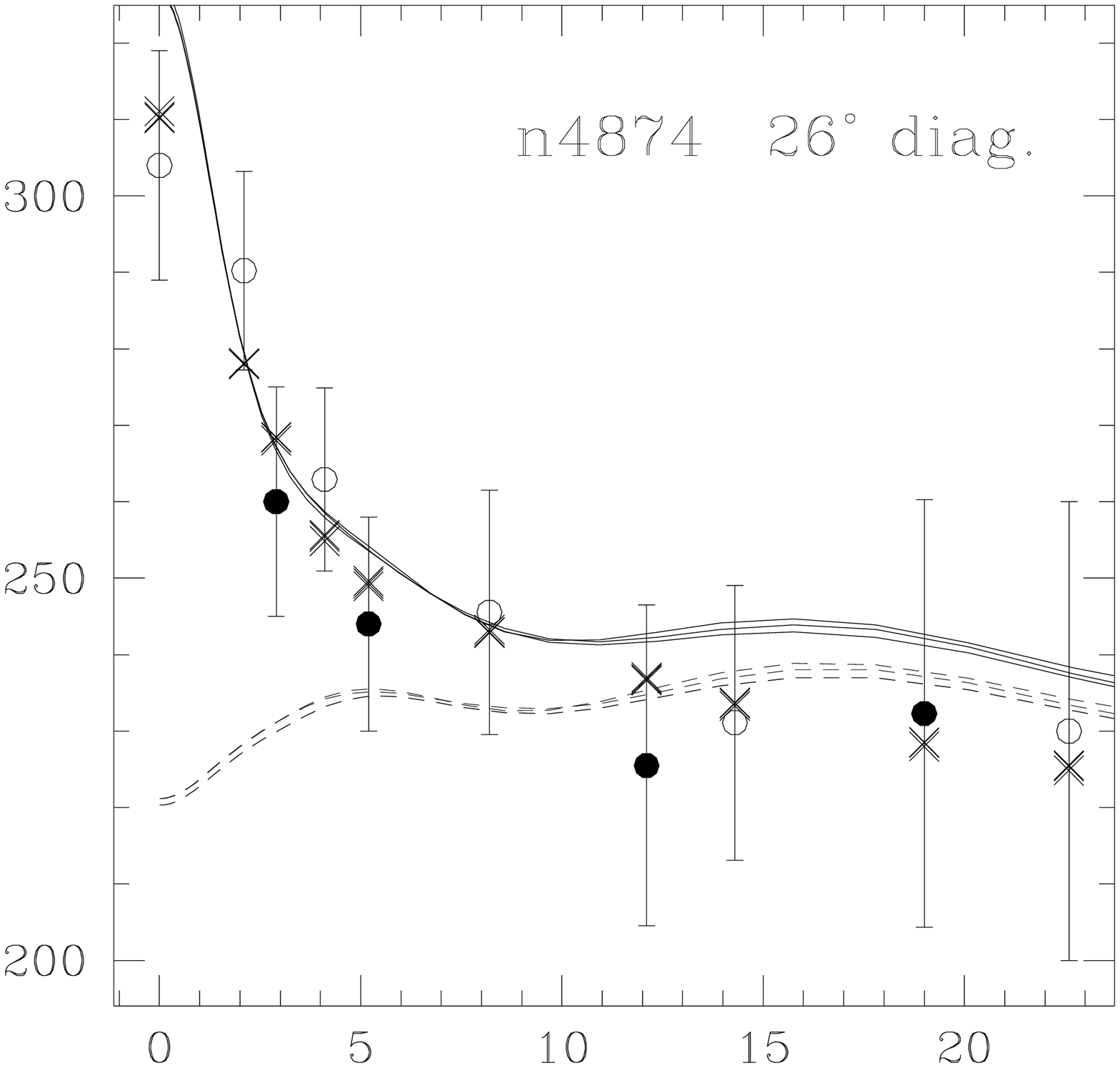,width=0.25\hsize}
\psfig{file=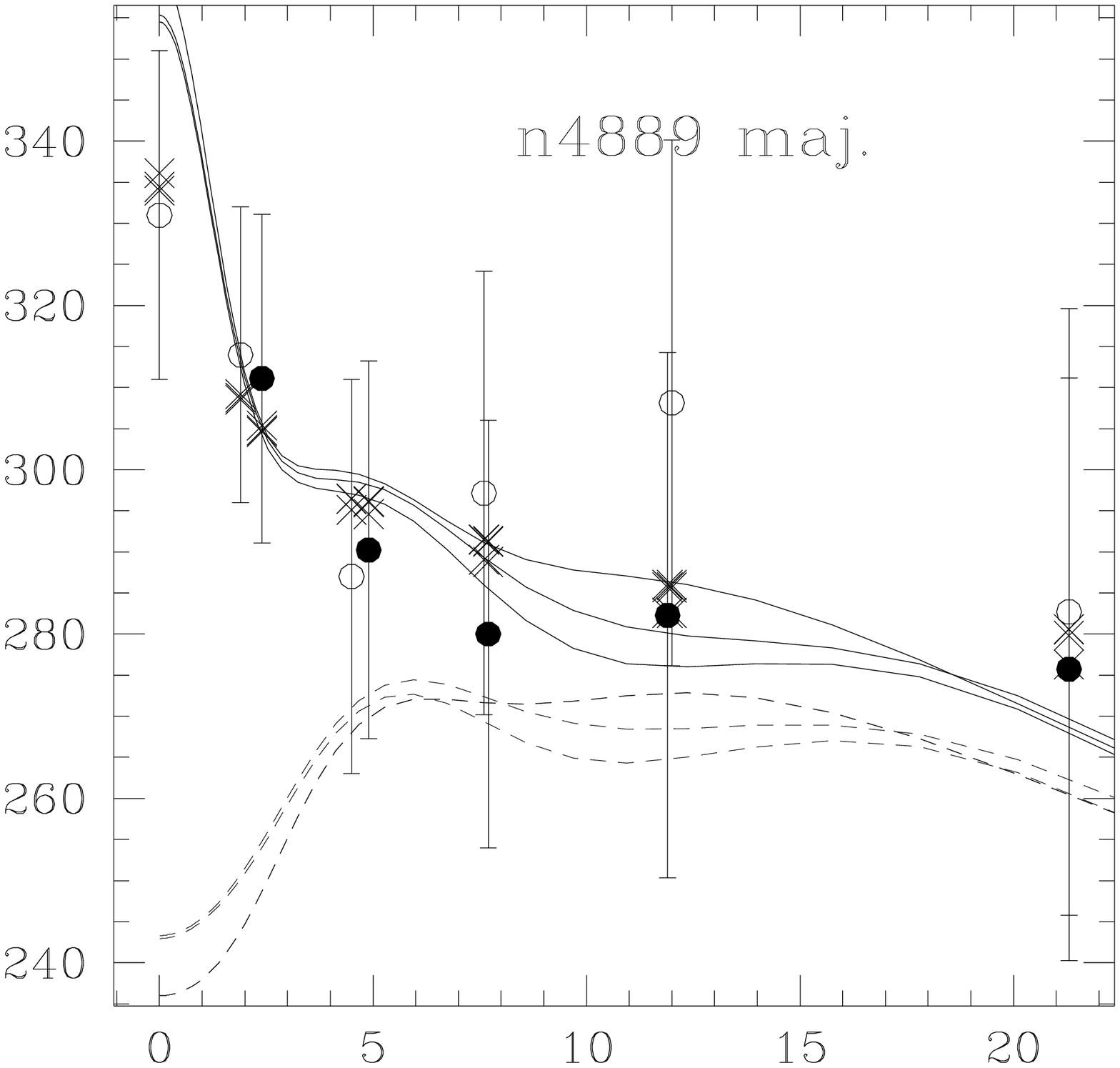,width=0.25\hsize}
\psfig{file=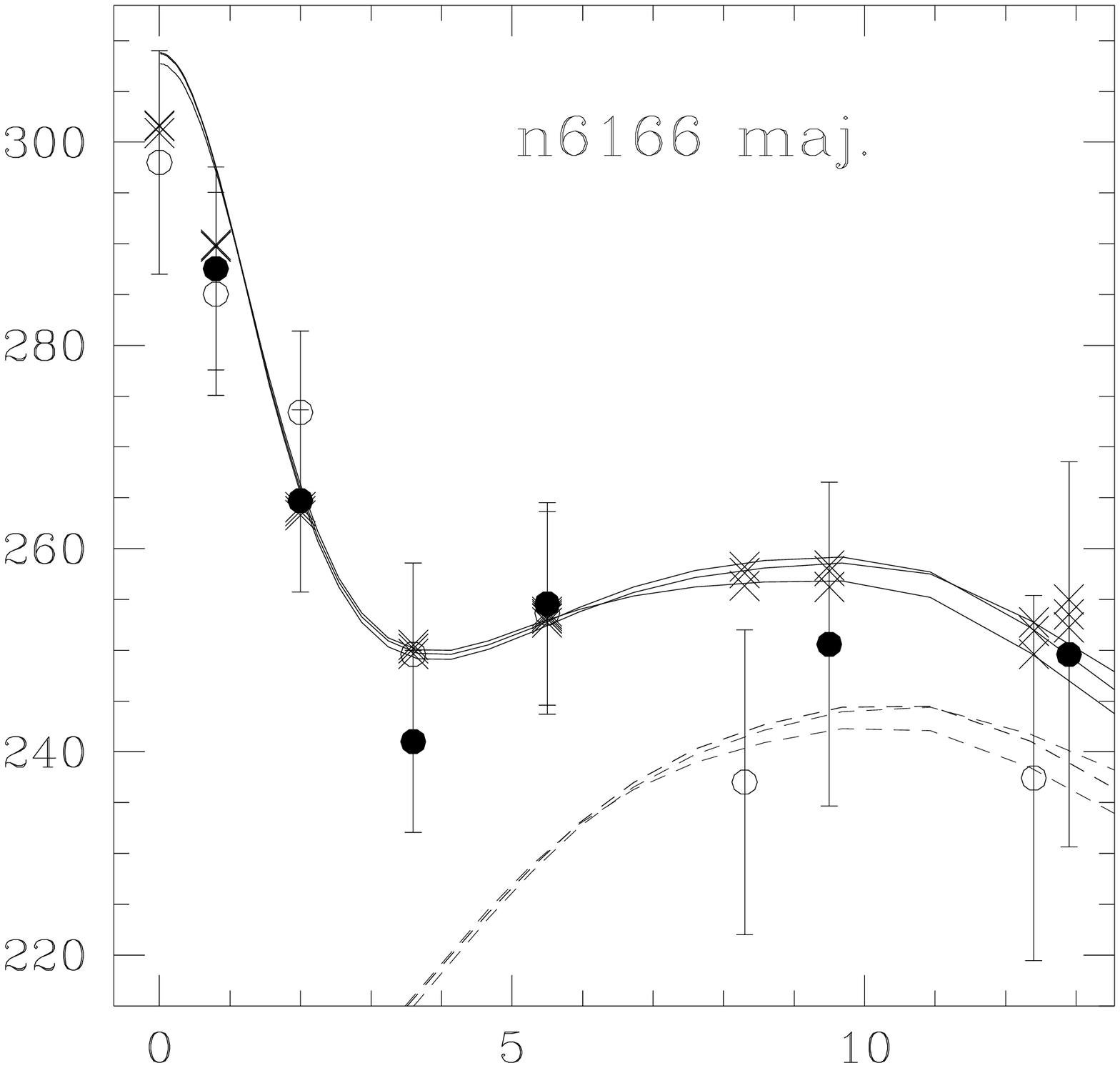,width=0.25\hsize}
\psfig{file=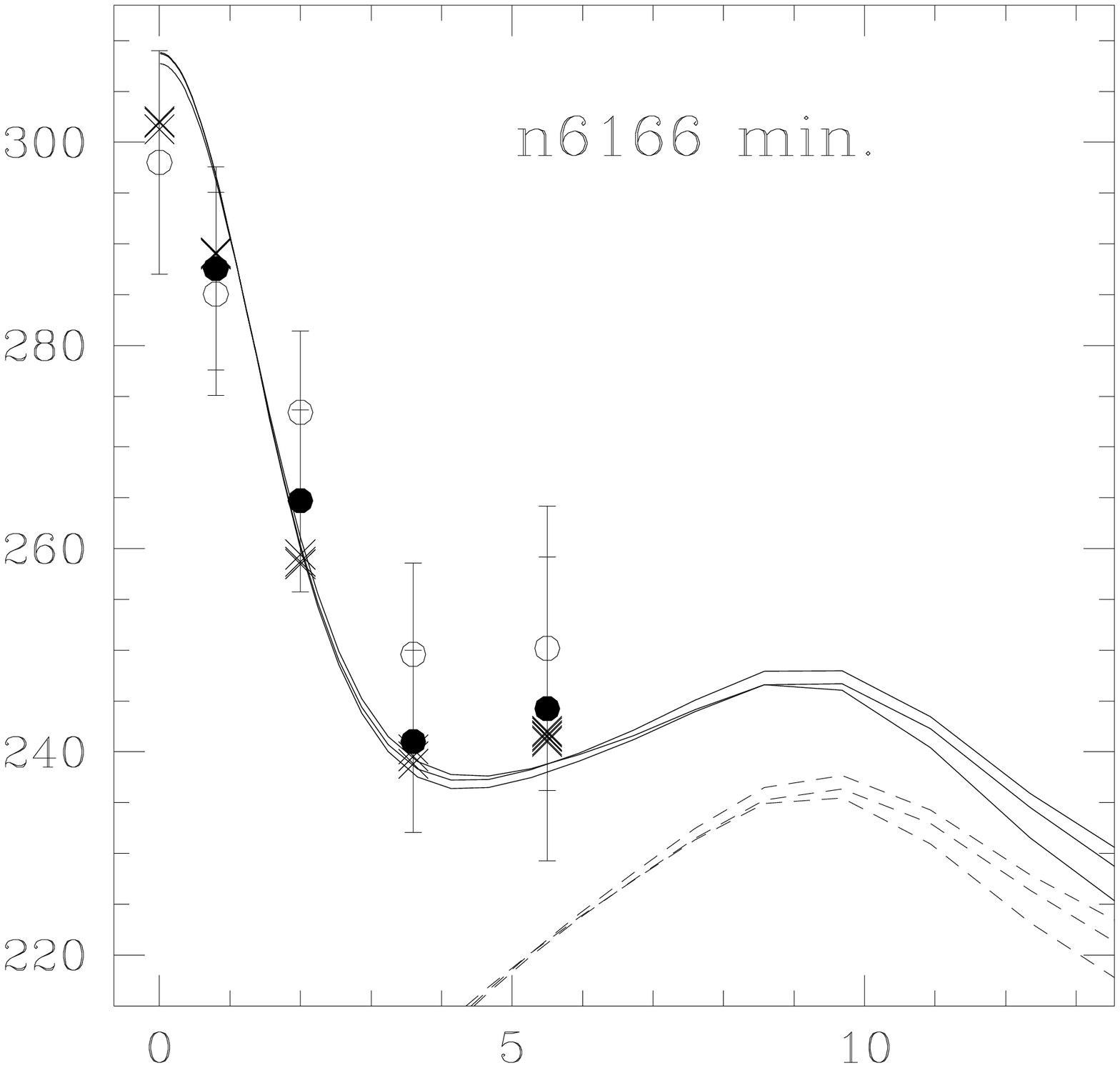,width=0.25\hsize}
}
\centerline{
\psfig{file=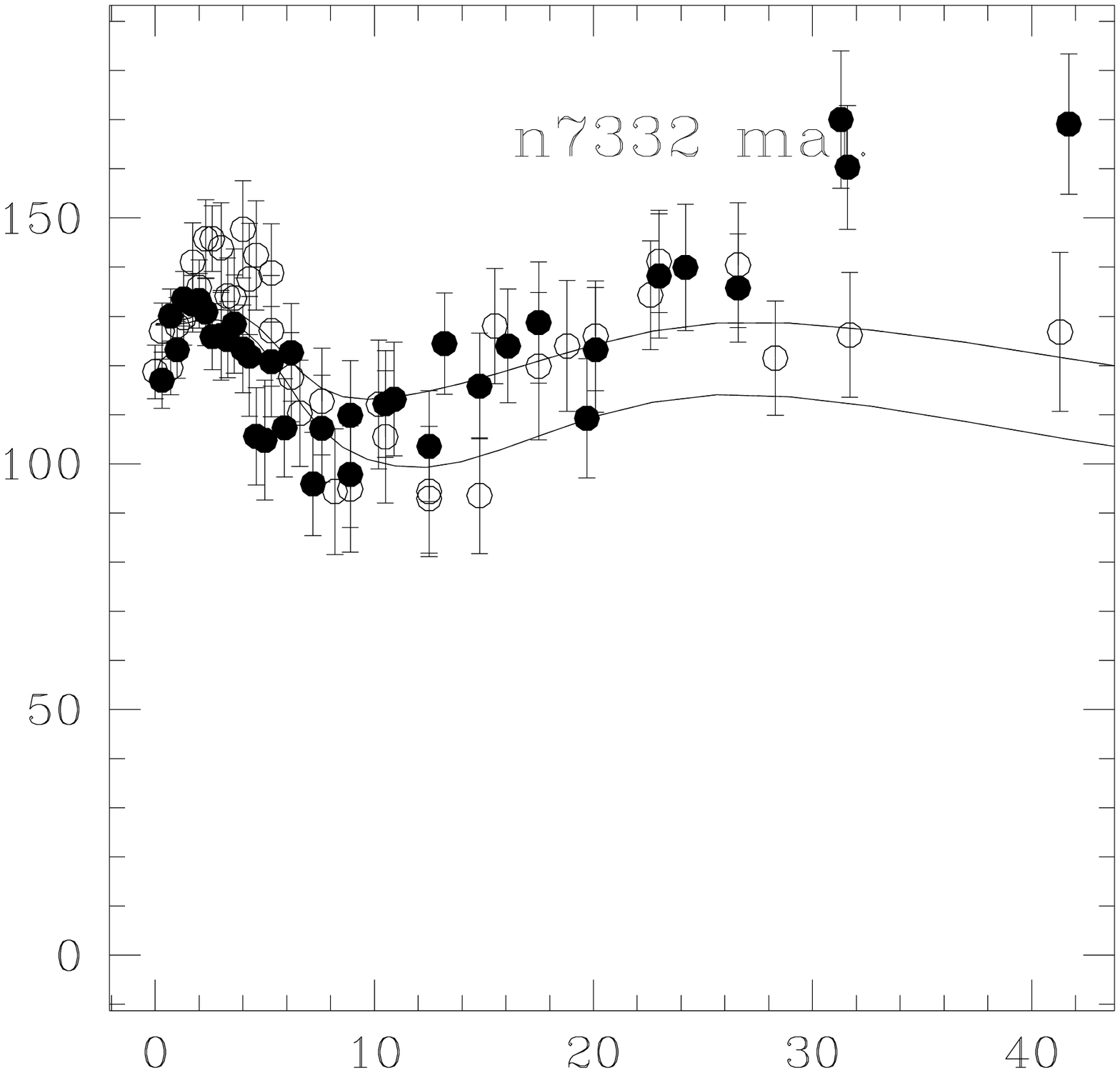,width=0.25\hsize}
\psfig{file=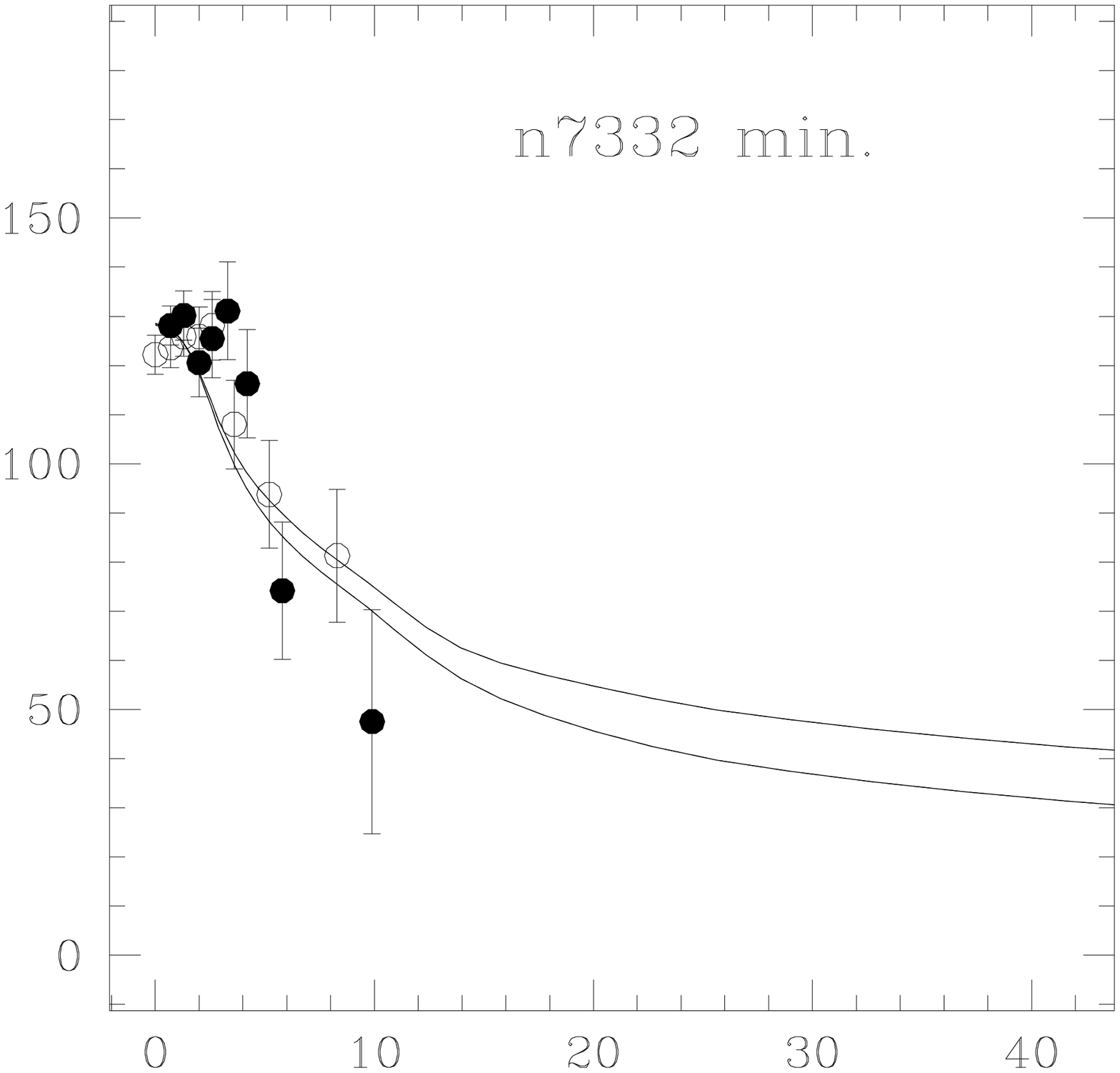,width=0.25\hsize}
\psfig{file=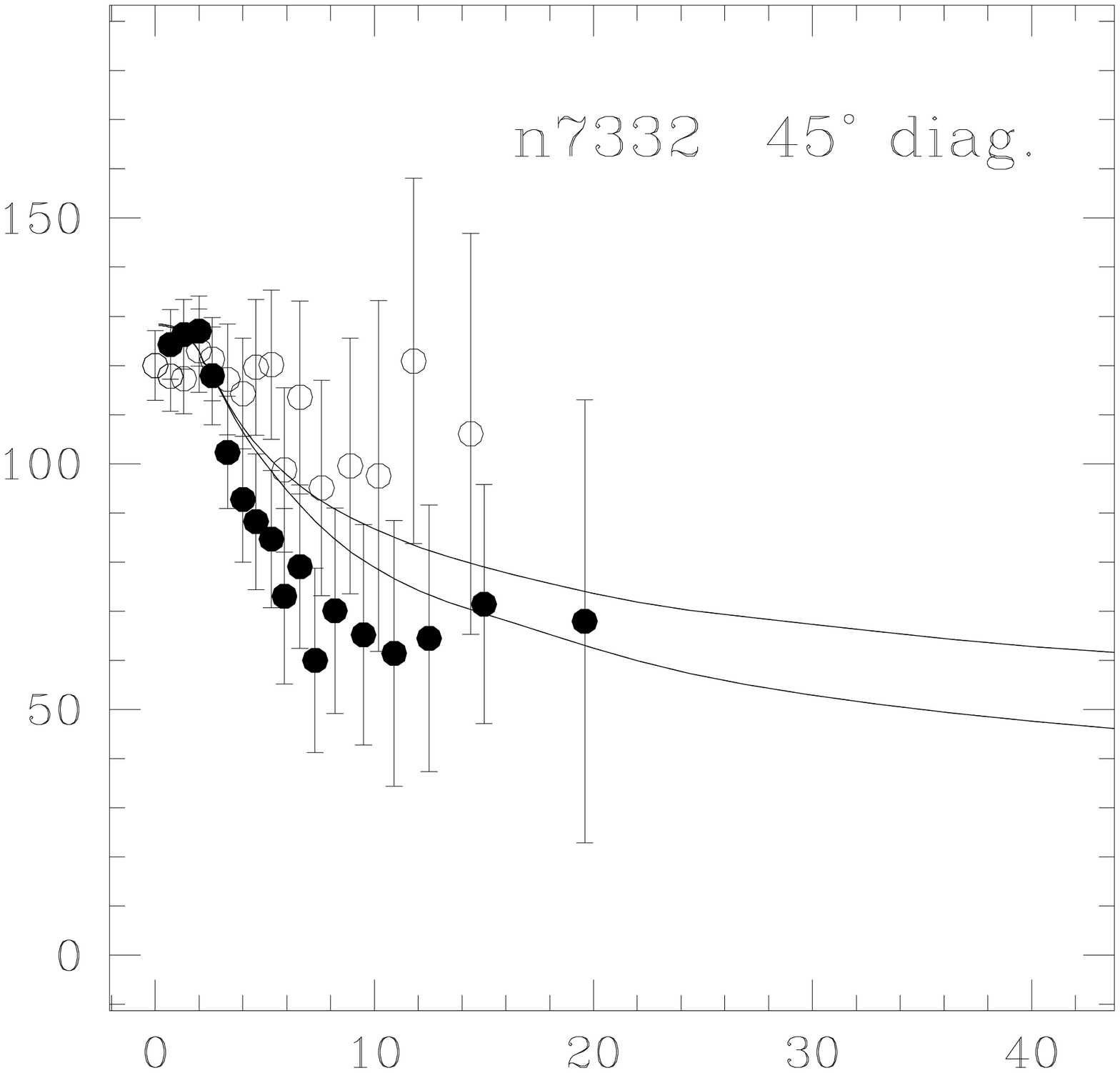,width=0.25\hsize}
\psfig{file=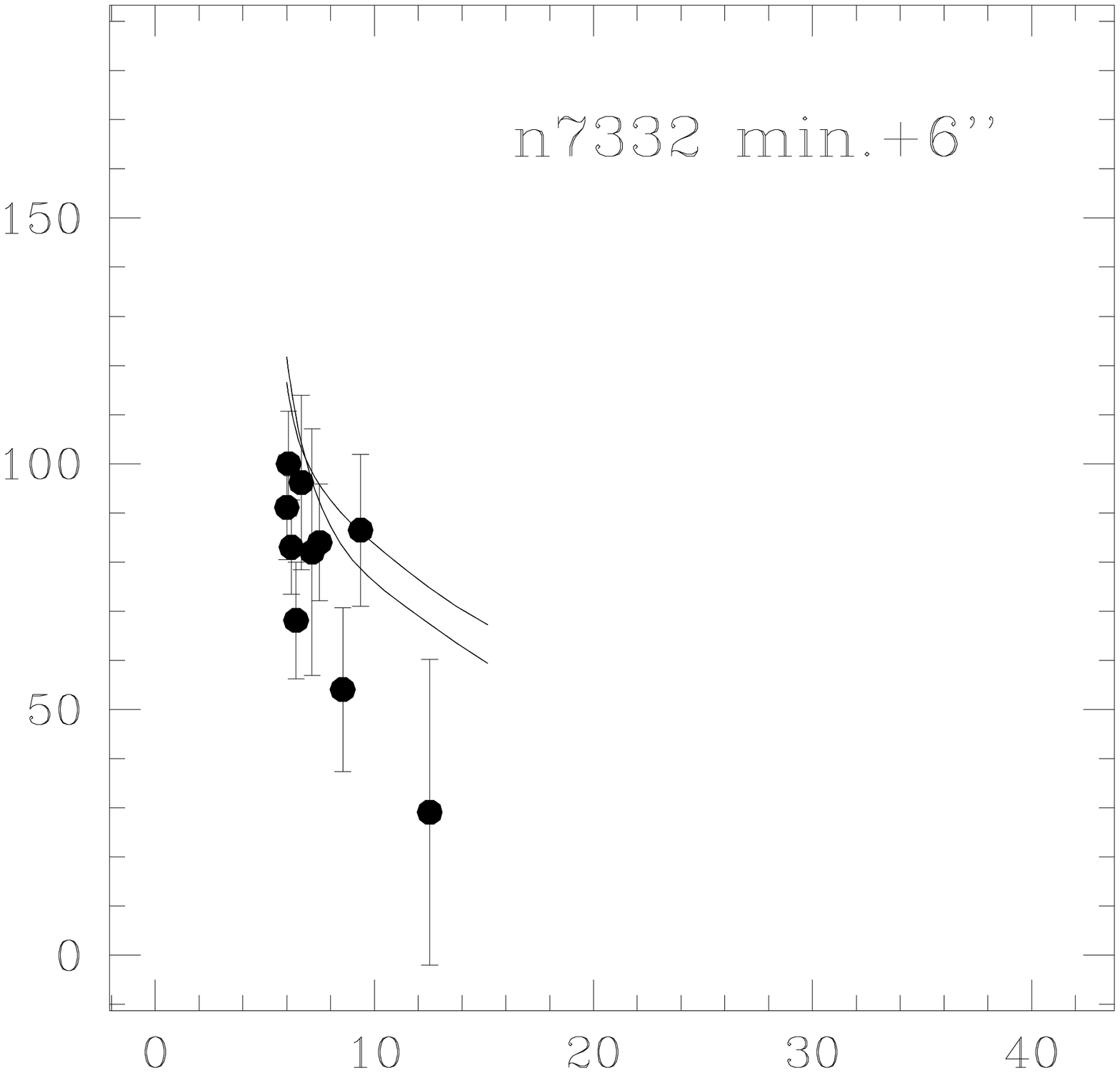,width=0.25\hsize}
}
\centerline{
\psfig{file=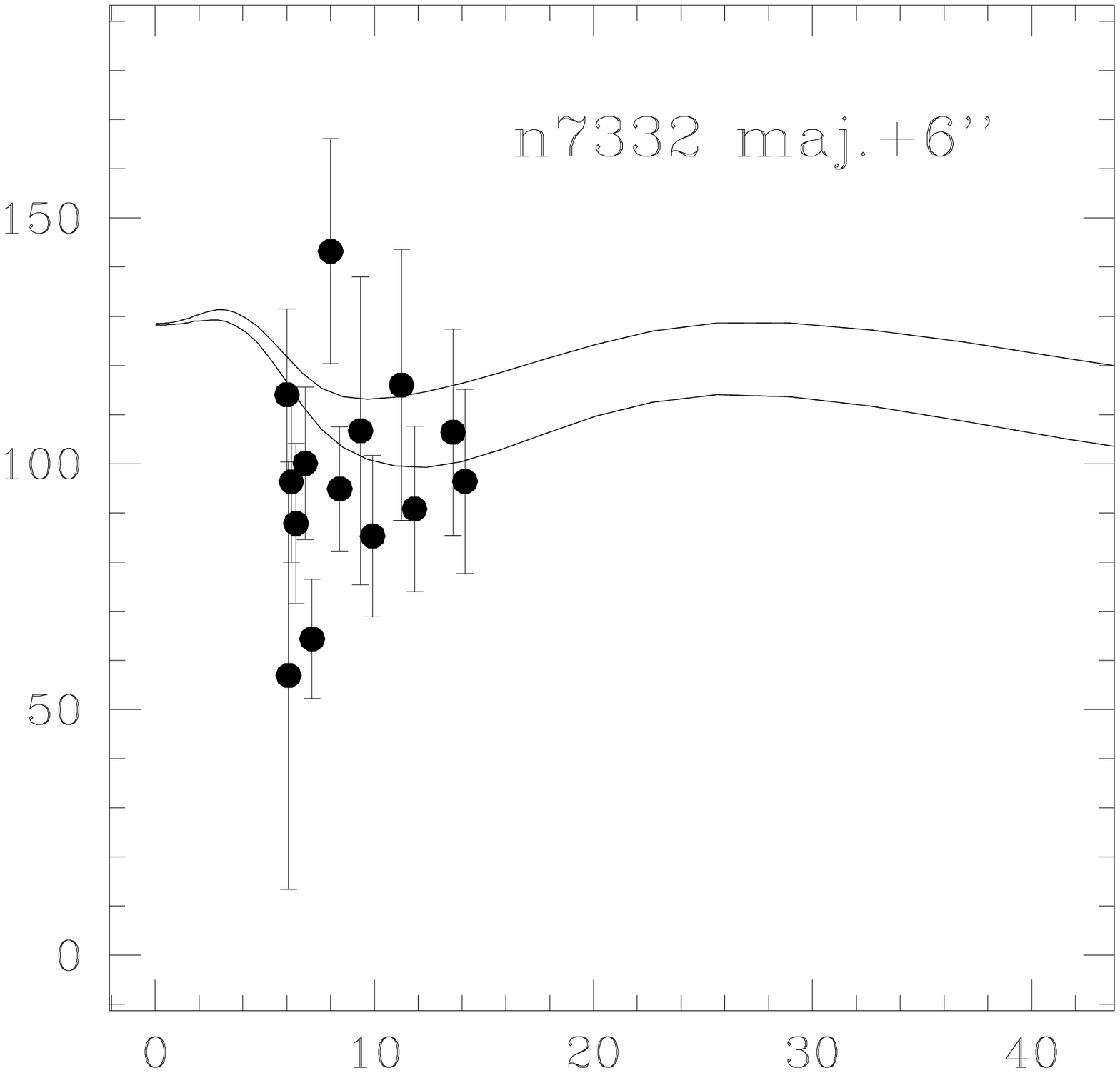,width=0.25\hsize}
\psfig{file=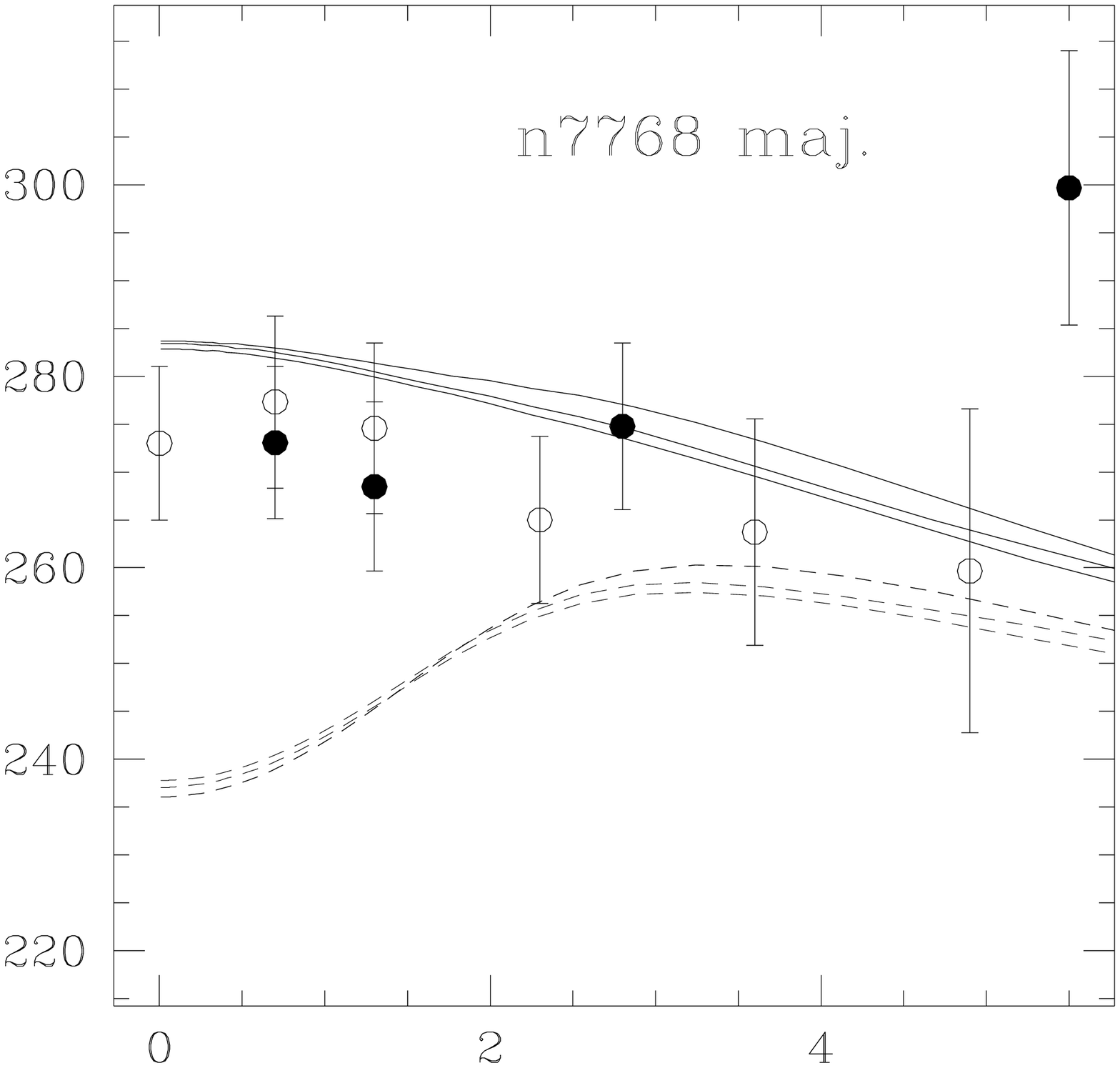,width=0.25\hsize}
\psfig{file=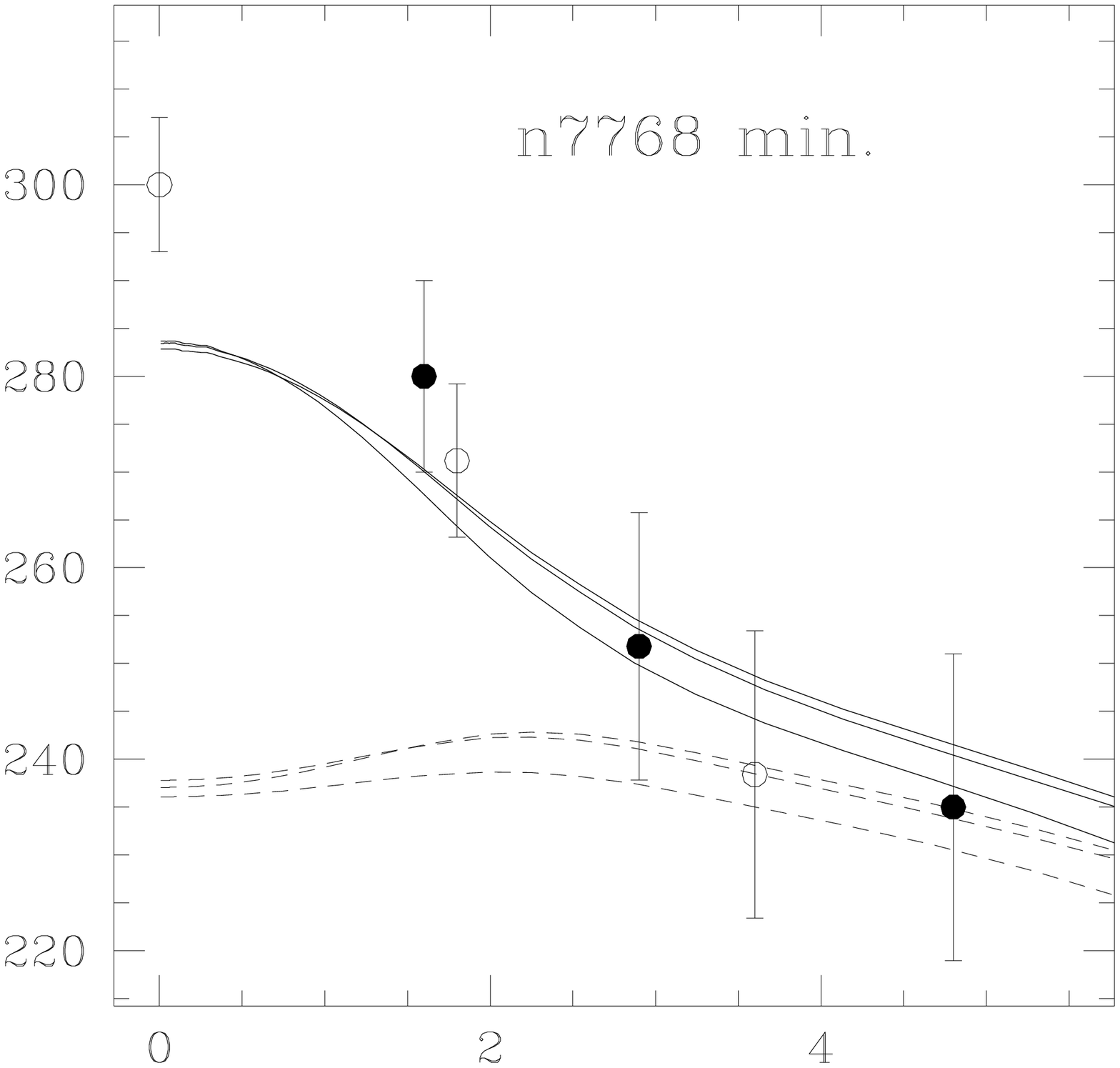,width=0.25\hsize}
\hbox to 0.25\hsize{}
}
}{\captioncont}

\vfill\supereject
\newfigure\figmdotest
\figure{\psfig{file=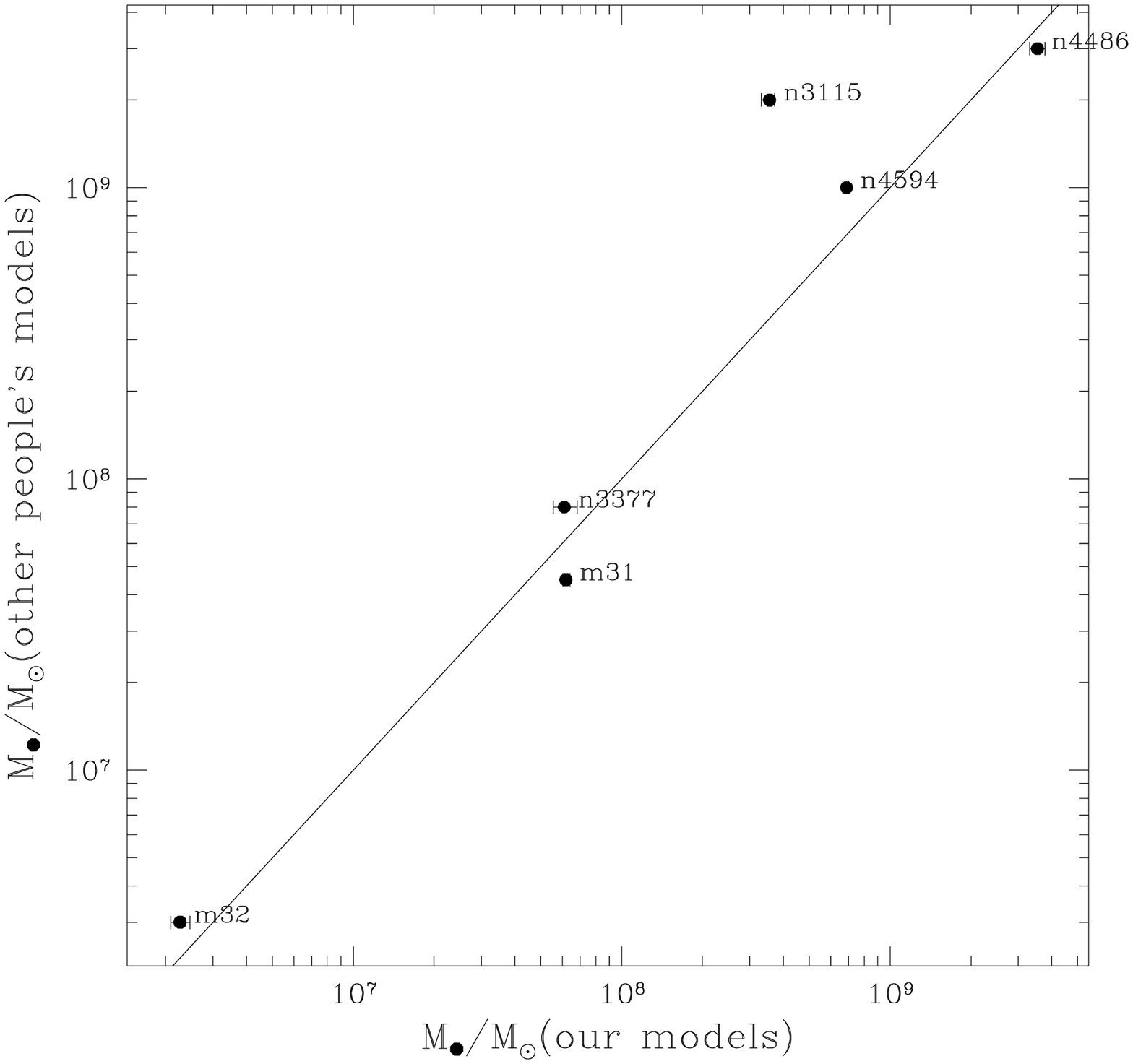,width=\hsize}}
{{\bf Figure~\ref{figmdotest}.} The correlation between the MDO masses
predicted by our models and those predicted by other methods.  The
errorbars give the 68\% confidence limits on $M_\bullet$.  Sources for
the other models are as follows: M31 -- Richstone et al.\ (1990); M32
-- van der Marel et al.\ (1997); NGC~3377 -- Richstone et al.\ (1997);
NGC~3115 -- Kormendy et al.\ (1996a);
NGC~4594 -- Kormendy et al.\ (1996b); NGC~4486 -- Harms et al.\ (1994).  }

\vfill\supereject
\newfigure\figcorrel
\figure{\centerline{\psfig{file=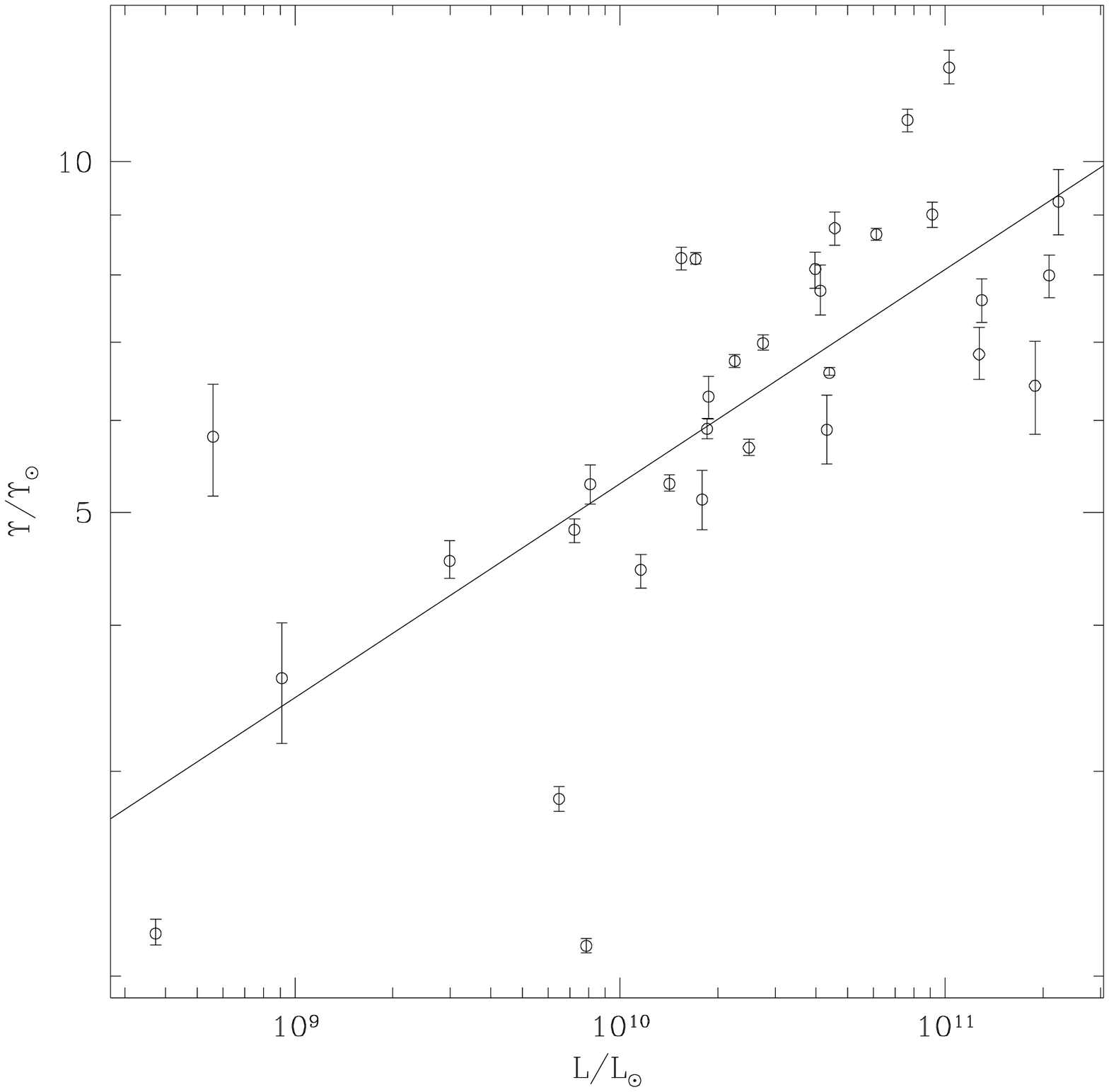,width=.5\hsize}
\psfig{file=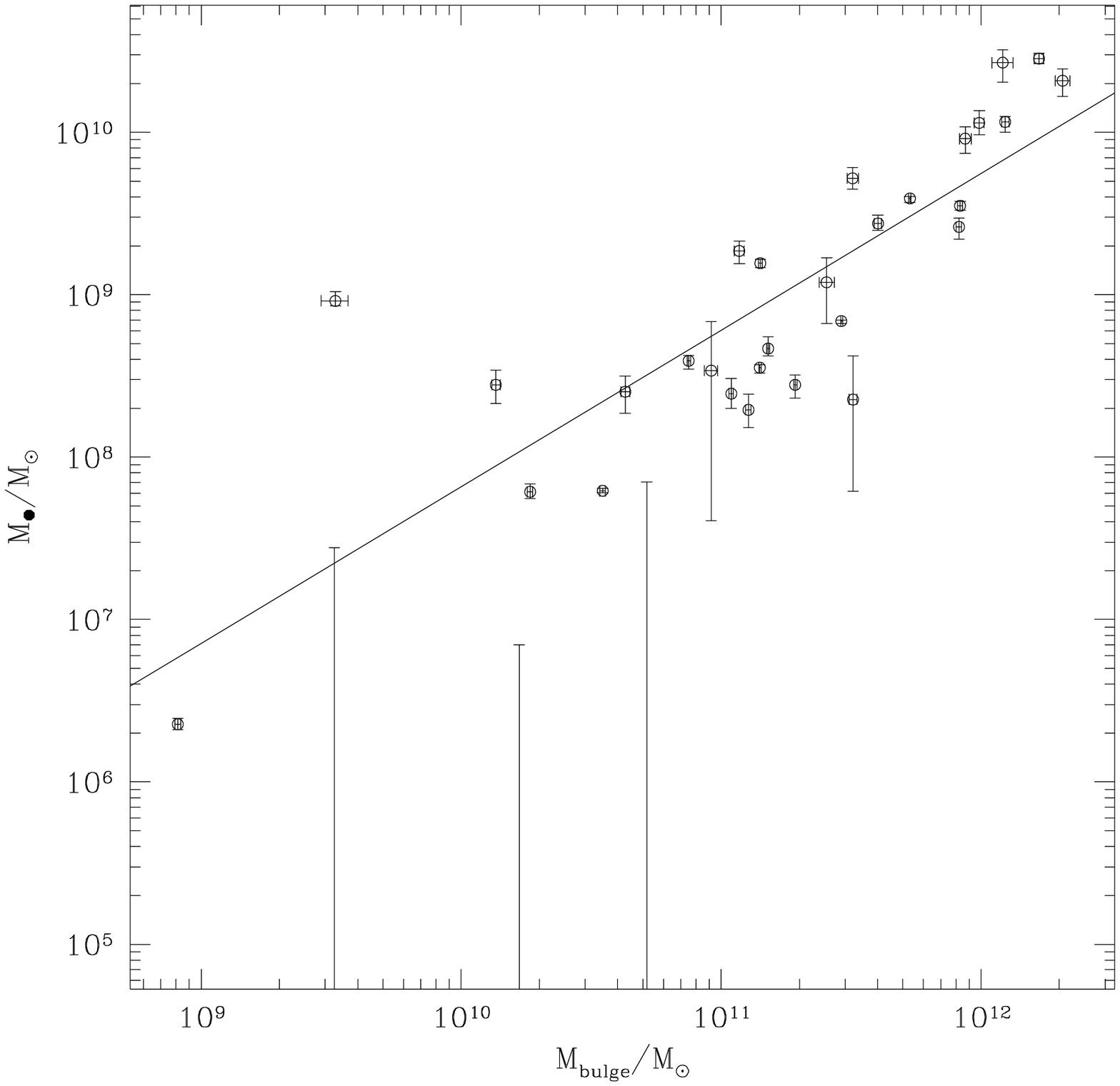,width=.5\hsize}}}
{{\bf Figure \ref{figcorrel}(a) and (b).} The correlations between stellar
mass-to-light ratio $\Upsilon$ and bulge luminosity~$L$ (left panel)
and between MDO mass $M_\bullet$ and $M_{\rm bulge}$ (right panel)
produced by our models.  The error bars give 68\% confidence
intervals.  The solid lines plot $\Upsilon_{\rm fit}$ and
$M_{\bullet\rm, fit}$ as described in the text (equations \ref{eqfita}
and~\ref{eqfitb}).}
\figure{\centerline{\psfig{file=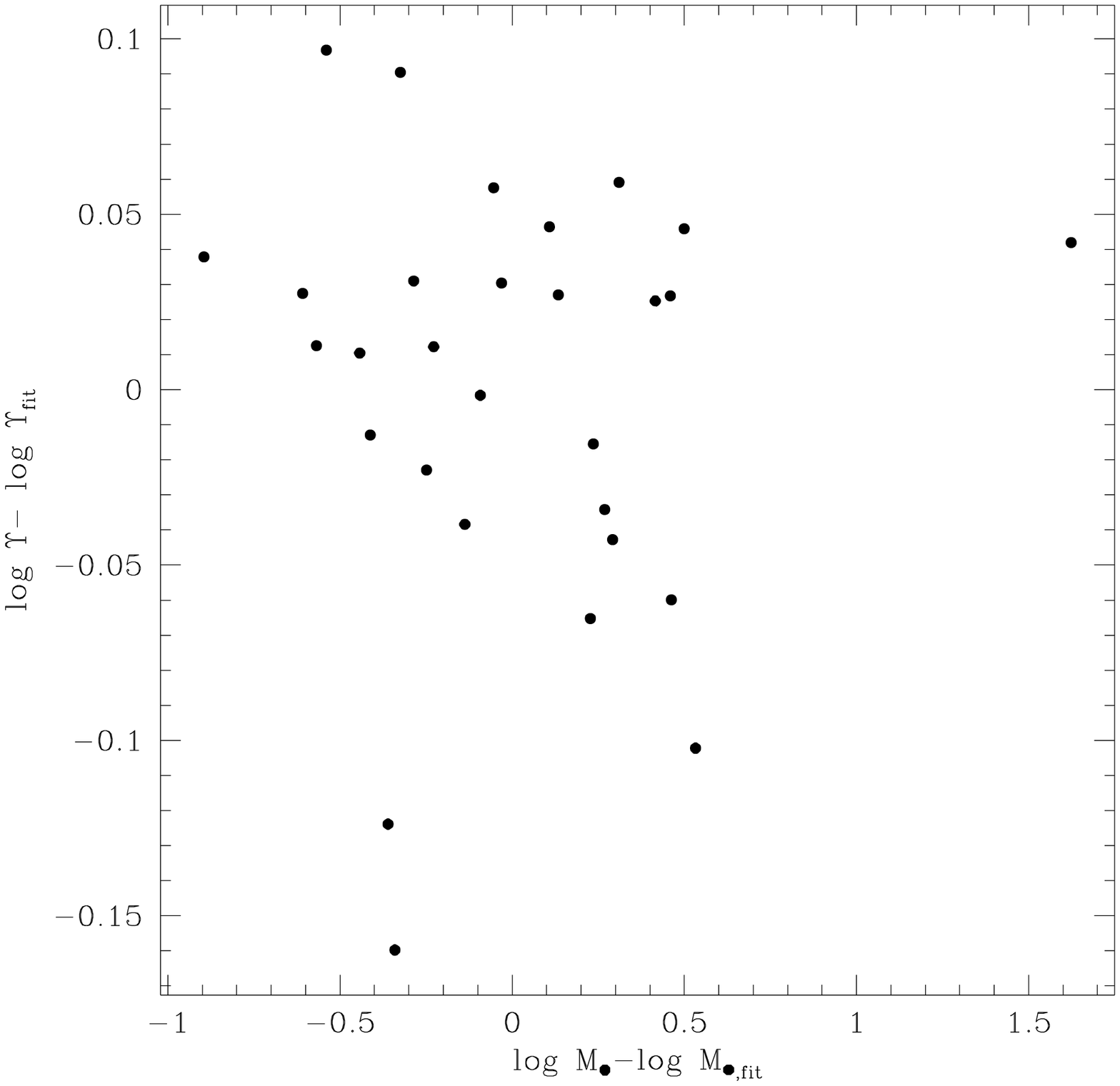,width=.5\hsize}}}
{{\bf Figure \ref{figcorrel}(c).} The correlation of the residuals in
the $\Upsilon$-versus-$L$ and $M_\bullet$-versus-$M_{\rm bulge}$
fits.}

\newfigure\figpost
\topinsert
\centerline{\psfig{file=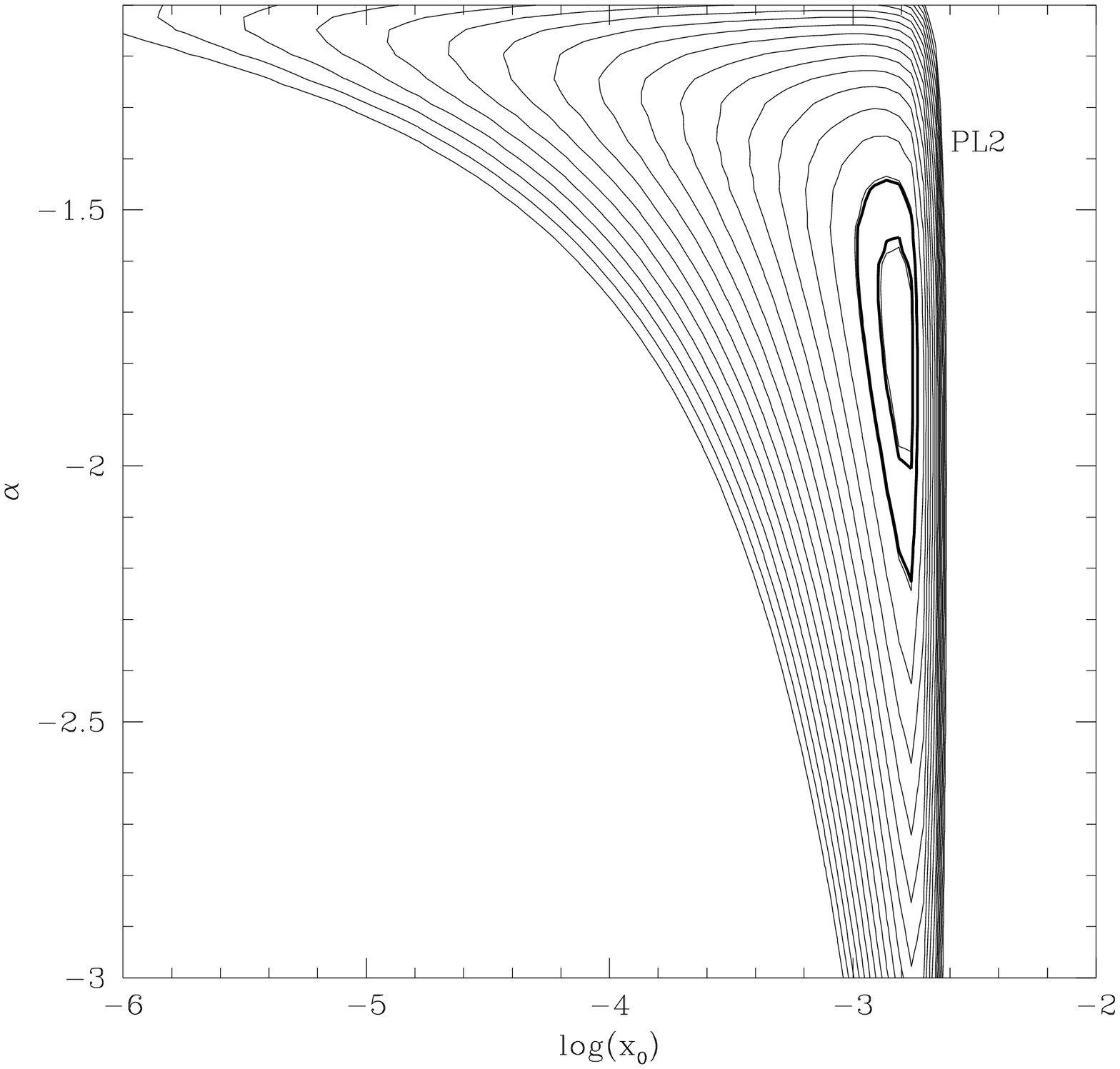,width=.4\hsize}
\psfig{file=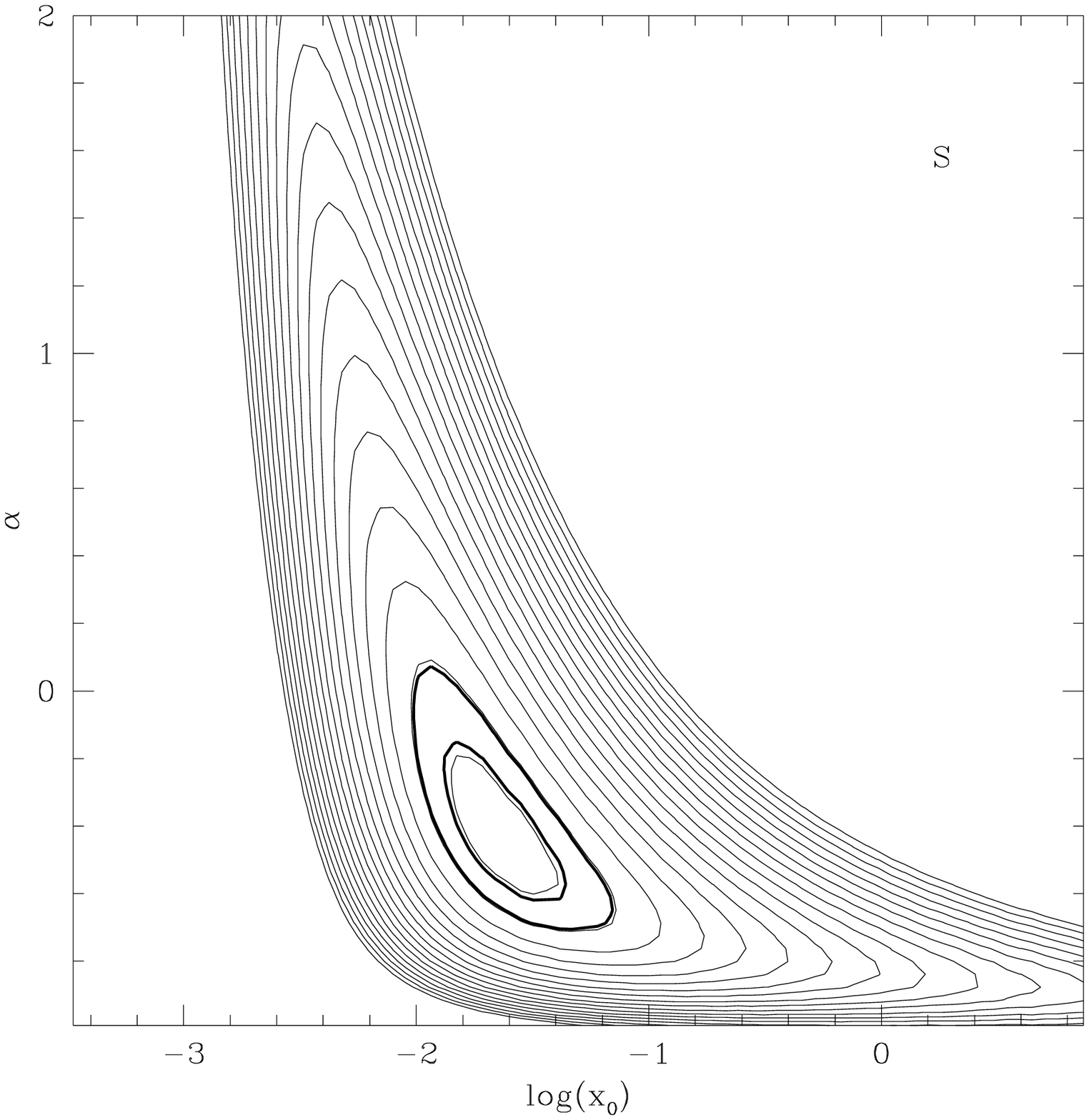,width=.4\hsize}}
\centerline{\psfig{file=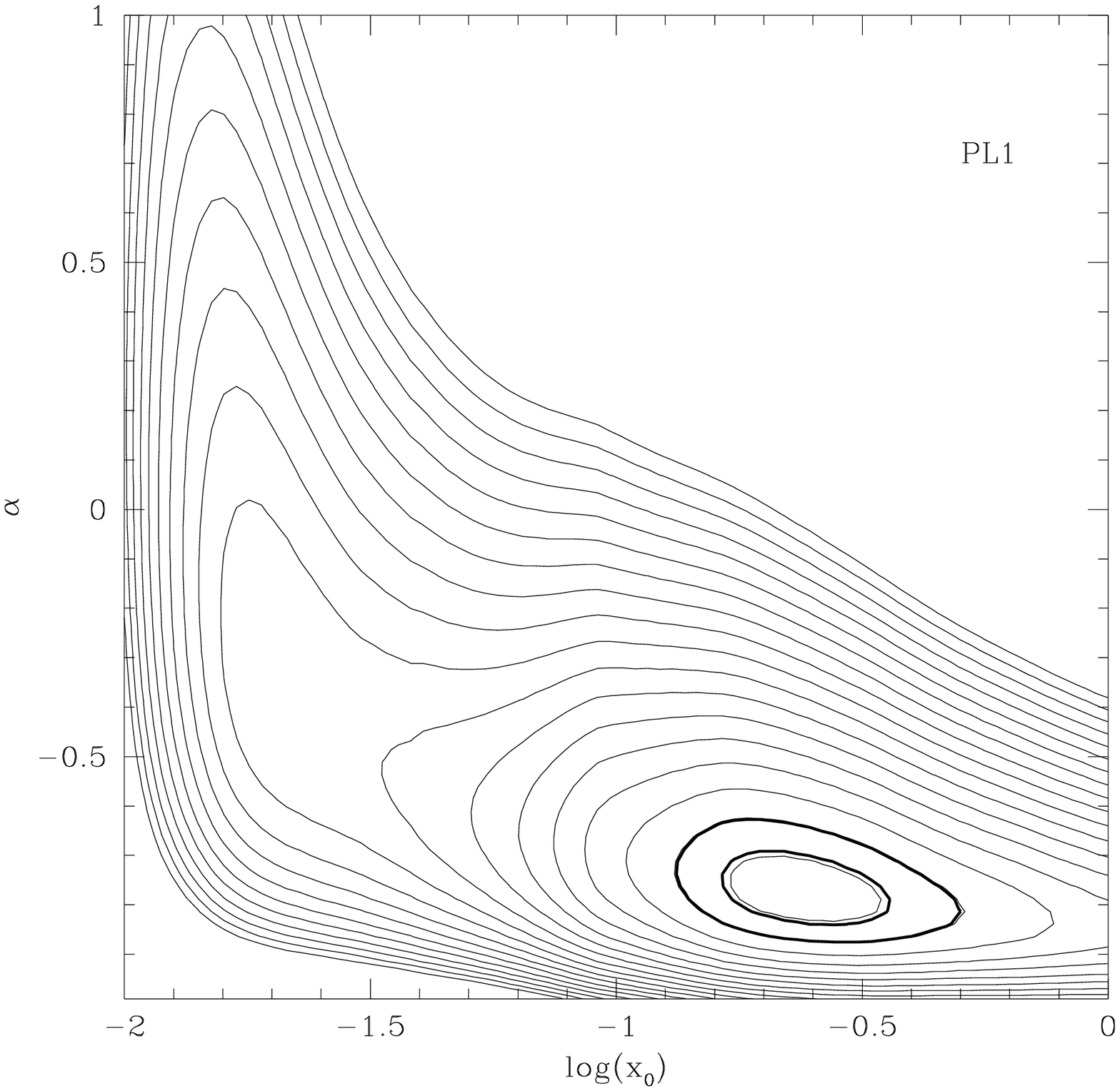,width=.4\hsize}
\psfig{file=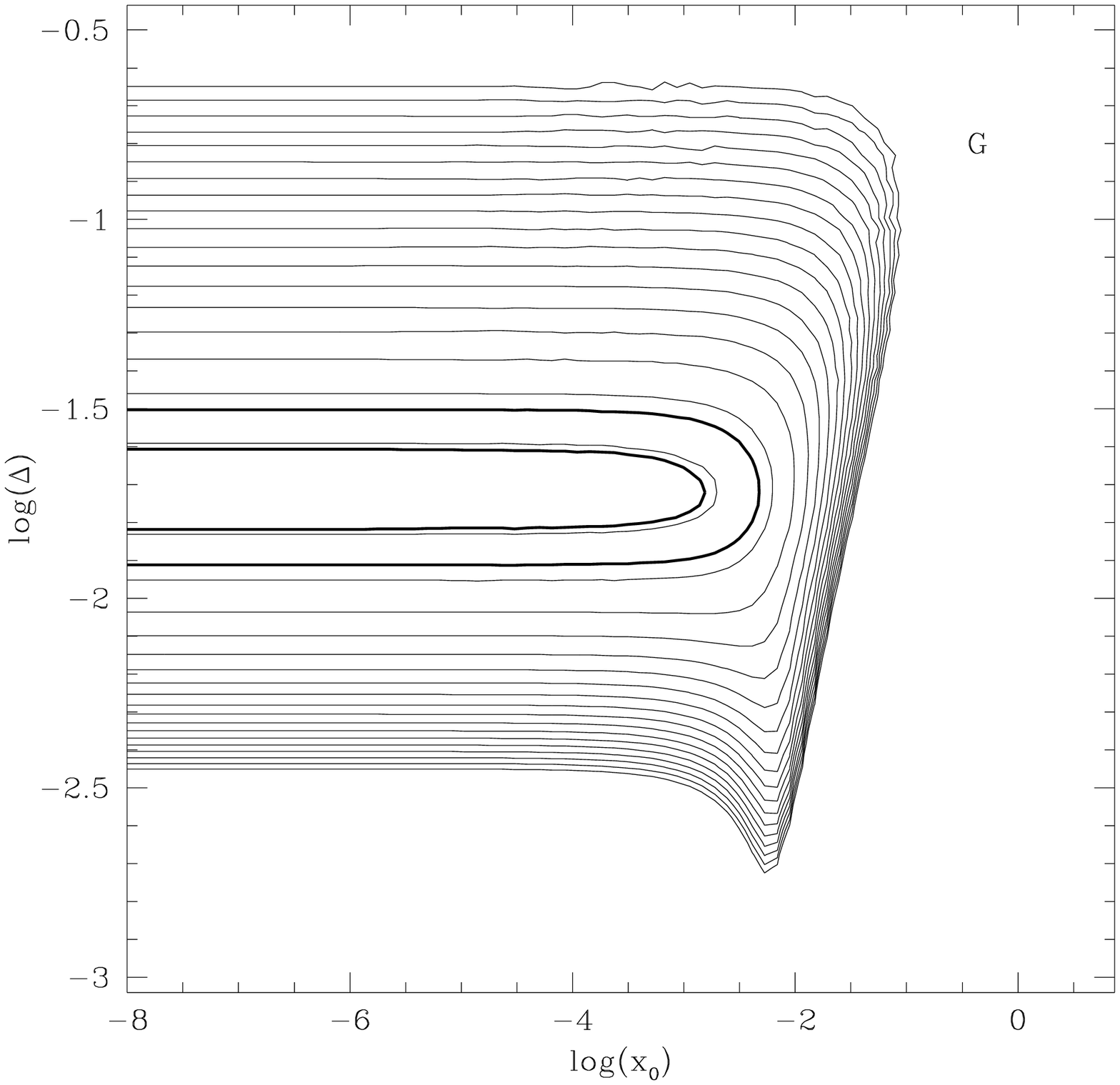,width=.4\hsize}}
\centerline{{\vbox{\hsize .4\hsize\eightpoint {\bf Figure~\ref{figpost}.}
\overfullrule=0pt
The posterior distributions $\pr(\omega\mid D,P)$ marginalized
over $f$ for (clockwise from above) $P_{\rm PL1}$, $P_{\rm PL2}$,
$P_{\rm S}$, $P_{\rm G}$ and $P_{\rm LG}$.  Successive light contours
correspond to a factor of 10 change in $\pr(\omega\mid D,P)$.  The heavy
contours enclose the 68\% and 95\% confidence areas of the
parameters~$\omega$.  The most likely value of~$f$ within the 95\%
confidence area in all five cases is $\sim0.96$.}}
\psfig{file=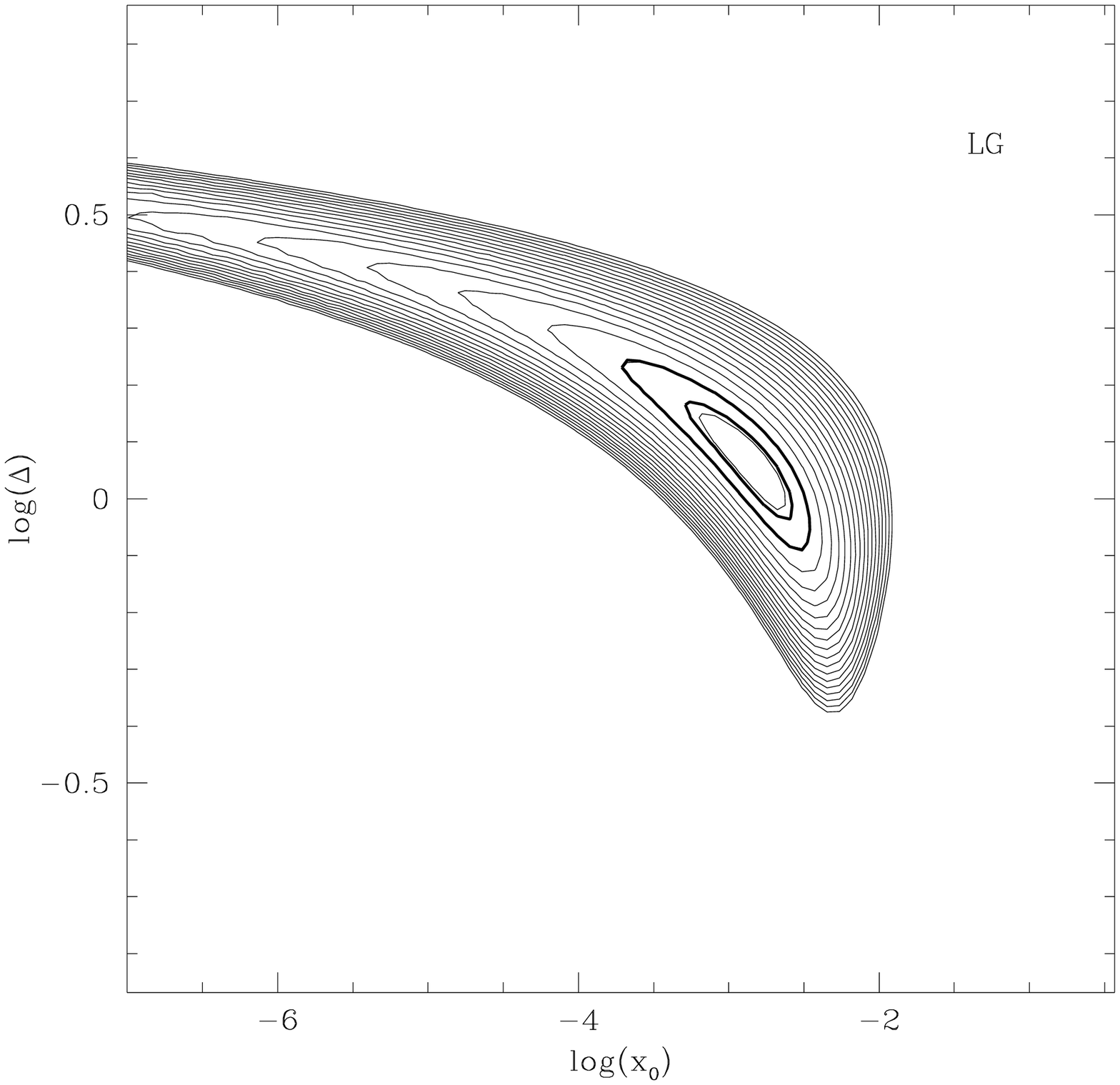,width=.4\hsize}}
\endinsert

\newfigure\figdist
\topinsert
\centerline{\psfig{file=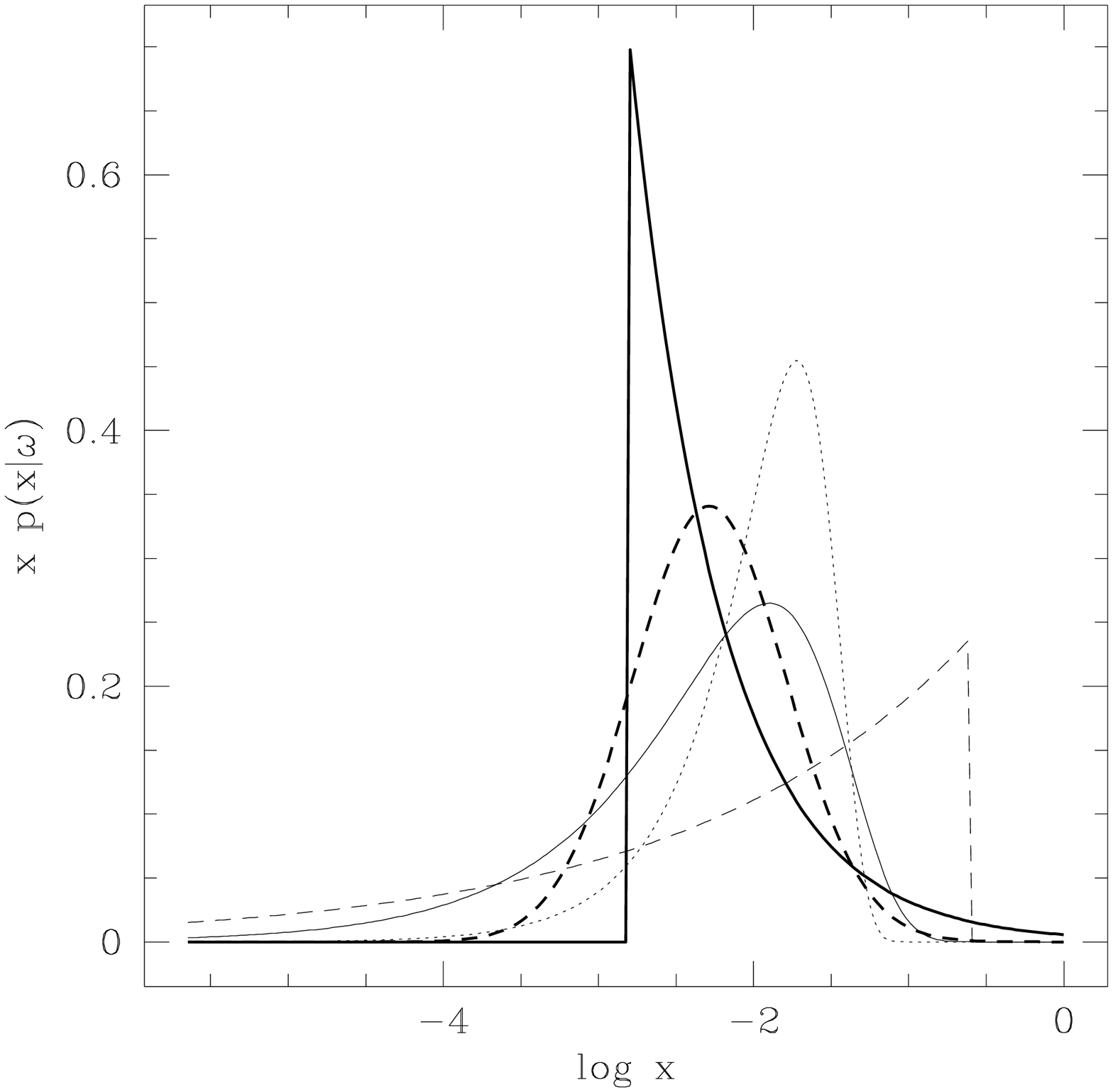,width=\hhsize}
\psfig{file=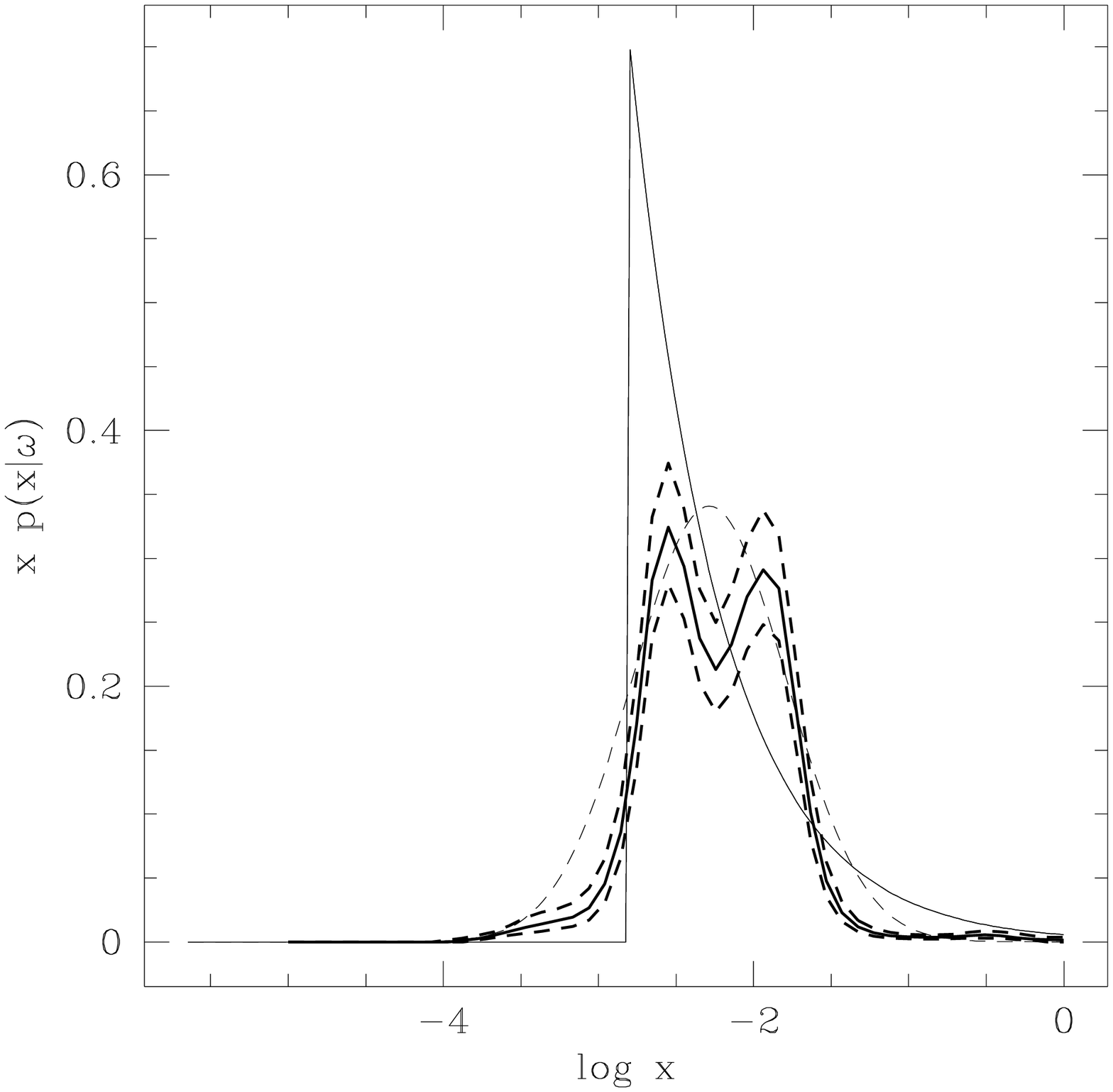,width=\hhsize}}
\eightpoint{\bf Figure \ref{figdist}(a).} (Left panel)
The probability distributions $\pr(x\mid\omega,P)$ for the
 best-fitting parameters $\omega$.  The heavy solid and dashed curves
 show results for $P_{\rm PL2}$ and $P_{\rm LG}$, the two best-fitting
 cases.  The lighter solid, dashed and dotted curves are for $P_{\rm
 S}$, $P_{\rm PL1}$ and $P_{\rm G}$ respectively.  {\bf (b).} (Right
 panel) The ``non-parametric'' probability distribution $\pr(x)$
 (heavy solid curve) and its 68\% confidence limits (heavy dashed
 curves) obtained using the Metropolis algorithm with $\lambda=5$.
 The rise in $\pr(x)$ at small~$x$ is caused by those galaxies without
 an MDO.  The best-fitting parameterized distributions $P_{\rm PL2}$
 and $P_{\rm LG}$ are overlaid as the lighter solid and dashed curves
 respectively.
\endinsert

\newfigure\figcompa
\topinsert
\centerline{\psfig{file=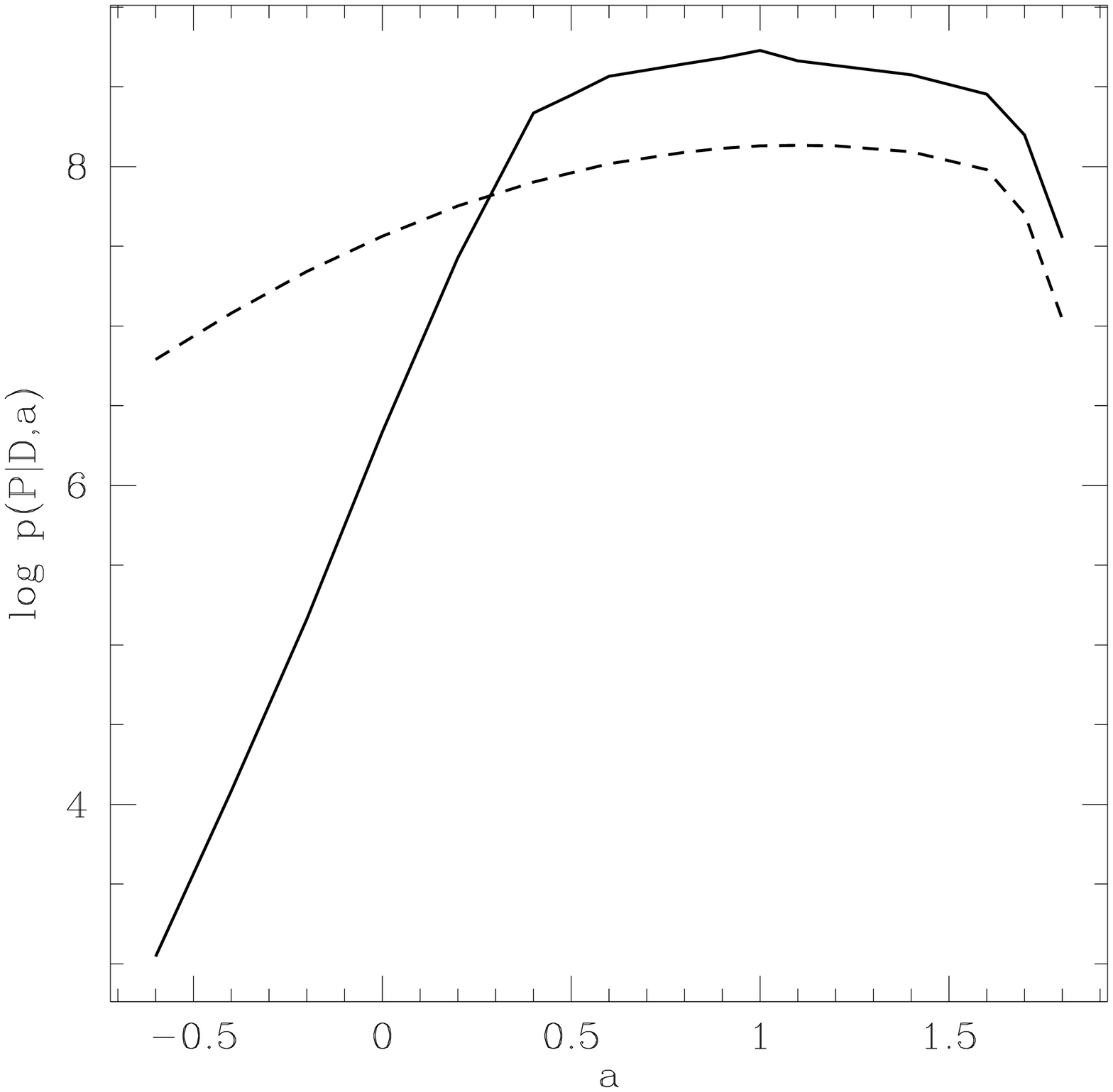,width=\hhsize}}
\eightpoint{\bf Figure \ref{figcompa}.}
Variation of $\pr(P\mid D,a)$ with $a$, where $a$ is defined in
equation~\ref{aparam}.  The solid and dashed curves show $P=P_{\rm
PL2}$ and $P=P_{\rm LG}$ respectively.  The vertical scale does not
extend down far enough to show the results for the other three
parameterizations.
\endinsert

\dosupereject

\begingroup\eightpoint
\begingroup\offinterlineskip
\halign{\strut\quad\bf# &\vrule#&\quad\hfil#\hfil & \hfil#\hfil & \hfil#\hfil & \hfil#\hfil\quad&\vrule# &\quad  \hfil#\hfil & \hfil#\quad\hfil & \hfil#\quad\hfil & \hfil#\hfil &\hfil#\hfil\cr
\omit&height1pt&\multispan4&\cr
Galaxy & & type & $M_V$ & $R_{\rm min}$ & Outer & & Slit & seeing & max. radius& no. of & Kin.\cr
&&&&(arcsec)&Photometry&&pos$^n$&(arcsec)&(arcsec)&bins&Source\cr
\omit&height1pt&\multispan4&\cr
&& (1) & (2) & (3) & (4) && (5) & (6) & (7) & (8) & (9)\cr
\omit&height1pt&\multispan4&\cr
\noalign{\hrule}
\omit&height3pt&\multispan4&\cr
m31 && S$\bigcap$ & $-19.82$ & -- & Ke87 & &maj. (bulge)   & 0.6 & 8.9 & 34 & KB97\cr
\omit&height0pt&\multispan4&\cr
&&&&&&&maj. (bulge)   & 0.75 & 8.2 & 18 & vdM94b\cr
\omit&height0pt&\multispan4&\cr
&&&&&&&min. (bulge)   & 1.2 & 8.3 & 17 & vdM94b\cr
\omit&height0pt&\multispan4&\cr
&&&&&&&maj. (nucl.)   & 1.16 & 10 & 19 & vdM94b\cr
\omit&height0pt&\multispan4&\cr
&&&&&&&min. (nucl.)   & 0.75 & 8.3 & 17 & vdM94b\cr
\omit&height0pt&\multispan4&\cr
\omit&height4pt&\multispan4&\cr
m32 && E$\setminus$ & $-16.60$ & -- & L92b & &maj. & 0.52 & 4 & 16 & BKD\cr
\omit&height0pt&\multispan4&\cr
&&&&&&&maj. & 0.75 & 7.5 & 13 & vdM94b\cr
\omit&height0pt&\multispan4&\cr
&&&&&&&min. & 0.83 & 11.3 & 9 & vdM94b\cr
\omit&height0pt&\multispan4&\cr
&&&&&&&$45^\circ$ diag & 0.83 & 12 & 18 & vdM94b\cr
\omit&height4pt&\multispan4&\cr
n821 && E$\setminus$ & $-20.64$ & -- & BDM & &maj. & 2.0 & 11.3 & 8 & BSG\cr
\omit&height0pt&\multispan4&\cr
&&&&&&&maj. & 2.0 & 27.5 & 24 & G93\cr
\omit&height0pt&\multispan4&\cr
&&&&&&&min. & 2.0 & 17.5 & 16 & G93\cr
\omit&height4pt&\multispan4&\cr
n1399 && E$\bigcap$ & $-21.71$ & -- & FIH & &maj. & 2.0 & 37.3 & 12 & FIH\cr
\omit&height0pt&\multispan4&\cr
&&&&&&&min. & 2.0 & 38.5 & 12 & FIH\cr
\omit&height4pt&\multispan4&\cr
n1600 && E$\bigcap$ & $-22.70$ & -- & BDM & &maj. & 2.0 & 24.9 & 23 & JS\cr
\omit&height0pt&\multispan4&\cr
&&&&&&&maj. & 2.0 & 27 & 22 & G93\cr
\omit&height0pt&\multispan4&\cr
&&&&&&&min. & 2.0 & 18.4 & 18 & JS\cr
\omit&height0pt&\multispan4&\cr
&&&&&&&min. & 2.0 & 18.5 & 16 & G93\cr
\omit&height4pt&\multispan4&\cr
n1700 && E$\setminus$ & $-21.65$ & -- & FIH & &maj. & 2.0 & 23.8 & 8 & BSG\cr
\omit&height0pt&\multispan4&\cr
&&&&&&&maj. & 2.0 & 20 & 11 & FIH\cr
\omit&height0pt&\multispan4&\cr
&&&&&&&min. & 2.0 & 20.5 & 7 & BSG\cr
\omit&height0pt&\multispan4&\cr
&&&&&&&min. & 2.0 & 19.6 & 11 & FIH\cr
\omit&height4pt&\multispan4&\cr
n2300 && E$\bigcap$ & $-21.82$ & -- & BDM & &maj. & 2.0 & 15.2 & 8 & BSG\cr
\omit&height0pt&\multispan4&\cr
&&&&&&&maj. & 2.0 & 21.5 & 18 & G93\cr
\omit&height0pt&\multispan4&\cr
&&&&&&&min. & 2.0 & 29.6 & 8 & BSG\cr
\omit&height0pt&\multispan4&\cr
&&&&&&&min. & 2.0 & 20 & 19 & G93\cr
\omit&height4pt&\multispan4&\cr
n2778 && E$\setminus$ & $-20.33$ & -- & PDIDC & &maj. & 2.0 & 5.5 & 7 & FIF95\cr
\omit&height0pt&\multispan4&\cr
&&&&&&&maj. & 2.0 & 11.5 & 10 & G93\cr
\omit&height0pt&\multispan4&\cr
&&&&&&&min. & 2.0 & 3.9 & 9 & JS\cr
\omit&height0pt&\multispan4&\cr
&&&&&&&min. & 2.0 & 10 & 10 & G93\cr
\omit&height4pt&\multispan4&\cr
n2832 && E$\bigcap$ & $-22.95$ & -- & PDIDC & &maj. & 2.0 & 25.6 & 13 & FIF95\cr
\omit&height4pt&\multispan4&\cr
n3115 && S0$\setminus$ & $-20.75$ & -- & BDM & &maj.   & 0.57 & 43.8 & 56 & K96a\cr
\omit&height0pt&\multispan4&\cr
&&&&&&&maj. & 2.0 & 21.2 & 56 & BSG\cr
\omit&height0pt&\multispan4&\cr
&&&&&&&maj. & 1.0 & 30.7 & 25 & KR92\cr
\omit&height0pt&\multispan4&\cr
&&&&&&&maj. & 1.0 & 28.4 & 20 & KR92\cr
\omit&height0pt&\multispan4&\cr
&&&&&&&maj. & 1.0 & 24.7 & 23 & KR92\cr
\omit&height0pt&\multispan4&\cr
&&&&&&&min. & 1.0 & 6.7 & 10 & KR92\cr
\omit&height0pt&\multispan4&\cr
\omit&height4pt&\multispan4&\cr
n3377 && E$\setminus$ & $-19.70$ & -- & SB & &maj.   & 0.59 & 22.6 & 30 & K97b\cr
\omit&height0pt&\multispan4&\cr
&&&&&&&maj. & 0.47 & 1.5 & 18 & K97b\cr
\omit&height0pt&\multispan4&\cr
&&&&&&&min. & 2.0 & 28.5 & 26 & G93\cr
\omit&height4pt&\multispan4&\cr
n3379 && E$\bigcap$ & $-20.55$ & -- & PDIDC & &maj. & 1.5 & 11.5 & 9 & G97\cr
\omit&height0pt&\multispan4&\cr
&&&&&&&$18^\circ$ diag. & 1.5 & 11.5 & 8 & G97\cr
\omit&height0pt&\multispan4&\cr
&&&&&&&$28^\circ$ diag. & 1.5 & 11.6 & 8 & G97\cr
\omit&height0pt&\multispan4&\cr
&&&&&&&$72^\circ$ diag. & 1.5 & 11.5 & 8 & G97\cr
\omit&height0pt&\multispan4&\cr
\omit&height4pt&\multispan4&\cr
n3608 && E$\bigcap$ & $-20.84$ & -- & BDM & &maj.   & 2.0 & 37.3 & 21 & JS\cr
\omit&height0pt&\multispan4&\cr
&&&&&&&maj. & 2.0 & 24.5 & 20 & G93\cr
\omit&height0pt&\multispan4&\cr
&&&&&&&min.   & 2.0 & 25.9 & 17 & JS\cr
\omit&height0pt&\multispan4&\cr
&&&&&&&min. & 2.0 & 24 & 19 & G93\cr
\omit&height4pt&\multispan4&\cr
n4168 && E$\bigcap$ & $-21.76$ & -- & BDM & &maj. & 2.0 & 30.6 & 11 & BSG\cr
\omit&height0pt&\multispan4&\cr
&&&&&&&maj. & 1.5 & 7.8 & 6 & BN\cr
\omit&height4pt&\multispan4&\cr
n4278 && E$\bigcap$ & $-21.16$ & 0.1 & PDIDC & &maj. & 2.0 & 31 & 36 & G93\cr
\omit&height0pt&\multispan4&\cr
&&&&&&&min. & 2.0 & 30 & 30 & G93\cr
\omit&height4pt&\multispan4&\cr
n4291 && E$\bigcap$ & $-20.85$ & -- & BDM & &maj. & 2.0 & 28 & 21 & JS\cr
\omit&height0pt&\multispan4&\cr
&&&&&&&maj. & 2.0 & 19.5 & 13 & BSG\cr
\omit&height0pt&\multispan4&\cr
&&&&&&&min. & 2.0 & 24.1 & 17 & JS\cr
\omit&height0pt&\multispan4&\cr
&&&&&&&min. & 2.0 & 16.5 & 12 & BSG\cr
\omit&height4pt&\multispan4&\cr
n4365 && E$\bigcap$ & $-22.06$ & -- & $R_{\rm eff}=57''$\rlap{(l)} & &maj. & 2.0 & 7.5 & 7 & BSG\cr
\omit&height0pt&\multispan4&\cr
&&&&&&&min. & 2.0 & 6.3 & 6 & BSG\cr
\omit&height4pt&\multispan4&\cr
n4467 && E$\setminus$ & $-17.04$ & -- & $R_{\rm eff}=10''$\rlap{(f)} & &maj. & 1.5 & 7.5 & 6 & BN\cr
\omit&height4pt&\multispan4&\cr
n4472 && E$\bigcap$ & $-22.57$ & -- & PDIDC & &maj. & 2.0 & 23.9 & 18 & BSG\cr
\omit&height0pt&\multispan4&\cr
&&&&&&&min. & 2.0 & 27.9 & 14 & BSG\cr
\omit&height4pt&\multispan4&\cr
n4473 && E$\bigcap$ & $-20.80$ & -- & BDM & &maj. & 2.0 & 20.4 & 11 & BSG\cr
\omit&height4pt&\multispan4&\cr
n4486 && E$\bigcap$ & $-22.38$ & -- & PDIDC & &maj. & 0.79 & 5.8 & 27 & vdM94a\cr
\omit&height4pt&\multispan4&\cr
n4486b && E$\bigcap$ & $-17.57$ & -- & $R_{\rm eff}=1''$\rlap{(f)} & &maj. & 0.66 & 5.5 & 15 & K97a\cr
\omit&height0pt&\multispan4&\cr
&&&&&&&maj. & 0.52 & 1.1 & 11 & K97a\cr
\omit&height0pt&\multispan4&\cr
&&&&&&&min. & 2.0 & 7.4 & 5 & BN\cr
\omit&height4pt&\multispan4&\cr
n4494 && E$\setminus$ & $-21.14$ & -- & BDM & &maj. & 2.0 & 32.3 & 14 & BSG\cr
\omit&height0pt&\multispan4&\cr
&&&&&&&min. & 2.0 & 36.8 & 20 & JS\cr
\omit&height4pt&\multispan4&\cr
n4552 && E$\bigcap$ & $-21.05$ & 0.1 & BDM & &maj. & 3.0 & 24.9 & 8 & BSG\cr
\omit&height0pt&\multispan4&\cr
&&&&&&&maj. & 3.0 & 34.5 & 42 & G93\cr
\omit&height0pt&\multispan4&\cr
&&&&&&&min. & 3.0 & 33.5 & 36 & G93\cr
\omit&height4pt&\multispan4&\cr
n4564 && E$\setminus$ & $-19.94$ & -- & BDM & &maj. & 2.0 & 33.2 & 16 & BSG\cr
\omit&height4pt&\multispan4&\cr
n4589 && E$\bigcap$ & $-21.69$ & -- & $R_{\rm eff}=30''$\rlap{(f)} & &maj. & 2.0 & 8.1 & 5 & BSG\cr
\omit&height4pt&\multispan4&\cr
n4594 && S0$\setminus$ & $-21.78$ & 0.1 & K88 & &maj. & 0.93 & 4.7 & 19 & K88\cr
\omit&height0pt&\multispan4&\cr
&&&&&&&maj. & 0.93 & 4.9 & 11 & K88\cr
\omit&height0pt&\multispan4&\cr
&&&&&&&maj. & 0.93 & 5 & 15 & vdM94b\cr
\omit&height0pt&\multispan4&\cr
&&&&&&&maj. & 0.93 & 4.2 & 7 & vdM94b\cr
\omit&height0pt&\multispan4&\cr
&&&&&&&min. & 0.93 & 8.2 & 10 & K88\cr
\omit&height4pt&\multispan4&\cr
n4621 && E$\setminus$ & $-21.27$ & -- & BDM & &maj. & 2.0 & 33.6 & 19 & BSG\cr
\omit&height0pt&\multispan4&\cr
&&&&&&&min. & 2.0 & 27.1 & 16 & BSG\cr
\omit&height4pt&\multispan4&\cr
n4636 && E$\bigcap$ & $-21.67$ & -- & BDM & &maj. & 3.0 & 33.5 & 15 & BSG\cr
\omit&height4pt&\multispan4&\cr
n4649 && E$\bigcap$ & $-22.14$ & -- & BDM & &maj. & 2.0 & 24.2 & 13 & BSG\cr
\omit&height0pt&\multispan4&\cr
&&&&&&&maj. & 2.0 & 37.5 & 35 & G93\cr
\omit&height0pt&\multispan4&\cr
&&&&&&&min. & 2.0 & 27.7 & 16 & BSG\cr
\omit&height0pt&\multispan4&\cr
&&&&&&&min. & 2.0 & 34.5 & 52 & G93\cr
\omit&height4pt&\multispan4&\cr
n4660 && E$\setminus$ & $-18.86$ & -- & SB & &maj. & 2.0 & 8.1 & 8 & BSG\cr
\omit&height0pt&\multispan4&\cr
&&&&&&&min. & 2.0 & 5.5 & 6 & BSG\cr
\omit&height4pt&\multispan4&\cr
n4874 && E$\bigcap$ & $-23.54$ & -- & PDIDC & &$26^\circ$ diag.   & 2.0 & 22.5 & 10 & FIF95\cr
\omit&height4pt&\multispan4&\cr
n4889 && E$\bigcap$ & $-23.36$ & -- & BDM & &maj.   & 2.0 & 21.3 & 11 & FIF95\cr
\omit&height4pt&\multispan4&\cr
n6166 && E$\bigcap$ & $-23.47$ & 0.2 & $R_{\rm eff}=56''$\rlap{(l)} & &maj.   & 2.0 & 12.9 & 13 & FIF95\cr
\omit&height0pt&\multispan4&\cr
&&&&&&&min.   & 2.0 & 5.5 & 9 & FIF95\cr
\omit&height4pt&\multispan4&\cr
n7332 && S0$\setminus$ & $-19.91$ & -- & FIF94 & &maj. & 2.0 & 31.3 & 33 & FIF94\cr
\omit&height0pt&\multispan4&\cr
&&&&&&&maj. & 2.0 & 41.7 & 41 & FIF94\cr
\omit&height0pt&\multispan4&\cr
&&&&&&&min. & 2.0 & 9.9 & 16 & FIF94\cr
\omit&height0pt&\multispan4&\cr
&&&&&&&$45^\circ$ diag. & 2.0 & 19.6 & 33 & FIF94\cr
\omit&height0pt&\multispan4&\cr
&&&&&&&min.+6'' & 2.0 & 12.5 & 10 & FIF94\cr
\omit&height0pt&\multispan4&\cr
&&&&&&&maj.+6'' & 2.0 & 14.1 & 14 & FIF94\cr
\omit&height4pt&\multispan4&\cr
n7768 && E$\bigcap$ & $-22.93$ & 0.4 & $R_{\rm eff}=30''$\rlap{(l)} & &maj. & 2.0 & 5.5 & 10 & FIF95\cr
\omit&height0pt&\multispan4&\cr
&&&&&&&min. & 2.0 & 4.8 & 6 & FIF95\cr
\omit&height4pt&\multispan4&\cr
}\endgroup

{\eightpoint {\bf Table 1.} The galaxy sample.  Column (1) gives the
galaxy type: ``S''=spiral bulge, ``S0''=lenticular, ``E''=elliptical;
``$\bigcap$''=cored, ``$\setminus$''=power law (Lauer et al.\ 1995).
The absolute $V$ magnitudes of the bulge or other hot component in
column (2) are taken from Faber et al.\ (1997), and assume
$H_0=80\hbox{ km s}^{-1}\hbox{ Mpc}^{-1}$.  $R_{\rm min}$ in column
(3) is the radius inside which we believe the galaxy light may be
contaminated by non-stellar radiation.  Column (4) gives the source of
the outer photometry used (if available), otherwise it gives the
effective radius $R_{\rm eff}$ used for the outward extrapolation.
Values of $R_{\rm eff}$ obtained from the literature are followed by
an '(l)', while an '(f)' follows those obtained by fitting to the HST
photometry.  Columns (5) to (9) list the kinematical data used.  For
each exposure along each slit position, columns (5) and (6) give the
position and FWHM of the seeing respectively.  The maximum radius and
the number of bins used by our models are given in columns (7) and
(8).  Finally, column (9) gives the source of the kinematical data.  }
\endgroup

\vfill\eject
\begingroup\offinterlineskip
\halign{ \qquad\strut# & \vrule # & \hfil\quad $#$\quad & \hfil\quad $#$\quad & \hfil\quad $#$\quad & \hfil\quad $#$ \quad & \hfil\quad $#$ \cr\omit&height3pt&\cr
 \bf Galaxy && D/\Mpc & \log(L/L_\odot) & \log(\Upsilon/\Upsilon_\odot) & \log(M_\bullet/M_\odot) & \log x\hfil\cr\omit&height3pt&\cr
\noalign{\hrule}
\omit&height3pt&\cr
m31 &&    0.8 &  9.860 &  0.684^{+0.010}_{-0.011} &  7.792^{+0.016}_{-0.011} & -2.752^{+0.021}_{-0.017}\cr
\omit&height3pt&\cr
m32 &&    0.8 &  8.572 &  0.338^{+0.012}_{-0.010} &  6.355^{+0.036}_{-0.034} & -2.553^{+0.040}_{-0.046}\cr
\omit&height3pt&\cr
n821 &&   19.5 & 10.188 &  0.918^{+0.009}_{-0.011} &  8.291^{+0.097}_{-0.107} & -2.796^{+0.085}_{-0.137}\cr
\omit&height3pt&\cr
n1399 &&   17.9 & 10.616 &  0.889^{+0.022}_{-0.021} &  9.718^{+0.065}_{-0.068} & -1.785^{+0.075}_{-0.091}\cr
\omit&height3pt&\cr
n1600 &&   50.2 & 11.012 &  1.081^{+0.015}_{-0.014} & 10.065^{+0.033}_{-0.064} & -2.046^{+0.060}_{-0.059}\cr
\omit&height3pt&\cr
n2300 &&   31.8 & 10.660 &  0.943^{+0.014}_{-0.015} &  9.438^{+0.051}_{-0.043} & -2.161^{+0.058}_{-0.060}\cr
\omit&height3pt&\cr
n2778 &&   33.6 & 10.064 &  0.650^{+0.013}_{-0.016} & < 7.849 & <-2.850\cr
\omit&height3pt&\cr
n2832 &&   90.2 & 11.112 &  0.881^{+0.018}_{-0.019} & 10.058^{+0.076}_{-0.072} & -1.935^{+0.089}_{-0.091}\cr
\omit&height3pt&\cr
n3115 &&    8.4 & 10.232 &  0.917^{+0.005}_{-0.004} &  8.551^{+0.019}_{-0.032} & -2.602^{+0.025}_{-0.030}\cr
\omit&height3pt&\cr
n3377 &&    9.9 &  9.812 &  0.453^{+0.010}_{-0.011} &  7.786^{+0.049}_{-0.041} & -2.469^{+0.045}_{-0.060}\cr
\omit&height3pt&\cr
n3379 &&    9.9 & 10.152 &  0.724^{+0.008}_{-0.006} &  8.595^{+0.031}_{-0.053} & -2.284^{+0.037}_{-0.056}\cr
\omit&height3pt&\cr
n3608 &&   20.3 & 10.268 &  0.771^{+0.009}_{-0.008} &  8.392^{+0.091}_{-0.091} & -2.638^{+0.088}_{-0.108}\cr
\omit&height3pt&\cr
n4168 &&   36.4 & 10.636 &  0.770^{+0.030}_{-0.030} &  9.077^{+0.151}_{-0.253} & -2.356^{+0.194}_{-0.273}\cr
\omit&height3pt&\cr
n4278 &&   17.5 & 10.396 &  0.755^{+0.007}_{-0.007} &  9.194^{+0.024}_{-0.027} & -1.959^{+0.030}_{-0.032}\cr
\omit&height3pt&\cr
n4291 &&   28.6 & 10.272 &  0.798^{+0.018}_{-0.019} &  9.271^{+0.060}_{-0.079} & -1.807^{+0.081}_{-0.091}\cr
\omit&height3pt&\cr
n4467 &&   15.3 &  8.748 &  0.764^{+0.045}_{-0.051} & < 7.442 & <-2.025\cr
\omit&height3pt&\cr
n4472 &&   15.3 & 10.960 &  0.955^{+0.010}_{-0.011} &  9.417^{+0.055}_{-0.074} & -2.509^{+0.074}_{-0.072}\cr
\omit&height3pt&\cr
n4473 &&   15.8 & 10.252 &  0.710^{+0.025}_{-0.026} &  8.533^{+0.301}_{-0.923} & -2.456^{+0.329}_{-1.338}\cr
\omit&height3pt&\cr
n4486 &&   15.3 & 10.884 &  1.036^{+0.009}_{-0.010} &  9.549^{+0.028}_{-0.029} & -2.377^{+0.041}_{-0.031}\cr
\omit&height3pt&\cr
n4486b &&   15.3 &  8.960 &  0.557^{+0.048}_{-0.056} &  8.963^{+0.055}_{-0.033} & -0.541^{+0.083}_{-0.100}\cr
\omit&height3pt&\cr
n4552 &&   15.3 & 10.352 &  0.829^{+0.006}_{-0.005} &  8.669^{+0.072}_{-0.045} & -2.495^{+0.059}_{-0.068}\cr
\omit&height3pt&\cr
n4564 &&   15.3 &  9.908 &  0.723^{+0.017}_{-0.017} &  8.404^{+0.097}_{-0.132} & -2.240^{+0.118}_{-0.136}\cr
\omit&height3pt&\cr
n4594 &&    9.2 & 10.644 &  0.819^{+0.005}_{-0.002} &  8.838^{+0.006}_{-0.015} & -2.631^{+0.013}_{-0.012}\cr
\omit&height3pt&\cr
n4621 &&   15.3 & 10.440 &  0.844^{+0.007}_{-0.006} &  8.445^{+0.061}_{-0.083} & -2.842^{+0.066}_{-0.086}\cr
\omit&height3pt&\cr
n4636 &&   15.3 & 10.600 &  0.908^{+0.014}_{-0.016} &  8.356^{+0.267}_{-0.566} & -3.154^{+0.273}_{-0.652}\cr
\omit&height3pt&\cr
n4649 &&   15.3 & 10.788 &  0.938^{+0.005}_{-0.005} &  9.594^{+0.011}_{-0.023} & -2.143^{+0.024}_{-0.016}\cr
\omit&height3pt&\cr
n4660 &&   15.3 &  9.476 &  0.657^{+0.017}_{-0.015} &  8.446^{+0.090}_{-0.115} & -1.699^{+0.110}_{-0.120}\cr
\omit&height3pt&\cr
n4874 &&   93.3 & 11.348 &  0.966^{+0.028}_{-0.028} & 10.319^{+0.071}_{-0.097} & -2.000^{+0.093}_{-0.119}\cr
\omit&height3pt&\cr
n4889 &&   93.3 & 11.276 &  0.808^{+0.038}_{-0.042} & 10.429^{+0.079}_{-0.119} & -1.678^{+0.127}_{-0.140}\cr
\omit&height3pt&\cr
n6166 &&  112.5 & 11.320 &  0.902^{+0.018}_{-0.019} & 10.454^{+0.034}_{-0.029} & -1.767^{+0.048}_{-0.047}\cr
\omit&height3pt&\cr
n7332 &&   20.3 &  9.896 &  0.327^{+0.006}_{-0.006} & < 6.845 & <-3.373\cr
\omit&height3pt&\cr
n7768 &&  103.1 & 11.104 &  0.835^{+0.023}_{-0.021} &  9.961^{+0.072}_{-0.089} & -1.991^{+0.101}_{-0.103}\cr
}\endgroup
{\eightpoint {\bf Table 2.} The best-fitting parameters
($\Upsilon$, $M_\bullet$ and $x\equiv M_\bullet/M_{\rm bulge}$) with
their 68\% confidence intervals for the 32 galaxies that our models
describe well.  The assumed galaxy distance $D$ and the luminosity $L$
of the bulge or other hot stellar component are also listed.}

\vfill\eject
\begingroup\offinterlineskip
\halign{\quad\strut\bf# &\vrule#&\hfil\quad\quad#\quad & \hfil\quad#\quad & \hfil\quad#\quad &\vrule# & \hfil\qquad#\quad&\hfil\quad#\quad\cr
\omit&height3pt&\multispan3&\cr
\hfil$P$\hfil & & $f$\hfil & $\log x_0$\hfil & $\alpha$ or $\log\Delta$ & &$\log \langle x\rangle$\hfil & $\langle\log x\rangle$\hfil\cr
\omit&height3pt&\multispan3&\cr
\noalign{\hrule}
\omit&height3pt&\multispan3&\cr
$P_{\rm PL1}$ && $ 1.000^{+0.000}_{-0.057}$ & $-0.608^{+0.094}_{-0.125}$ & $-0.765^{+0.037}_{-0.056}$ && $-1.347^{+0.115}_{-0.111}$ &  \cr
\omit&height3pt&\multispan3&\cr
$P_{\rm PL2}$ && $ 0.950^{+0.032}_{-0.065}$ & $-2.815^{+0.063}_{-0.038}$ & $-1.725^{+0.131}_{-0.161}$ &&   & $-2.266^{+0.097}_{-0.089}$\cr
\omit&height3pt&\multispan3&\cr
$P_{\rm S}$ && $ 1.000^{+0.000}_{-0.067}$ & $-1.705^{+0.204}_{-0.109}$ & $-0.456^{+0.178}_{-0.112}$ && $-1.879^{+0.117}_{-0.107}$ & $-2.334^{+0.153}_{-0.187}$\cr
\omit&height3pt&\multispan3&\cr
$P_{\rm G}$ && $ 0.940^{+0.042}_{-0.067}$ & $-4.747^{+0.222}_{-3.253}$ & $-1.717^{+0.098}_{-0.090}$ && $-1.809^{+0.106}_{-0.096}$ & $-1.993^{+0.106}_{-0.091}$\cr
\omit&height3pt&\multispan3&\cr
$P_{\rm LG}$ && $ 0.970^{+0.030}_{-0.057}$ & $-2.842^{+0.206}_{-0.226}$ & $ 0.074^{+0.065}_{-0.065}$ && $-1.964^{+0.149}_{-0.119}$ & $-2.274^{+0.104}_{-0.111}$\cr
\omit&height3pt&\multispan3&\cr
}\endgroup
{\eightpoint {\bf Table 4.} The best-fitting parameters $\omega$ and
their 68\% confidence limits for each assumed distribution
$\pr(x\mid\omega,P)$.  By definition $0\le f\le1$.  The last two
columns give the logarithm of the expectation value of $x\equiv
M_\bullet/M_{\rm bulge}$ and the expectation value of $\log x$ for
those galaxies with $M_\bullet\ne0$ (both calculated from
$\pr{}_+(x\mid\omega,P)$).  The mean $\langle x\rangle$ does not exist
for $P_{\rm PL2}$, while $\langle\log x\rangle$ does not exist for
$P_{\rm PL1}$.  }

\bye